\documentclass[11pt]{cernrep}
\usepackage{graphicx}

\usepackage{epsf}
\usepackage{epsfig}

\renewcommand{\theequation}{\arabic{section}.\arabic{equation}}
\renewcommand{\thesection}{\arabic{section}}
\renewcommand{\thesubsection}{\arabic{section}.\arabic{subsection}}
\renewcommand{\thesubsubsection}{\arabic{section}.\arabic{subsection}.\arabic{subsubsection}}

\newcommand{\lsim}{
\mathrel{\hbox{\rlap{\hbox{\lower4pt\hbox{$\sim$}}}\hbox{$<$}}}}

\newcommand{\gsim}{
\mathrel{\hbox{\rlap{\hbox{\lower4pt\hbox{$\sim$}}}\hbox{$>$}}}}

\begin{document}


\thispagestyle{empty}

\begin{flushright}
CERN-PH-TH/2008-034
\end{flushright}

\vspace*{2.0truecm}

\begin{center}
\boldmath
\large\bf  Flavour Physics and CP Violation: Expecting the LHC
\unboldmath
\end{center}

\vspace{0.9truecm}
\begin{center}
Robert Fleischer\\[0.1cm]
{\sl CERN, Department of Physics, Theory Division\\
CH-1211 Geneva 23, Switzerland}
\end{center}

\vspace{1.3truecm}

\begin{center}
{\bf Abstract}
\end{center}

{\small
\vspace{0.2cm}\noindent
The starting point of these lectures is an introduction to the weak 
interactions of quarks and the Standard-Model description of CP violation, 
where the central r\^ole is played by the Cabibbo--Kobayashi--Maskawa matrix 
and the corresponding unitarity triangles. Since the $B$-meson system governs 
the stage of (quark) flavour physics and CP violation, it is our main focus: we  
shall classify $B$-meson decays, introduce the 
theoretical tools to deal with them, investigate the requirements for non-vanishing 
CP-violating asymmetries, and discuss the main strategies to explore
CP violation and the preferred avenues for physics beyond the Standard Model
to enter. This formalism allows us then to discuss important benchmark modes, 
where we will also address the question of how much space for new-physics effects 
in the $B$ studies at the LHC is left by the recent experimental results 
from the $B$ factories and the Tevatron. 
}

\vspace{1.5truecm}

\begin{center}
Lectures given at the {\sl 4th CERN--CLAF School of High-Energy Physics,}\\
Vi\~na del Mar (Valparaiso Region), Chile, 18 February -- 3 March 2007\\
To appear in the Proceedings (CERN Report)
\end{center}

\vfill
\noindent
February 2008

\newpage
\thispagestyle{empty}
\vbox{}
\newpage
 
\setcounter{page}{1}


\pagestyle{plain}

\setcounter{page}{1}
\pagenumbering{roman}

\tableofcontents

\newpage

\setcounter{page}{1}
\pagenumbering{arabic}

\title{FLAVOUR PHYSICS AND CP VIOLATION: EXPECTING THE LHC}
\author{Robert Fleischer}
\institute{CERN, Geneva, Switzerland}
\maketitle
\begin{abstract}
The starting point of these lectures is an introduction to the weak 
interactions of quarks and the Standard-Model description of CP violation, 
where the central r\^ole is played by the Cabibbo--Kobayashi--Maskawa matrix 
and the corresponding unitarity triangles. Since the $B$-meson system governs 
the stage of (quark) flavour physics and CP violation, it is our main focus: we  
shall classify $B$-meson decays, introduce the 
theoretical tools to deal with them, investigate the requirements for non-vanishing 
CP-violating asymmetries, and discuss the main strategies to explore
CP violation and the preferred avenues for physics beyond the Standard Model
to enter. This formalism allows us then to discuss important benchmark modes, 
where we will also address the question of how much space for new-physics effects 
in the $B$ studies at the LHC is left by the recent experimental results 
from the $B$ factories and the Tevatron. 
\end{abstract}

\pagestyle{plain}

\pagestyle{plain}

\section{INTRODUCTION}\label{sec:intro}
\setcounter{equation}{0}
The history of CP violation, i.e.\ the non-invariance of the weak interactions
with respect to a combined charge-conjugation (C) and parity (P) 
transformation, goes back to the year 1964, where this phenomenon was
discovered through the observation of $K_{\rm L}\to\pi^+\pi^-$ decays
\cite{CP-obs}. This 
surprising effect is a manifestation of {\it indirect} CP violation, which arises 
from the fact that the mass eigenstates $K_{\rm L,S}$ of the neutral kaon 
system, which shows $K^0$--$\bar K^0$ mixing, are not eigenstates of the 
CP operator. In particular, the $K_{\rm L}$ state is governed by the CP-odd 
eigenstate, but has also a tiny admixture of the CP-even eigenstate, which 
may decay through CP-conserving interactions into the $\pi^+\pi^-$ final state. 
These CP-violating effects are described by the following observable:
\begin{equation}\label{epsK}
\varepsilon_K=(2.280\pm0.013)\times10^{-3}\times e^{i\pi/4}.
\end{equation}
On the other hand, CP-violating effects may also arise directly at the decay-amplitude
level, thereby yielding {\it direct} CP violation. This phenomenon, which leads to a
non-vanishing value of a quantity Re$(\varepsilon_K'/\varepsilon_K)$, could 
eventually be established in 1999 through the NA48 (CERN) and KTeV 
(FNAL) collaborations \cite{eps-prime}; the final results of the corresponding 
measurements are given by
\begin{equation}\label{epsp-eps-final}
\mbox{Re}(\varepsilon_K'/\varepsilon_K)=\left\{\begin{array}{ll}
(14.7\pm2.2)\times10^{-4}&\mbox{(NA48 \cite{NA48-final})}\\
(20.7\pm2.8)\times10^{-4}&\mbox{(KTeV \cite{KTeV-final}).}
\end{array}
\right.
\end{equation}

In this decade, there are huge experimental efforts to further 
explore CP violation and the quark-flavour sector of the Standard Model 
(SM). In these studies, the main actor is
the $B$-meson system, where we distinguish between charged and neutral 
$B$ mesons, which are characterized by the following valence-quark contents:
\begin{equation}\label{B-valence}
\begin{array}{c}
B^+\sim u \bar b, \quad B^+_c\sim c \bar b, \quad B^0_d\sim d \bar b, \quad
B^0_s\sim s \bar b, \\
B^-\sim \bar u b, \quad B^-_c\sim \bar c b, \quad \bar B^0_d\sim \bar d  b, \quad
\bar B^0_s\sim \bar s  b.
\end{array}
\end{equation}
In contrast to the charged $B$ mesons, their neutral counterparts
$B_q$ ($q\in \{d,s\}$) show -- in analogy to $K^0$--$\bar K^0$ 
mixing -- the phenomenon of 
$B_q^0$--$\bar B_q^0$ mixing. Decays of $B$ mesons are studied at two kinds 
of experimental facilities. The first are the ``$B$ factories" at SLAC and KEK with
the BaBar and Belle experiments, respectively. These machines are asymmetric 
$e^+e^-$ colliders that have by now produced altogether ${\cal O}(10^9)$ $B\bar B$ pairs, establishing CP violation in the $B$ system through the ``golden"
$B^0_d\to J/\psi K_{\rm S}$ channel in 2001 \cite{CP-B-obs}, and leading to 
many other interesting results. There are currently discussions of a ``super-$B$ 
factory'', with an increase of luminosity by two orders of magnitude
with respect to the current machines \cite{superB}. 
Since the $B$ factories are operated at the $\Upsilon(4S)$ resonance, only 
$B^0_d\bar B^0_d$ and $B^+_uB^-_u$ pairs are produced. On the other hand, 
hadron colliders produce, in addition to $B_d$ and $B_u$,  also $B_s$ 
mesons,\footnote{Recently, data were taken by Belle at $\Upsilon(5S)$, allowing also access to $B_s$ decays \cite{Belle-U5S}.} as well as $B_c$ 
and $\Lambda_b$ hadrons, and the Tevatron experiments CDF and D0 have 
reported first $B_{(s)}$-decay results. The physics potential of the $B_s$-meson 
system can be fully exploited at the LHC, starting operation in the summer of 2008. 
Here the general purpose experiments ATLAS and CMS can also address some 
$B$-physics topics. However, 
these studies are the main target of the dedicated LHCb 
experiment~\cite{nakada}, which will
allow us to enter a new territory in the exploration of CP violation. Concerning
the kaon system, there are plans to measure the ``rare" kaon decays 
$K^+\to\pi^+\nu\bar\nu$ and $K_{\rm L}\to\pi^0\nu\bar\nu$, which are absent at 
the tree level in the SM and exhibt extremely tiny branching ratios at the
$10^{-10}$ level, at CERN and J-PARC (for a recent overview, see 
Ref.~\cite{WG2-rep}).

The main interest in the study of CP violation and flavour physics in general is 
due to the fact that ``new physics" (NP) typically leads to new patterns in the 
flavour sector \cite{wagner}. This is actually the case in several specific extensions of the 
SM, such as supersymmetry (SUSY) scenarios, left--right-symmetric 
models, models with extra 
$Z'$ bosons, scenarios with extra dimensions, or ``little Higgs" models. Moreover, 
also the evidence for non-zero neutrino masses points towards an
origin lying beyond the SM \cite{barenboim}, raising questions of having CP violation 
in the neutrino sector and about connections between lepton- and quark-flavour physics.
 
Interestingly, CP violation offers also a link to cosmology. One of the key features 
of our Universe is the
cosmological baryon asymmetry of ${\cal O}(10^{-10})$ \cite{ellis}. As was pointed
out by Sakharov \cite{sach}, the necessary conditions for the generation of 
such an asymmetry include also the requirement that elementary interactions
violate CP (and C). Model calculations of the baryon asymmetry indicate, however,
that the CP violation present in the SM seems to be too small to generate
the observed asymmetry  \cite{shapos}. On the one hand, the required new sources 
of CP violation could be associated with very high energy scales, as in 
``leptogenesis", where new CP-violating effects appear in decays of heavy
Majorana neutrinos \cite{LG-rev}. On the other hand, new sources of
CP violation could also be accessible in the laboratory, as they arise 
naturally when going beyond the SM, as we have noted above. 

Before searching for NP at flavour factories, it is essential to understand first 
the picture of flavour physics and CP violation arising in the framework of the 
SM, where the Cabibbo--Kobayashi--Maskawa (CKM) matrix -- the
quark-mixing matrix -- plays the central r\^ole \cite{cab,KM}. The 
corresponding phenomenology is extremely rich \cite{CKM-book}. In general,
the key problem for the theoretical interpretation of experimental results 
is related to strong interactions, i.e.\ to ``hadronic" uncertainties. A famous example is
the observable $\mbox{Re}(\varepsilon_K'/\varepsilon_K)$, where we
have to deal with a subtle interplay between different contributions
which largely cancel \cite{epsp-rev}. Although the non-vanishing value of this
quantity has unambiguously ruled out ``superweak" models of
CP violation \cite{superweak}, it does currently not allow a stringent
test of the SM. 

In the $B$-meson system, there are various strategies to eliminate
the hadronic uncertainties in the exploration of CP violation.
Moreover, we may also search for relations and/or correlations that 
hold in the SM but could well be spoiled by NP contributions. 
These topics will be the focus of this lecture, which is 
organized as follows: in Section~\ref{sec:CP-SM}, we discuss the quark 
mixing in the SM by having a closer look at the CKM matrix and the
associated unitarity triangles. In Section~\ref{sec:Bdecays}, we make
first contact with weak decays of $B$ mesons, and introduce the
theoretical tool of low-energy effective Hamiltonians that is used for the 
analysis of non-leptonic $B$-meson decays, representing the key players for the 
exploration of CP violation. We will discuss the challenges in these studies, and 
will classify the main strategies to deal with them. Here we will encounter two major 
avenues: the use of amplitude relations and the study of CP violation through neutral 
$B$ decays. In Section~\ref{sec:A-REL}, we illustrate the former 
kind of methods, whereas we discuss the features of neutral $B_q$ mesons 
and $B^0_q$--$\bar B^0_q$ mixing ($q\in\{d,s\}$) in Section~\ref{sec:mix}.
In Section~\ref{sec:NP}, we address the question of how NP could enter
the $B$-physics landscape, while we turn to puzzling patterns in the current
$B$-factory data in Section~\ref{sec:puzzle}. Finally, in Section~\ref{sec:LHC}, 
we have a detailed look 
at the key targets of the $B$-physics programme at the LHC, which is characterized 
by high statistics and the complementarity to the studies at the $e^+e^-$ 
$B$ factories. The conclusions and a brief outlook are given in Section~\ref{sec:concl}.

For more detailed discussions and textbooks dealing with flavour physics and 
CP violation, the reader is referred to Refs.~\cite{BF-rev}--\cite{mannel-book}, 
alternative lecture notes can be found in Refs.~\cite{buras-spain}--\cite{nir-lecture}, 
and a selection of more compact recent reviews is given in 
Refs.~\cite{ali-rev}--\cite{HoLi-rev}.

\section{CP VIOLATION IN THE STANDARD MODEL}\label{sec:CP-SM}
\setcounter{equation}{0}
\subsection{Weak Interactions of Quarks and the Quark-Mixing Matrix}
In the framework of the Standard Model of electroweak interactions 
\cite{SM,pich}, which is based on the spontaneously broken gauge group
\begin{equation}
SU(2)_{\rm L}\times U(1)_{\rm Y}
\stackrel{{\rm SSB}}
{\longrightarrow}U(1)_{\rm em},
\end{equation}
CP-violating effects may originate from the charged-current 
interactions of quarks, having the structure
\begin{equation}\label{cc-int}
D\to U W^-.
\end{equation}
Here $D\in\{d,s,b\}$ and $U\in\{u,c,t\}$ denote down- and up-type quark 
flavours, respectively, whereas the $W^-$ is the usual $SU(2)_{\rm L}$ 
gauge boson. From a phenomenological point of view, it is convenient to 
collect the generic ``coupling strengths'' $V_{UD}$ of the charged-current 
processes in (\ref{cc-int}) in the form of the following matrix:
\begin{equation}\label{ckm0}
\hat V_{\rm CKM}=
\left(\begin{array}{ccc}
V_{ud}&V_{us}&V_{ub}\\
V_{cd}&V_{cs}&V_{cb}\\
V_{td}&V_{ts}&V_{tb}
\end{array}\right),
\end{equation}
which is referred to as the Cabibbo--Kobayashi--Maskawa (CKM) matrix
\cite{cab,KM}. 

From a theoretical point of view, this matrix connects the electroweak 
states $(d',s',b')$ of the down, strange and bottom quarks with their 
mass eigenstates $(d,s,b)$ through the following unitary transformation 
\cite{pich}:
\begin{equation}\label{ckm}
\left(\begin{array}{c}
d'\\
s'\\
b'
\end{array}\right)=\left(\begin{array}{ccc}
V_{ud}&V_{us}&V_{ub}\\
V_{cd}&V_{cs}&V_{cb}\\
V_{td}&V_{ts}&V_{tb}
\end{array}\right)\cdot
\left(\begin{array}{c}
d\\
s\\
b
\end{array}\right).
\end{equation}
Consequently, $\hat V_{\rm CKM}$ is actually a {\it unitary} matrix.
This feature ensures the absence of flavour-changing neutral-current 
(FCNC) processes at the tree level in the SM, and is hence at the basis 
of the famous Glashow--Iliopoulos--Maiani (GIM) mechanism \cite{GIM}. 
We shall return to the unitarity of the CKM matrix in
Subsection~\ref{ssec:UT}, discussing the ``unitarity triangles''.
If we express the non-leptonic charged-current interaction Lagrangian 
in terms of the mass eigenstates appearing in (\ref{ckm}), we arrive at 
\begin{equation}\label{cc-lag2}
{\cal L}_{\mbox{{\scriptsize int}}}^{\mbox{{\scriptsize CC}}}=
-\frac{g_2}{\sqrt{2}}\left(\begin{array}{ccc}
\bar u_{\mbox{{\scriptsize L}}},& \bar c_{\mbox{{\scriptsize L}}},
&\bar t_{\mbox{{\scriptsize L}}}\end{array}\right)\gamma^\mu\,\hat
V_{\mbox{{\scriptsize CKM}}}
\left(\begin{array}{c}
d_{\mbox{{\scriptsize L}}}\\
s_{\mbox{{\scriptsize L}}}\\
b_{\mbox{{\scriptsize L}}}
\end{array}\right)W_\mu^\dagger\,\,+\,\,\mbox{h.c.,}
\end{equation}
where the gauge coupling $g_2$ is related to the gauge group 
$SU(2)_{\mbox{{\scriptsize L}}}$, and the $W_\mu^{(\dagger)}$ field 
corresponds to the charged $W$ bosons. Looking at the
interaction vertices following from (\ref{cc-lag2}), we observe 
that the elements of the CKM matrix describe in fact the generic
strengths of the associated charged-current processes, as we have 
noted above.

\begin{figure}
\centerline{
\epsfysize=3.3truecm
\epsffile{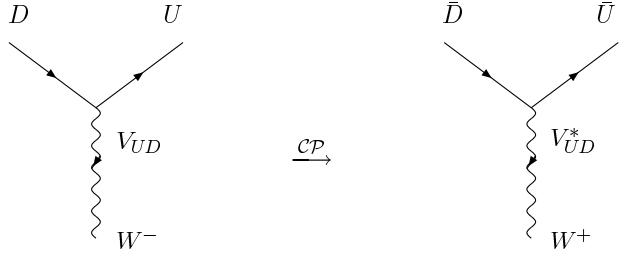}}
\caption{CP-conjugate charged-current quark-level interaction processes
in the SM.}\label{fig:CC} 
\end{figure}

In Fig.~\ref{fig:CC}, we show the $D\to U W^-$ vertex and its CP 
conjugate. Since the corresponding CP transformation involves the 
replacement
\begin{equation}\label{CKM-CP}
V_{UD}\stackrel{{\cal CP}}{\longrightarrow}V_{UD}^\ast,
\end{equation}
CP violation could -- in principle -- be accommodated in the SM through 
complex phases in the CKM matrix. The crucial question in this context
is, of course, whether we may actually have physical complex phases in 
that matrix.

\subsection{Phase Structure of the CKM Matrix}
We have the freedom of redefining the up- and down-type quark fields
in the following way:
\begin{equation}
U\to \exp(i\xi_U)U,\quad D\to \exp(i\xi_D)D. 
\end{equation}
If we perform such transformations in (\ref{cc-lag2}), the invariance 
of the charged-current interaction Lagrangian implies the 
following phase transformations of the CKM matrix elements:
\begin{equation}\label{CKM-trafo}
V_{UD}\to\exp(i\xi_U)V_{UD}\exp(-i\xi_D).
\end{equation}
Using these transformations to eliminate unphysical phases, it can be shown 
that the parametrization of the general $N\times N$ quark-mixing matrix, 
where $N$ denotes the number of fermion generations, involves the following 
parameters:
\begin{equation}
\underbrace{\frac{1}{2}N(N-1)}_{\mbox{Euler angles}} \, + \,
\underbrace{\frac{1}{2}(N-1)(N-2)}_{\mbox{complex phases}}=
(N-1)^2.
\end{equation}

If we apply this expression to the case of $N=2$ generations, we observe
that only one rotation angle -- the Cabibbo angle $\theta_{\rm C}$
\cite{cab} -- is required for the parametrization of the $2\times2$
quark-mixing matrix, which can be written in the following form:
\begin{equation}\label{Cmatrix}
\hat V_{\rm C}=\left(\begin{array}{cc}
\cos\theta_{\rm C}&\sin\theta_{\rm C}\\
-\sin\theta_{\rm C}&\cos\theta_{\rm C}
\end{array}\right),
\end{equation}
where $\sin\theta_{\rm C}=0.22$ can be determined from $K\to\pi\ell\bar\nu$ 
decays. On the other hand, in the case of $N=3$ generations, the 
parametrization of the corresponding $3\times3$ quark-mixing matrix involves 
three Euler-type angles and a single {\it complex} phase. This complex phase 
allows us to accommodate CP violation in the SM, as was pointed out by 
Kobayashi and Maskawa in 1973 \cite{KM}. The corresponding picture
is referred to as the Kobayashi--Maskawa (KM) mechanism of CP violation.

In the ``standard parametrization'' advocated by the Particle Data Group
(PDG) \cite{PDG}, the three-generation CKM matrix takes the following 
form:
\begin{equation}\label{standard}
\hat V_{\rm CKM}=\left(\begin{array}{ccc}
c_{12}c_{13}&s_{12}c_{13}&s_{13}e^{-i\delta_{13}}\\ -s_{12}c_{23}
-c_{12}s_{23}s_{13}e^{i\delta_{13}}&c_{12}c_{23}-
s_{12}s_{23}s_{13}e^{i\delta_{13}}&
s_{23}c_{13}\\ s_{12}s_{23}-c_{12}c_{23}s_{13}e^{i\delta_{13}}&-c_{12}s_{23}
-s_{12}c_{23}s_{13}e^{i\delta_{13}}&c_{23}c_{13}
\end{array}\right),
\end{equation}
where $c_{ij}\equiv\cos\theta_{ij}$ and $s_{ij}\equiv\sin\theta_{ij}$. 
Performing appropriate redefinitions of the quark-field phases, the real 
angles $\theta_{12}$, $\theta_{23}$ and $\theta_{13}$ can all be made to
lie in the first quadrant. The advantage of this parametrization is that
the generation labels $i,j=1,2,3$ are introduced in such a way that
the mixing between two chosen generations vanishes if the corresponding
mixing angle $\theta_{ij}$ is set to zero. In particular, for 
$\theta_{23}=\theta_{13}=0$, the third generation decouples, and the
$2\times2$ submatrix describing the mixing between the first and 
second generations takes the same form as (\ref{Cmatrix}).

Let us finally note that physical observables, for instance CP-violating
asymmetries, {\it cannot} depend on the chosen parametrization of the CKM 
matrix, i.e.\ have to be invariant under the phase transformations specified 
in (\ref{CKM-trafo}).

\subsection{Further Requirements for CP Violation}
As we have just seen, in order to be able to accommodate CP violation within 
the framework of the SM through a complex phase in the CKM matrix, at least 
three generations are required. However, this feature is not sufficient for 
observable CP-violating effects. To this end, further conditions have to 
be satisfied, which can be summarized as follows \cite{jarlskog,BBG}:
\begin{equation}\label{CP-req}
(m_t^2-m_c^2)(m_t^2-m_u^2)(m_c^2-m_u^2)
(m_b^2-m_s^2)(m_b^2-m_d^2)(m_s^2-m_d^2)\times J_{\rm CP}\,\not=\,0,
\end{equation}
where
\begin{equation}\label{JCP}
J_{\rm CP}=|\mbox{Im}(V_{i\alpha}V_{j\beta}V_{i\beta}^\ast 
V_{j\alpha}^\ast)|\quad(i\not=j,\,\alpha\not=\beta)\,.
\end{equation}

The mass factors in (\ref{CP-req}) are related to the fact that the 
CP-violating phase of the CKM matrix could be eliminated through an 
appropriate unitary transformation of the quark fields if any two quarks 
with the same charge had the same mass. Consequently, the origin 
of CP violation is closely related to the ``flavour problem'' in
elementary particle physics, and cannot be understood in a deeper 
way, unless we have fundamental insights into the hierarchy of quark 
masses and the number of fermion generations.

The second element of (\ref{CP-req}), the ``Jarlskog parameter'' 
$J_{\rm CP}$ \cite{jarlskog}, can be interpreted as a measure of the 
strength of CP violation in the SM. It does not depend on the chosen 
quark-field parametrization, i.e.\ it is invariant under (\ref{CKM-trafo}), 
and the unitarity of the CKM matrix implies that all combinations 
$|\mbox{Im}(V_{i\alpha}V_{j\beta}V_{i\beta}^\ast V_{j\alpha}^\ast)|$ 
are equal to one another. Using the standard parametrization of the
CKM matrix introduced in (\ref{standard}), we obtain
\begin{equation}\label{JCP-PDG}
J_{\rm CP}=s_{12}s_{13}s_{23}c_{12}c_{23}c_{13}^2\sin\delta_{13}.
\end{equation}
The  experimental information on the CKM parameters implies 
$J_{\rm CP}={\cal O}(10^{-5})$, so that
CP-violating phenomena are hard to observe.

\boldmath\subsection{Experimental Information on $|V_{\rm CKM}|$}
\begin{figure}
\centerline{
\epsfysize=3.8truecm
\epsffile{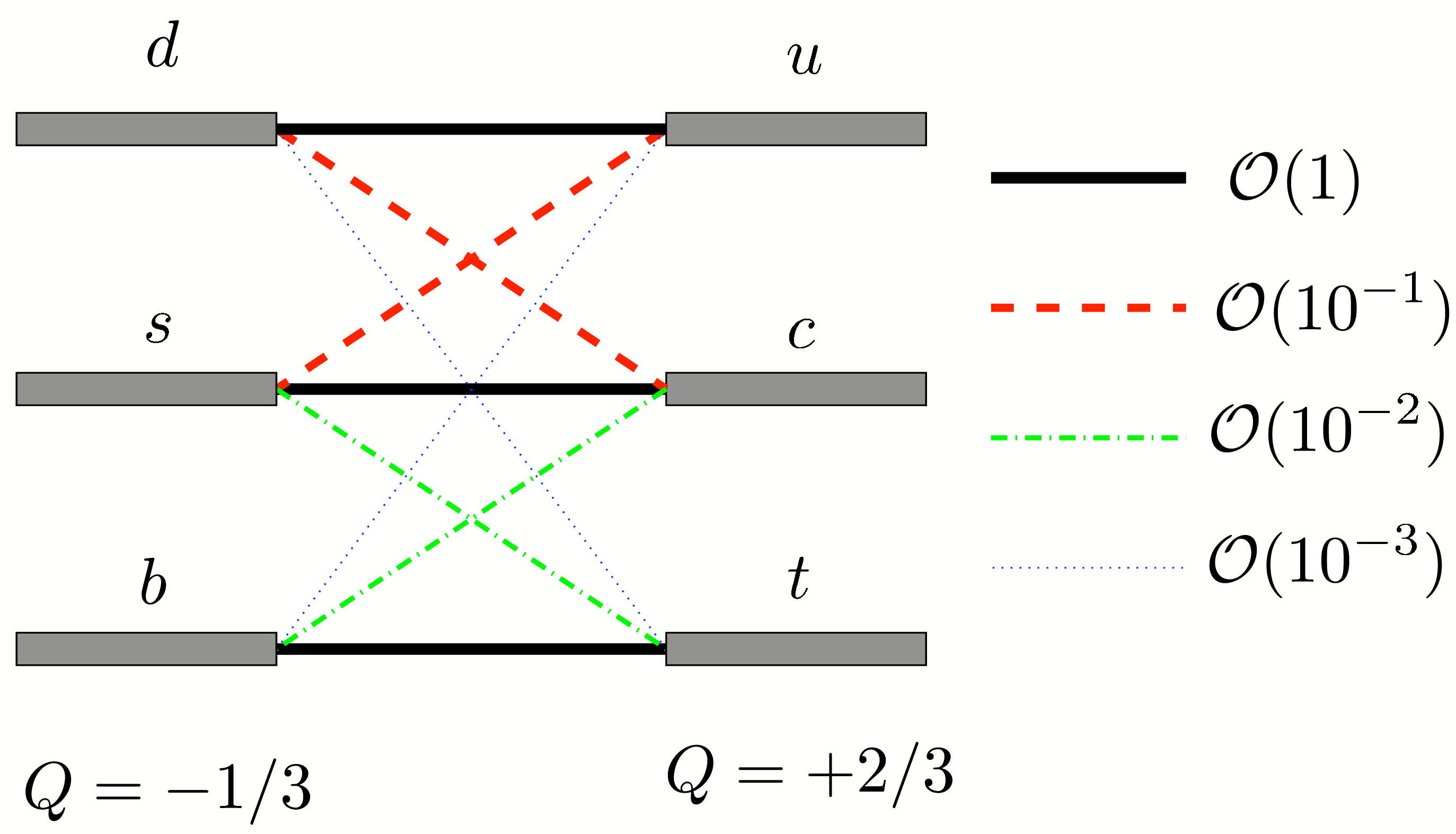}}
\vspace*{-0.2truecm}
\caption{Hierarchy of the quark transitions mediated through 
charged-current processes.}\label{fig:term}
\end{figure}

\unboldmath
In order to determine the magnitudes $|V_{ij}|$ of the elements of the
CKM matrix, we may use the following tree-level processes:
\begin{itemize}
\item Nuclear beta decays, neutron decays $\Rightarrow$ $|V_{ud}|$.
\item $K\to\pi\ell\bar\nu$ decays $\Rightarrow$ $|V_{us}|$.
\item $\nu$ production of charm off valence $d$ quarks
$\Rightarrow$ $|V_{cd}|$.
\item Charm-tagged $W$ decays (as well as $\nu$ production and 
semileptonic $D$ decays)  $\Rightarrow$ $|V_{cs}|$.
\item Exclusive and inclusive $b\to c \ell \bar\nu$ decays 
$\Rightarrow$ $|V_{cb}|$.
\item Exclusive and inclusive 
$b\to u \ell \bar \nu$ decays $\Rightarrow$ $|V_{ub}|$.
\item $\bar t\to \bar b \ell \bar\nu$ processes $\Rightarrow$ (crude direct 
determination of) $|V_{tb}|$.
\end{itemize}
If we use the corresponding experimental information, together with the 
CKM unitarity condition, and assume that there are only three generations, 
the following 90\% C.L. limits for the $|V_{ij}|$ emerge \cite{PDG,PDG-n}:
\begin{equation}\label{CKM-mag}
|\hat V_{\rm CKM}|=
\left(\begin{array}{ccc}
$0.9739\mbox{--}0.9751$ & $0.221\mbox{--}0.227$ & $0.0029\mbox{--}0.0045$\\ 
$0.221\mbox{--}0.227$ & $0.9730\mbox{--}0.9744$ & $0.039\mbox{--}0.044$\\ 
$0.0048\mbox{--}0.014$ & $0.037\mbox{--}0.043$ & $0.9990\mbox{--}0.9992$\\ 
\end{array}\right).
\end{equation}
In Fig.~\ref{fig:term}, we have illustrated the resulting hierarchy 
of the strengths of the charged-current quark-level processes:
transitions within the same generation are governed by 
CKM matrix elements of ${\cal O}(1)$, those between the first and the second 
generation are suppressed by CKM factors of ${\cal O}(10^{-1})$, those 
between the second and the third generation are suppressed by 
${\cal O}(10^{-2})$, and the transitions between the first and the third 
generation are even suppressed by CKM factors of ${\cal O}(10^{-3})$. 
In the standard parametrization (\ref{standard}), this hierarchy is 
reflected by 
\begin{equation}
s_{12}=0.22 \,\gg\, s_{23}={\cal O}(10^{-2}) \,\gg\, 
s_{13}={\cal O}(10^{-3}). 
\end{equation}

\subsection{Wolfenstein Parametrization of the CKM Matrix}
For phenomenological applications, it would be useful to have a 
parametrization of the CKM matrix available that makes the 
hierarchy arising in (\ref{CKM-mag}) -- and illustrated in
Fig.~\ref{fig:term} -- explicit \cite{wolf}. In order to derive such 
a parametrization, we introduce a set of new parameters, 
$\lambda$, $A$, $\rho$ and $\eta$, by imposing the following 
relations \cite{blo}:
\begin{equation}\label{set-rel}
s_{12}\equiv\lambda=0.22,\quad s_{23}\equiv A\lambda^2,\quad 
s_{13}e^{-i\delta_{13}}\equiv A\lambda^3(\rho-i\eta).
\end{equation}
If we now go back to the standard parametrization (\ref{standard}), we 
obtain an {\it exact} parametrization of the CKM matrix as a function of 
$\lambda$ (and $A$, $\rho$, $\eta$), allowing us to expand each CKM 
element in powers of the small parameter $\lambda$. If we neglect terms of 
${\cal O}(\lambda^4)$, we arrive at the famous ``Wolfenstein 
parametrization'' \cite{wolf}:
\begin{equation}\label{W-par}
\hat V_{\mbox{{\scriptsize CKM}}} =\left(\begin{array}{ccc}
1-\frac{1}{2}\lambda^2 & \lambda & A\lambda^3(\rho-i\eta) \\
-\lambda & 1-\frac{1}{2}\lambda^2 & A\lambda^2\\
A\lambda^3(1-\rho-i\eta) & -A\lambda^2 & 1
\end{array}\right)+{\cal O}(\lambda^4),
\end{equation}
which makes the hierarchical structure of the CKM matrix very transparent 
and is an important tool for phenomenological considerations, as we 
will see throughout this lecture. 

For several applications, next-to-leading order corrections in $\lambda$ 
play an important r\^ole. Using the exact parametrization following from 
(\ref{standard}) and (\ref{set-rel}), they can be calculated straightforwardly 
by expanding each CKM element to the desired accuracy in 
$\lambda$ \cite{blo,Brev01}:
\begin{displaymath}
V_{ud}=1-\frac{1}{2}\lambda^2-\frac{1}{8}\lambda^4+{\cal O}(\lambda^6),\quad
V_{us}=\lambda+{\cal O}(\lambda^7),\quad
V_{ub}=A\lambda^3(\rho-i\,\eta),
\end{displaymath}
\begin{displaymath}
V_{cd}=-\lambda+\frac{1}{2}A^2\lambda^5\left[1-2(\rho+i\eta)\right]+
{\cal O}(\lambda^7),
\end{displaymath}
\begin{equation}\label{NLO-wolf}
V_{cs}=1-\frac{1}{2}\lambda^2-\frac{1}{8}\lambda^4(1+4A^2)+
{\cal O}(\lambda^6),
\end{equation}
\begin{displaymath}
V_{cb}=A\lambda^2+{\cal O}(\lambda^8),\quad
V_{td}=A\lambda^3\left[1-(\rho+i\eta)\left(1-\frac{1}{2}\lambda^2\right)
\right]+{\cal O}(\lambda^7),
\end{displaymath}
\begin{displaymath}
V_{ts}=-A\lambda^2+\frac{1}{2}A(1-2\rho)\lambda^4-i\eta A\lambda^4
+{\cal O}(\lambda^6),\quad
V_{tb}=1-\frac{1}{2}A^2\lambda^4+{\cal O}(\lambda^6).
\end{displaymath}
It should be noted that 
\begin{equation}
V_{ub}\equiv A\lambda^3(\rho-i\eta)
\end{equation}
receives {\it by definition} no power corrections in $\lambda$ within
this prescription. If we follow Ref.~\cite{blo} and introduce the generalized
Wolfenstein parameters
\begin{equation}\label{rho-eta-bar}
\bar\rho\equiv\rho\left(1-\frac{1}{2}\lambda^2\right),\quad
\bar\eta\equiv\eta\left(1-\frac{1}{2}\lambda^2\right),
\end{equation}
we may simply write, up to corrections of ${\cal O}(\lambda^7)$,
\begin{equation}\label{Vtd-expr}
V_{td}=A\lambda^3(1-\bar\rho-i\,\bar\eta).
\end{equation}
Moreover, we have to an excellent accuracy
\begin{equation}\label{Def-A}
V_{us}=\lambda\quad \mbox{and}\quad 
V_{cb}=A\lambda^2,
\end{equation}
as these quantities receive only corrections at the $\lambda^7$ and
$\lambda^8$ levels, respectively. In comparison with other generalizations
of the Wolfenstein parametrization found in the literature, the advantage
of (\ref{NLO-wolf}) is the absence of relevant corrections to $V_{us}$
and $V_{cb}$, and that $V_{ub}$ and $V_{td}$ take forms similar to those 
in (\ref{W-par}). As far as the Jarlskog parameter introduced in
(\ref{JCP}) is concerned, we obtain the simple expression
\begin{equation}
J_{\rm CP}=\lambda^6A^2\eta,
\end{equation}
which should be compared with (\ref{JCP-PDG}).

\subsection{Unitarity Triangles of the CKM Matrix}\label{ssec:UT}
The unitarity of the CKM matrix, which is described by
\begin{equation}
\hat V_{\mbox{{\scriptsize CKM}}}^{\,\,\dagger}\cdot\hat 
V_{\mbox{{\scriptsize CKM}}}=
\hat 1=\hat V_{\mbox{{\scriptsize CKM}}}\cdot\hat V_{\mbox{{\scriptsize 
CKM}}}^{\,\,\dagger},
\end{equation}
leads to a set of 12 equations, consisting of 6 normalization 
and 6 orthogonality relations. The latter can be represented as 6 
triangles in the complex plane \cite{AKL}, all having the same area, 
$2 A_{\Delta}=J_{\rm CP}$ \cite{JS}. Let us now have a closer look at 
these relations: those describing the orthogonality of different columns 
of the CKM matrix are given by
\begin{eqnarray}
\underbrace{V_{ud}V_{us}^\ast}_{{\cal O}(\lambda)}+
\underbrace{V_{cd}V_{cs}^\ast}_{{\cal O}(\lambda)}+
\underbrace{V_{td}V_{ts}^\ast}_{{\cal O}(\lambda^5)} & = &
0\\
\underbrace{V_{us}V_{ub}^\ast}_{{\cal O}(\lambda^4)}+
\underbrace{V_{cs}V_{cb}^\ast}_{{\cal O}(\lambda^2)}+
\underbrace{V_{ts}V_{tb}^\ast}_{{\cal O}(\lambda^2)} & = &
0\\
\underbrace{V_{ud}V_{ub}^\ast}_{(\rho+i\eta)A\lambda^3}+
\underbrace{V_{cd}V_{cb}^\ast}_{-A\lambda^3}+
\underbrace{V_{td}V_{tb}^\ast}_{(1-\rho-i\eta)A\lambda^3} & = &
0,\label{UT1}
\end{eqnarray}
whereas those associated with the orthogonality of different rows 
take the following form:
\begin{eqnarray}
\underbrace{V_{ud}^\ast V_{cd}}_{{\cal O}(\lambda)}+
\underbrace{V_{us}^\ast V_{cs}}_{{\cal O}(\lambda)}+
\underbrace{V_{ub}^\ast V_{cb}}_{{\cal O}(\lambda^5)} & = &
0\\
\underbrace{V_{cd}^\ast V_{td}}_{{\cal O}(\lambda^4)}+
\underbrace{V_{cs}^\ast V_{ts}}_{{\cal O}(\lambda^2)}+
\underbrace{V_{cb}^\ast V_{tb}}_{{\cal O}(\lambda^2)} & = & 0\\
\underbrace{V_{ud}^\ast V_{td}}_{(1-\rho-i\eta)A\lambda^3}+
\underbrace{V_{us}^\ast V_{ts}}_{-A\lambda^3}+
\underbrace{V_{ub}^\ast V_{tb}}_{(\rho+i\eta)A\lambda^3}
& = & 0.\label{UT2}
\end{eqnarray}
Here we have also indicated the structures that arise if we apply the 
Wolfenstein parametrization by keeping just the leading, non-vanishing 
terms. We observe that only in (\ref{UT1}) and (\ref{UT2}), which 
describe the orthogonality of the first and third columns 
and of the first and third rows, respectively, all three sides are 
of comparable magnitude, ${\cal O}(\lambda^3)$, while in the 
remaining relations, one side is suppressed with respect to the others 
by factors of ${\cal O}(\lambda^2)$ or ${\cal O}(\lambda^4)$. Consequently,
we have to deal with only {\it two} non-squashed unitarity triangles 
in the complex plane. However, as we have already indicated in (\ref{UT1}) 
and (\ref{UT2}), the corresponding orthogonality relations agree with each 
other at the $\lambda^3$ level, yielding
\begin{equation}\label{UTLO}
\left[(\rho+i\eta)+(-1)+(1-\rho-i\eta)\right]A\lambda^3=0.
\end{equation}
Consequently, they describe the same triangle, which is usually referred 
to as {\it the} unitarity triangle of the CKM matrix 
\cite{JS,ut}.

\begin{figure}[t]
\centerline{
\begin{tabular}{ll}
   {\small(a)} & {\small(b)} \\
   \qquad \includegraphics[width=6.3truecm]{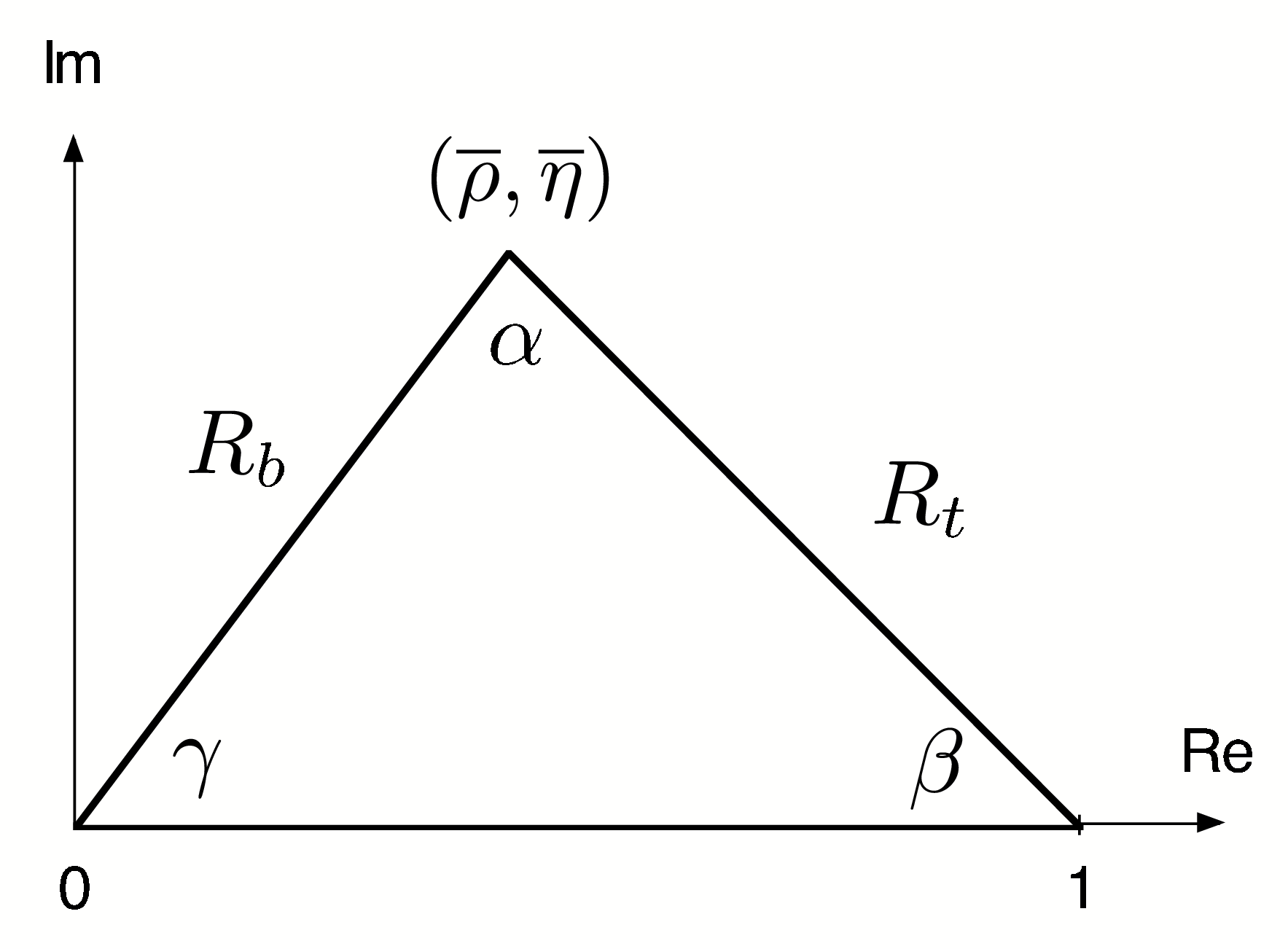}
&
\qquad \includegraphics[width=6.3truecm]{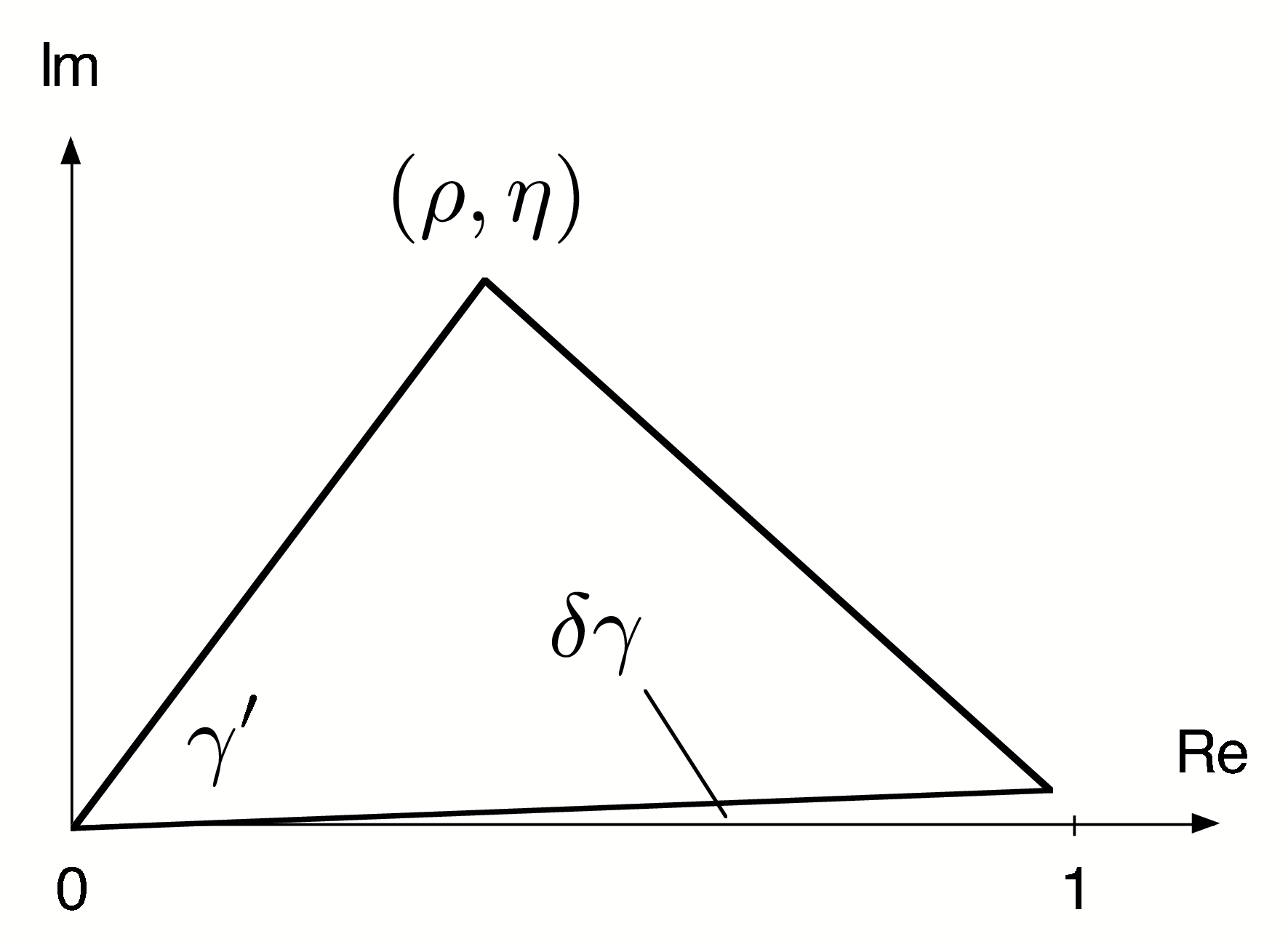}
 \end{tabular}}
 \vspace*{-0.2truecm}
\caption{The two non-squashed unitarity triangles of the CKM matrix, as
explained in the text: (a) and (b) correspond to the orthogonality 
relations (\ref{UT1}) and (\ref{UT2}), respectively. In Asia, the notation 
$\phi_1\equiv\beta$,
$\phi_2\equiv\alpha$ and $\phi_3\equiv\gamma$ is used for the angles of the
triangle shown in (a).}
\label{fig:UT}
\end{figure}

Concerning the $B$-decay studies in the LHC era, we have to take 
the next-to-leading order terms of the Wolfenstein expansion
into account, and have to distinguish between the unitarity triangles 
following from (\ref{UT1}) and (\ref{UT2}). Let us first have a closer
look at the former relation. Including terms of ${\cal O}(\lambda^5)$, 
we obtain the following generalization of (\ref{UTLO}):
\begin{equation}\label{UT1-NLO}
\left[(\bar\rho+i\bar\eta)+(-1)+(1-\bar\rho-
i\bar\eta)\right]A\lambda^3 +{\cal O}(\lambda^7)=0, 
\end{equation}
where $\bar\rho$ and $\bar\eta$ are as defined in (\ref{rho-eta-bar}). 
If we divide this relation by the overall normalization factor $A\lambda^3$, 
and introduce
\begin{equation}\label{Rb-def}
R_b\equiv\sqrt{\overline{\rho}^2+\overline{\eta}^2}=\left(1-\frac{\lambda^2}{2}
\right)\frac{1}{\lambda}\left|\frac{V_{ub}}{V_{cb}}\right|
\end{equation}
\begin{equation}\label{Rt-def}
R_t\equiv\sqrt{(1-\overline{\rho})^2+\overline{\eta}^2}=
\frac{1}{\lambda}\left|\frac{V_{td}}{V_{cb}}\right|,
\end{equation}
we arrive at the unitarity triangle illustrated in Fig.\ \ref{fig:UT} (a). 
It is a straightforward generalization of the leading-order
case described by (\ref{UTLO}): instead of $(\rho,\eta)$, the apex
is now simply given by $(\bar\rho,\bar\eta)$ \cite{blo}. The two UT sides 
$R_b$ and $R_t$ as well as the UT angles will show up at several places 
throughout this lecture. Moreover, the relations
\begin{equation}
V_{ub}=A\lambda^3\left(\frac{R_b}{1-\lambda^2/2}\right)e^{-i\gamma},\quad 
V_{td}=A\lambda^3 R_t e^{-i\beta}
\end{equation}
are also useful for phenomenological applications, since they make the
dependences of $\gamma$ and $\beta$ explicit; they correspond to the
phase convention chosen both in the standard parametrization 
(\ref{standard}) and in the generalized Wolfenstein parametrization 
(\ref{NLO-wolf}). Finally, if we take also (\ref{set-rel}) into account, 
we obtain
\begin{equation}
\delta_{13}=\gamma.
\end{equation}

Let us now turn to (\ref{UT2}). Here we arrive at an expression 
that is more complicated than (\ref{UT1-NLO}): 
\begin{equation}
\left[\left\{1-\frac{\lambda^2}{2}-
(1-\lambda^2)\rho-i(1-\lambda^2)\eta
\right\}\!+\!\left\{-1+\left(\frac{1}{2}-\rho\right)
\lambda^2-i\eta\lambda^2\right\}
\!+\!\left\{\rho+i\eta\right\}\right]A\lambda^3+{\cal O}(\lambda^7)=0.
\end{equation}
If we divide again by $A\lambda^3$, we obtain the unitarity triangle
sketched in Fig.\ \ref{fig:UT} (b), where the apex is given by $(\rho,\eta)$ 
and {\it not} by $(\bar\rho,\bar\eta)$. On the other hand, we encounter a 
tiny angle 
\begin{equation}
\delta\gamma\equiv\lambda^2\eta={\cal O}(1^\circ)
\end{equation}
between real axis and basis of the triangle, which satisfies 
\begin{equation}
\gamma=\gamma'+\delta\gamma, 
\end{equation}
where $\gamma$ coincides with the corresponding angle in  
Fig.\ \ref{fig:UT} (a).

\begin{figure}[t]
\centerline{
\begin{tabular}{ll}
  \includegraphics[width=7.5truecm]{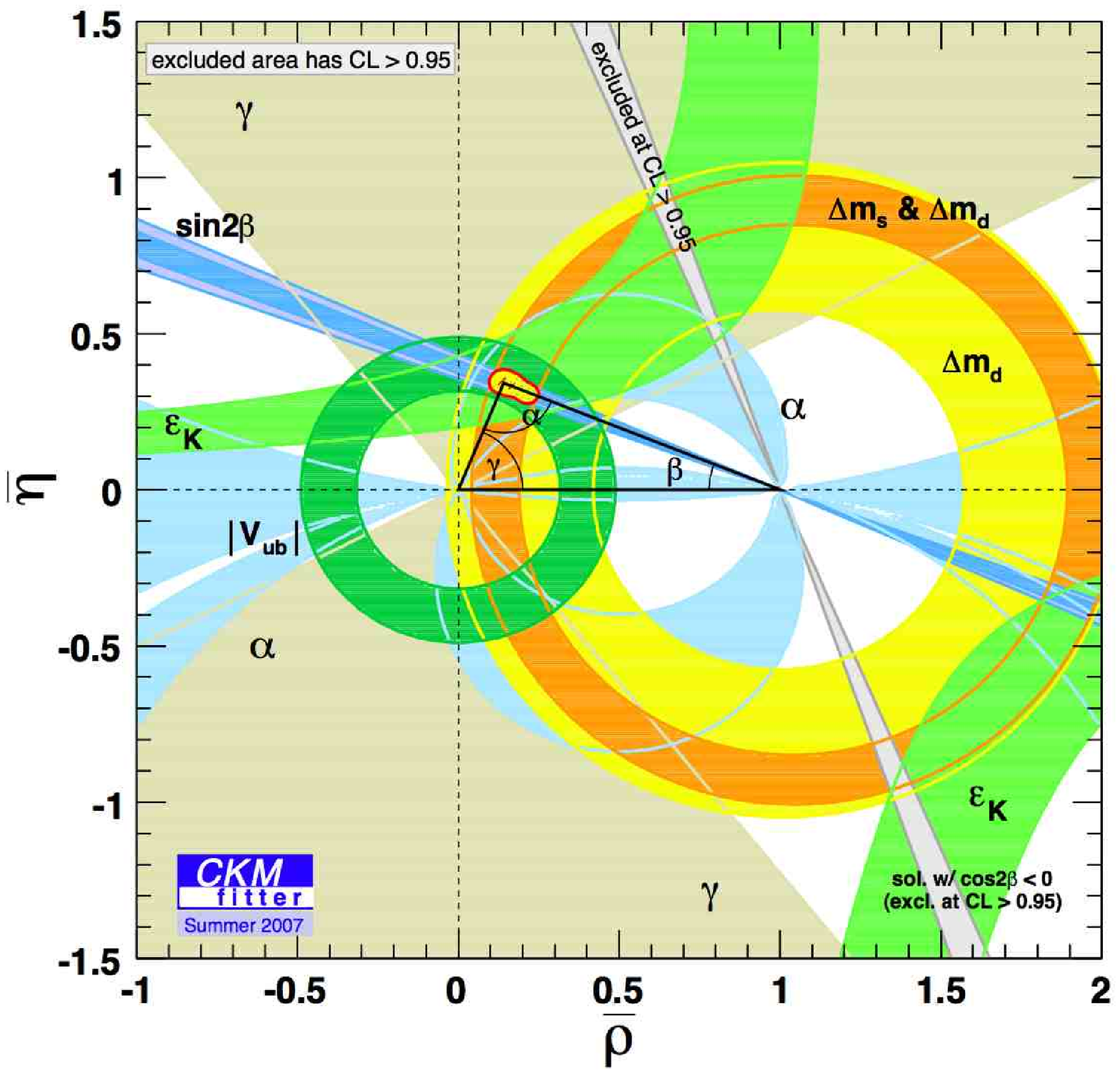} &
\includegraphics[width=8.2truecm]{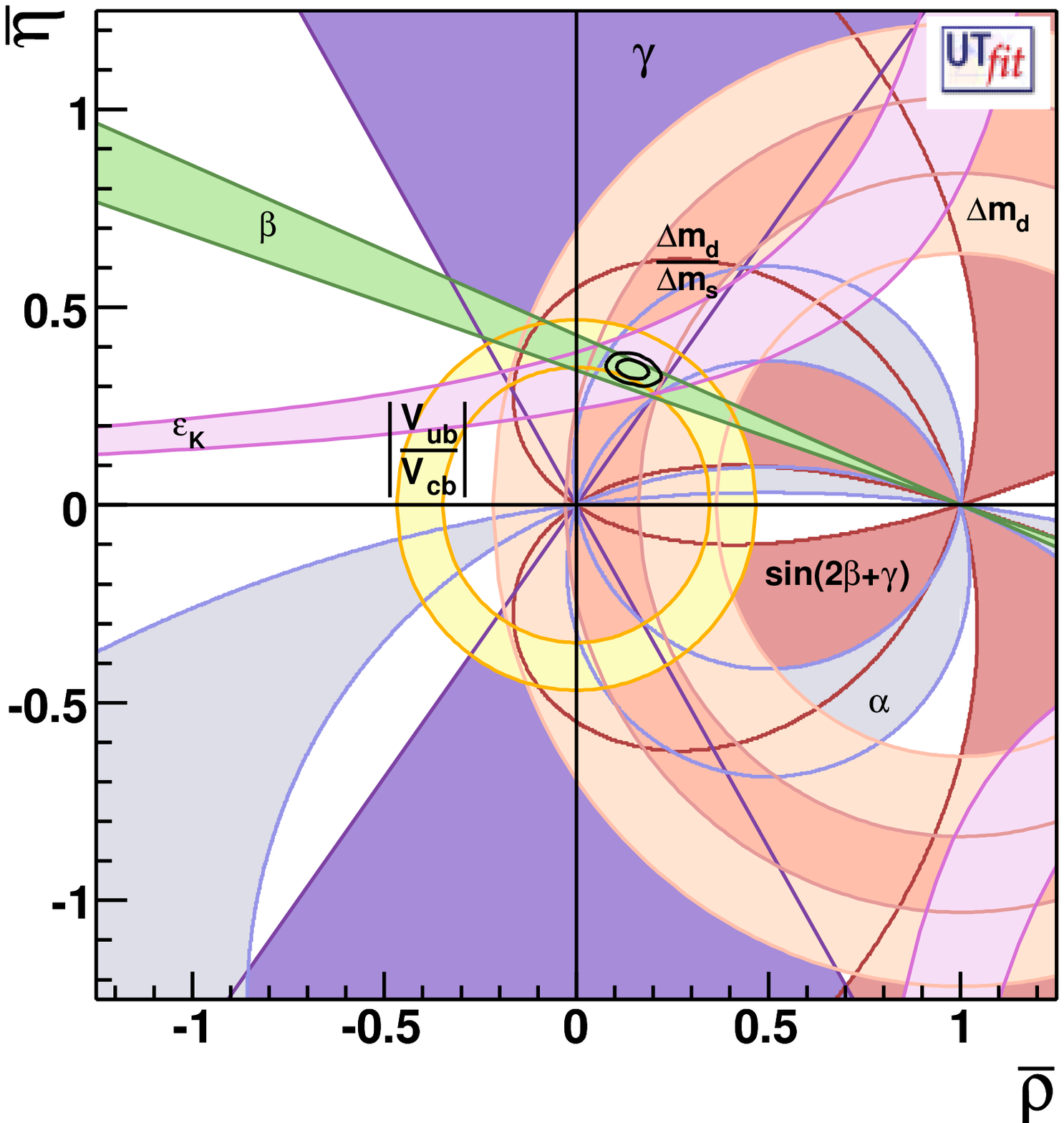}
 \end{tabular}}
 \vspace*{-0.5truecm}
\caption{Analyses of the CKMfitter (left panel) and 
UTfit (right panel) collaborations \cite{CKMfitter,UTfit}.}\label{fig:UTfits}
\end{figure}

Whenever referring to a ``unitarity triangle'' (UT) in the following 
discussion, we mean the one illustrated in Fig.\ \ref{fig:UT} (a), which
is the generic generalization of the leading-order case described
by (\ref{UTLO}). The UT is a central target for the experimental testing 
of the SM description of CP violation. 
Interestingly, also the tiny angle $\delta\gamma$ can be probed directly 
through certain CP-violating effects that can be explored at the
LHCb experiment, as we will see in Section~\ref{sec:LHC}.

\subsection{The Determination of the Unitarity Triangle}\label{subsec:CKM-fits}
The next obvious question is how the UT can be determined. There are two 
conceptually different avenues that we may follow to this end:
\begin{itemize}
\item[(i)] In the ``CKM fits'', theory is used to convert 
experimental data into contours in the $\bar\rho$--$\bar\eta$ plane. In particular, 
semileptonic $b\to u \ell \bar\nu_\ell$, $c \ell \bar\nu_\ell$ decays and 
$B^0_q$--$\bar B^0_q$ mixing ($q\in\{d,s\}$) allow us to determine the UT sides 
$R_b$ and $R_t$, respectively, i.e.\ to fix two circles in the $\bar\rho$--$\bar\eta$ 
plane. Furthermore, the indirect CP violation in the neutral kaon system
described by $\varepsilon_K$ can be transformed into a hyperbola. 
\item[(ii)] Theoretical considerations allow us to convert measurements of 
CP-violating effects in $B$-meson decays into direct information on the UT angles. 
The most prominent example is the determination of $\sin2\beta$ through 
CP violation in $B^0_d\to J/\psi K_{\rm S}$ decays, but several other strategies 
were proposed and can be confronted with the experimental data. 
\end{itemize}
The goal is to ``overconstrain'' the UT as much as possible. Additional 
contours can be fixed in the $\bar\rho$--$\bar\eta$ plane through 
the measurement of rare decays \cite{BF-rev}. 

In Fig.~\ref{fig:UTfits}, we show examples of the comprehensive
analyses of the UT that are performed -- and continuously updated --
by the ``CKM Fitter Group'' \cite{CKMfitter}
and the ``UTfit collaboration''~\cite{UTfit}. In these figures, we can nicely see the
circles that are determined through the semileptonic $B$ decays and the 
$\varepsilon_K$ hyperbolas. Moreover, also the straight lines following from the 
direct measurement of $\sin 2\beta$ with the help of $B^0_d\to J/\psi K_{\rm S}$ 
modes are shown. We observe that the global consistency is very good. However,
looking closer, we also see that the average for 
$(\sin 2\beta)_{\psi K_{\rm S}}$ is now on the lower side, so that the situation in 
the $\bar\rho$--$\bar\eta$ plane is no longer fully ``perfect". Furthermore, as we
shall discuss in detail in Section~\ref{sec:puzzle}, there are certain puzzling
patterns in the $B$-factory data, and various key aspects could not yet be addressed
experimentally and are hence still essentially unexplored. Consequently, still
a lot of space is left for the detection of possible, unambiguous inconsistencies
with respect to the SM picture of CP violation and quark-flavour physics. Since 
weak decays  of $B$ mesons play a key r\^ole in this adventure, let us next have 
a closer look at them.

\begin{figure}
\vskip -0.2truein
\begin{center}
\leavevmode
\epsfysize=3.8truecm 
\epsffile{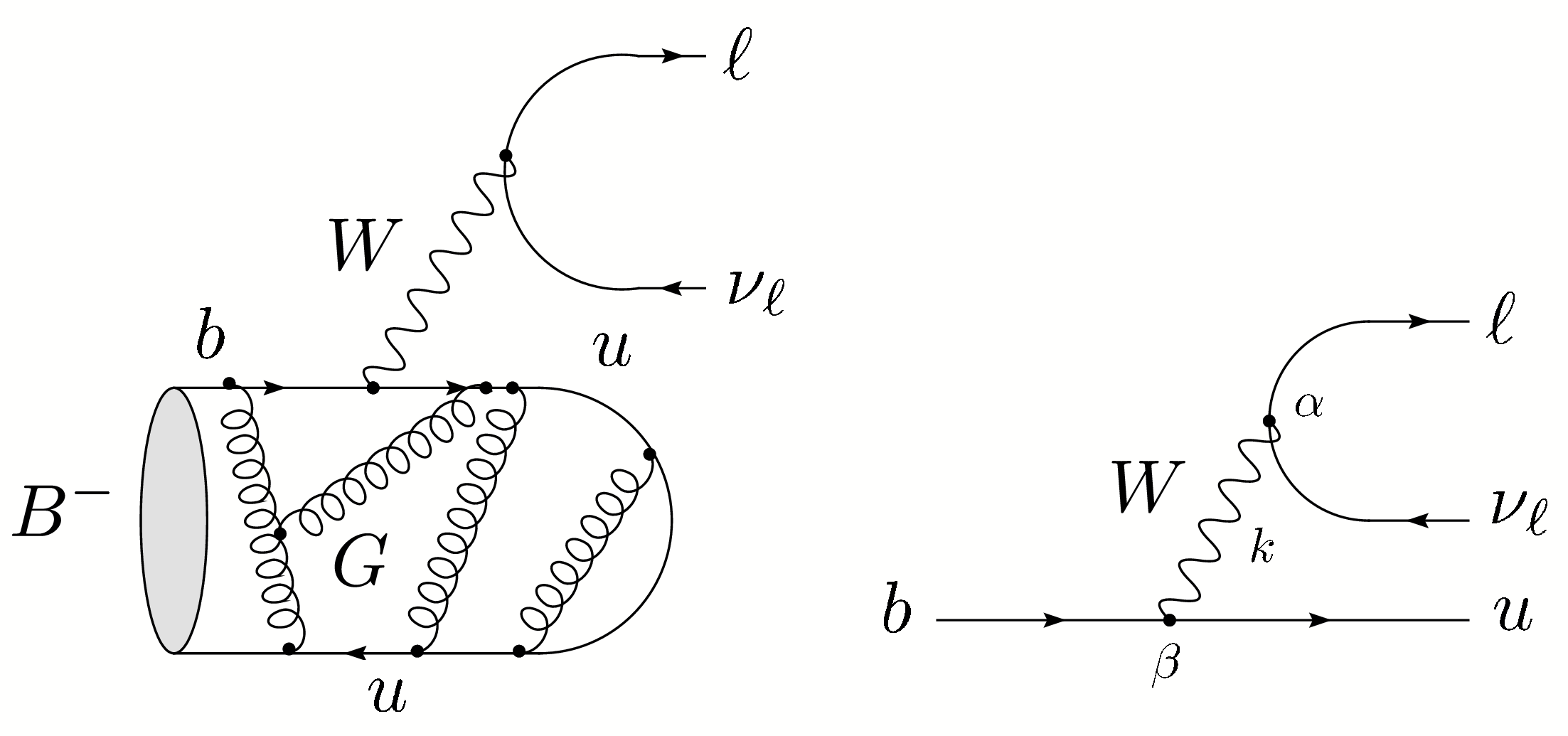} 
\end{center}
\vspace*{-0.8truecm}
\caption{Feynman diagrams contributing to the leptonic decay
$B^-\to \ell\bar\nu_\ell$.}\label{fig:lep}
\end{figure}

\section{WEAK DECAYS OF {\boldmath$B$\unboldmath} MESONS}\label{sec:Bdecays}
\setcounter{equation}{0}
The $B$-meson system consists of charged and neutral $B$ mesons, which are 
characterized by the valence quark contents in (\ref{B-valence}).
The characteristic feature
of the neutral $B_q$ ($q\in \{d,s\}$) mesons is the phenomenon
of $B_q^0$--$\bar B_q^0$ mixing, which will be discussed in 
Section~\ref{sec:mix}. As far as the weak decays of 
$B$ mesons are concerned, we distinguish between leptonic, 
semileptonic and non-leptonic transitions.

\subsection{Leptonic Decays}
The simplest $B$-meson decay class is given by leptonic decays 
of the kind $B^-\to \ell\bar\nu_\ell$, as illustrated in Fig.~\ref{fig:lep}.
If we evaluate the corresponding Feynman diagram, we arrive at the 
following transition amplitude:
\begin{equation}\label{Tfi-lept}
T_{fi}=-\,\frac{g_2^2}{8} V_{ub}
\underbrace{\left[\bar u_\ell\gamma^\alpha(1-\gamma_5)v_\nu
\right]}_{\mbox{Dirac spinors}}
\left[\frac{g_{\alpha\beta}}{k^2-M_W^2}\right]
\underbrace{\langle 0|\bar u\gamma^\beta
(1-\gamma_5)b|B^-\rangle}_{\mbox{hadronic ME}},
\end{equation}
where $g_2$ is the $SU(2)_{\rm L}$ gauge coupling, $V_{ub}$ the corresponding
element of the CKM matrix, $\alpha$ and $\beta$ are Lorentz indices,
and $M_W$ denotes the mass of the $W$ gauge boson. Since the four-momentum
$k$ that is carried by the $W$ satisfies $k^2=M_B^2\ll M_W^2$, we may
write
\begin{equation}\label{W-int-out}
\frac{g_{\alpha\beta}}{k^2-M_W^2}\quad\longrightarrow\quad
-\,\frac{g_{\alpha\beta}}{M_W^2}\equiv-\left(\frac{8G_{\rm F}}{\sqrt{2}g_2^2}
\right)g_{\alpha\beta},
\end{equation}
where $G_{\rm F}$ is Fermi's constant. Consequently, we may ``integrate out'' 
the $W$ boson in (\ref{Tfi-lept}), which yields
\begin{equation}\label{Tfi-lept-2}
T_{fi}=\frac{G_{\rm F}}{\sqrt{2}}V_{ub}\left[\bar u_\ell\gamma^\alpha
(1-\gamma_5)v_\nu\right]\langle 0|\bar u\gamma_\alpha(1-\gamma_5)b
|B^-\rangle.
\end{equation}
In this simple expression, {\it all} the hadronic physics is encoded 
in the {\it hadronic matrix element} 
\begin{displaymath}
\langle 0|\bar u\gamma_\alpha(1-\gamma_5)b
|B^-\rangle,
\end{displaymath}
i.e.\ there are no other strong-interaction QCD effects (for a detailed discussion
of QCD, see Ref.~\cite{kop}). Since the $B^-$ meson is a pseudoscalar particle, 
we have
\begin{equation}\label{ME-rel1}
\langle 0|\overline{u}\gamma_\alpha b|B^-\rangle=0,
\end{equation}
and may write
\begin{equation}\label{ME-rel2}
\langle 0|\bar u\gamma_\alpha\gamma_5 b|B^-(q)\rangle =
i f_B q_\alpha,
\end{equation}
where $f_B$ is the $B$-meson {\it decay constant}, which is an important
input for phenomenological studies. In order to determine this
quantity, which is a very challenging task, non-perturbative 
techniques, such as QCD sum-rule analyses \cite{khod} or 
lattice studies, where a numerical evaluation of the QCD path integral is 
performed with the help of a space-time lattice \cite{luscher}--\cite{delmo}, 
are required.  If we use (\ref{Tfi-lept-2}) with (\ref{ME-rel1}) and 
(\ref{ME-rel2}), and perform the corresponding phase-space integrations, 
we obtain the following decay rate:
\begin{equation}
\Gamma(B^-\to\ell \bar \nu_\ell)=\frac{G_{\rm F}^2}{8\pi}
M_Bm_\ell^2\left(1-\frac{m_\ell^2}{M_B^2}\right)^2f_B^2|V_{ub}|^2 ,
\end{equation}
where $M_B$ and $m_\ell$ denote the masses of the $B^-$ and $\ell$,
respectively. Because of the tiny value of $|V_{ub}|\propto\lambda^3$ 
and a helicity-suppression mechanism, we obtain unfortunately very small 
branching ratios of ${\cal O}(10^{-10})$ and ${\cal O}(10^{-7})$ for 
$\ell=e$ and $\ell=\mu$, respectively \cite{fulvia}. 

The helicity suppression is not effective for $\ell=\tau$, but -- because of the 
required $\tau$ reconstruction -- these modes are also very challenging from 
an experimental point of view. Nevertheless, the Belle experiment has recently
reported the first evidence for the purely leptonic decay 
$B^- \to \tau^- \bar\nu_\tau$, with the following branching ratio \cite{Belle-leptonic}:
\begin{equation}
\mbox{BR}(B^- \to \tau^- \bar\nu_\tau) = \left[1.79^{+0.56}_{-0.49} \,
\mbox{(stat)} \,  ^{+0.46}_{-0.51} \, \mbox{(syst)}\right]\times 10^{-4},
\end{equation}
which corresponds to a significance of about 3.5 standard deviations. On the other
hand, BaBar gives an upper limit of $\mbox{BR}(B^- \to \tau^- \bar\nu_\tau)<
1.8\times10^{-4}$ (90\% C.L.), as well as the following value \cite{Babar-leptonic}:
\begin{equation}
\mbox{BR}(B^- \to \tau^- \bar\nu_\tau) = \left[0.88^{+0.68}_{-0.67} \,
\mbox{(stat)} \,  \pm0.11 \, \mbox{(syst)}\right]\times 10^{-4}.
\end{equation}

\begin{figure}
   \centering
   \includegraphics[width=5.5truecm]{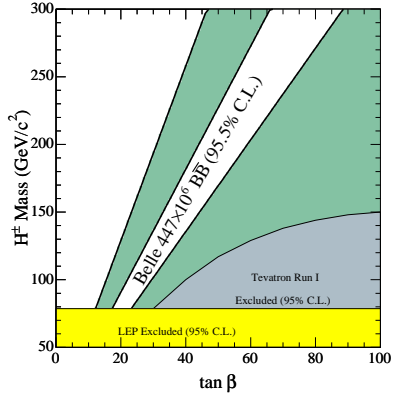}
   \vspace*{-0.2truecm}
   \caption{Constrains on the charged Higgs parameter space 
   \cite{browder}.}\label{fig:2HDlimit}
\end{figure}

Using the
SM expression for this branching ratio and the measured values of
$G_{\rm F}$, $M_B$, $m_\tau$ and the $B$-meson lifetime, the 
product of the $B$-meson decay constant $f_B$ and the magnitude of the
CKM matrix element $|V_{ub}|$ is obtained as
\begin{equation}
f_B|V_{ub}|=\left[10.1^{+1.6}_{-1.4}\,
\mbox{(stat)} \,  ^{+1.3}_{-1.4} \, \mbox{(syst)}\right]\times 10^{-4} \, \mbox{GeV}
\end{equation}
from the Belle result. The determination of this quantity is very interesting, as 
knowledge of $|V_{ub}|$ (see Subsection~\ref{subsec:semi-lept}) allows us 
to extract $f_B$, thereby providing tests of 
non-perturbative calculations of this important parameter. On the other hand,
when going beyond the SM, the $B^- \to \tau^- \bar\nu_\tau$ decay is 
a sensitive probe of effects from charged Higgs bosons; the corresponding
Feynman diagram can easily be obtained from Fig.~\ref{fig:lep} by replacing
the $W$ boson through a charged Higgs $H$. The SM expression for the branching
ratio is then simply modified by the following factor  \cite{hou}:
\begin{equation}
r_H=\left[1-\left(\frac{M_B}{M_H}\tan\beta\right)^2\right]^2
\stackrel{\rm Belle}{\longrightarrow}1.13\pm0.53,
\end{equation}
where $\tan\beta\equiv v_2/v_1$ is defined through the ratio of vacuum expectation
values and does {\it not} involve the UT angle $\beta$. Using information on $f_B$ 
and $|V_{ub}|$, constraints on the charged Higgs parameter space can be obtained
from the measured $B^- \to \tau^- \bar\nu_\tau$ branching ratio, as shown
in Fig.~\ref{fig:2HDlimit}.

Before discussing the determination of  $|V_{ub}|$ from semileptonic $B$ decays 
in the next subsection, let us have a look at the leptonic $D$-meson decay
$D^+\to \mu^+\nu$. It is governed by the CKM factor 
\begin{equation}
|V_{cd}|=|V_{us}|+{\cal O}(\lambda^5)=
\lambda[1+{\cal O}(\lambda^4)],
\end{equation}
whereas $B^-\to \mu^-\bar \nu$ involves $|V_{ub}|=\lambda^3 R_b$.
Consequently, we win a factor of ${\cal O}(\lambda^4)$ in the decay rate,
so that $D^+\to \mu^+\nu$ is accessible at the CLEO-c experiment \cite{CLEO-c}. 
Since the corresponding CKM factor is well known, the decay constant  $f_{D^+}$
defined in analogy to (\ref{ME-rel2}) can be extracted,  
allowing another interesting testing ground for lattice QCD calculations. Thanks to
recent progress in these techniques \cite{davies-eps}, the ``quenched" 
approximation, which had to be applied for many many years and ingnores 
quark loops, is no longer required for the calculation of $f_{D^+}$. In the 
summer of 2005, there was a first show down between the corresponding 
theoretical prediction and experiment:
the lattice result of $f_{D^+}=(201\pm3\pm17)\mbox{MeV}$ was reported
\cite{fD-lat}, while CLEO-c announced the measurement of
$f_{D^+}=(222.6 \pm 16.7 ^{+2.8}_{-3.4})\,\mbox{MeV}$ \cite{fD-cleoc}.
Both numbers agree well within the uncertainties. For a review of recent
developments and other results on decay constants of pseudoscalar mesons, see
Ref.~\cite{ro-st}.

\begin{figure}
\vskip -0.2truein
\begin{center}
\leavevmode
\epsfysize=3.8truecm 
\epsffile{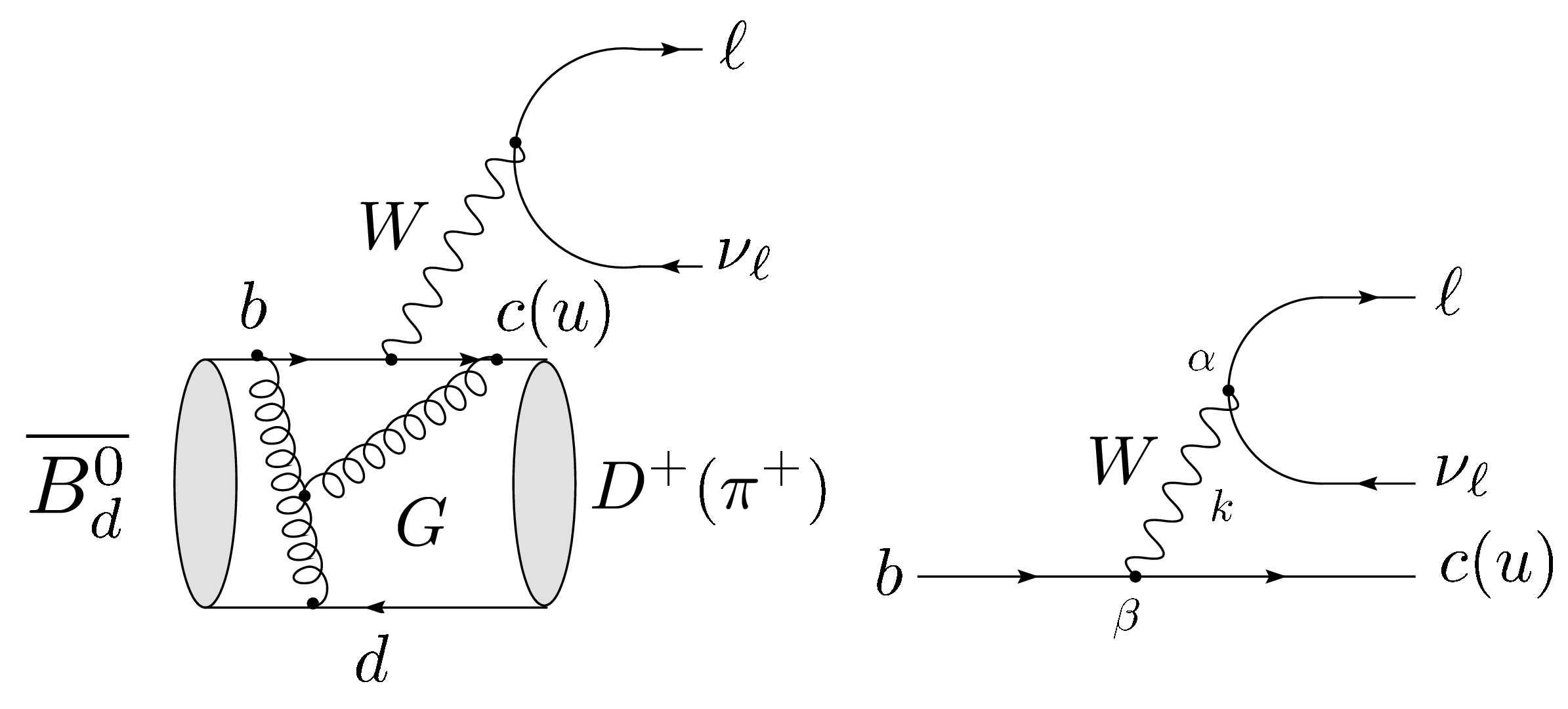} 
\end{center}
\vspace*{-0.8truecm}
\caption{Feynman diagrams contributing to semileptonic 
$\bar B^0_d\to D^+ (\pi^+) \ell \bar \nu_\ell$ decays.}\label{fig:semi}
\end{figure}

\subsection{Semileptonic Decays}\label{subsec:semi-lept}
\subsubsection{General Structure}
Semileptonic $B$-meson decays of the kind shown in Fig.~\ref{fig:semi}
have a structure that is more complicated than the one of the 
leptonic transitions. If we evaluate the corresponding Feynman diagram
for the $b\to c$ case, we obtain
\begin{equation}\label{Tfi-semi-full}
T_{fi}=-\,\frac{g_2^2}{8} V_{cb}
\underbrace{\left[\bar u_\ell\gamma^\alpha(1-\gamma_5)v_\nu
\right]}_{\mbox{Dirac spinors}}
\left[\frac{g_{\alpha\beta}}{k^2-M_W^2}\right]
\underbrace{\langle D^+|\bar c\gamma^\beta
(1-\gamma_5)b|\bar B^0_d\rangle}_{\mbox{hadronic ME}}.
\end{equation}
Because of $k^2\sim M_B^2\ll M_W^2$, we may again -- as in (\ref{Tfi-lept}) --
integrate out the $W$ boson with the help of (\ref{W-int-out}), which
yields
\begin{equation}\label{Tfi-semi}
T_{fi}=\frac{G_{\rm F}}{\sqrt{2}}V_{cb}\left[\bar u_\ell\gamma^\alpha
(1-\gamma_5)v_\nu\right]\langle  D^+|\bar c\gamma_\alpha(1-\gamma_5)b
|\bar B^0_d\rangle,
\end{equation}
where {\it all} the hadronic physics is encoded in the hadronic
matrix element
\begin{displaymath}
\langle D^+|\bar c\gamma_\alpha
(1-\gamma_5)b|\bar B^0_d\rangle,
\end{displaymath}
i.e.\ there are {\it no} other QCD effects.
Since the $\bar B^0_d$ and $D^+$ are pseudoscalar mesons, we have
\begin{equation}
\langle D^+|\bar c\gamma_\alpha\gamma_5b|
\bar B^0_d\rangle=0,
\end{equation}
and may write
\begin{equation}\label{BD-ME}
\langle D^+(k)|\bar c\gamma_\alpha b|\bar B^0_d(p)
\rangle=F_1(q^2)\left[(p+k)_\alpha -
\left(\frac{M_B^2-M_D^2}{q^2}\right)q_\alpha\right]
+F_0(q^2)\left(\frac{M_B^2-M_D^2}{q^2}\right)q_\alpha, 
\end{equation}
where $q\equiv p-k$, and the $F_{1,0}(q^2)$ denote the {\it form factors}
of the $\bar B\to D$ transitions. Consequently, in contrast to the simple 
case of the leptonic transitions, semileptonic decays involve {\it two} 
hadronic form factors instead of the decay constant $f_B$. In order to 
calculate these parameters, which depend on the momentum transfer $q$, 
again non-perturbative techniques (QCD sum rules, lattice,  etc.) are 
required.

\subsubsection{Aspects of the Heavy-Quark Effective Theory}
If the mass $m_Q$ of a quark $Q$ is much larger than the QCD scale parameter
$\Lambda_{\rm QCD}={\cal O}(100\,\mbox{MeV})$, it is referred to as 
a ``heavy'' quark. Since the bottom and charm quarks have masses at the 
level of $5\,\mbox{GeV}$ and $1\,\mbox{GeV}$, respectively, they belong 
to this important category. As far as the extremely heavy top quark, 
with $m_t\sim 170\,\mbox{GeV}$ is concerned, it decays unfortunately 
through weak interactions before a hadron can be formed. Let us now 
consider a heavy quark that is bound inside a hadron, i.e.\ a bottom 
or a charm quark. The heavy quark then moves almost with the 
hadron's four velocity $v$ and is almost on-shell, so that
\begin{equation}
p_Q^\mu=m_Qv^\mu + k^\mu,
\end{equation}
where $v^2=1$ and $k\ll m_Q$ is the ``residual'' momentum. Owing to 
the interactions of the heavy quark with the light degrees of freedom of 
the hadron, the residual momentum may only change by 
$\Delta k\sim\Lambda_{\rm QCD}$, and $\Delta v \to 0$ for 
$\Lambda_{\rm QCD}/m_Q\to 0$. 

It is now instructive to have a look at the elastic scattering process 
$\bar B(v)\to \bar B(v')$ in the limit of $\Lambda_{\rm QCD}/m_b \to 0$, 
which is characterized by the following matrix element:
\begin{equation}\label{BB-ME}
\frac{1}{M_B}\langle\bar B(v')|\bar b_{v'}\gamma_\alpha b_v
|\bar B(v)\rangle=\xi(v'\cdot v)(v+v')_\alpha.
\end{equation}
Since the contraction of this matrix element with $(v-v')^\alpha$ has to 
vanish because of $\not \hspace*{-0.1truecm}v b_v= b_v$ and 
$\overline{b}_{v'} \hspace*{-0.2truecm}\not \hspace*{-0.1truecm}v' = 
\overline{b}_{v'}$, no $(v-v')_\alpha$ term arises in the parametrization
in (\ref{BB-ME}). On the other hand, the $1/M_B$ factor is related
to the normalization of states, i.e.\ the right-hand side of
\begin{equation}
\left(\frac{1}{\sqrt{M_B}}\langle\bar B(p')|\right)
\left(|\bar B(p)\rangle\frac{1}{\sqrt{M_B}}\right)=2v^0(2\pi)^3
\delta^3(\vec p-\vec p')
\end{equation}
does not depend on $M_B$. Finally, current conservation implies 
the following normalization condition:
\begin{equation}
\xi(v'\cdot v=1)=1,
\end{equation}
where  the ``Isgur--Wise'' function $\xi(v'\cdot v)$ does 
{\it not} depend on the flavour of the heavy quark (heavy-quark symmetry)
\cite{IW}. Consequently, for $\Lambda_{\rm QCD}/m_{b,c}\to 0$, we may write
\begin{equation}\label{BD-ME-HQ}
\frac{1}{\sqrt{M_D M_B}}\langle D(v')|\bar c_{v'}\gamma_\alpha b_v
|\bar B(v)\rangle=\xi(v'\cdot v)(v+v')_\alpha,
\end{equation}
and observe that this transition amplitude is governed -- in the 
heavy-quark limit -- by {\it one} hadronic form factor $\xi(v'\cdot v)$, 
which satisfies $\xi(1)=1$. If we now compare (\ref{BD-ME-HQ})
with (\ref{BD-ME}), we obtain 
\begin{equation}
F_1(q^2)=\frac{M_D+M_B}{2\sqrt{M_DM_B}}\xi(w)
\end{equation}
\begin{equation}
F_0(q^2)=\frac{2\sqrt{M_DM_B}}{M_D+M_B}\left[\frac{1+w}{2}\right]\xi(w),
\end{equation}
with
\begin{equation}
w\equiv v_D\cdot v_B=\frac{M_D^2+M_B^2-q^2}{2M_DM_B}.
\end{equation}
Similar relations hold  for the $\bar B\to D^\ast$ form factors
because of the heavy-quark spin symmetry, since the $D^\ast$ is 
related to the $D$ by a rotation of the heavy-quark spin. A detailed 
discussion of these interesting features and the associated ``heavy-quark 
effective theory'' (HQET) is beyond the scope of this lecture. For
a detailed overview, we refer the reader to Ref.~\cite{neubert-rev}, where
also a comprehensive list of original references can be found. 
For a more phenomenological discussion, also Ref.~\cite{BaBar-book} is very
useful.

\subsubsection{Applications}
An important application of the formalism sketched above is the
extraction of the CKM element $|V_{cb}|$. To this end, 
$\bar B\to D^*\ell\bar \nu$ decays are particularly promising. 
The corresponding rate can be written as 
\begin{equation}\label{BD-rate}
\frac{{\rm d}\Gamma}{{\rm d}w}=G_{\rm F}^2 K(M_B,M_{D^\ast},w) 
F(w)^2 |V_{cb}|^2,
\end{equation}
where $K(M_B,M_{D^\ast},w)$ is a known kinematic function, and 
$F(w)$ agrees with the Isgur--Wise function, up to perturbative
QCD corrections and $\Lambda_{\rm QCD}/m_{b,c}$ terms. The form 
factor $F(w)$ is a non-perturbative quantity. However, it satisfies 
the following normalization condition:
\begin{equation}\label{F1-norm}
F(1)=\eta_A(\alpha_s)\left[1+\frac{0}{m_c}+
\frac{0}{m_b}+
{\cal O}(\Lambda_{\rm QCD}^2/m_{b,c}^2)\right],
\end{equation} 
where $\eta_A(\alpha_s)$ is a perturbatively calculable short-distance
QCD factor, and the $\Lambda_{\rm QCD}/m_{b,c}$ corrections {\it vanish}
\cite{neubert-rev,neu-BDast}. The important latter feature is an 
implication of Luke's theorem \cite{luke}. Consequently, 
if we extract $F(w)|V_{cb}|$ from a measurement of (\ref{BD-rate}) 
as a function of $w$ and extrapolate to the ``zero-recoil point'' $w=1$
(where the rate vanishes), we may determine $|V_{cb}|$. In the case of 
$\bar B\to D\ell\bar \nu$ decays, we have 
${\cal O}(\Lambda_{\rm QCD}/m_{b,c})$ corrections to the 
corresponding rate ${\rm d}\Gamma/{\rm d}w$ at $w=1$. 
In order to determine $|V_{cb}|$, inclusive $B\to X_c\ell\bar \nu$ decays 
offer also very attractive avenues. As becomes obvious from (\ref{Def-A})
and the considerations in Subsection~\ref{ssec:UT}, $|V_{cb}|$ fixes 
the normalization of the UT. Moreover, this quantity is an important 
input parameter for various theoretical calculations. The CKM matrix
element $|V_{cb}|$ is currently known with about $2\%$ precision;  performing 
an analysis of leptonic and hadronic moments in inclusive $b\to c \ell \bar \nu$ 
processes \cite{Gambino}, the following value was extracted from the $B$-factory 
data \cite{OBuchmuller}:
\begin{equation}\label{Vcb}
|V_{cb}| = (42.0\pm 0.7)\times 10^{-3},
\end{equation}
which agrees with that from exclusive decays. 

Let us now turn to $\bar B\to \pi\ell\bar\nu, \rho\ell\bar\nu$ decays, 
which originate from $b\to u\ell \bar\nu$ quark-level processes, as
can be seen in Fig.~\ref{fig:semi}, and provide access to $|V_{ub}|$. 
If we complement this CKM matrix element with $|V_{cb}|$, we may determine 
the UT side $R_b$ with the help of (\ref{Rb-def}). The 
determination of $|V_{ub}|$ is hence a very important aspect of
flavour physics. Since the $\pi$ and $\rho$ are ``light'' mesons, 
the HQET symmetry relations 
cannot be applied to the $\bar B\to \pi\ell\bar\nu, \rho\ell\bar\nu$ modes. 
Consequently, in order to determine $|V_{ub}|$ from these exclusive 
channels, the corresponding heavy-to-light form factors have to be
described by models. An important alternative is provided by inclusive 
decays. The corresponding decay rate takes the following form:
\begin{equation}\label{inclusive-rate}
\Gamma(\bar B\to X_u \ell \bar \nu)=
\frac{G_{\rm F}^2|V_{ub}|^2}{192\pi^3}m_b^5
\left[1-2.41\frac{\alpha_s}{\pi}+\frac{\lambda_1-9\lambda_2}{2m_b^2}
+\ldots\right],
\end{equation}
where $\lambda_1$ and $\lambda_2$ are non-perturbative parameters, 
which describe the hadronic matrix elements of certain ``kinetic'' 
and ``chromomagnetic'' operators appearing within the framework of
the HQET. Using the heavy-quark expansions
\begin{equation}\label{mass-exp}
M_B=m_b+\bar\Lambda-\frac{\lambda_1+3\lambda_2}{2m_b}+\ldots, \quad
M_{B^\ast}=m_b+\bar\Lambda-\frac{\lambda_1-\lambda_2}{2m_b}+\ldots
\end{equation}
for the $B^{(\ast)}$-meson masses, where $\bar\Lambda\sim\Lambda_{\rm QCD}$ 
is another non-perturbative parameter that is related to the light degrees
of freedom, the parameter $\lambda_2$ can be determined from the measured
values of the $M_{B^{(\ast)}}$. The strong dependence of 
(\ref{inclusive-rate}) on $m_b$ is a significant
source of uncertainty. On the other hand, the $1/m_b^2$ corrections
can be better controlled than in the exclusive case (\ref{F1-norm}), 
where we have, moreover, to deal with $1/m_c^2$ corrections. From an 
experimental point of view, we have to struggle with large backgrounds, 
which originate from $b\to c \ell \bar\nu$ processes and require also 
a model-dependent treatment. The determination of $|V_{ub}|$ from 
$B$-meson decays caused by $b\to u\ell \bar\nu$ 
quark-level processes is therefore a very challenging issue, and the situation 
is less favourable than with $|V_{cb}|$ \cite{BF-DMs}. In particular, the 
values from inclusive and exclusive transitions differ at the 1$\,\sigma$ 
level \cite{HFAG}:
\begin{equation}\label{Vub-det}
|V_{ub}|_{\rm incl} = (4.4\pm 0.3)\times 10^{-3}\,,\quad 
|V_{ub}|_{\rm excl} = (3.8\pm 0.6)\times 10^{-3}\,,
\end{equation}
which has to be fully settled in the future. 
The error on $|V_{ub}|_{\rm excl}$ is dominated by the theoretical
uncertainty of lattice and light-cone sum rule calculations of $B\to\pi$ and
$B\to\rho$ transition form factors \cite{Vublatt,LCSR}, whereas for
$|V_{ub}|_{\rm incl}$ experimental and theoretical errors are at par.
Using the values of $|V_{cb}|$ and $|V_{ub}|$ given above
and $\lambda =0.225\pm0.001$ \cite{blucher}, we obtain
\begin{equation}\label{Rb}
R_b^{\rm incl} = 0.45\pm 0.03\,,\qquad R_b^{\rm excl} = 0.39\pm
0.06\,,
\end{equation}
where the labels ``incl" and ``excl" refer to the determinations of $|V_{ub}|$
through inclusive and exclusive $b\to u\ell\bar \nu_\ell$ transitions, 
respectively.

For a much more detailed discussion of the determinations of $|V_{cb}|$ 
and $|V_{ub}|$, we refer the reader to 
Refs.~\cite{WG2-rep,CKM-book,BaBar-book}, where 
also the references to the vast original literature can be found.

\begin{figure}
   \centerline{
   \begin{tabular}{lc}
     {\small(a)} & \\
    &     \includegraphics[width=3.3truecm]{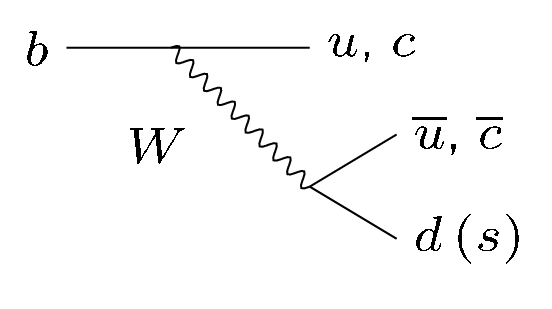}\\
     {\small(b)} & \\
    &  \includegraphics[width=5.0truecm]{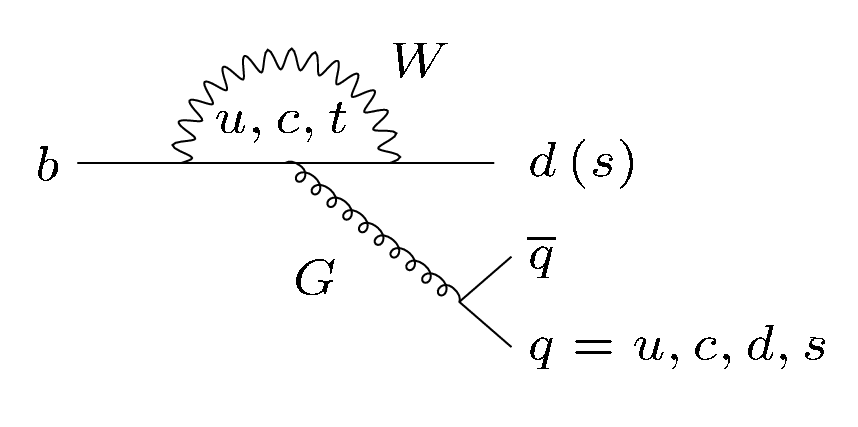}\\
     {\small(c)} & \\
    &     \includegraphics[width=8.3truecm]{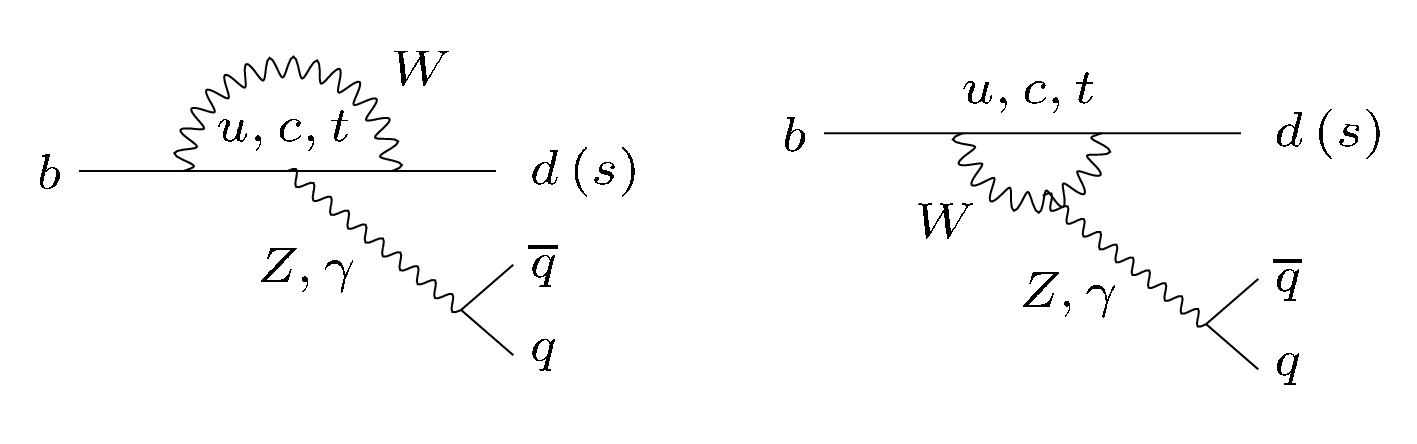} 
     \end{tabular}}
     \caption{Feynman diagrams of the topologies characterizing non-leptonic 
     $B$-meson decays: trees (a), QCD penguins (b), and electroweak penguins 
     (c).}\label{fig:topol}
\end{figure}

\subsection{Non-Leptonic Decays}\label{subsec:non-lept}
\subsubsection{Classification}\label{sec:class}
The most complicated $B$ decays are the non-leptonic transitions, 
which are mediated by 
$b\to q_1\,\bar q_2\,d\,(s)$ quark-level processes, with 
$q_1,q_2\in\{u,d,c,s\}$. There are two kinds of 
topologies contributing to such decays: tree-diagram-like and ``penguin'' 
topologies. The latter consist of gluonic (QCD) and electroweak (EW) 
penguins. In Fig.~\ref{fig:topol}, the corresponding 
leading-order Feynman diagrams are shown. Depending
on the flavour content of their final states, we may classify 
$b\to q_1\,\bar q_2\,d\,(s)$ decays as follows:
\begin{itemize}
\item $q_1\not=q_2\in\{u,c\}$: {\it only} tree diagrams contribute.
\item $q_1=q_2\in\{u,c\}$: tree {\it and} penguin diagrams contribute.
\item $q_1=q_2\in\{d,s\}$: {\it only} penguin diagrams contribute.
\end{itemize}

\subsubsection{Low-Energy Effective Hamiltonians}\label{subsec:ham}
In order to analyse non-leptonic $B$ decays theoretically, we use 
low-energy effective Hamiltonians, which are calculated by making use 
of the ``operator product expansion'', yielding transition 
matrix elements of the following structure:
\begin{equation}\label{ee2}
\langle f|{\cal H}_{\rm eff}|i\rangle=\frac{G_{\rm F}}{\sqrt{2}}
\lambda_{\rm CKM}\sum_k C_k(\mu)\langle f|Q_k(\mu)|i\rangle\,.
\end{equation}
The technique of the operator product expansion allows us to separate 
the short-distance contributions to this transition amplitude from the 
long-distance ones, which are described by perturbative quantities 
$C_k(\mu)$ (``Wilson coefficient functions'') and non-perturbative 
quantities $\langle f|Q_k(\mu)|i\rangle$ (``hadronic matrix elements''), 
respectively. As before, $G_{\rm F}$ is the Fermi constant, whereas
$\lambda_{\rm CKM}$ is a CKM factor and $\mu$ denotes an appropriate 
renormalization scale. The $Q_k$ are local operators, which 
are generated by electroweak interactions and QCD, and govern ``effectively'' 
the decay in question. The Wilson coefficients $C_k(\mu)$ can be 
considered as scale-dependent couplings related to the vertices described
by the $Q_k$.

\begin{figure}
\begin{center}
\leavevmode
\begin{tabular}{cc}
\epsfysize=4.0truecm 
\epsffile{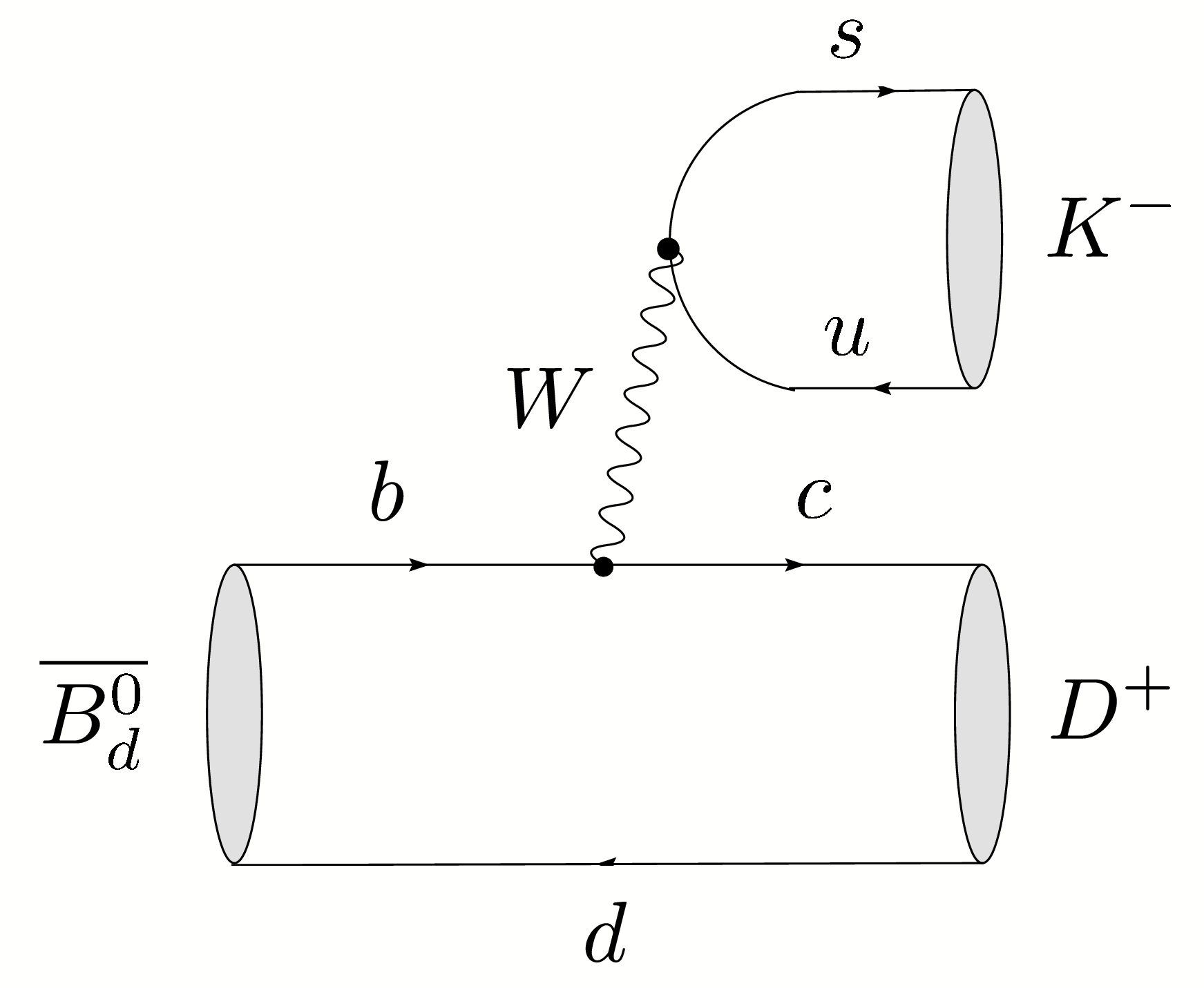} &
\epsfysize=2.0truecm 
\epsffile{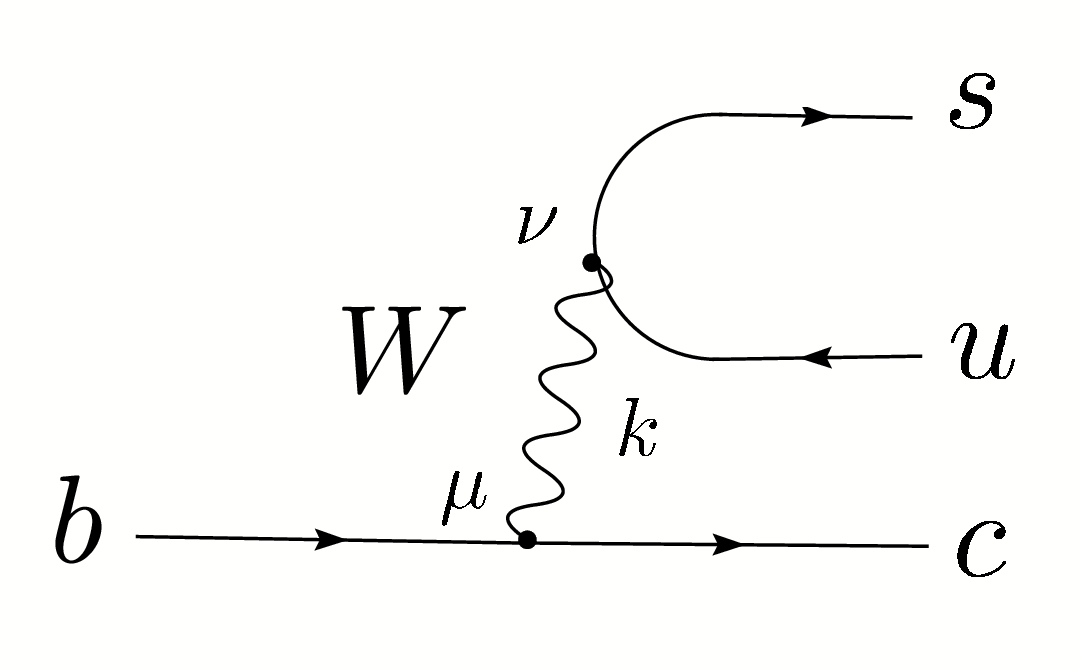} 
\end{tabular}
\end{center}
\vspace*{-0.8truecm}
\caption{Feynman diagrams contributing to the non-leptonic 
$\bar B^0_d\to D^+K^-$ decay.}\label{fig:non-lept-ex}
\end{figure}

\begin{figure}
\begin{center}
\leavevmode
\epsfysize=2.0truecm 
\epsffile{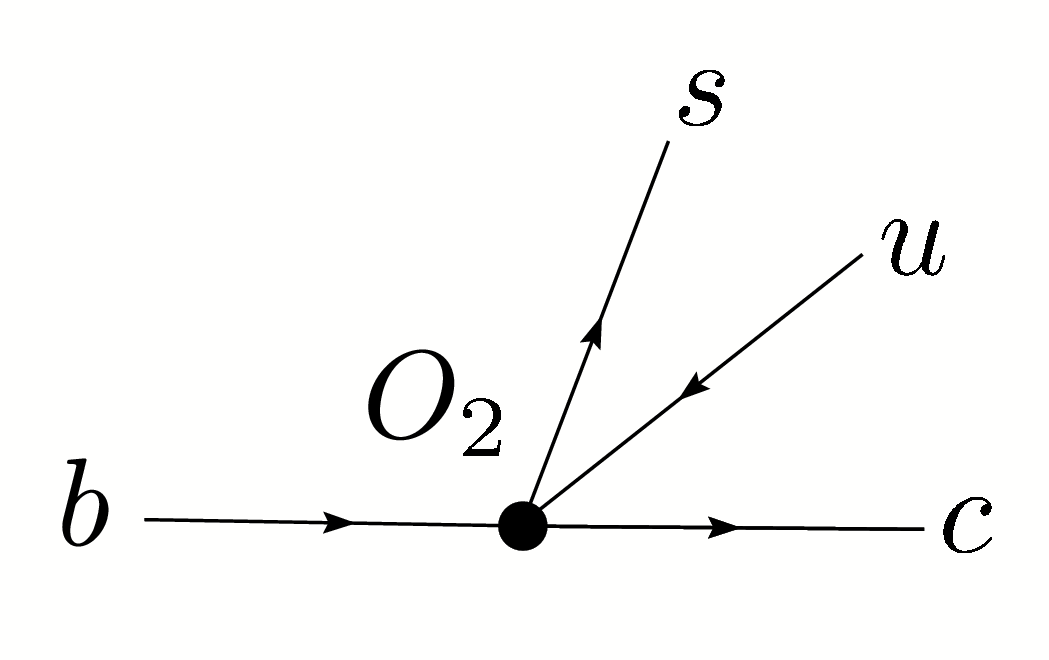} 
\end{center}
\vspace*{-0.8truecm}
\caption{The description of the $b\to d \bar u s$ process through the four-quark
operator $O_2$ in the effective theory after the $W$ boson has been integrated 
out.}\label{fig:non-lept-eff}
\end{figure}

In order to illustrate this rather abstract formalism, let us consider
the decay $\bar B^0_d\to D^+K^-$, which allows a transparent discussion 
of the evaluation of the corresponding low-energy effective Hamiltonian.
Since this transition originates from a $b\to c \bar u s$ quark-level 
process, it is -- as we have just seen -- a pure ``tree'' decay, i.e.\ we do not have
to deal with penguin topologies, which simplifies the analysis
considerably. The leading-order Feynman diagram contributing to
$\bar B^0_d\to D^+K^-$ can straightforwardly be obtained from 
Fig.~\ref{fig:semi} by substituting $\ell$ and $\nu_\ell$ by $s$ and $u$, 
respectively, as can be seen in Fig.~\ref{fig:non-lept-ex}. Consequently, 
the lepton current is simply replaced by a 
quark current, which will have important implications shown below. 
Evaluating the corresponding Feynman diagram yields
\begin{equation}\label{trans-ampl}
-\,\frac{g_2^2}{8}V_{us}^\ast V_{cb}
\left[\bar s\gamma^\nu(1-\gamma_5)u\right]
\left[\frac{g_{\nu\mu}}{k^2-M_W^2}\right]
\left[\bar c\gamma^\mu(1-\gamma_5)b\right].
\end{equation}
Because of $k^2\sim m_b^2\ll M_W^2$, we may -- as in (\ref{Tfi-semi-full}) -- 
``integrate out'' the $W$ boson with the help of (\ref{W-int-out}),
and arrive at
\begin{eqnarray}
\lefteqn{{\cal H}_{\rm eff}=\frac{G_{\rm F}}{\sqrt{2}}V_{us}^\ast V_{cb}
\left[\bar s_\alpha\gamma_\mu(1-\gamma_5)u_\alpha\right]
\left[\bar c_\beta\gamma^\mu(1-\gamma_5)b_\beta\right]}\nonumber\\
&&=\frac{G_{\rm F}}{\sqrt{2}}V_{us}^\ast V_{cb}
(\bar s_\alpha u_\alpha)_{\mbox{{\scriptsize 
V--A}}}(\bar c_\beta b_\beta)_{\mbox{{\scriptsize V--A}}}
\equiv\frac{G_{\rm F}}{\sqrt{2}}V_{us}^\ast V_{cb}O_2\,,
\end{eqnarray}
where $\alpha$ and $\beta$ denote the colour indices of the $SU(3)_{\rm C}$
gauge group of QCD. Effectively, our $b\to c \bar u s$ decay process 
is now described by the ``current--current'' operator $O_2$, as is illustrated
in Fig.~\ref{fig:non-lept-eff}.

\begin{figure}
\begin{center}
\leavevmode
\begin{tabular}{ccc}
\epsfysize=2.7truecm 
\epsffile{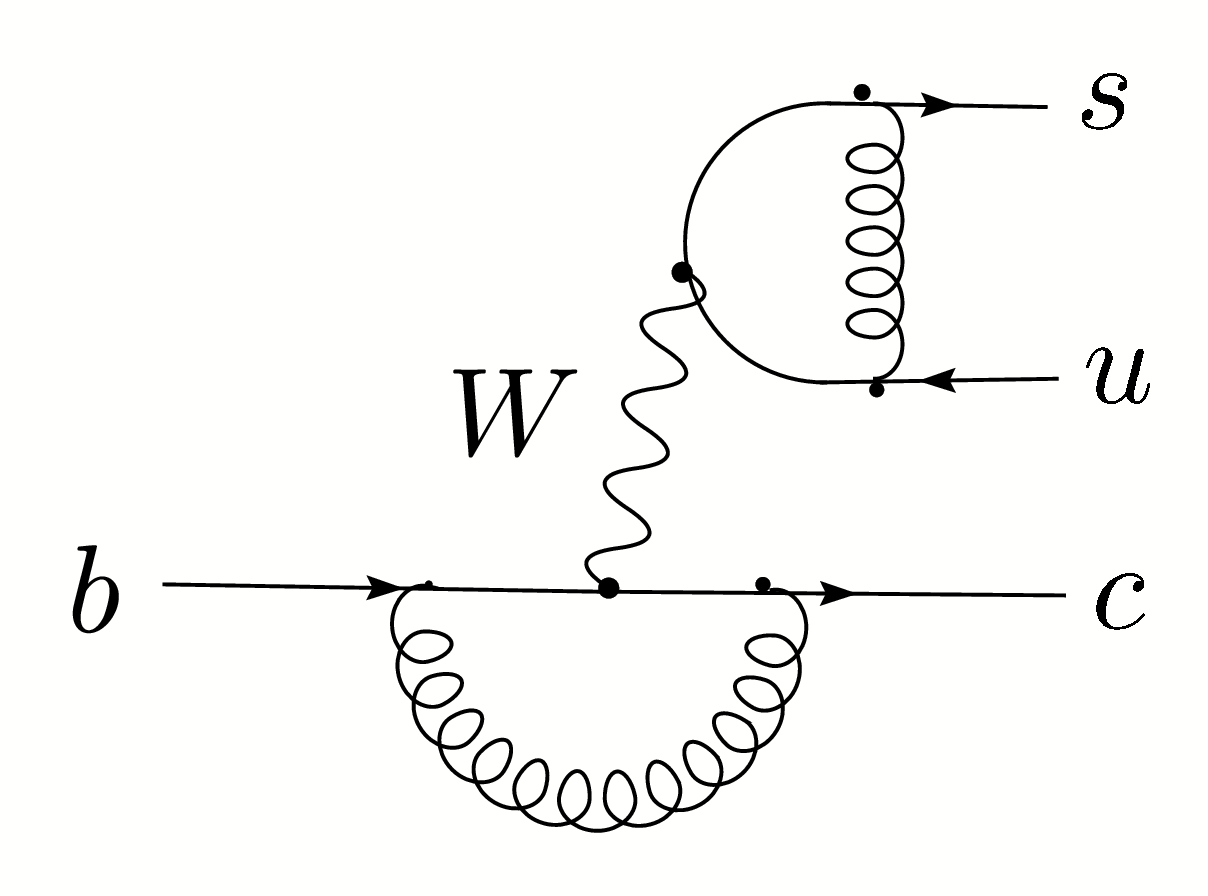} & \mbox{} &
\epsfysize=2.7truecm 
\epsffile{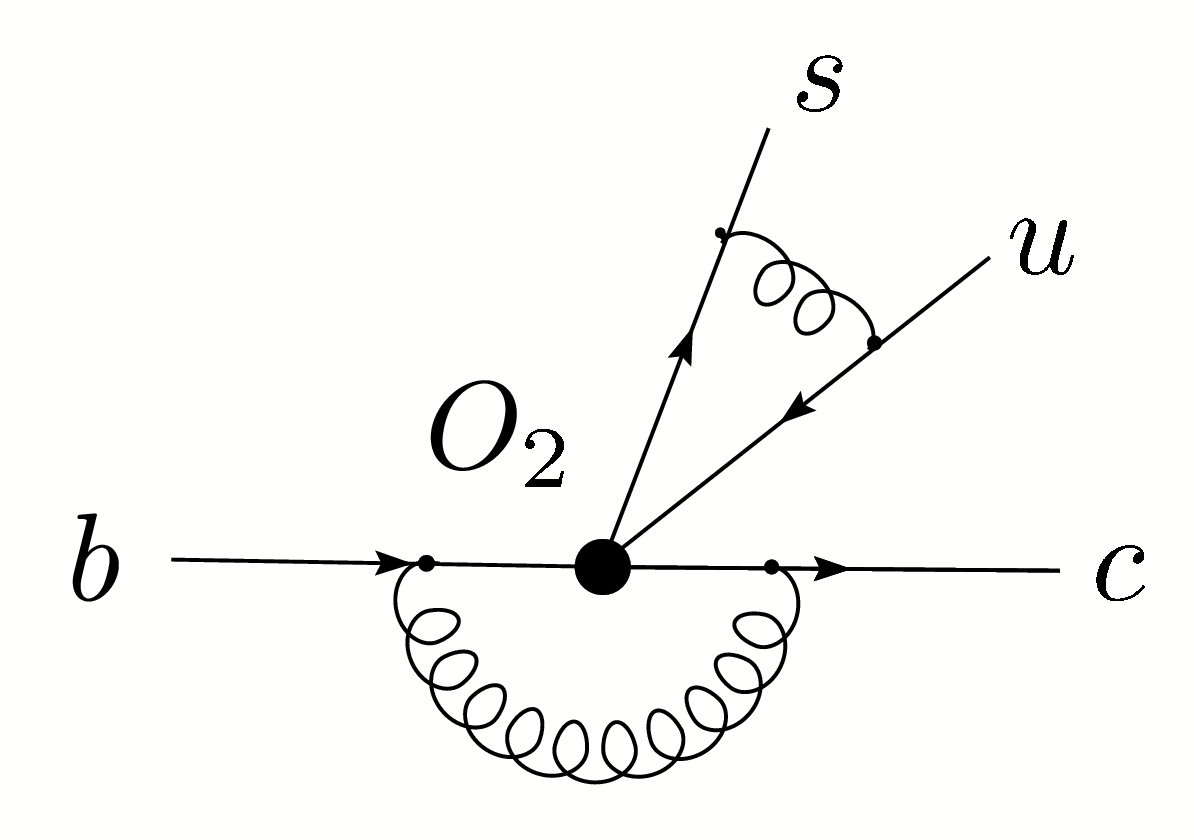}
\end{tabular}
\end{center}
\vspace*{-0.5truecm}
\caption{Factorizable QCD corrections in the full (left panel) and effective 
(right panel) theories.}\label{fig:QCD-fact}
\end{figure}

\begin{figure}
\begin{center}
\leavevmode
\begin{tabular}{ccc}
\epsfysize=1.8truecm 
\epsffile{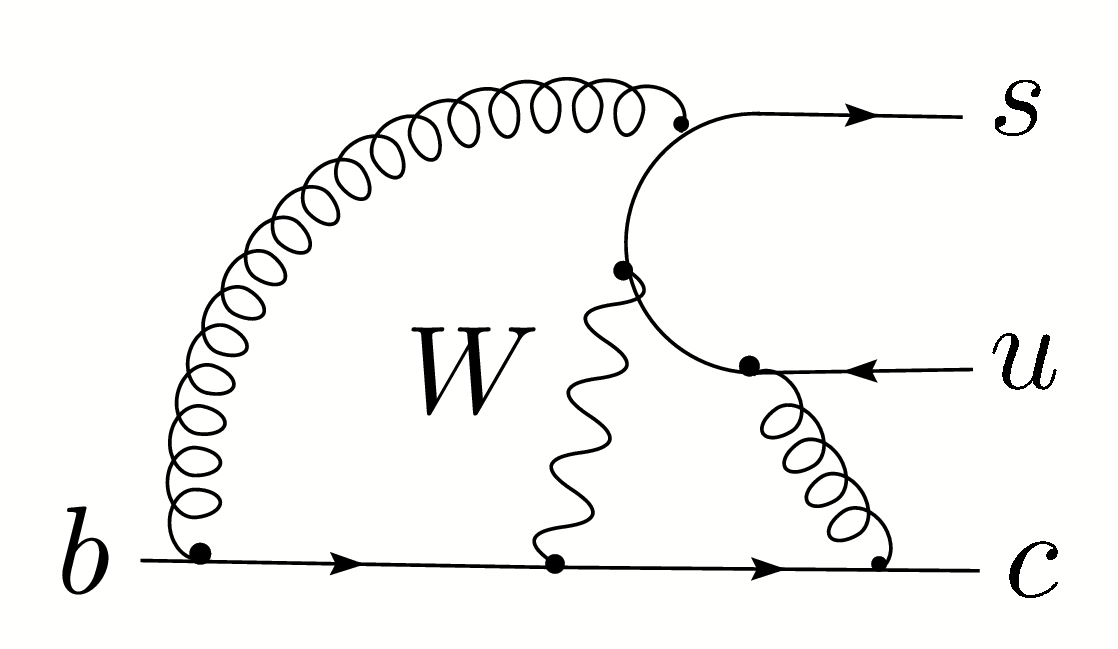} &   \mbox{} &
\epsfysize=1.8truecm 
\epsffile{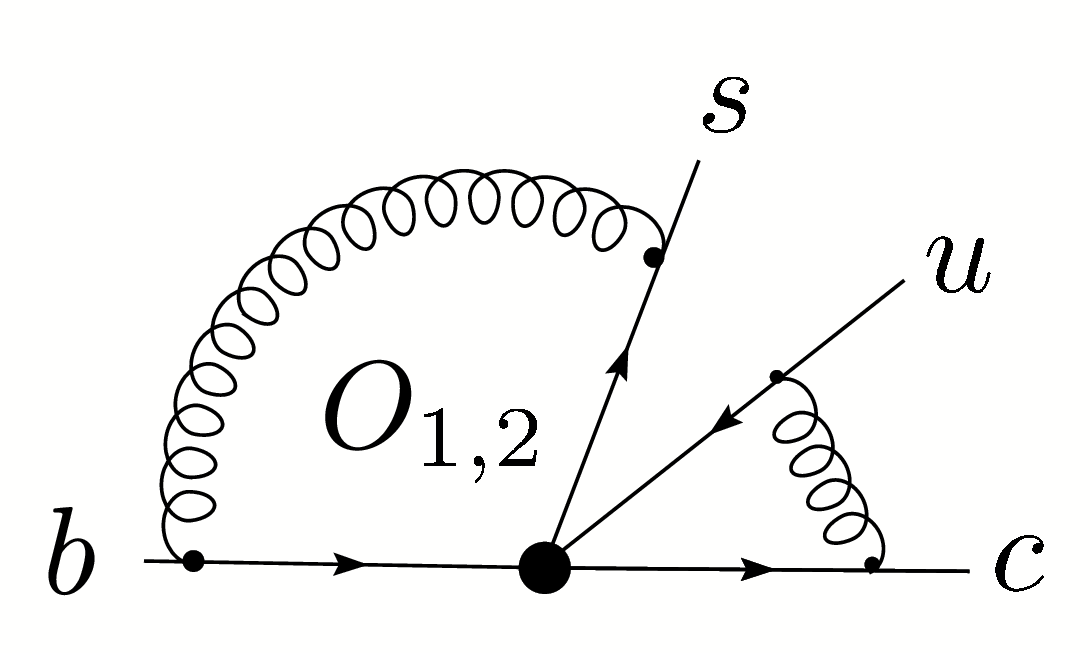}
\end{tabular}
\end{center}
\vspace*{-0.5truecm}
\caption{Non-factorizable QCD corrections in the full (left panel) and effective 
(right panel) theories.}\label{fig:QCD-nonfact}
\end{figure}

So far, we have neglected QCD corrections. Their important impact is
twofold: thanks to {\it factorizable} QCD corrections as shown in Fig.~\ref{fig:QCD-fact}, 
the Wilson coefficient $C_2$ acquires a renormalization-scale dependence,
i.e.\ $C_2(\mu)\not=1$. On the other hand,  {\it non-factorizable} QCD corrections as
illustrated in Fig.~\ref{fig:QCD-nonfact} generate a second current--current operator
through ``operator mixing", which is given by
\begin{equation}
O_1\equiv\left[\bar s_\alpha\gamma_\mu(1-\gamma_5)u_\beta\right]
\left[\bar c_\beta\gamma^\mu(1-\gamma_5)b_\alpha\right].
\end{equation}
Consequently, we eventually arrive at  a low-energy effective Hamiltonian of the
following structure:
\begin{equation}\label{Heff-example}
{\cal H}_{\rm eff}=\frac{G_{\rm F}}{\sqrt{2}}V_{us}^\ast V_{cb}
\left[C_1(\mu)O_1+C_2(\mu)O_2\right].
\end{equation}
In order to evaluate the Wilson  
coefficients $C_1(\mu)\not=0$ and $C_2(\mu)\not=1$ \cite{HEFF-TREE}, 
we must first calculate the QCD corrections to the decay processes 
both in the full theory, i.e. with $W$ exchange, and in the effective 
theory, where the $W$ is integrated out (see Figs.~\ref{fig:QCD-fact} and 
\ref{fig:QCD-nonfact}), and have then to express 
the QCD-corrected transition amplitude in terms of QCD-corrected matrix 
elements and Wilson coefficients as in (\ref{ee2}). This procedure is 
called ``matching'' between the full and the effective theory. 
The results for the $C_k(\mu)$ thus obtained contain 
terms of $\mbox{log}(\mu/M_W)$, which become large for $\mu={\cal O}(m_b)$, 
the scale governing the hadronic matrix elements of the $O_k$. Making use of 
the renormalization group, which exploits the fact that the transition 
amplitude (\ref{ee2}) cannot depend on the chosen renormalization scale 
$\mu$, we may sum up the following terms of the Wilson coefficients:
\begin{equation}
\alpha_s^n\left[\log\left(\frac{\mu}{M_W}\right)\right]^n 
\,\,\mbox{(LO)},\quad\,\,\alpha_s^n\left[\log\left(\frac{\mu}{M_W}\right)
\right]^{n-1}\,\,\mbox{(NLO)},\quad ...\quad ;
\end{equation}
detailed discussions of these rather technical aspects can be found in
Ref.~\cite{BBL-rev}.

For the exploration of CP violation, the class of non-leptonic $B$ decays 
that receives contributions both from tree and from penguin topologies plays 
a central r\^ole. In this important case, the operator basis is much larger 
than in our example (\ref{Heff-example}), where we considered a pure 
``tree'' decay. If we apply the relation
\begin{equation}\label{CKM-UT-Rel}
V_{ur}^\ast V_{ub}+V_{cr}^\ast V_{cb}+V_{tr}^\ast V_{tb}=0
\quad (r\in\{d,s\}),
\end{equation}
which follows from the unitarity of the CKM matrix, and ``integrate out''
the top quark (which enters through the penguin loop processes) and 
the $W$ boson, we may write 
\begin{equation}\label{e4}
{\cal H}_{\mbox{{\scriptsize eff}}}=\frac{G_{\mbox{{\scriptsize 
F}}}}{\sqrt{2}}\left[\sum\limits_{j=u,c}V_{jr}^\ast V_{jb}\left\{
\sum\limits_{k=1}^2C_k(\mu)\,Q_k^{jr}
+\sum\limits_{k=3}^{10}C_k(\mu)\,Q_k^{r}\right\}\right].
\end{equation}
Here we have introduced another quark-flavour label $j\in\{u,c\}$,
and the $Q_k^{jr}$ can be divided as follows:
\begin{itemize}
\item Current--current operators:
\begin{equation}
\begin{array}{rcl}
Q_{1}^{jr}&=&(\bar r_{\alpha}j_{\beta})_{\mbox{{\scriptsize V--A}}}
(\bar j_{\beta}b_{\alpha})_{\mbox{{\scriptsize V--A}}}\\
Q_{2}^{jr}&=&(\bar r_\alpha j_\alpha)_{\mbox{{\scriptsize 
V--A}}}(\bar j_\beta b_\beta)_{\mbox{{\scriptsize V--A}}}.
\end{array}
\end{equation}
\item QCD penguin operators:
\begin{equation}\label{qcd-penguins}
\begin{array}{rcl}
Q_{3}^r&=&(\bar r_\alpha b_\alpha)_{\mbox{{\scriptsize V--A}}}\sum_{q'}
(\bar q'_\beta q'_\beta)_{\mbox{{\scriptsize V--A}}}\\
Q_{4}^r&=&(\bar r_{\alpha}b_{\beta})_{\mbox{{\scriptsize V--A}}}
\sum_{q'}(\bar q'_{\beta}q'_{\alpha})_{\mbox{{\scriptsize V--A}}}\\
Q_{5}^r&=&(\bar r_\alpha b_\alpha)_{\mbox{{\scriptsize V--A}}}\sum_{q'}
(\bar q'_\beta q'_\beta)_{\mbox{{\scriptsize V+A}}}\\
Q_{6}^r&=&(\bar r_{\alpha}b_{\beta})_{\mbox{{\scriptsize V--A}}}
\sum_{q'}(\bar q'_{\beta}q'_{\alpha})_{\mbox{{\scriptsize V+A}}}.
\end{array}
\end{equation}
\item EW penguin operators (the $e_{q'}$ denote the
electrical quark charges):
\begin{equation}
\begin{array}{rcl}
Q_{7}^r&=&\frac{3}{2}(\bar r_\alpha b_\alpha)_{\mbox{{\scriptsize V--A}}}
\sum_{q'}e_{q'}(\bar q'_\beta q'_\beta)_{\mbox{{\scriptsize V+A}}}\\
Q_{8}^r&=&
\frac{3}{2}(\bar r_{\alpha}b_{\beta})_{\mbox{{\scriptsize V--A}}}
\sum_{q'}e_{q'}(\bar q_{\beta}'q'_{\alpha})_{\mbox{{\scriptsize V+A}}}\\
Q_{9}^r&=&\frac{3}{2}(\bar r_\alpha b_\alpha)_{\mbox{{\scriptsize V--A}}}
\sum_{q'}e_{q'}(\bar q'_\beta q'_\beta)_{\mbox{{\scriptsize V--A}}}\\
Q_{10}^r&=&
\frac{3}{2}(\bar r_{\alpha}b_{\beta})_{\mbox{{\scriptsize V--A}}}
\sum_{q'}e_{q'}(\bar q'_{\beta}q'_{\alpha})_{\mbox{{\scriptsize V--A}}}.
\end{array}
\end{equation}
\end{itemize}
The current--current, QCD and EW penguin operators are related to the tree, 
QCD and EW penguin processes shown in 
Fig.~\ref{fig:topol}. At a renormalization scale
$\mu={\cal O}(m_b)$, the Wilson coefficients of the current--current operators
are $C_1(\mu)={\cal O}(10^{-1})$ and $C_2(\mu)={\cal O}(1)$, whereas those
of the penguin operators are ${\cal O}(10^{-2})$ \cite{BBL-rev}. 
Note that penguin 
topologies with internal charm- and up-quark exchanges \cite{BSS}
are described in this framework by penguin-like matrix elements of 
the corresponding current--current operators \cite{RF-DIPL}, and 
may also have important phenomenological consequences \cite{BF-PEN,CHARM-PEN}.

Since the ratio $\alpha/\alpha_s={\cal O}(10^{-2})$ of the QED and QCD 
couplings is very small, we would expect na\"\i vely that EW penguins 
should play a minor r\^ole in comparison with QCD penguins. This would 
actually be the case if the top quark was not ``heavy''. However, since 
the Wilson coefficient $C_9$ increases strongly with $m_t$, we obtain 
interesting EW penguin effects in several $B$ decays: $B\to K\phi$ 
modes are affected significantly by EW penguins, whereas $B\to\pi\phi$ 
and $B_s\to\pi^0\phi$ transitions are even {\it dominated} by such 
topologies \cite{RF-EWP,RF-rev}. As we will see in Subsection~\ref{ssec:BpiK},
EW penguins have also an important impact on the 
$B\to\pi K$ system \cite{EWP-BpiK,PAPIII}.

The low-energy effective Hamiltonians discussed above apply to all 
$B$ decays that are caused by the same quark-level transition, i.e.\ 
they are ``universal''. Consequently, the differences between the 
various exclusive modes of a given decay class arise within this formalism 
only through the hadronic matrix elements of the relevant four-quark 
operators. Unfortunately, the evaluation of such matrix elements is 
associated with large uncertainties and is a very challenging task. In this 
context, ``factorization'' is a widely used concept, which is our next 
topic.

\subsubsection{Factorization of Hadronic Matrix Elements}\label{ssec:ME-fact}
In order to discuss ``factorization'', let us consider once more 
the decay $\bar B^0_d\to D^+K^-$. Evaluating the corresponding 
transition amplitude, we encounter the hadronic matrix elements of the 
$O_{1,2}$ operators between the $\langle K^-D^+|$ final and the 
$|\bar B^0_d\rangle$ initial states. If we use the well-known 
$SU(N_{\rm C})$ colour-algebra relation
\begin{equation}
T^a_{\alpha\beta}T^a_{\gamma\delta}=\frac{1}{2}\left(\delta_{\alpha\delta}
\delta_{\beta\gamma}-\frac{1}{N_{\rm C}}\delta_{\alpha\beta}
\delta_{\gamma\delta}\right)
\end{equation}
to rewrite the operator $O_1$, we obtain
\begin{displaymath}
\langle K^-D^+|{\cal H}_{\rm eff}|\bar B^0_d\rangle=
\frac{G_{\rm F}}{\sqrt{2}}V_{us}^\ast V_{cb}\Bigl[a_1\langle K^-D^+|
(\bar s_\alpha u_\alpha)_{\mbox{{\scriptsize V--A}}}
(\bar c_\beta b_\beta)_{\mbox{{\scriptsize V--A}}}
|\bar B^0_d\rangle
\end{displaymath}
\vspace*{-0.3truecm}
\begin{equation}\label{ME-rewritten}
+2\,C_1\langle K^-D^+|
(\bar s_\alpha\, T^a_{\alpha\beta}\,u_\beta)_{\mbox{{\scriptsize 
V--A}}}(\bar c_\gamma 
\,T^a_{\gamma\delta}\,b_\delta)_{\mbox{{\scriptsize V--A}}}
|\bar B^0_d\rangle\Bigr],\nonumber
\end{equation}
with
\begin{equation}\label{a1-def}
a_1=C_1/N_{\rm C}+C_2 \sim 1.
\end{equation}
It is now straightforward to ``factorize'' the hadronic matrix elements
in (\ref{ME-rewritten}):
\begin{eqnarray}
\lefteqn{\left.\langle K^-D^+|
(\bar s_\alpha u_\alpha)_{\mbox{{\scriptsize 
V--A}}}(\bar c_\beta b_\beta)_{\mbox{{\scriptsize V--A}}}
|\bar B^0_d\rangle\right|_{\rm fact}}\nonumber\\
&&=\langle K^-|\left[\bar s_\alpha\gamma_\mu(1-\gamma_5)u_\alpha\right]
|0\rangle\langle D^+|\left[\bar c_\beta\gamma^\mu
(1-\gamma_5)b_\beta\right]|\bar B^0_d\rangle\nonumber\\
&&=\underbrace{i f_K}_{\mbox{decay constant}} \, \times \, 
\underbrace{F^{(BD)}_0(M_K^2)}_{\mbox{$B\to D$ form factor}} 
\, \times \,\underbrace{(M_B^2-M_D^2),}_{\mbox{kinematical factor}}
\end{eqnarray}
\begin{equation}
\left.\langle K^-D^+|
(\bar s_\alpha\, T^a_{\alpha\beta}\,u_\beta)_{\mbox{{\scriptsize 
V--A}}}(\bar c_\gamma 
\,T^a_{\gamma\delta}\,b_\delta)_{\mbox{{\scriptsize V--A}}}
|\bar B^0_d\rangle\right|_{\rm fact}=0.
\end{equation}
The quantity $a_1$ is a phenomenological ``colour factor'', 
which governs ``colour-allowed'' decays; the
decay $\bar B^0_d\to D^+K^-$ belongs to this category, since the 
colour indices of the $K^-$ meson and the $\bar B^0_d$--$D^+$ system 
run independently from each other in the corresponding leading-order 
diagram shown in Fig.~\ref{fig:non-lept-ex}. On the other hand, in the case 
of ``colour-suppressed'' modes, for instance $\bar B^0_d\to \pi^0D^0$, where only 
one colour index runs through the whole diagram, we have to deal with the combination
\begin{equation}\label{a2-def}
a_2=C_1+C_2/N_{\rm C}\sim0.25.
\end{equation}

The concept of factorizing the hadronic matrix elements of four-quark 
operators into the product of hadronic matrix elements of quark currents 
has a long history \cite{Neu-Ste}, and can be justified, for example, 
in the large-$N_{\rm C}$ limit \cite{largeN}. Interesting more recent 
developments are the following:
\begin{itemize}
\item ``QCD factorization'' \cite{BBNS}, which is in 
accordance with the old picture that factorization should 
hold for certain decays in the limit of $m_b\gg\Lambda_{\rm QCD}$ 
\cite{QCDF-old}, provides a formalism to calculate the 
relevant amplitudes at the leading order of a $\Lambda_{\rm QCD}/m_b$ 
expansion. The resulting expression for the transition amplitudes 
incorporates elements both of the na\"\i ve factorization approach 
sketched above and of the hard-scattering picture. Let us consider a 
decay $\bar B\to M_1M_2$, where $M_1$ picks up the spectator quark. 
If $M_1$ is either a heavy ($D$) or a light ($\pi$, $K$) meson, and 
$M_2$ a light ($\pi$, $K$) meson, QCD factorization gives a transition 
amplitude of the following structure:
\begin{equation}
A(\bar B\to M_1M_2)=\left[\mbox{``na\"\i ve factorization''}\right]
\times\left[1+{\cal O}(\alpha_s)+{\cal O}(\Lambda_{\rm QCD}/m_b)\right].
\end{equation}
While the ${\cal O}(\alpha_s)$ terms, i.e.\ the radiative
non-factorizable corrections, can be calculated systematically, 
the main limitation of the theoretical accuracy originates from 
the ${\cal O}(\Lambda_{\rm QCD}/m_b)$ terms. 

\item Another QCD approach to deal with non-leptonic $B$-meson decays -- 
the ``perturbative hard-scattering approach '' (PQCD) -- was developed 
independently in Ref.~\cite{PQCD}, and differs from the QCD factorization 
formalism in some technical aspects.

\item An interesting technique for ``factorization proofs'' is provided 
by the framework of the ``soft collinear effective theory'' (SCET) 
\cite{SCET}, which led to various applications.

\item Non-leptonic $B$ decays can also be studied within 
QCD light-cone sum-rule approaches \cite{sum-rules}.
\end{itemize}
A detailed presentation of these topics would be very technical and is
beyond the scope of this lecture. However, for the discussion of the
CP-violating effects in the $B$-meson system, we must only be familiar 
with the general structure of the non-leptonic $B$ decay amplitudes and 
not enter the details of the techniques to deal with the
corresponding hadronic matrix elements. Let us finally note that the 
$B$-decay data will eventually decide how well factorization and 
the new concepts sketched above are actually working. For example, 
data on the $B\to\pi\pi$ system point towards 
large non-factorizable corrections \cite{BFRS2}--\cite{CGRS}.

\subsection{Towards Studies of CP Violation}\label{To-CP}
As we have seen above, leptonic and semileptonic $B$-meson decays
involve only a single weak (CKM) amplitude. On the other hand, the 
structure of non-leptonic transitions is considerably more complicated. 
Let us consider a non-leptonic decay $\bar B\to\bar f$ that is described by
the low-energy effective Hamiltonian  in (\ref{e4}). The corresponding
decay amplitude is then given as follows:
\begin{eqnarray}
\lefteqn{\hspace*{-1.3truecm}A(\bar B\to \bar f)=\langle \bar f\vert
{\cal H}_{\mbox{{\scriptsize eff}}}\vert\bar B\rangle}\nonumber\\
&&\hspace*{-1.3truecm}=\frac{G_{\mbox{{\scriptsize F}}}}{\sqrt{2}}\left[
\sum\limits_{j=u,c}V_{jr}^\ast V_{jb}\left\{\sum\limits_{k=1}^2
C_{k}(\mu)\langle \bar f\vert Q_{k}^{jr}(\mu)\vert\bar B\rangle
+\sum\limits_{k=3}^{10}C_{k}(\mu)\langle \bar f\vert Q_{k}^r(\mu)
\vert\bar B\rangle\right\}\right].~~~\mbox{}\label{Bbarfbar-ampl}
\end{eqnarray}
Concerning the CP-conjugate process $B\to\ f$, we have
\begin{eqnarray}
\lefteqn{\hspace*{-1.3truecm}A(B \to f)=\langle f|
{\cal H}_{\mbox{{\scriptsize 
eff}}}^\dagger|B\rangle}\nonumber\\
&&\hspace*{-1.3truecm}=\frac{G_{\mbox{{\scriptsize F}}}}{\sqrt{2}}
\left[\sum\limits_{j=u,c}V_{jr}V_{jb}^\ast \left\{\sum\limits_{k=1}^2
C_{k}(\mu)\langle f\vert Q_{k}^{jr\dagger}(\mu)\vert B\rangle
+\sum\limits_{k=3}^{10}C_{k}(\mu)\langle f\vert Q_k^{r\dagger}(\mu)
\vert B\rangle\right\}\right].~~~\mbox{}\label{Bf-ampl}
\end{eqnarray}
If we use now that strong interactions are invariant under CP transformations, 
insert $({\cal CP})^\dagger({\cal CP})=\hat 1$ both after the $\langle f|$ and in 
front of the $|B\rangle$, and take the relation 
\begin{equation}
({\cal CP})Q_k^{jr\dagger}({\cal CP})^\dagger=Q_k^{jr}
\end{equation}
into account, we arrive at
\begin{eqnarray}
\lefteqn{\hspace*{-1.3truecm}A(B \to f)=
e^{i[\phi_{\mbox{{\scriptsize CP}}}(B)-\phi_{\mbox{{\scriptsize CP}}}(f)]}}\nonumber\\
&&\hspace*{-1.3truecm}\times\frac{G_{\mbox{{\scriptsize F}}}}{\sqrt{2}}
\left[\sum\limits_{j=u,c}V_{jr}V_{jb}^\ast\left\{\sum\limits_{k=1}^2
C_{k}(\mu)\langle \bar f\vert Q_{k}^{jr}(\mu)\vert\bar B\rangle
+\sum\limits_{k=3}^{10}C_{k}(\mu)
\langle \bar f\vert Q_{k}^r(\mu)\vert\bar B\rangle\right\}\right],
\end{eqnarray}
where the convention-dependent phases $\phi_{\mbox{{\scriptsize CP}}}(B)$ 
and $\phi_{\mbox{{\scriptsize CP}}}(f)$ are defined through
\begin{equation}\label{CP-phase-def}
({\cal CP})\vert B\rangle=
e^{i\phi_{\mbox{{\scriptsize CP}}}(B)}
\vert\bar B\rangle, \quad 
({\cal CP})\vert f\rangle=
e^{i\phi_{\mbox{{\scriptsize CP}}}(f)}
\vert\bar f\rangle.
\end{equation}
Consequently, we may write
\begin{eqnarray}
A(\bar B\to\bar f)&=&e^{+i\varphi_1}
|A_1|e^{i\delta_1}+e^{+i\varphi_2}|A_2|e^{i\delta_2}\label{par-ampl}\\
A(B\to f)&=&e^{i[\phi_{\mbox{{\scriptsize CP}}}(B)-\phi_{\mbox{{\scriptsize CP}}}(f)]}
\left[e^{-i\varphi_1}|A_1|e^{i\delta_1}+e^{-i\varphi_2}|A_2|e^{i\delta_2}
\right].\label{par-ampl-CP}
\end{eqnarray}
Here the CP-violating phases $\varphi_{1,2}$ originate from the CKM factors 
$V_{jr}^\ast V_{jb}$, and the CP-conserving ``strong'' amplitudes
$|A_{1,2}|e^{i\delta_{1,2}}$ involve the hadronic matrix elements of the 
four-quark operators. In fact, these expressions are the most general forms
of any non-leptonic $B$-decay amplitude in the SM, i.e.\ they do not only
refer to the $\Delta C=\Delta U=0$ case described by (\ref{e4}). 
Using (\ref{par-ampl}) and (\ref{par-ampl-CP}), we obtain
the following CP asymmetry:
\begin{eqnarray}
{\cal A}_{\rm CP}&\equiv&\frac{\Gamma(B\to f)-
\Gamma(\bar B\to\bar f)}{\Gamma(B\to f)+\Gamma(\bar B
\to \bar f)}=\frac{|A(B\to f)|^2-|A(\bar B\to \bar f)|^2}{|A(B\to f)|^2+
|A(\bar B\to \bar f)|^2}\nonumber\\
&=&\frac{2|A_1||A_2|\sin(\delta_1-\delta_2)
\sin(\varphi_1-\varphi_2)}{|A_1|^2+2|A_1||A_2|\cos(\delta_1-\delta_2)
\cos(\varphi_1-\varphi_2)+|A_2|^2},\label{direct-CPV}
\end{eqnarray}
where the convention-dependent phase in (\ref{par-ampl-CP}) cancels.
We observe that a non-vanishing value can be generated through the 
interference between the two weak amplitudes, provided both a non-trivial 
weak phase difference $\varphi_1-\varphi_2$ and a non-trivial strong phase 
difference $\delta_1-\delta_2$ are present. This kind of
CP violation is referred to as ``direct'' CP violation, as it originates 
directly at the amplitude level of the considered decay. It is the 
$B$-meson counterpart of the effects that are probed through 
$\mbox{Re}(\varepsilon'/\varepsilon)$ in the neutral kaon 
system,\footnote{In order to calculate this quantity, an approriate 
low-energy effective Hamiltonian having the same structure as (\ref{e4}) 
is used. The large theoretical uncertainties mentioned in Section~\ref{sec:intro} 
originate from a strong cancellation 
between the contributions of the QCD and EW penguins (caused by the
large top-quark mass) and the associated
hadronic matrix elements.} and could first be established with the help of 
$B_d\to\pi^\mp K^\pm$ decays \cite{CP-B-dir}.

Since $\varphi_1-\varphi_2$ is in general given by one of the UT angles  -- 
usually $\gamma$ -- the goal is to extract this quantity from the measured 
value of ${\cal A}_{\rm CP}$. Unfortunately, hadronic uncertainties affect this
determination through the poorly known hadronic matrix elements in 
(\ref{Bbarfbar-ampl}). In order to deal with this problem, we may proceed along 
one of the following two avenues:
\begin{itemize}
\item[(i)] Amplitude relations can be used to eliminate the hadronic matrix 
elements. In these strategies, we distinguish between exact relations, 
using pure ``tree'' decays  of the kind $B^\pm\to K^\pm D$ \cite{gw,ADS} or 
$B_c^\pm\to D^\pm_s D$ \cite{fw}, and relations, which follow from the flavour symmetries 
of strong interactions, i.e.\ isospin or $SU(3)_{\rm F}$, and involve 
$B_{(s)}\to\pi\pi,\pi K,KK$ modes~\cite{GHLR}. 
\item[(ii)] In decays of neutral $B_q$ mesons, interference effects 
between $B^0_q$--$\bar B^0_q$ mixing and decay processes may induce
 ``mixing-induced CP violation''. If a single CKM amplitude governs the decay, 
 the hadronic matrix elements cancel in the corresponding
CP asymmetries (otherwise we have to use again amplitude relations \cite{RF-BsKK}).
The most important example is the decay $B^0_d\to J/\psi K_{\rm S}$ \cite{bisa}.
\end{itemize}
Before discussing the features of neutral $B_q$ mesons and  
$B^0_q$--$\bar B^0_q$ mixing in detail in Section~\ref{sec:mix}, let us illustrate
the use of amplitude relations for clean extractions of the UT angle $\gamma$
from decays of charged $B_u$ and $B_c$ mesons.

\begin{figure}
\begin{center}
\leavevmode
\epsfysize=3.9truecm 
\epsffile{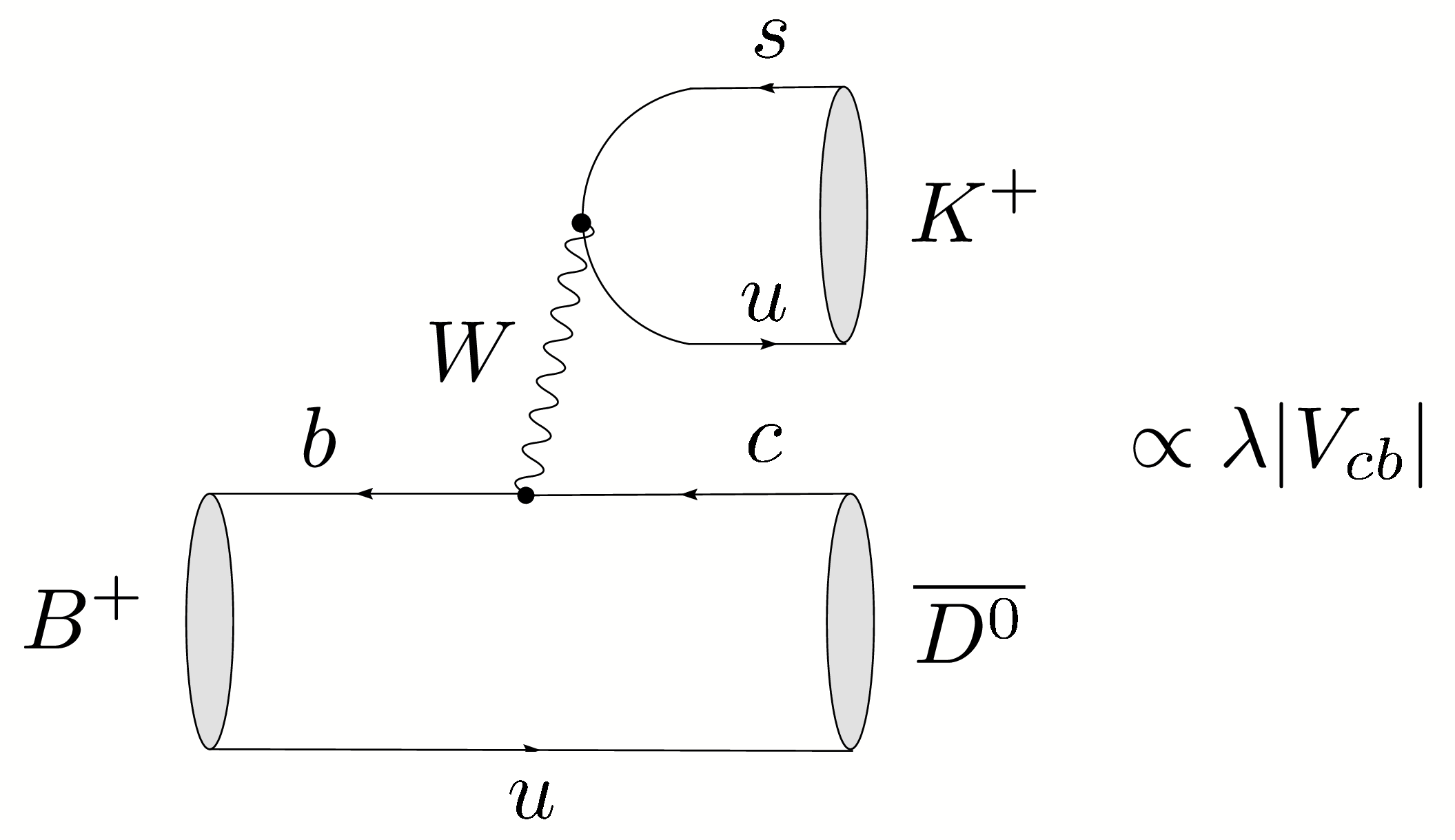} \hspace*{1truecm}
\epsfysize=4.3truecm 
\epsffile{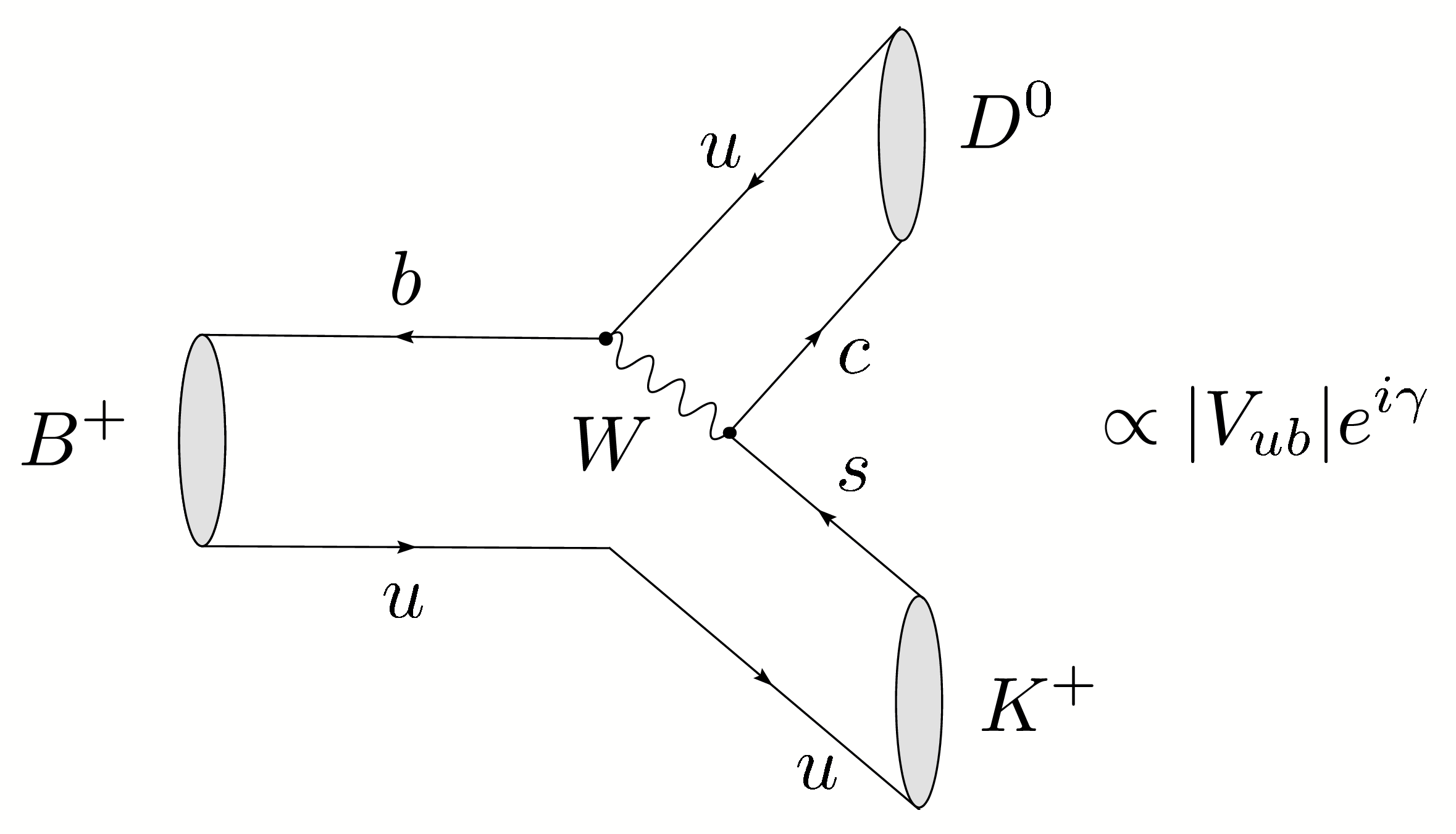}
\end{center}
\vspace*{-0.4truecm}
\caption{Feynman diagrams contributing to $B^+\to K^+\bar D^0$ and 
$B^+\to K^+D^0$. }\label{fig:BDK}
\end{figure}

\begin{figure}
\vspace*{0.3truecm}
\begin{center}
\leavevmode
\epsfysize=2.7truecm 
\epsffile{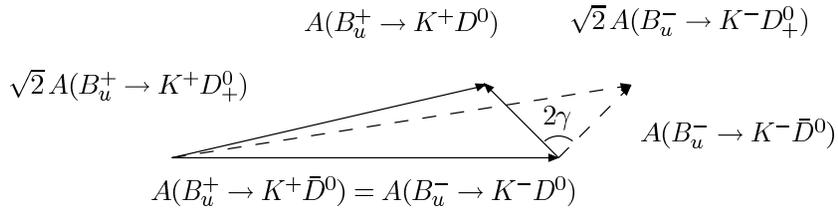} 
\end{center}
\vspace*{-0.6truecm}
\caption{The extraction of $\gamma$ from 
$B^\pm\to K^\pm\{D^0,\bar D^0,D^0_+\}$ 
decays.}\label{fig:BDK-triangle}
\end{figure}

\section{AMPLITUDE RELATIONS}\label{sec:A-REL}
\setcounter{equation}{0}
\boldmath
\subsection{$B^\pm\to K^\pm D$}
\unboldmath
The prototype of the strategies using theoretically clean amplitude 
relations is provided by $B^\pm \to K^\pm D$ decays \cite{gw}. Looking at 
Fig.~\ref{fig:BDK}, we observe that $B^+\to K^+\bar D^0$ and $B^+\to K^+D^0$ 
are pure ``tree'' decays. If we consider, in addition, the transition 
$B^+\to D^0_+K^+$, where $D^0_+$ denotes the CP 
eigenstate of the neutral $D$-meson system with eigenvalue $+1$,
\begin{equation}\label{ED85}
|D^0_+\rangle=\frac{1}{\sqrt{2}}\left[|D^0\rangle+
|\bar D^0\rangle\right],
\end{equation}
we obtain interference effects, which are described by
\begin{eqnarray}
\sqrt{2}A(B^+\to K^+D^0_+)&=&A(B^+\to K^+D^0)+
A(B^+\to K^+\bar D^0)\\
\sqrt{2}A(B^-\to K^-D^0_+)&=&A(B^-\to K^-\bar D^0)+
A(B^-\to K^-D^0).
\end{eqnarray}
These relations can be represented as two triangles in 
the complex plane. Since we have only to deal with tree-diagram-like 
topologies, we have moreover
\begin{eqnarray}
A(B^+\to K^+\bar D^0)&=&A(B^-\to K^-D^0)\\
A(B^+\to K^+D^0)&=&A(B^-\to K^-\bar D^0)\times e^{2i\gamma},
\end{eqnarray}
allowing a {\it theoretically clean} extraction of $\gamma$, as shown 
in Fig.~\ref{fig:BDK-triangle}. Unfortunately, these triangles are 
very squashed, since $B^+\to K^+D^0$ is colour-suppressed 
with respect to $B^+\to K^+\bar D^0$:
\begin{equation}\label{BDK-suppr}
\left|\frac{A(B^+\to K^+D^0)}{A(B^+\to K^+\bar D^0}\right|=
\left|\frac{A(B^-\to K^-\bar D^0)}{A(B^-\to K^-D^0}\right|\approx
\frac{1}{\lambda}\frac{|V_{ub}|}{|V_{cb}|}\times\frac{a_2}{a_1}
\approx 0.4\times0.3={\cal O}(0.1),
\end{equation}
where the phenomenological ``colour'' factors were introduced in
Subsection~\ref{ssec:ME-fact}. 

Another -- more subtle -- problem is related to the measurement of
$\mbox{BR}(B^+\to K^+D^0)$. From the theoretical point of view, 
$D^0\to K^-\ell^+\nu$ would be ideal to measure this tiny 
branching ratio. However, because of the huge background from 
semileptonic $B$ decays, we must rely on Cabibbo-allowed hadronic 
$D^0\to f_{\rm NE}$ decays, such as $f_{\rm NE}=\pi^+K^-$, $\rho^+K^-$,
$\ldots$, i.e.\ have to measure 
\begin{equation}\label{chain1}
B^+\to K^+D^0 \,[\to f_{\rm NE}].
\end{equation}
Unfortunately, we then encounter another decay path into the {\it same} 
final state $K^+ f_{\rm NE}$ through 
\begin{equation}\label{chain2}
B^+\to K^+\bar D^0 \,[\to f_{\rm NE}], 
\end{equation}
where BR$(B^+\to K^+\bar D^0)$ is {\it larger} than BR$(B^+\to K^+D^0)$
by a factor of ${\cal O}(10^2)$, while $\bar D^0\to f_{\rm NE}$ is doubly 
Cabibbo-suppressed, i.e.\ the corresponding branching ratio is suppressed
with respect to the one of $D^0\to f_{\rm NE}$ by a factor of 
${\cal O}(10^{-2})$. Consequently, we obtain interference effects of 
${\cal O}(1)$ between the decay chains in (\ref{chain1}) and (\ref{chain2}). 
However, if two different final states $f_{\rm NE}$ are considered, 
$\gamma$  can be extracted \cite{ADS}, although this determination is  
then more involved than the original triangle approach presented in 
\cite{gw}. 

The angle $\gamma$ can actually be determined in a variety of ways
through CP-violating effects in pure
tree decays of type $B\to D^{(*)} K^{(*)}$ \cite{WG-sum}. Using the
present $B$-factory data, the following results were obtained through a
combination of various methods:
\begin{equation}\label{gam-DK}
\left.\gamma\right|_{D^{(*)} K^{(*)}} = \left\{
\begin{array}{ll}
(77^{+30}_{-32})^\circ & \mbox{(CKMfitter \cite{CKMfitter}),}\\[5pt] 
(88\pm 16)^\circ & \mbox{(UTfit \cite{UTfit}).}
\end{array}
\right.
\end{equation}
Here we have discarded a second solution given by $180^\circ+\left.
\gamma\right|_{D^{(*)} K^{(*)}}$
in the third quadrant of the $\bar\rho$--$\bar\eta$ plane, as it is disfavoured
by the global fits of
the UT, and by the data for mixing-induced CP violation in pure tree decays
of type
$B_d\to D^{\pm}\pi^\mp, D^{\ast\pm}\pi^\mp, ...$ \cite{RF-gam-ca}. A similar
comment applies
to the information from $B\to\pi\pi, \pi K$ modes \cite{BFRS-5}.

\boldmath
\subsection{$B_c^\pm\to D_s^\pm D$}
\unboldmath
In addition to the ``conventional'' $B_u^\pm$ mesons, there is yet another 
species of charged $B$ mesons, the $B_c$-meson system, which consists of
$B_c^+\sim c\overline{b}$ and $B_c^-\sim b\overline{c}$. These mesons were 
observed by the CDF collaboration through their decay 
$B_c^+\to J/\psi \ell^+ \nu$, with the following mass and lifetime 
\cite{CDF-Bc}:
\begin{equation}
M_{B_c}=(6.40\pm0.39\pm0.13)\,\mbox{GeV}, \quad
\tau_{B_c}=(0.46^{+0.18}_{-0.16}\pm 0.03)\,\mbox{ps}.
\end{equation}
Meanwhile, the D0 collaboration observed the $B_c^+\to J/\psi\,\mu^+ X$ mode
\cite{D0-Bc}, which led to the following $B_c$ mass and lifetime determinations:
\begin{equation}
M_{B_c}=(5.95^{+0.14}_{-0.13}\pm0.34)\,\mbox{GeV}, \quad
\tau_{B_c}=(0.448^{+0.123}_{-0.096}\pm 0.121)\,\mbox{ps},
\end{equation}
and CDF reported evidence for the $B_c^+\to J/\psi \pi^+$ channel \cite{CDF-Bc-nl},
implying
\begin{equation}
M_{B_c}= (6.2870 \pm 0.0048  \pm 0.0011)\,\mbox{GeV}.
\end{equation}

\begin{figure}
\begin{center}
\leavevmode
\epsfysize=4.0truecm 
\epsffile{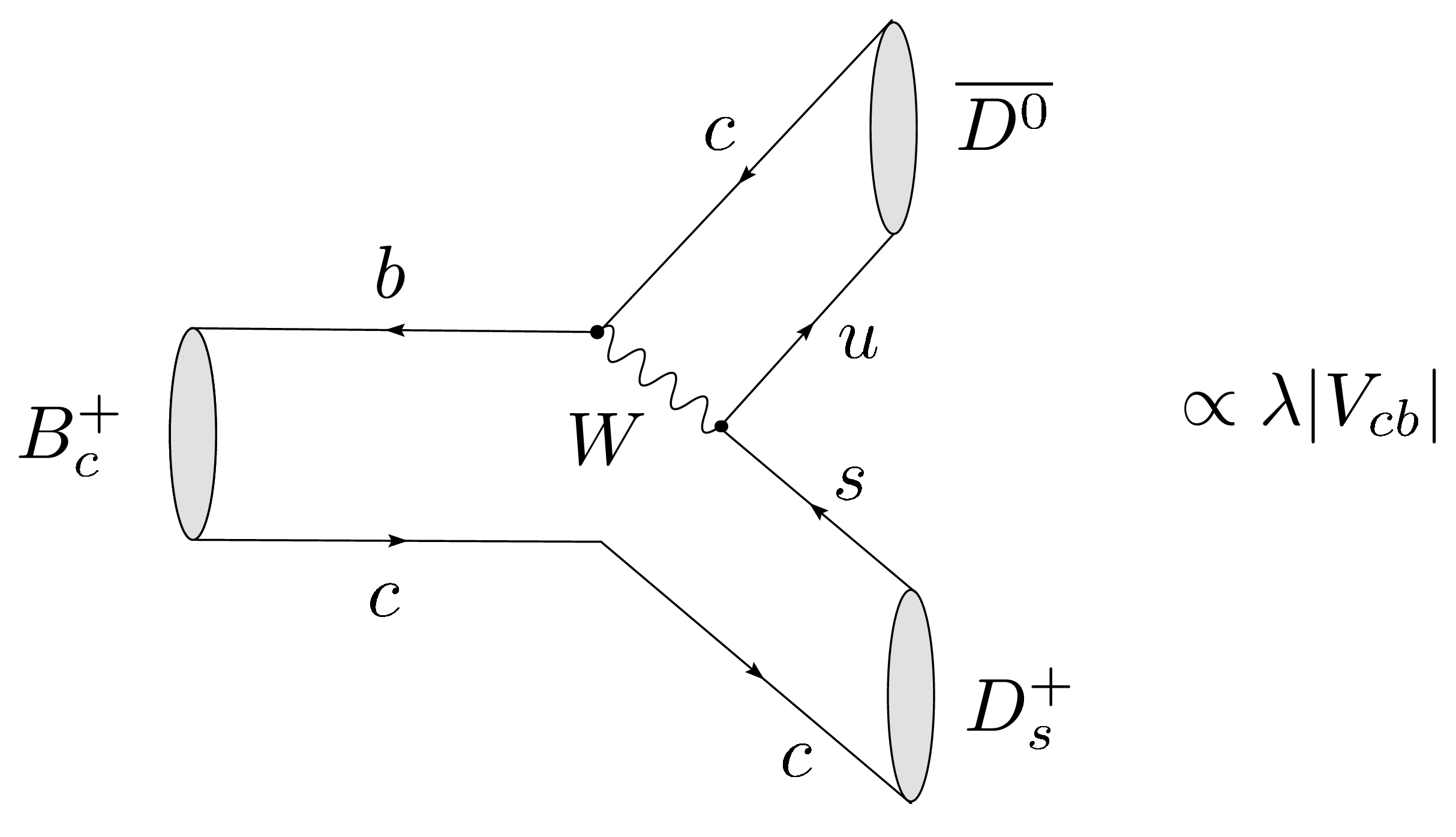} \hspace*{1truecm}
\epsfysize=3.8truecm 
\epsffile{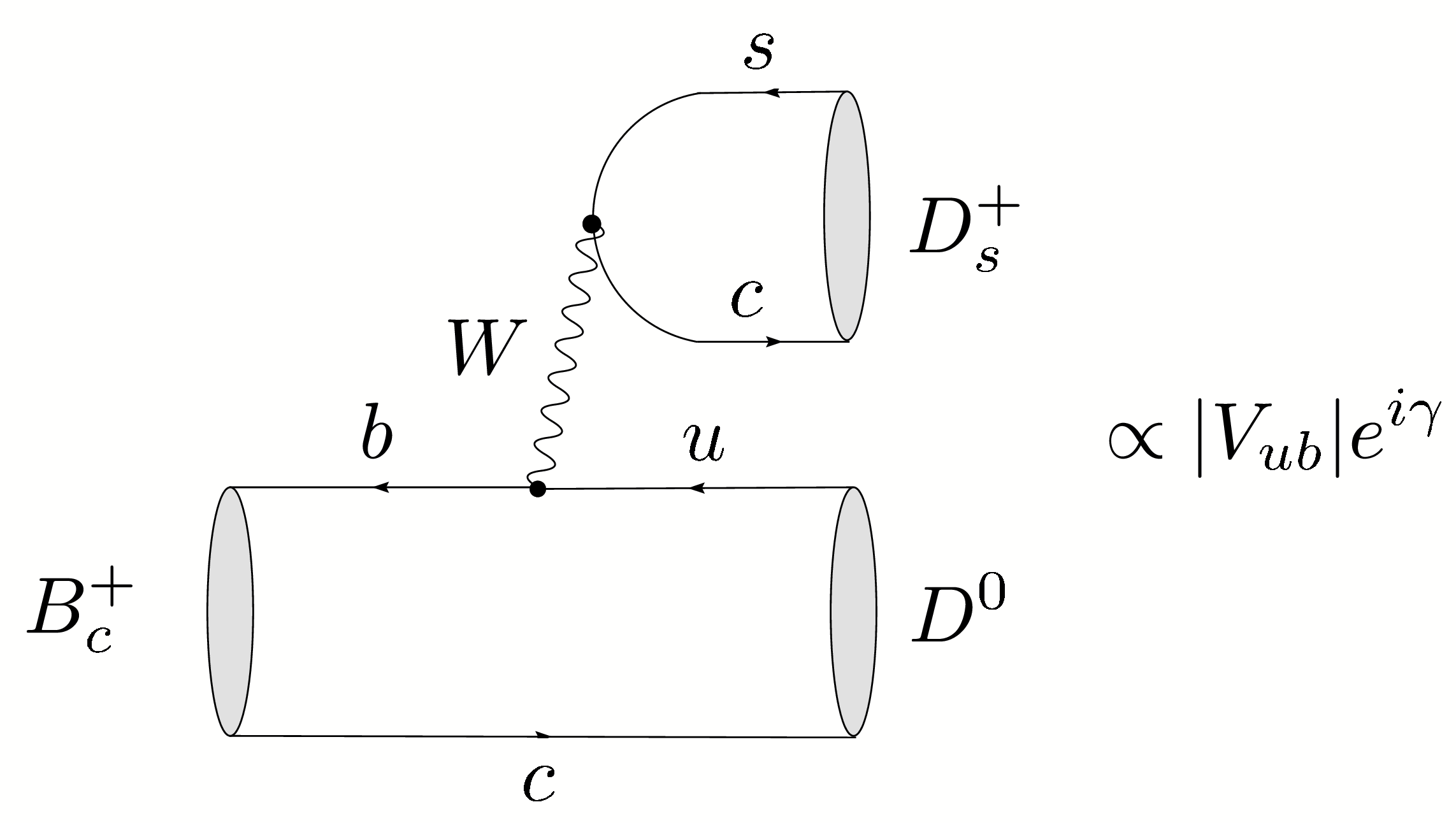}
\end{center}
\vspace*{-0.4truecm}
\caption{Feynman diagrams contributing to $B^+_c\to D_s^+\bar D^0$ and 
$B^+\to D_s^+D^0$. }\label{fig:BcDsD}
\end{figure}

\begin{figure}
\vspace*{0.3truecm}
\begin{center}
\leavevmode
\epsfysize=4.7truecm 
\epsffile{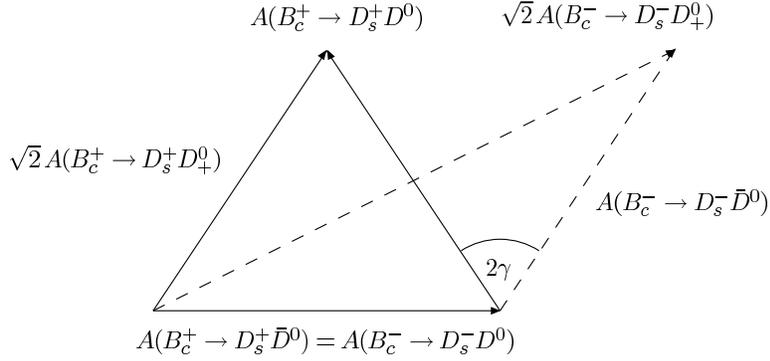} 
\end{center}
\vspace*{-0.6truecm}
\caption{The extraction of $\gamma$ from 
$B_c^\pm\to D^\pm_s\{D^0,\bar D^0,D^0_+\}$ decays.}\label{fig:triangles}
\end{figure}

Since a huge number of $B_c$ mesons will be produced at the LHC, the 
natural question of how to explore CP violation with charged $B_c$ decays arises,
in particular whether an extraction of $\gamma$ with the help of the
triangle approach is possible.  Such a determination is actually offered by 
$B_c^\pm\to D_s^\pm D$ decays, which are the $B_c$ counterparts 
of the $B_u^\pm\to K^\pm D$ modes (see Fig.\ \ref{fig:BcDsD}), 
and satisfy the following amplitude relations \cite{masetti}:
\begin{eqnarray}
\sqrt{2}A(B_c^+\to D_s^+D^0_+)&=&A(B_c^+\to D_s^+D^0)+
A(B_c^+\to D_s^+\bar D^0)\\
\sqrt{2}A(B_c^-\to D_s^-D^0_+)&=&A(B_c^-\to D_s^-\bar D^0)+
A(B_c^-\to D_s^-D^0),
\end{eqnarray}
with
\begin{eqnarray}
A(B^+_c\to D_s^+\bar D^0)&=&A(B^-_c\to D_s^-D^0)\\
A(B_c^+\to D_s^+D^0)&=&A(B_c^-\to D_s^-\bar D^0)\times e^{2i\gamma}.
\end{eqnarray}
At first sight, everything is completely analogous to the $B_u^\pm\to K^\pm D$
case. However, there is an important difference \cite{fw}, 
which becomes obvious by comparing the Feynman diagrams shown in 
Figs.~\ref{fig:BDK} and \ref{fig:BcDsD}: in the $B_c^\pm\to D_s^\pm D$ 
system, the amplitude with the rather small CKM matrix element $V_{ub}$ 
is not colour-suppressed, while the larger element $V_{cb}$ comes with 
a colour-suppression factor. Therefore, we obtain
\begin{equation}\label{Bc-ratio1}
\left|\frac{A(B^+_c\to D_s^+ D^0)}{A(B^+_c\to D_s^+ 
\bar D^0)}\right|=\left|\frac{A(B^-_c\to D_s^-\bar D^0)}{A(B^-_c\to D_s^- 
D^0)}\right|\approx\frac{1}{\lambda}\frac{|V_{ub}|}{|V_{cb}|}
\times\frac{a_1}{a_2}\approx0.4\times 3 = {\cal O}(1),
\end{equation}
and conclude that the two amplitudes are similar in size. In contrast 
to this favourable situation, in the decays $B_u^{\pm}\to K^{\pm}D$, 
the matrix element $V_{ub}$ comes with the colour-suppression factor, 
resulting in a very stretched triangle. The extraction of $\gamma$ from 
the $B_c^\pm\to D_s^\pm D$ triangles is illustrated in 
Fig.~\ref{fig:triangles}, which should be compared with the
squashed $B^\pm_u\to K^\pm D$ triangles shown in 
Fig.\ \ref{fig:BDK-triangle}. Another important advantage is that 
the interference effects arising from $D^0,\bar D^0\to\pi^+K^-$ are 
practically unimportant for the measurement of BR$(B^+_c\to D_s^+ D^0)$ 
and BR$(B^+_c\to D_s^+ \bar D^0)$ since the $B_c$-decay amplitudes are 
of the same order of magnitude. Consequently, the $B_c^\pm\to D_s^\pm D$
decays provide -- from the theoretical point of view -- the ideal
realization of the ``triangle'' approach to determine $\gamma$. 
However, the practical implementation at LHCb still appears to 
be challenging.  The corresponding branching ratios
were estimated in Ref.~\cite{IKP}, with a pattern in accordance 
with (\ref{Bc-ratio1}).

\begin{figure}
\centerline{
 \includegraphics[width=5.9truecm]{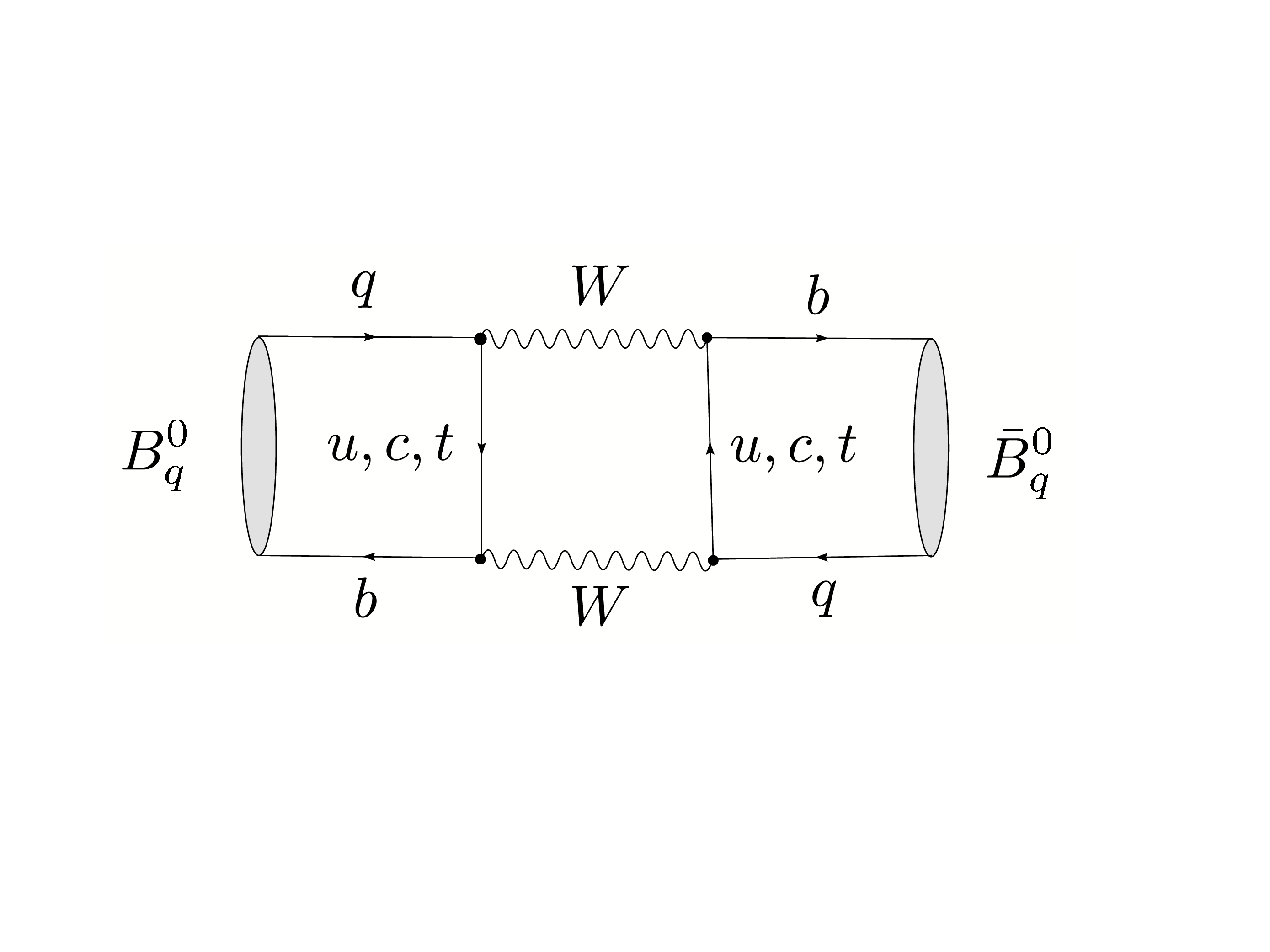}
 \hspace*{0.5truecm}
 \includegraphics[width=5.9truecm]{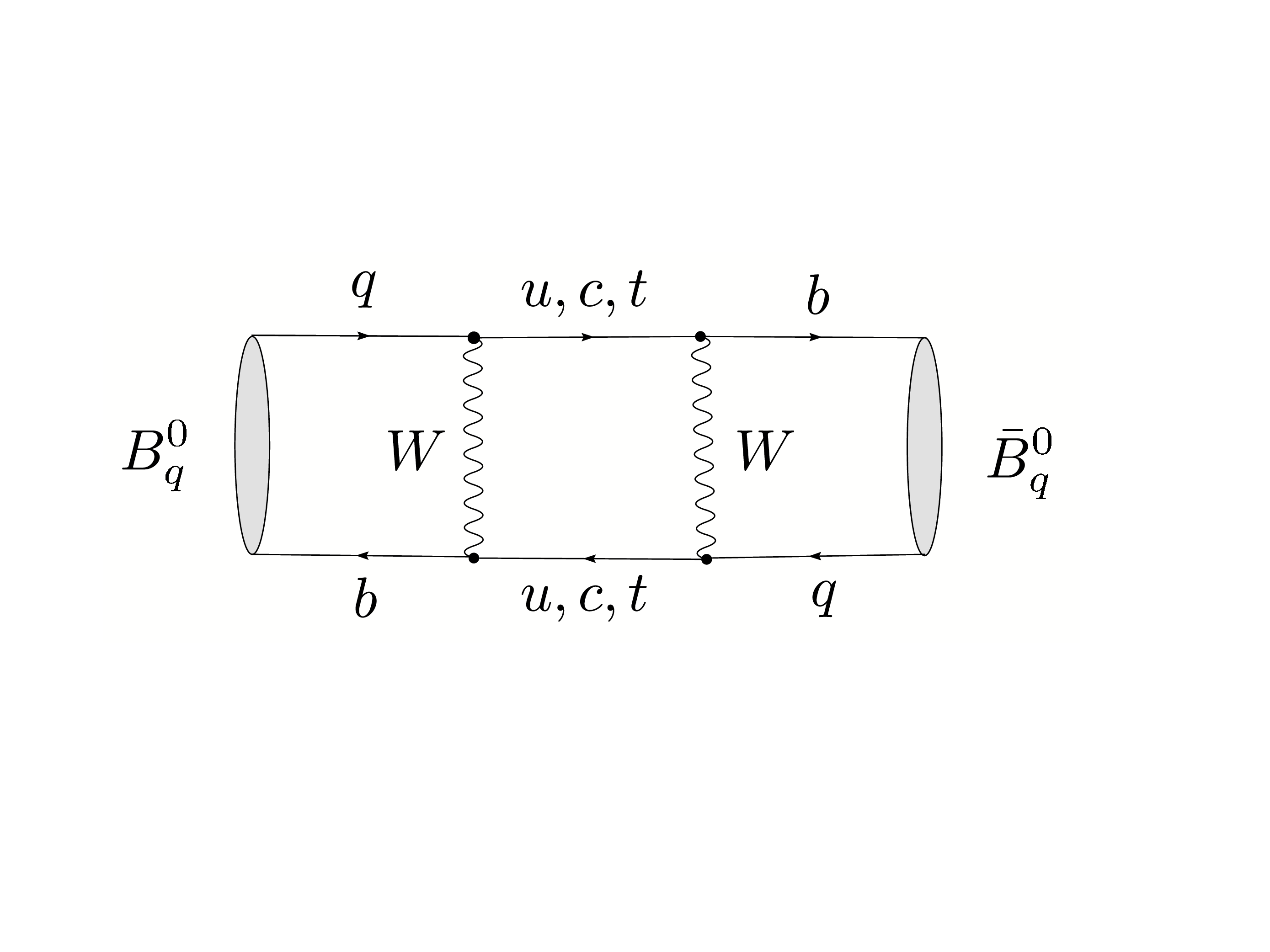}  
 }
 \vspace*{-0.4truecm}
 \caption{Box diagrams contributing to $B^0_q$--$\bar B^0_q$ mixing in the
 SM ($q\in\{d,s\}$).}
   \label{fig:boxes}
\end{figure}

\section{THE NEUTRAL {\boldmath$B$\unboldmath}-MESON SYSTEM}\label{sec:mix}
\setcounter{equation}{0}
\boldmath\subsection{Schr\"odinger Equation for $B^0_q$--$\bar B^0_q$ 
Mixing}\unboldmath\label{ssec:BBbar-mix}
Within the SM, $B^0_q$--$\bar B^0_q$ mixing arises from
the box diagrams shown in Fig.~\ref{fig:boxes}. Thanks to 
this phenomenon, an initially, i.e.\ at time $t=0$, present 
$B^0_q$-meson state evolves into a time-dependent linear combination of 
$B^0_q$ and $\bar B^0_q$ states:
\begin{equation}
|B_q(t)\rangle=a(t)|B^0_q\rangle + b(t)|\bar B^0_q\rangle,
\end{equation}
where $a(t)$ and $b(t)$ are governed by a Schr\"odinger equation of 
the following form:
\begin{equation}\label{SG-OSZ}
i\,\frac{{\rm d}}{{\rm d} t}\left(\begin{array}{c} a(t)\\ b(t)
\end{array}
\right)= H \cdot\left(\begin{array}{c}
a(t)\\ b(t)\nonumber
\end{array}
\right) \equiv
\Biggl[\underbrace{\left(\begin{array}{cc}
M_{0}^{(q)} & M_{12}^{(q)}\\ M_{12}^{(q)\ast} & M_{0}^{(q)}
\end{array}\right)}_{\mbox{mass matrix}}-
\frac{i}{2}\underbrace{\left(\begin{array}{cc}
\Gamma_{0}^{(q)} & \Gamma_{12}^{(q)}\\
\Gamma_{12}^{(q)\ast} & \Gamma_{0}^{(q)}
\end{array}\right)}_{\mbox{decay matrix}}\Biggr]
\cdot\left(\begin{array}{c}
a(t)\\ b(t)\nonumber
\end{array}
\right).
\end{equation}
The special form $H_{11}=H_{22}$ of the Hamiltonian $H$ is an implication 
of the CPT theorem, i.e.\ of the invariance under combined CP and 
time-reversal (T) transformations. 

It is straightforward to calculate the eigenstates 
$\vert B_{\pm}^{(q)}\rangle$ and eigenvalues $\lambda_{\pm}^{(q)}$ 
of (\ref{SG-OSZ}):
\begin{equation}
\vert B_{\pm}^{(q)} \rangle  =
\frac{1}{\sqrt{1+\vert \alpha_q\vert^{2}}}
\left(\vert B^{0}_q\rangle\pm\alpha_q\vert\bar B^{0}_q\rangle\right)
\end{equation}
\begin{equation}\label{lam-pm}
\lambda_{\pm}^{(q)}  =
\left(M_{0}^{(q)}-\frac{i}{2}\Gamma_{0}^{(q)}\right)\pm
\left(M_{12}^{(q)}-\frac{i}{2}\Gamma_{12}^{(q)}\right)\alpha_q,
\end{equation}
where
\begin{equation}\label{alpha-q-expr}
\alpha_q e^{+i\left(\Theta_{\Gamma_{12}}^{(q)}+n'
\pi\right)}=
\sqrt{\frac{4\vert M_{12}^{(q)}\vert^{2}
e^{-i2\delta\Theta_{M/\Gamma}^{(q)}}+\vert\Gamma_{12}^{(q)}\vert^{2}}{4\vert 
M_{12}^{(q)}\vert^{2}+\vert\Gamma_{12}^{(q)}\vert^{2}- 
4\vert M_{12}^{(q)}\vert\vert\Gamma_{12}^{(q)}\vert
\sin\delta\Theta_{M/\Gamma}^{(q)}}}.
\end{equation}
Here we have written
\begin{equation}
M_{12}^{(q)}\equiv e^{i\Theta_{M_{12}}^{(q)}}\vert
M_{12}^{(q)}\vert,\quad \Gamma_{12}^{(q)}\equiv
e^{i\Theta_{\Gamma_{12}}^{(q)}}\vert\Gamma_{12}^{(q)}\vert,\quad
\delta\Theta_{M/\Gamma}^{(q)}\equiv
\Theta_{M_{12}}^{(q)}-\Theta_{\Gamma_{12}}^{(q)},
\end{equation}
and have introduced the quantity $n'\in\{0,1\}$ to parametrize the 
sign of the square root in (\ref{alpha-q-expr}). 

Evaluating the dispersive parts of the box diagrams shown in
Fig.~\ref{fig:boxes}, which are dominated by internal top-quark 
exchanges, yields (for a more detailed discussion, see Ref.~\cite{BF-rev}):
\begin{equation}\label{M12-calc}
M_{12}^{(q)}=
\frac{G_{\rm F}^2 M_{W}^{2}}{12\pi^{2}}
\eta_B M_{B_q}f_{B_q}^{2}\hat B_{B_q}\left(V_{tq}^\ast V_{tb}\right)^2 
S_0(x_{t}) e^{i(\pi-\phi_{\mbox{{\scriptsize CP}}}(B_q))},
\end{equation}
where $\phi_{\mbox{{\scriptsize CP}}}(B_q)$ is a convention-dependent
phase, which is defined in analogy to (\ref{CP-phase-def}).
The short-distance physics is encoded in the ``Inami--Lim'' function 
$S_0(x_{t}\equiv m_t^2/M_W^2)$ \cite{IL}, and in the perturbative 
QCD correction factor  $\eta_B$,
which does {\it not} depend on $q\in\{d,s\}$, i.e.\ is the same for $B_d$ and $B_s$ 
mesons. On the other hand, the non-perturbative physics is described by the 
quantities $f_{B_q}\hat B_{B_q}^{1/2}$, involving -- in addition to the $B_q$ decay 
constant $f_{B_q}$ -- the ``bag'' parameter $\hat B_{B_q}$, which is related to the hadronic matrix element $\langle \bar B^0_q|(\bar bq)_{{\rm V}-{\rm A}}
(\bar bq)_{{\rm V}-{\rm A}}|B^0_q\rangle$. These non-perturbative parameters
can be determined through QCD sum-rule calculations
or lattice studies. 

If we calculate also the absorptive parts of the box diagrams in
Fig.~\ref{fig:boxes}, we obtain
\begin{equation}
\frac{\Gamma_{12}^{(q)}}{M_{12}^{(q)}}\approx
-\frac{3\pi}{2S_0(x_{t})}\left(\frac{m_b^2}{M_W^2}\right)
={\cal O}(m_b^2/m_t^2)\ll 1.
\end{equation}
Consequently, we may expand (\ref{alpha-q-expr}) in 
$\Gamma_{12}^{(q)}/M_{12}^{(q)}$. Neglecting second-order 
terms, we arrive at
\begin{equation}
\alpha_q=\left[1+\frac{1}{2}\left|
\frac{\Gamma_{12}^{(q)}}{M_{12}^{(q)}}\right|\sin\delta
\Theta_{M/\Gamma}^{(q)}\right]e^{-i\left(\Theta_{M_{12}}^{(q)}+n'\pi\right)}.
\end{equation}

The deviation of $|\alpha_q|$ from 1 measures CP violation in 
$B^0_q$--$\bar B^0_q$ oscillations, and can be probed through
the following ``wrong-charge'' lepton asymmetries:
\begin{equation}
{\cal A}^{(q)}_{\mbox{{\scriptsize SL}}}\equiv
\frac{\Gamma(B^0_q(t)\to \ell^-\bar\nu X)-\Gamma(\bar B^0_q(t)\to
\ell^+\nu X)}{\Gamma(B^0_q(t)\to \ell^-\bar \nu X)+
\Gamma(\bar B^0_q(t)\to \ell^+\nu X)}
=\frac{|\alpha_q|^4-1}{|\alpha_q|^4+1}\approx\left|
\frac{\Gamma_{12}^{(q)}}{M_{12}^{(q)}}\right|
\sin\delta\Theta^{(q)}_{M/\Gamma}.
\end{equation}
Because of $|\Gamma_{12}^{(q)}|/|M_{12}^{(q)}|\propto
m_b^2/m_t^2$ and $\sin\delta\Theta^{(q)}_{M/\Gamma}\propto m_c^2/m_b^2$,
the asymmetry ${\cal A}^{(q)}_{\mbox{{\scriptsize SL}}}$ is suppressed by 
a factor of $m_c^2/m_t^2={\cal O}(10^{-4})$ and is hence tiny in the SM.
However, this observable may be enhanced through NP effects, thereby
representing an interesing probe for physics beyond the SM 
\cite{LLNP,BBLN-CFLMT}.  The current experimental average for the $B_d$-meson
system compiled by the ``Heavy Flavour Averaging Group" \cite{HFAG} reads
as follows:
\begin{equation}
{\cal A}^{(d)}_{\mbox{{\scriptsize SL}}}=0.0005 \pm 0.0056,
\end{equation}
and does not indicate any non-vanishing effect.

\subsection{Mixing Parameters}\label{ssec:Mix-Par}
Let us denote the masses of the eigenstates of (\ref{SG-OSZ}) by 
$M^{(q)}_{\rm H}$ (``heavy'') and $M^{(q)}_{\rm L}$ (``light''). 
It is then useful to introduce 
\begin{equation}
M_q\equiv\frac{M^{(q)}_{\rm H}+M^{(q)}_{\rm L}}{2}=
M^{(q)}_0,
\end{equation}
as well as the mass difference
\begin{equation}\label{DeltaMq-def}
\Delta M_q\equiv M_{\rm H}^{(q)}-M_{\rm L}^{(q)}=2|M_{12}^{(q)}|>0,
\end{equation}
which is by definition {\it positive}. While $B^0_d$--$\bar B^0_d$ mixing is well established and
\begin{equation}\label{DMd-exp}
\Delta M_d = (0.507\pm 0.005)\,{\rm ps}^{-1}
\end{equation}
known with impressive
experimental accuracy \cite{HFAG},  only lower bounds on $\Delta M_s$ 
were available,  for many years, from the LEP (CERN) 
experiments and SLD (SLAC) \cite{LEPBOSC}. In 2006, the value 
of $\Delta M_s$ could eventually be pinned down at the Tevatron \cite{DMs-obs}.
The most recent results read as follows:
\begin{equation}\label{MDs}
\Delta M_s=\left\{
\begin{array}{ll}
(18.56 \pm0.87){\rm ps}^{-1}  & \mbox{(D0 \cite{D0-DMs})} \\
(17.77\pm0.10 \pm 0.07){\rm ps}^{-1} & 
\mbox{(CDF \cite{CDF-DMs})}.
\end{array}
\right.
\end{equation}
We shall return to the theoretical interpretation of these measurements 
in Subsection~\ref{ssec:Bs-prelim}. 

The decay widths $\Gamma_{\rm H}^{(q)}$ and 
$\Gamma_{\rm L}^{(q)}$ of the mass eigenstates, which correspond to 
$M^{(q)}_{\rm H}$ and $M^{(q)}_{\rm L}$, respectively, satisfy 
\begin{equation}
\Delta\Gamma_q\equiv\Gamma_{\rm H}^{(q)}-\Gamma_{\rm L}^{(q)}=
\frac{4\mbox{\,Re}\left[M_{12}^{(q)}\Gamma_{12}^{(q)\ast}\right]}{\Delta M_q},
\end{equation}
whereas 
\begin{equation}
\Gamma_q\equiv\frac{\Gamma^{(q)}_{\rm H}+\Gamma^{(q)}_{\rm L}}{2}=
\Gamma^{(q)}_0.
\end{equation}
There is the following interesting relation:
\begin{equation}\label{DGoG}
\frac{\Delta\Gamma_q}{\Gamma_q}\approx-\frac{3\pi}{2S_0(x_t)}
\left(\frac{m_b^2}{M_W^2}\right)x_q=-{\cal O}(10^{-2})\times x_q,
\end{equation}
where
\begin{equation}\label{mix-par}
x_q\equiv\frac{\Delta M_q}{\Gamma_q}=\left\{\begin{array}{cc}
0.776\pm0.008&(q=d)\\
{\cal O}(20)& (q=s)
\end{array}\right.
\end{equation}
is often referred to as {\it the} 
$B^0_q$--$\bar B^0_q$ ``mixing parameter''.\footnote{Note that
$\Delta\Gamma_q/\Gamma_q$ is negative in the SM because of the
minus sign in (\ref{DGoG}).}
Consequently, we observe that $\Delta\Gamma_d/\Gamma_d\sim 10^{-2}$ is 
negligibly small, while $\Delta\Gamma_s/\Gamma_s\sim 10^{-1}$ may
be sizeable. In fact, the current state-of-the-art calculations of these quantities give 
the following numbers \cite{lenz}:
\begin{equation}\label{DGam-numbers}
\frac{|\Delta\Gamma_d|}{\Gamma_d}=(3\pm1.2)\times 10^{-3}, \quad
\frac{|\Delta\Gamma_s|}{\Gamma_s}=0.147 \pm 0.060.
\end{equation}
Recently, results for 
$\Delta\Gamma_s$ were reported from the Tevatron, using the 
$B^0_s\to J/\psi\phi$ channel \cite{DDF}:
\begin{equation}\label{DG-det}
\Delta\Gamma_s=\left\{
\begin{array}{ll}
(0.17\pm0.09\pm0.02)\mbox{ps}^{-1}  & \mbox{(D0 \cite{D0-DG})}\\
(0.076^{+0.059}_{-0.063}\pm0.006)\mbox{ps}^{-1} & \mbox{(CDF \cite{CDF-DG})}.
\end{array}
\right.
\end{equation}
It will be interesting to follow the evolution of these data. At LHCb, we expect a 
precision of $\sigma(\Delta\Gamma_s)=0.027\mbox{ps}^{-1}$ already with
$0.5\,\mbox{fb}^{-1}$ data, which is expected to be available by the end of 2009
\cite{nakada}; ATLAS expects a relative accuracy of $13\%$ with 
$30\,\mbox{fb}^{-1}$ of data taken at low luminosity \cite{smsp}.

\subsection{Time-Dependent Decay Rates}
The time evolution of initially, i.e.\ at $t=0$, pure $B^0_q$- and 
$\bar B^0_q$-meson states is given by
\begin{equation}
|B^0_q(t)\rangle=f_+^{(q)}(t)|B^{0}_q\rangle
+\alpha_qf_-^{(q)}(t)|\bar B^{0}_q\rangle
\end{equation}
and
\begin{equation}
|\bar B^0_q(t)\rangle=\frac{1}{\alpha_q}f_-^{(q)}(t)
|B^{0}_q\rangle+f_+^{(q)}(t)|\bar B^{0}_q\rangle,
\end{equation}
respectively, with
\begin{equation}\label{f-functions}
f_{\pm}^{(q)}(t)=\frac{1}{2}\left[e^{-i\lambda_+^{(q)}t}\pm
e^{-i\lambda_-^{(q)}t}\right].
\end{equation}
These time-dependent state vectors allow the calculation of the 
corresponding transition rates. To this end, it is useful to introduce
\begin{equation}\label{g-funct-1}
|g^{(q)}_{\pm}(t)|^2=\frac{1}{4}\left[e^{-\Gamma_{\rm L}^{(q)}t}+
e^{-\Gamma_{\rm H}^{(q)}t}\pm2\,e^{-\Gamma_q t}\cos(\Delta M_qt)\right]
\end{equation}
\begin{equation}\label{g-funct-2}
g_-^{(q)}(t)\,g_+^{(q)}(t)^\ast=\frac{1}{4}\left[e^{-\Gamma_{\rm L}^{(q)}t}-
e^{-\Gamma_{\rm H}^{(q)}t}+2\,i\,e^{-\Gamma_q t}\sin(\Delta M_qt)\right],
\end{equation}
as well as
\begin{equation}\label{xi-def}
\xi_f^{(q)}=e^{-i\Theta_{M_{12}}^{(q)}}
\frac{A(\bar B_q^0\to f)}{A(B_q^0\to f)},\quad
\xi_{\bar f}^{(q)}=e^{-i\Theta_{M_{12}}^{(q)}}
\frac{A(\bar B_q^0\to \bar f)}{A(B_q^0\to \bar f)}.
\end{equation}
Looking at (\ref{M12-calc}), we find
\begin{equation}\label{theta-def}
\Theta_{M_{12}}^{(q)}=\pi+2\mbox{arg}(V_{tq}^\ast V_{tb})-
\phi_{\mbox{{\scriptsize CP}}}(B_q),
\end{equation}
and observe that this phase depends on the chosen CKM and 
CP phase conventions specified in (\ref{CKM-trafo}) and (\ref{CP-phase-def}), 
respectively. However, these dependences are cancelled through the 
amplitude ratios in (\ref{xi-def}), so that $\xi_f^{(q)}$ and 
$\xi_{\bar f}^{(q)}$ are {\it convention-independent} observables. 
Whereas $n'$ enters the functions in (\ref{f-functions}) through 
(\ref{lam-pm}), the dependence on this parameter is cancelled in 
(\ref{g-funct-1}) and (\ref{g-funct-2}) through the introduction of 
the {\it positive} mass difference $\Delta M_q$ (see (\ref{DeltaMq-def})). 
Combining the formulae listed above, we eventually arrive at the 
following transition rates for decays of initially, i.e.\ at $t=0$, 
present $B^0_q$ or $\bar B^0_q$ mesons:
\begin{equation}\label{rates}
\Gamma(\stackrel{{\mbox{\tiny (---)}}}{B^0_q}(t)\to f)
=\left[|g_\mp^{(q)}(t)|^2+|\xi_f^{(q)}|^2|g_\pm^{(q)}(t)|^2-
2\mbox{\,Re}\left\{\xi_f^{(q)}
g_\pm^{(q)}(t)g_\mp^{(q)}(t)^\ast\right\}
\right]\tilde\Gamma_f,
\end{equation}
where the time-independent rate $\tilde\Gamma_f$ corresponds to the 
``unevolved'' decay amplitude $A(B^0_q\to f)$, and can be calculated by 
performing the usual phase-space integrations. The rates into the 
CP-conjugate final state $\bar f$ can straightforwardly be obtained from 
(\ref{rates}) by making the substitutions
\begin{equation}
\tilde\Gamma_f  \,\,\,\to\,\,\, 
\tilde\Gamma_{\bar f},
\quad\,\,\xi_f^{(q)} \,\,\,\to\,\,\, 
\xi_{\bar f}^{(q)}.
\end{equation}

\subsection{``Untagged'' Rates}
The expected sizeable width difference $\Delta\Gamma_s$ may provide interesting studies of CP 
violation through ``untagged'' $B_s$ rates 
(see Refs.~\cite{DDF} and \cite{dun}--\cite{DFN}), which are defined as 
\begin{equation}
\langle\Gamma(B_s(t)\to f)\rangle
\equiv\Gamma(B^0_s(t)\to f)+\Gamma(\bar B^0_s(t)\to f),
\end{equation}
and are characterized by the feature that we do not distinguish between
initially, i.e.\ at time $t=0$, present $B^0_s$ or $\bar B^0_s$ mesons. 
If we consider a final state $f$ to which both a $B^0_s$ and a $\bar B^0_s$ 
may decay, and use the expressions in (\ref{rates}), we find
\begin{equation}\label{untagged-rate}
\hspace*{-0.7truecm}\langle\Gamma(B_s(t)\to f)\rangle
\propto \left[\cosh(\Delta\Gamma_st/2)-{\cal A}_{\Delta\Gamma}(B_s\to f)
\sinh(\Delta\Gamma_st/2)\right]e^{-\Gamma_s t},
\end{equation}
with
\begin{equation}\label{ADGam}
{\cal A}_{\rm \Delta\Gamma}(B_s\to f)\equiv
\frac{2\,\mbox{Re}\,\xi^{(s)}_f}{1+\bigl|\xi^{(s)}_f
\bigr|^2}.
\end{equation}
We observe that the rapidly oscillating 
$\Delta M_st$ terms cancel, and that we may obtain information about the 
phase structure of the observable $\xi_f^{(s)}$, thereby providing valuable
insights into CP violation. 

Following these lines, for instance,  the untagged observables offered by 
the angular distribution of the $B_s\to K^{*+}K^{*-}, K^{*0}\bar K^{*0}$ decay 
products allow a determination of the UT angle $\gamma$, 
provided $\Delta\Gamma_s$ is 
actually sizeable \cite{FD-CP}. Untagged $B_s$-decay rates are interesting in 
terms of efficiency, acceptance and purity, and are already applied for the physics
analyses at the Tevatron \cite{D0-DG,CDF-DG}. Soon they will help us to fully 
exploit the physics potential of the $B_s$-meson system at the LHC.

\subsection{CP Asymmetries}\label{subsec:CPasym}
A particularly simple -- but also very interesting  -- situation arises 
if we restrict ourselves to decays of neutral $B_q$ mesons 
into final states $f$ that are eigenstates of the CP operator, i.e.\
satisfy the relation 
\begin{equation}\label{CP-eigen}
({\cal CP})|f\rangle=\pm |f\rangle. 
\end{equation}
Consequently, we have $\xi_f^{(q)}=\xi_{\bar f}^{(q)}$ in this case, 
as can be seen in (\ref{xi-def}). Using the decay rates in (\ref{rates}), 
we find that the corresponding time-dependent CP asymmetry is given by
\begin{eqnarray}
{\cal A}_{\rm CP}(t)&\equiv&\frac{\Gamma(B^0_q(t)\to f)-
\Gamma(\bar B^0_q(t)\to f)}{\Gamma(B^0_q(t)\to f)+
\Gamma(\bar B^0_q(t)\to f)}\nonumber\\
&=&\left[\frac{{\cal A}_{\rm CP}^{\rm dir}(B_q\to f)\,\cos(\Delta M_q t)+
{\cal A}_{\rm CP}^{\rm mix}(B_q\to f)\,\sin(\Delta 
M_q t)}{\cosh(\Delta\Gamma_qt/2)-{\cal A}_{\rm 
\Delta\Gamma}(B_q\to f)\,\sinh(\Delta\Gamma_qt/2)}\right],\label{ee6}
\end{eqnarray}
with
\begin{equation}\label{CPV-OBS}
{\cal A}^{\mbox{{\scriptsize dir}}}_{\mbox{{\scriptsize CP}}}(B_q\to f)\equiv
\frac{1-\bigl|\xi_f^{(q)}\bigr|^2}{1+\bigl|\xi_f^{(q)}\bigr|^2},\qquad
{\cal A}^{\mbox{{\scriptsize mix}}}_{\mbox{{\scriptsize
CP}}}(B_q\to f)\equiv\frac{2\,\mbox{Im}\,\xi^{(q)}_f}{1+
\bigl|\xi^{(q)}_f\bigr|^2}.
\end{equation}
Because of the relation
\begin{equation}
{\cal A}^{\mbox{{\scriptsize dir}}}_{\mbox{{\scriptsize CP}}}(B_q\to f)=
\frac{|A(B^0_q\to f)|^2-|A(\bar B^0_q\to \bar f)|^2}{|A(B^0_q\to f)|^2+
|A(\bar B^0_q\to \bar f)|^2},
\end{equation}
this observable measures the direct CP violation in the decay
$B_q\to f$, which originates from the interference between different
weak amplitudes, as we have seen in (\ref{direct-CPV}). On the other
hand, the interesting {\it new} aspect of (\ref{ee6}) is due to 
${\cal A}^{\mbox{{\scriptsize mix}}}_{\mbox{{\scriptsize
CP}}}(B_q\to f)$, which originates from interference effects between 
$B_q^0$--$\bar B_q^0$ mixing and decay processes, and describes
``mixing-induced'' CP violation. Finally, the width difference 
$\Delta\Gamma_q$, which may be sizeable in the $B_s$-meson system, 
provides access to ${\cal A}_{\Delta\Gamma}(B_q\to f)$  introduced 
in (\ref{ADGam}). However, this observable is not independent from 
${\cal A}^{\mbox{{\scriptsize dir}}}_{\mbox{{\scriptsize CP}}}(B_q\to f)$ and 
${\cal A}^{\mbox{{\scriptsize mix}}}_{\mbox{{\scriptsize CP}}}(B_q\to f)$,
satisfying 
\begin{equation}\label{Obs-rel}
\Bigl[{\cal A}_{\rm CP}^{\rm dir}(B_q\to f)\Bigr]^2+
\Bigl[{\cal A}_{\rm CP}^{\rm mix}(B_q\to f)\Bigr]^2+
\Bigl[{\cal A}_{\Delta\Gamma}(B_q\to f)\Bigr]^2=1.
\end{equation}

In order to calculate $\xi_f^{(q)}$, we use the general expressions
(\ref{par-ampl}) and (\ref{par-ampl-CP}), where 
$e^{-i\phi_{\mbox{{\scriptsize CP}}}(f)}=\pm1$ because of (\ref{CP-eigen}), and 
$\phi_{\mbox{{\scriptsize CP}}}(B)=\phi_{\mbox{{\scriptsize CP}}}(B_q)$. If we insert
these amplitude parametrizations into (\ref{xi-def}) and take (\ref{theta-def}) into 
account, we observe that the phase-convention-dependent 
quantity $\phi_{\mbox{{\scriptsize CP}}}(B_q)$ cancels, and finally 
arrive at
\begin{equation}\label{xi-re}
\xi_f^{(q)}=\mp\, e^{-i\phi_q}\left[
\frac{e^{+i\varphi_1}|A_1|e^{i\delta_1}+
e^{+i\varphi_2}|A_2|e^{i\delta_2}}{
e^{-i\varphi_1}|A_1|e^{i\delta_1}+
e^{-i\varphi_2}|A_2|e^{i\delta_2}}\right],
\end{equation}
where
\begin{equation}\label{phiq-def}
\phi_q\equiv 2\,\mbox{arg} (V_{tq}^\ast V_{tb})=\left\{\begin{array}{cl}
+2\beta&\mbox{($q=d$)}\\
-2\delta\gamma&\mbox{($q=s$)}\end{array}\right.
\end{equation}
is associated with the CP-violating weak $B_q^0$--$\bar B_q^0$ mixing
phase arising in the SM; $\beta$ and $\delta\gamma$ refer to the corresponding
angles in the unitarity triangles shown in Fig.\ \ref{fig:UT}.

In analogy to (\ref{direct-CPV}), the caclulation
of $\xi_f^{(q)}$ is -- in general -- also affected by large hadronic uncertainties. 
However, if one CKM amplitude plays the dominant r\^ole in the $B_q\to f$
transition, we obtain
\begin{equation}\label{xi-si}
\xi_f^{(q)}=\mp\, e^{-i\phi_q}\left[
\frac{e^{+i\phi_f/2}|M_f|e^{i\delta_f}}{e^{-i\phi_f/2}|M_f|e^{i\delta_f}}
\right]=\mp\, e^{-i(\phi_q-\phi_f)},
\end{equation}
and observe that the hadronic matrix element $|M_f|e^{i\delta_f}$ 
cancels in this expression. Since the requirements for 
direct CP violation discussed above are then no longer satisfied, direct CP violation 
vanishes in this important special case, i.e.\ 
${\cal A}^{\mbox{{\scriptsize dir}}}_{\mbox{{\scriptsize CP}}}
(B_q\to f)=0$. On the other hand, this is {\it not} the case for the mixing-induced 
CP asymmetry. In particular, 
\begin{equation}\label{Amix-simple}
{\cal A}^{\rm mix}_{\rm CP}(B_q\to f)=\pm\sin\phi
\end{equation}
is now governed by the CP-violating weak phase difference 
$\phi\equiv\phi_q-\phi_f$ and is not affected by hadronic 
uncertainties. The corresponding time-dependent CP asymmetry
takes then the simple form
\begin{equation}\label{Amix-t-simple}
\left.\frac{\Gamma(B^0_q(t)\to f)-
\Gamma(\bar B^0_q(t)\to \bar f)}{\Gamma(B^0_q(t)\to f)+
\Gamma(\bar B^0_q(t)\to \bar f)}\right|_{\Delta\Gamma_q=0}
=\pm\sin\phi\,\sin(\Delta M_q t),
\end{equation}
and allows an elegant determination of $\sin\phi$.

\begin{figure}[t]
\centerline{
 \includegraphics[width=5.7truecm]{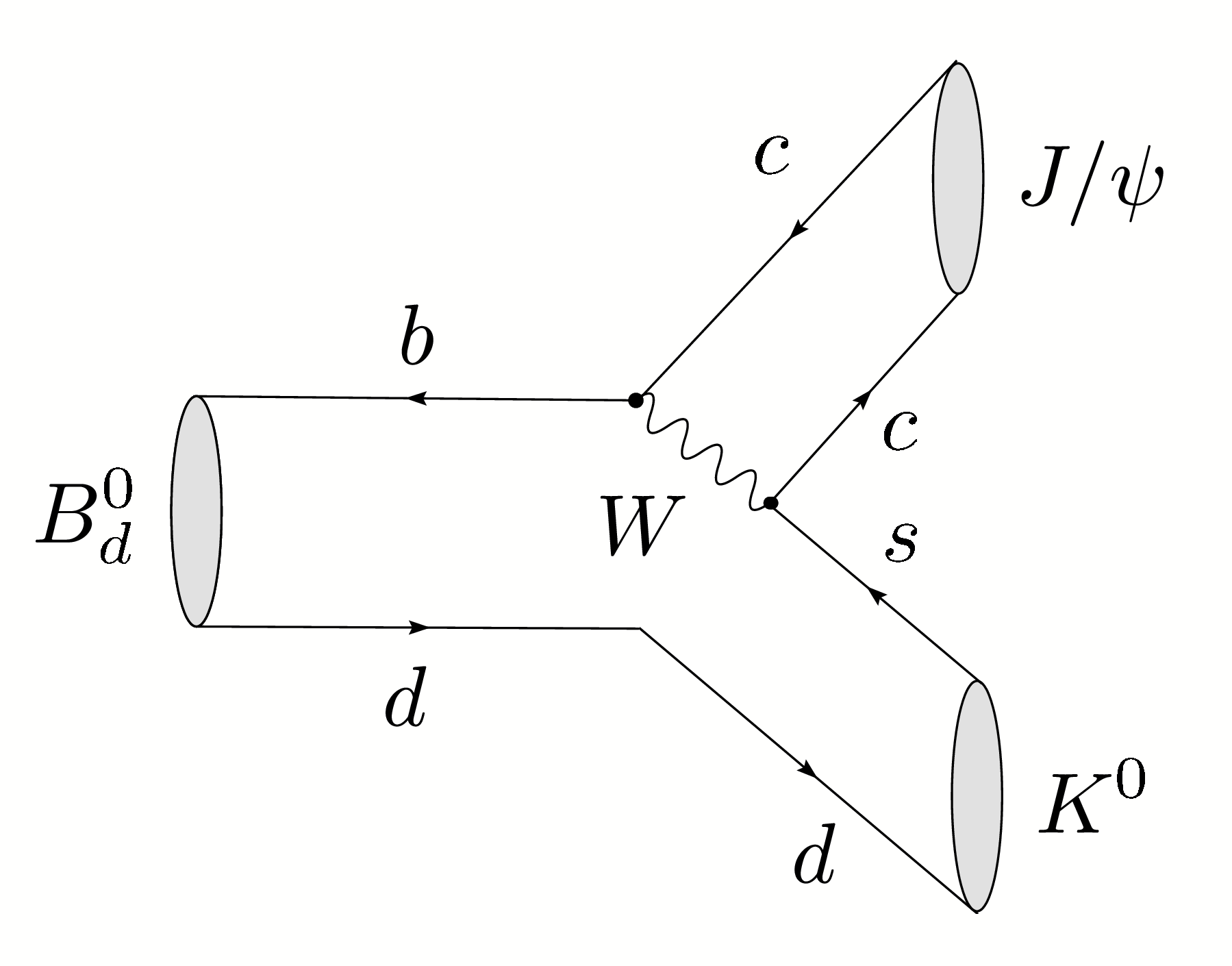}
 \hspace*{0.5truecm}
 \includegraphics[width=5.7truecm]{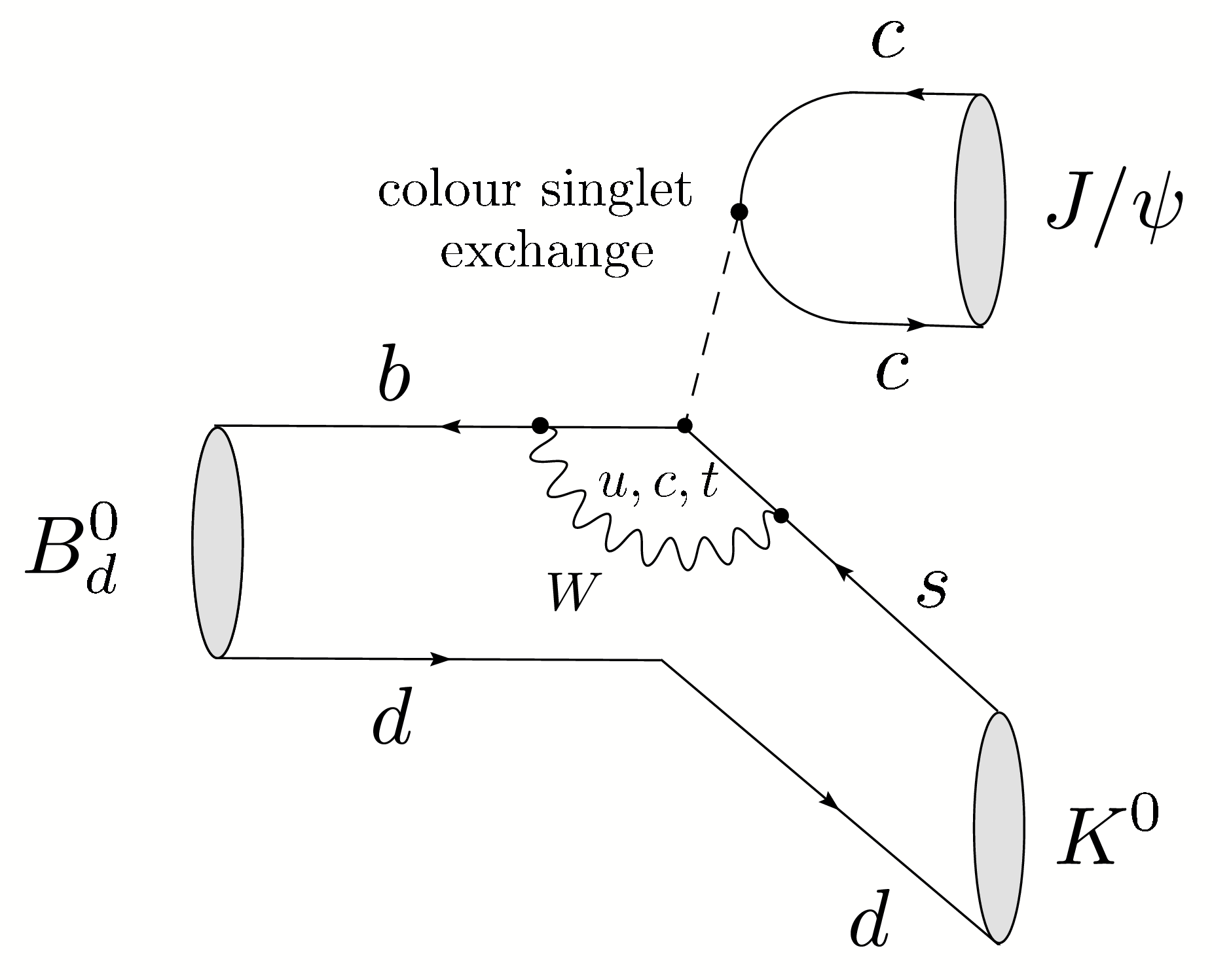}  
 }
 \vspace*{-0.3truecm}
\caption{Feynman diagrams contributing to the $B^0_d\to J/\psi K^0$ 
decay.}\label{fig:BpsiK-diag}
\end{figure}

\boldmath
\subsection{Key Application: $B^0_d\to J/\psi K_{\rm S}$}\label{ssec:BpsiK}
\unboldmath
This decay has a CP-odd final state, and originates from 
$\bar b\to\bar c c \bar s$ quark-level transitions, as can be 
seen in Fig.~\ref{fig:BpsiK-diag}. Consequently, it receives 
contributions both from tree and from penguin topologies 
(see Subsection~\ref{sec:class}). In the SM, the decay 
amplitude can hence be written as follows \cite{RF-BdsPsiK}:
\begin{equation}\label{Bd-ampl1}
A(B_d^0\to J/\psi K_{\rm S})=\lambda_c^{(s)}\left(A_{\rm T}^{c'}+
A_{\rm P}^{c'}\right)+\lambda_u^{(s)}A_{\rm P}^{u'}
+\lambda_t^{(s)}A_{\rm P}^{t'}.
\end{equation}
Here the
\begin{equation}\label{lamqs-def}
\lambda_q^{(s)}\equiv V_{qs}V_{qb}^\ast
\end{equation}
are CKM factors, $A_{\rm T}^{c'}$ is the CP-conserving strong tree amplitude, 
while the $A_{\rm P}^{q'}$ describe the penguin topologies with internal 
$q$ quarks ($q\in\{u,c,t\})$, including QCD and EW penguins; 
the primes remind us that we are dealing with a $\bar b\to\bar s$ 
transition. If we eliminate now $\lambda_t^{(s)}$ through (\ref{CKM-UT-Rel}) 
and apply the Wolfenstein parametrization, we obtain
\begin{equation}\label{BdpsiK-ampl2}
A(B_d^0\to J/\psi K_{\rm S})\propto\left[1+\lambda^2 a e^{i\theta}
e^{i\gamma}\right],
\end{equation}
where
\begin{equation}
a e^{i\vartheta}\equiv\left(\frac{R_b}{1-\lambda^2}\right)
\left[\frac{A_{\rm P}^{u'}-A_{\rm P}^{t'}}{A_{\rm T}^{c'}+
A_{\rm P}^{c'}-A_{\rm P}^{t'}}\right]
\end{equation}
is a hadronic parameter. Using now the formalism discussed in
Subsection~\ref{subsec:CPasym} yields
\begin{equation}\label{xi-BdpsiKS}
\xi_{\psi K_{\rm S}}^{(d)}=+e^{-i\phi_d}\left[\frac{1+
\lambda^2a e^{i\vartheta}e^{-i\gamma}}{1+\lambda^2a e^{i\vartheta}
e^{+i\gamma}}\right].
\end{equation}
Unfortunately, $a e^{i\vartheta}$, which is a measure for the ratio of the
$B_d^0\to J/\psi K_{\rm S}$ penguin to tree contributions,
can only be estimated with large hadronic uncertainties. However, since 
this parameter enters (\ref{xi-BdpsiKS}) in a doubly Cabibbo-suppressed way, its 
impact on the CP-violating observables is practically negligible. We can put 
this important statement on a more quantitative basis by making the plausible
assumption that $a={\cal O}(\bar\lambda)={\cal O}(0.2)={\cal O}(\lambda)$,
where $\bar\lambda$ is a ``generic'' expansion parameter \cite{FM-BpsiK}:
\begin{eqnarray}
{\cal A}^{\mbox{{\scriptsize dir}}}_{\mbox{{\scriptsize
CP}}}(B_d\to J/\psi K_{\mbox{{\scriptsize S}}})&=&0+
{\cal O}(\overline{\lambda}^3)\label{Adir-BdpsiKS}\\
{\cal A}^{\mbox{{\scriptsize mix}}}_{\mbox{{\scriptsize
CP}}}(B_d\to J/\psi K_{\mbox{{\scriptsize S}}})&=&-\sin\phi_d +
{\cal O}(\overline{\lambda}^3) \, \stackrel{\rm SM}{=} \, -\sin2\beta+
{\cal O}(\overline{\lambda}^3).\label{Amix-BdpsiKS}
\end{eqnarray}
Consequently, (\ref{Amix-BdpsiKS}) allows an essentially {\it clean}
determination of $\sin2\beta$ \cite{bisa}.

Since the CKM fits performed within the SM pointed to a large value of 
$\sin2\beta$, $B^0_d\to J/\psi K_{\rm S}$ offered the exciting perspective 
of exhibiting {\it large} mixing-induced CP violation. In 2001, the measurement of  
${\cal A}^{\mbox{{\scriptsize mix}}}_{\mbox{{\scriptsize CP}}}
(B_d\to J/\psi K_{\mbox{{\scriptsize S}}})$
allowed indeed the first observation of CP violation {\it outside} the 
$K$-meson system \cite{CP-B-obs}.
The most recent data are still not showing any signal for {\it direct} CP violation
in $B^0_d\to J/\psi K_{\rm S}$ within the current uncertainties, as is expected from 
(\ref{Adir-BdpsiKS}), and the world average reads \cite{HFAG}
\begin{equation}
{\cal A}_{\rm CP}^{\rm dir}(B_d\to J/\psi K_{\rm S})=0.002 \pm 0.021.
\end{equation}
As far as (\ref{Amix-BdpsiKS}) is concerned, we have
\begin{equation}\label{s2b-psiK-exp}
\hspace*{-2.0truecm}(\sin2\beta)_{\psi K_{\rm S}}\equiv 
-{\cal A}^{\mbox{{\scriptsize mix}}}_{\mbox{{\scriptsize
CP}}}(B_d\to J/\psi K_{\mbox{{\scriptsize S}}})
=\left\{
\begin{array}{ll}
0.714 \pm 0.032 \pm 0.018 & \mbox{(BaBar)}\\
0.651 \pm 0.034 & \mbox{(Belle),}
\end{array}
\right.
\end{equation}
where also other final states similar to $J/\psi K_{\rm S}$ were included  
\cite{HFAG}; the corresponding  world average is then given as follows:
\begin{equation}\label{s2b-average}
(\sin 2\beta)_{\psi K_{\rm S}}= 0.681 \pm 0.025.
\end{equation}
In the SM, the theoretical uncertainties are generically expected to be
below the 0.01 level  \cite{FM-BpsiK}; significantly smaller effects are found in 
Ref.~\cite{BMR}, 
whereas a fit performed in Ref.~\cite{CPS} yields a theoretical penguin uncertainty 
comparable to the present experimental systematic error. A possibility
to control these uncertainties is provided by $B^0_s\to J/\psi K_{\rm S}$  
\cite{RF-BdsPsiK}, which can be explored at the LHC \cite{LHC-Book}.

The average in (\ref{s2b-average}) yields a twofold solution for the phase
$(2\beta)_{\psi K_{\rm S}}$ itself:
\begin{equation}\label{phid-det}
(2\beta)_{\psi K_{\rm S}}=(43\pm2)^\circ \quad \lor \quad (137\pm2)^\circ .
\end{equation}
Here the latter solution would be in dramatic conflict with the CKM fits, and
would require a large NP contribution to $B^0_d$--$\bar B^0_d$ 
mixing \cite{FlMa,FIM}.  However, experimental information on the sign 
of $\cos2\beta$ rules out a negative value of this quantity with more than
95\% C.L.\ \cite{WG-sum}, so that we are left with the solution around $43^\circ$.

\begin{figure}
   \centering
   \includegraphics[width=9.0truecm]{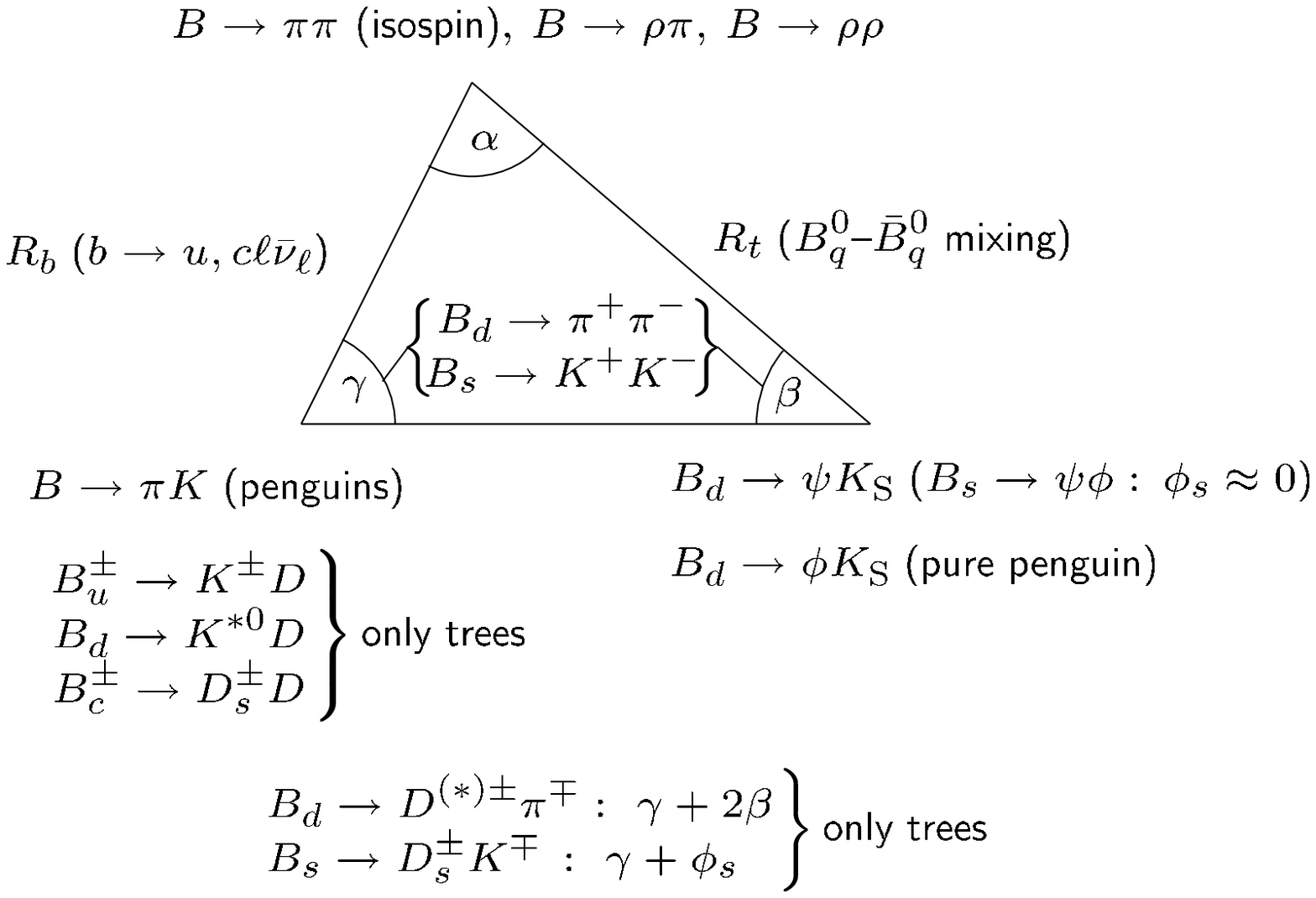} 
   \caption{A brief roadmap of $B$-decay strategies for the exploration of
   CP violation.}
   \label{fig:flavour-map}
\end{figure}

\section{IMPACT OF NEW PHYSICS}\label{sec:NP}
\setcounter{equation}{0}
\subsection{General Remarks}
Using the concept of the low-energy effective Hamiltonians introduced
in Subsection~\ref{subsec:ham}, we may address this important question
in a systematic way  \cite{buras-NP}:
\begin{itemize}
\item NP may modify the ``strength" of the SM operators through new
short-distance functions which depend on the NP parameters, such as the masses 
of charginos, squarks, charged Higgs particles and $\tan \beta\equiv v_2/v_1$ 
in the ``minimal supersymmetric SM'' (MSSM). The NP particles may enter in 
box and penguin topologies, and are ``integrated out'' as the $W$ boson and 
top quark in the SM. Consequently, the initial conditions for the
renormalization-group evolution take the following form:
\begin{equation}\label{WC-NP}
C_k \to C_k^{\rm SM} + C_k^{\rm NP}.
\end{equation}
It should be emphasized that the NP pieces $C_k^{\rm NP}$ may also involve 
new CP-violating phases which are {\it not} related to the CKM matrix.
\item NP may enhance the operator basis:
\begin{equation}
\{Q_k\} \to \{Q_k^{\rm SM}, Q_l^{\rm NP}\},
\end{equation}
so that operators which are not present (or strongly suppressed) in the 
SM may actually play an important r\^ole. In this case, we encounter, 
in general, also new sources for flavour and CP violation.
\end{itemize}

Concerning model-dependent NP analyses, in particular SUSY
scenarios have received a lot of attention. Examples of other fashionable 
NP frameworks are left--right-symmetric models, scenarios with extra dimensions, 
models with an extra $Z'$, ``little Higgs'' scenarios, and models with a 
fourth generation. For a recent overview and a guide to the literature, 
we refer to reader to Ref.~\cite{WG2-rep}.

The simplest extension of the SM is given by models with ``minimal flavour violation'' (MFV) \cite{MFV-1,MFV-2}. Simply speaking, in this class of
models, there are no new sources of flavour and CP violation, i.e.\ these
phenomena are still governed by the CKM matrix. Due to their simplicity, 
the extensions of the SM with MFV show several correlations between 
various observables, thereby allowing for powerful tests of this scenario. 
A comprehensive recent discussion can be found in Ref.~\cite{WG2-rep}. 

The $B$-meson system offers a variety of processes and strategies for the
exploration of CP violation \cite{CKM-book,RF-Phys-Rep}, as we have illustrated in 
Fig.~\ref{fig:flavour-map} through a collection of prominent examples. 
We see that there are processes with a very {\it different} dynamics that 
are -- in the SM -- sensitive to the {\it same} angles of the UT. 
Moreover, rare $B$- and $K$-meson decays \cite{rare}, 
which originate from loop effects in the SM, provide complementary insights 
into flavour physics and interesting correlations with the CP-B sector; key 
examples are $B\to X_s\gamma$ and the exclusive modes
$B\to K^\ast \gamma$, $B\to\rho\gamma$, as well as $B_{s,d}\to \mu^+\mu^-$ 
and $K^+\to\pi^+\nu\bar\nu$, $K_{\rm L}\to\pi^0\nu\bar\nu$. In the presence
of NP in the TeV regime, discrepancies with respect to the SM picture should emerge
at some level of accuracy. There are two promising avenues for NP to
enter $B$-physics obserevables, as we will discuss in the remainder of this
section.

\boldmath
\subsection{New Physics in $B$-Decay Amplitudes}
\unboldmath
NP has typically a small effect if SM tree processes
play the dominant r\^ole. However, NP could well have a significant impact on 
the FCNC sector: new particles may enter in penguin or box diagrams, or new 
FCNC contributions may even be generated at the tree level. In fact, sizeable 
contributions arise generically in field-theoretical estimates with 
$\Lambda_{\rm NP}\sim\mbox{TeV}$ \cite{FM-BphiK}, as well as in specific 
NP models. In Section~\ref{sec:puzzle}, we will  have a closer look
at $B$ decays that may actually indicate NP effects at the decay-amplitude level.

In the case of the ``golden" decay $B^0_d\to J/\psi K_{\rm S}$, NP effects 
have to compete with a tree contribution and are hence not expected to play 
a significant r\^ole. This is indeed signalled by a set of observables that
were introduced in Ref.~\cite{FM-BpsiK} to search for NP contributions 
to the $B\to J/\psi K $ decay amplitudes \cite{RF-JPHYSG}. In the 
following discussion, we will therefore assume that these NP effects 
are negligible.

\begin{figure}
$$\epsfxsize=0.38\textwidth\epsffile{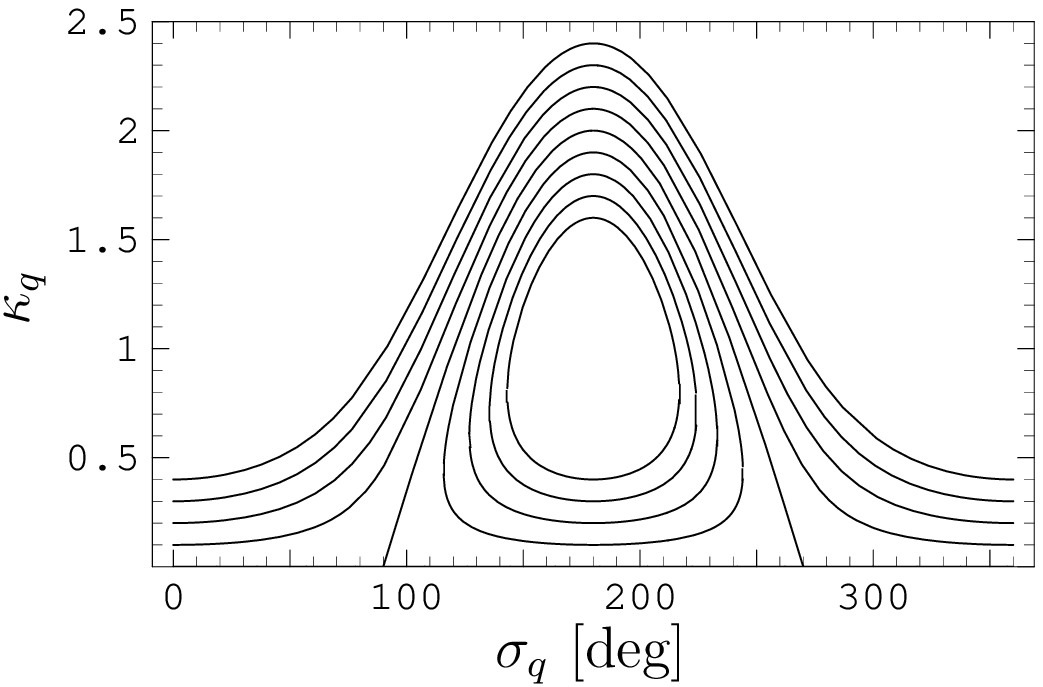}\quad
\epsfxsize=0.38\textwidth\epsffile{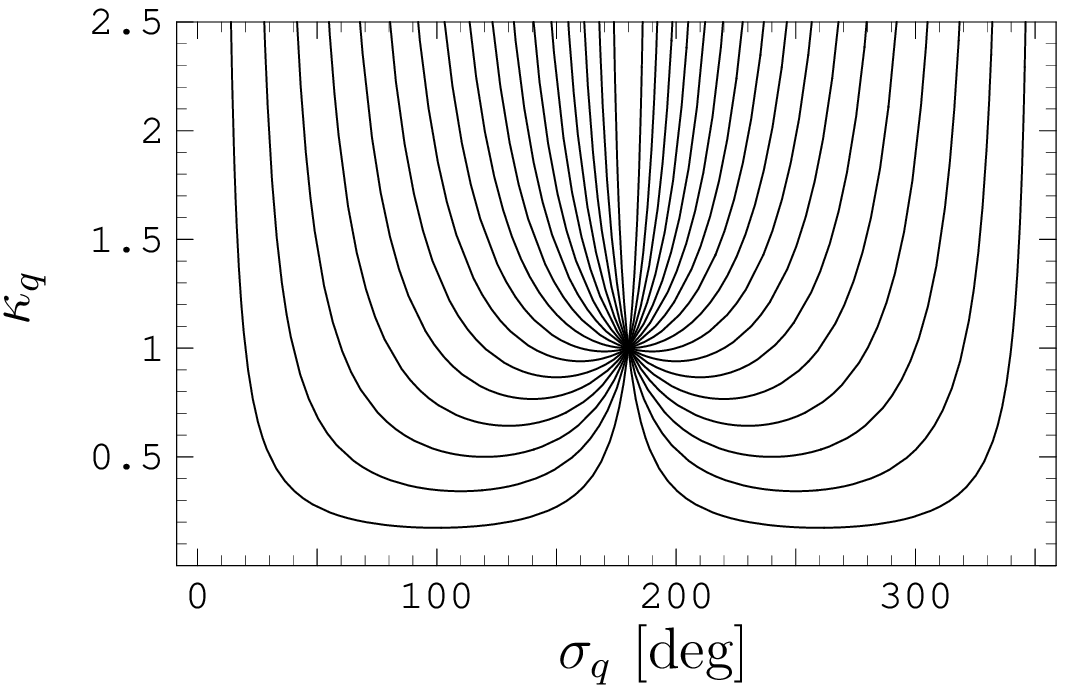}
$$
\vspace*{-1truecm}
\caption[]{Left panel: dependence of $\kappa_q$ on $\sigma_q$ for values of 
$\rho_q$ varied in steps of 0.1 between 1.4 (upper) and 0.6 (inner curve). Right 
panel: dependence of $\kappa_q$ on $\sigma_q$ for values of 
$\phi_q^{\rm NP}$ varied  in steps of $10^\circ$ between $\pm10^\circ$ 
(lower curves) and $\pm170^\circ$; the curves for $0^\circ<\sigma_q<180^\circ$
and $180^\circ<\sigma_q<360^\circ$ correspond to positive and negative values
of $\phi_q^{\rm NP}$, respectively.}\label{fig:kappa-rho}
\end{figure}

\begin{figure}
$$\epsfxsize=0.38\textwidth\epsffile{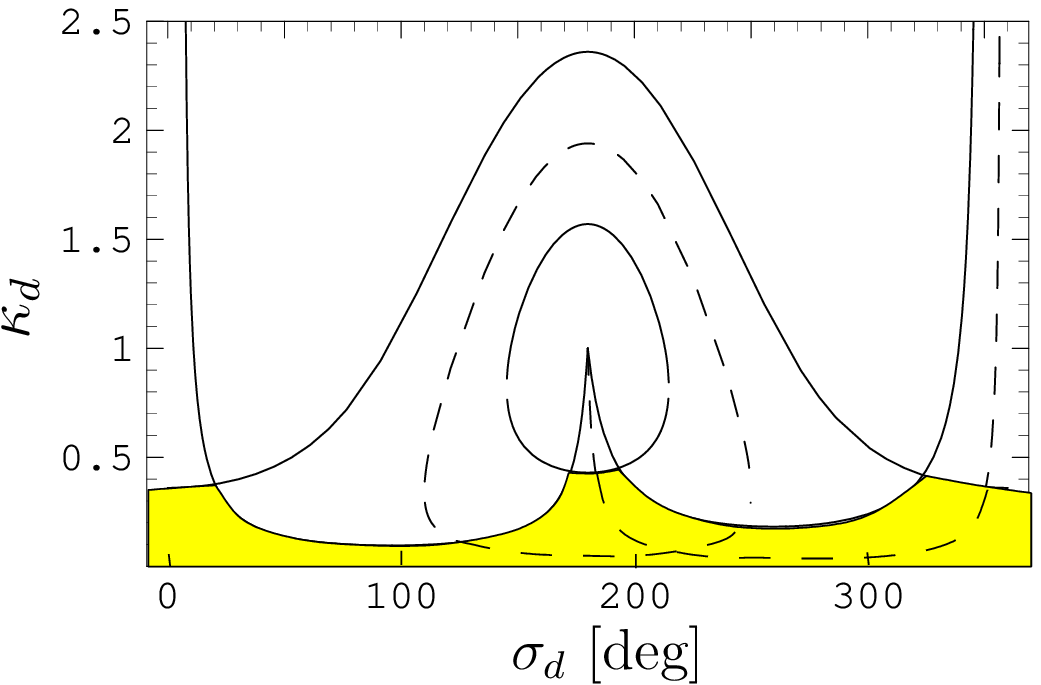}\quad
\epsfxsize=0.38\textwidth\epsffile{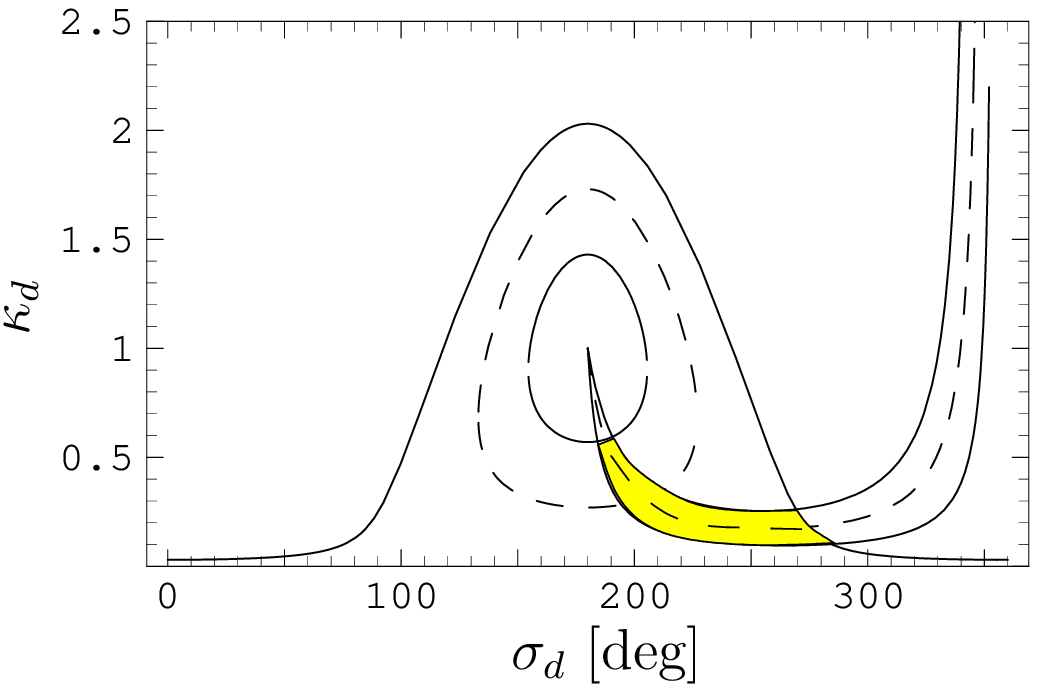}
$$
 \vspace*{-1truecm}
\caption[]{Left panel: allowed region (yellow/grey) in the $\sigma_d$--$\kappa_d$
  plane in a scenario with the JLQCD lattice results and the
  ``exclusive" value of $R_b$ in (\ref{Rb}). Right panel: ditto for the 
 scenario with the (HP+JL)QCD   lattice results
  and  the ``inclusive" value of $R_b$ in (\ref{Rb}), as discussed in the text.
  }\label{fig:res-k-sig-d}
\end{figure}

\boldmath
\subsection{New Physics in $B^0_q$--$\bar B^0_q$ Mixing}\label{ssec:NP-mix}
\unboldmath
Another attractive mechanism for NP to manifest itself in the $B$-physics landscape
is offered by  $B^0_q$--$\bar B^0_q$ mixing. Here NP could enter through the 
exchange of new particles in box diagrams, or through new contributions at the
tree level. In general, we may write
\begin{equation}
M_{12}^{(q)} =M_{12}^{q,{\rm SM}} \left(1 + \kappa_q e^{i\sigma_q}\right),
\end{equation}
where the expression for $M_{12}^{q,{\rm SM}}$ can be found 
in (\ref{M12-calc}). Consequently, we obtain
\begin{eqnarray}
\Delta M_q & = &\Delta M_q^{\rm SM}+\Delta M_q^{\rm NP} =
\Delta M_q^{\rm SM}\left| 1 + \kappa_q
  e^{i\sigma_q}\right|,\label{DMq-NP}\\
\phi_q & = & \phi_q^{\rm SM}+\phi_q^{\rm NP}=
\phi_q^{\rm SM} + \arg (1+\kappa_q e^{i\sigma_q}),\label{phiq-NP}
\end{eqnarray}
with $\Delta M_q^{\rm SM}$ and $\phi_q^{\rm SM}$ given in (\ref{DeltaMq-def}) and
(\ref{phiq-def}), respectively. 
Using dimensional arguments borrowed from effective field 
theory \cite{FM-BpsiK,FIM}, it can be shown that 
$\Delta M_q^{\rm NP}/\Delta M_q^{\rm SM}\sim1$ and
$\phi_q^{\rm NP}/\phi_q^{\rm SM}\sim1$ could -- in principle -- be possible
for a NP scale $\Lambda_{\rm NP}$ in the TeV regime; such a pattern may 
also arise in specific NP scenarios. 

Introducing 
\begin{equation}\label{rhoq-def}
\rho_q\equiv
\left|\frac{\Delta M_q}{\Delta M_q^{\rm SM}}\right|=
\sqrt{1+2\kappa_q\cos\sigma_q+\kappa_q^2}\,,
\end{equation}
the measured values of the mass differences $\Delta M_q$ can be converted
into constraints in NP parameter space through the contours shown in left
panel of Fig.~\ref{fig:kappa-rho}. Further constraints are implied by the NP
phases $\phi_q^{\rm NP}$, which can be probed by means of mixing-induced
CP asymmetries, through the curves in the right panel of 
Fig.~\ref{fig:kappa-rho}. Interestingly, 
$\kappa_q$ is bounded from below for any value of $\phi_q^{\rm NP}\not=0$. 
For example, even a small phase $|\phi_q^{\rm NP}|=10^\circ$ implies a 
clean lower bound of $\kappa_q\geq0.17$, i.e.\ NP contributions of at most 
17\% \cite{BF-DMs}.

\begin{figure}
\centerline{
 \includegraphics[width=5.5truecm]{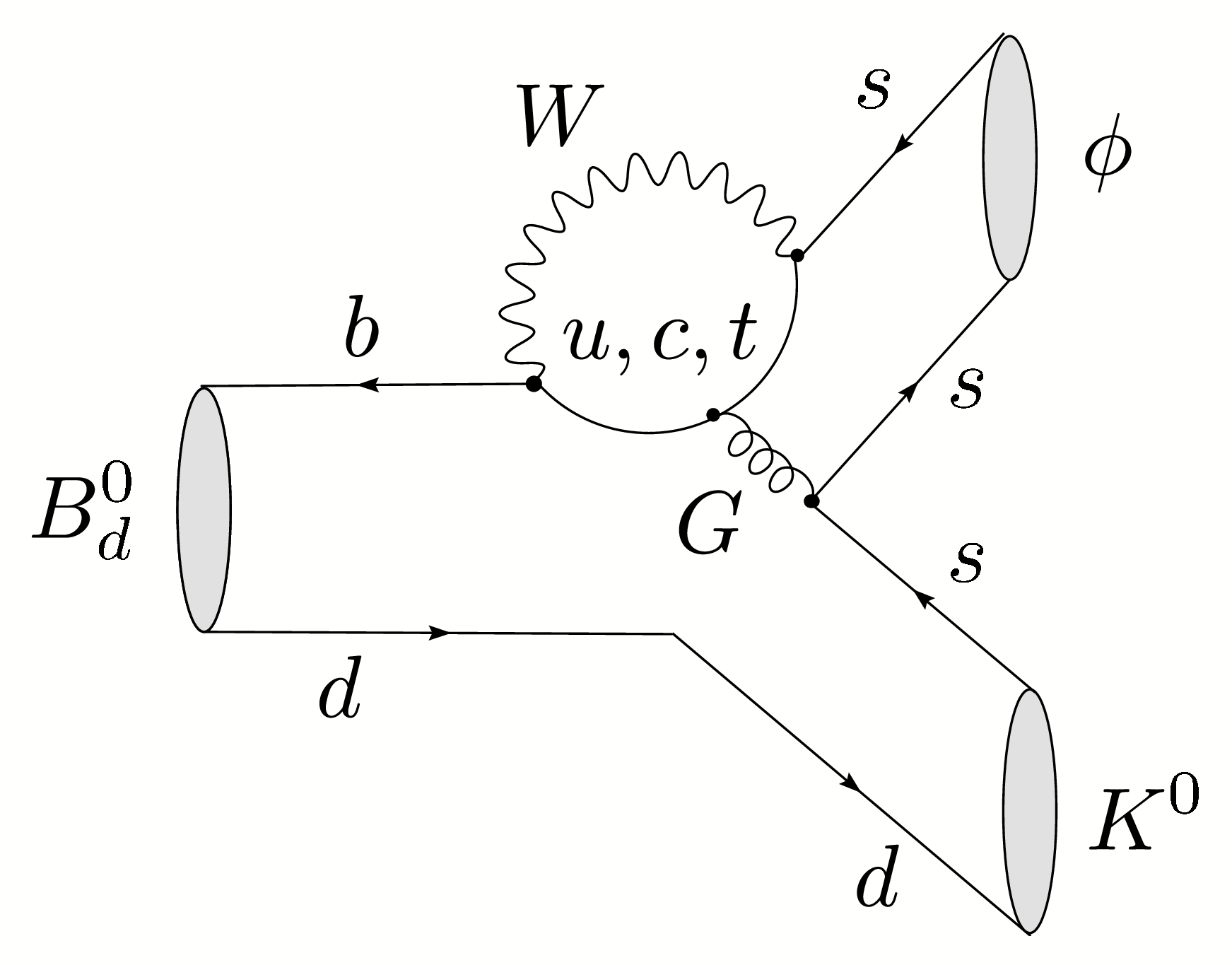} 
 }
 \vspace*{-0.3truecm}
\caption{Feynman diagrams contributing to the $B^0_d\to \phi K^0$ 
decay.}\label{fig:BphiK-diag}
\end{figure}

Consequently, using the $B$-factory data to measure $\Delta M_d$ and 
to extract the NP phase $\phi_d^{\rm NP}$, two sets of contours can be 
fixed in the $\sigma_d$--$\kappa_d$ plane. In the former case, the SM value 
$\Delta M_d^{\rm SM}$ is required. It involves the CKM parameter 
$|V_{td}^*V_{tb}|$, which is governed by $\gamma$ in the corresponding
numerical analysis if the unitarity of the CKM matrix is used. Moreover, information 
about the hadronic parameter $f_{B_d}\hat B_{B_d}^{1/2}$ which we
encountered in Subsection~\ref{ssec:BBbar-mix} is needed. For the 
purpose of comparison, we use two benchmark sets of such results for these 
quantities \cite{BF-DMs}: the JLQCD results for two flavours of dynamical light 
Wilson quarks \cite{JLQCD}, and a combination of $f_{B_d}$  as determined
by the HPQCD collaboration \cite{HPQCD} for three dynamical flavours with the 
JLQCD result for $\hat B_{B_d}$  [(HP+JL)QCD] \cite{Okamoto}.
For the determination of the NP phase $\phi_d^{\rm NP}=\phi_d-\phi_d^{\rm SM}$, 
we use $\phi_d=(43\pm2)^\circ$ (see (\ref{phid-det})),  and fix the ``true" value 
of $\phi_d^{\rm SM}=2\beta$ with the help of the data for tree processes.
This can simply be done through the relations 
\begin{equation}
\sin\beta=\frac{R_b\sin\gamma}{\sqrt{1-2R_b\cos\gamma+R_b^2}}\,, \quad
\cos\beta=\frac{1-R_b\cos\gamma}{\sqrt{1-2R_b\cos\gamma+R_b^2}}
\end{equation}
between the side $R_b\propto |V_{ub}/V_{cb}|$ of the UT and its angle
$\gamma$, which are determined through semileptonic $b\to u \ell \bar\nu_\ell$ 
decays and $B\to D K$ modes, respectively. A numerical analysis shows that the 
value of $\phi_d^{\rm NP}$ is actually governed by $R_b\propto |V_{ub}/V_{cb}|$, 
while $\gamma|_{{D^{(*)}K^{(*)}}}$, which suffers currently from large uncertainties 
as we saw in (\ref{gam-DK}), plays only a minor r\^ole, in contrast to the 
SM analysis of $\Delta M_d$ \cite{BF-DMs}. However, the values of $R_b$ in (\ref{Rb})
following from the analyses of inclusive and exclusive decays differ at the
$1\,\sigma$ level. We show the resulting situation in the $\sigma_d$--$\kappa_d$ 
plane in Fig.~\ref{fig:res-k-sig-d}, and observe that the measurement of CP violation in 
$B^0_d\to J/\psi K_{\rm S}$ and similar decays has a dramatic impact on the
allowed region in the NP parameter space; the right panel may indicate the presence 
of NP, although no definite conclusions can be drawn at the moment. It will
be interesting to monitor this effect in the future. In order to detect such CP-violating
NP contributions, which would immediately rule out MFV scenarios, things 
are much more promising in the $B_s$ system, as we will see in 
Subsection~\ref{ssec:Bspsiphi}.

\begin{figure}
$$\epsfxsize=0.48\textwidth\epsffile{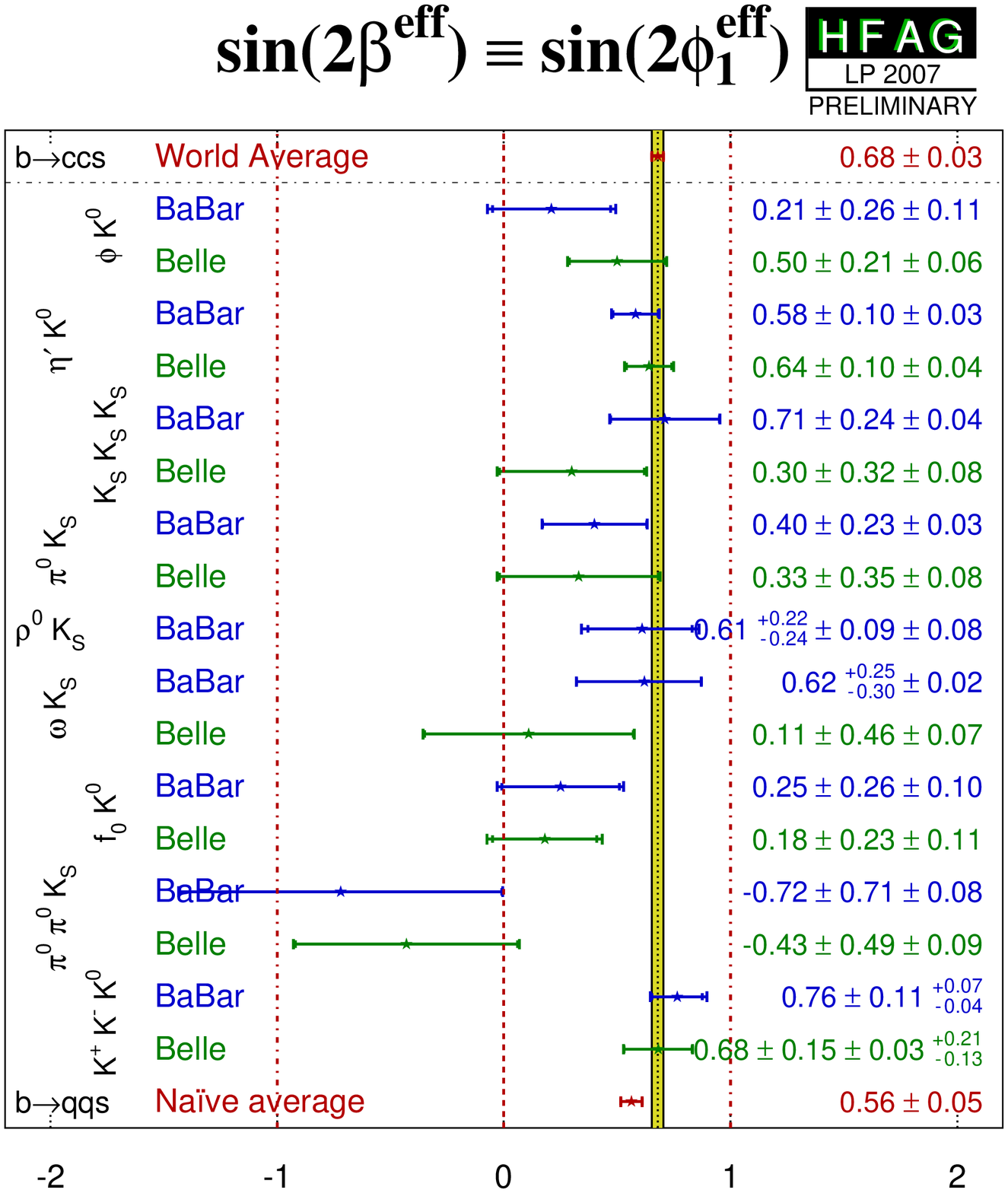}\quad
\epsfxsize=0.48\textwidth\epsffile{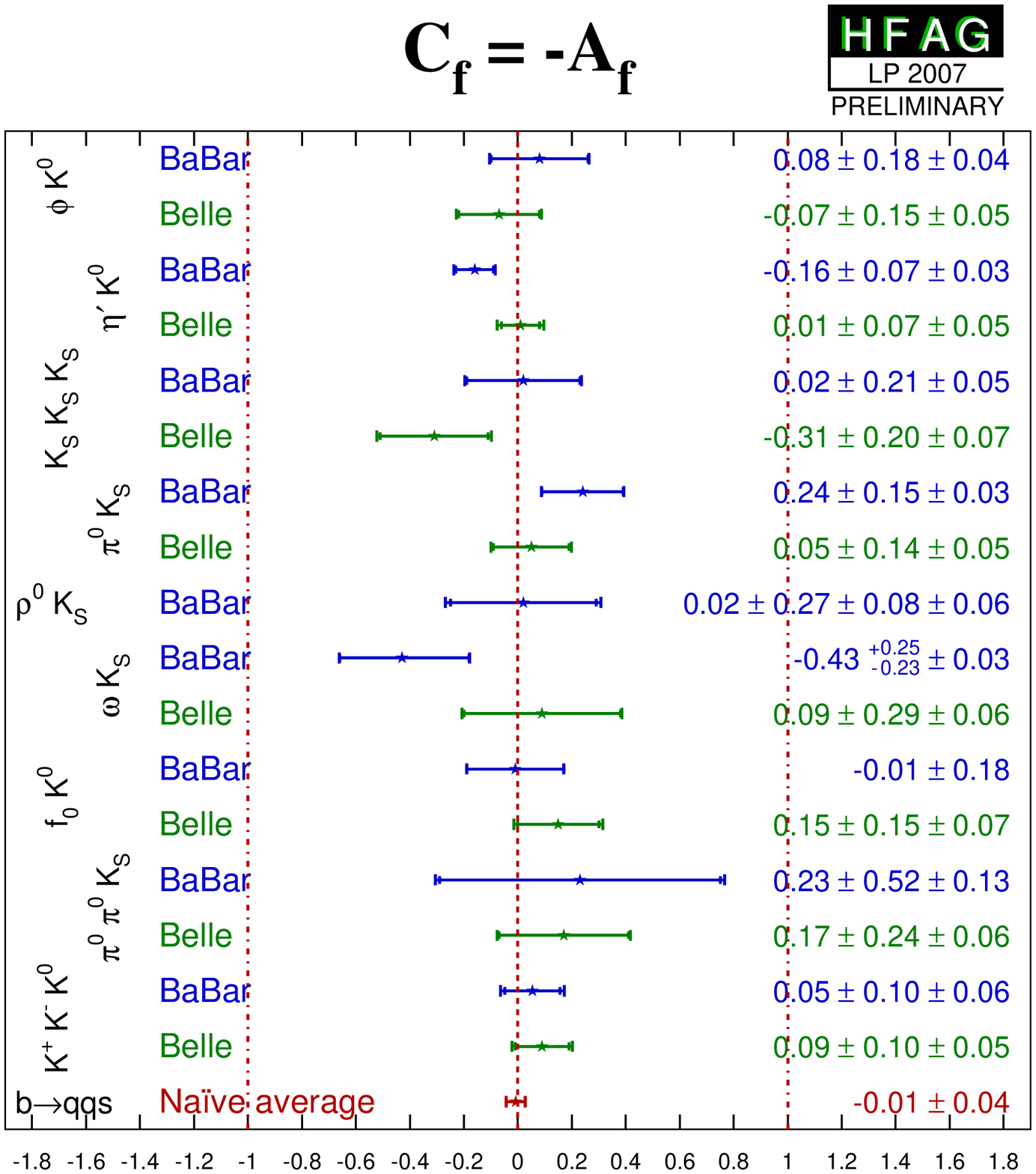}
$$
 \vspace*{-1truecm}
\caption[]{Compilation of the CP violation measurements in $B$ decays that are
dominated by $b\to s$ penguin processes \cite{HFAG}: mixing-induced 
CP asymmetries (left panel), and direct CP asymmetries, where
$C={\cal A}_{\rm CP}^{\rm dir}$ (right panel).}\label{fig:b-s}
\end{figure}

\section{PUZZLING PATTERNS IN THE  {\boldmath$B$\unboldmath}-FACTORY 
DATA}\label{sec:puzzle}
\setcounter{equation}{0}
\boldmath
\subsection{CP Violation in $b\to s$ Penguin Modes}\label{ssec:BphiK}
\unboldmath
A particularly interesting probe of NP is offered by the decay $B^0_d\to \phi K_{\rm S}$. 
As can be seen in Fig.~\ref{fig:BphiK-diag}, it originates from 
$\bar b\to \bar s s \bar s$ transitions and is, therefore, a 
pure penguin mode. This decay is described by the low-energy effective 
Hamiltonian in (\ref{e4}) with $r=s$, where the current--current operators 
may only contribute through penguin-like contractions, which describe 
penguin topologies with internal up- and charm-quark exchanges
\cite{RF-DIPL,RF-EWP}. The dominant
r\^ole is played by the QCD penguin operators \cite{BphiK-old}. However,
thanks to the large top-quark mass, EW penguins have a sizeable impact as 
well \cite{RF-EWP,DH-PhiK}. In the SM, we may write
\begin{equation}\label{B0phiK0-ampl}
A(B_d^0\to \phi K_{\rm S})=\lambda_u^{(s)}\tilde A_{\rm P}^{u'}
+\lambda_c^{(s)}\tilde A_{\rm P}^{c'}+\lambda_t^{(s)}\tilde A_{\rm P}^{t'},
\end{equation}
where we have applied the same notation as in Subsection~\ref{ssec:BpsiK}.
Eliminating the CKM factor $\lambda_t^{(s)}$ with the help of
(\ref{CKM-UT-Rel}) yields
\begin{equation}
A(B_d^0\to \phi K_{\rm S})\propto
\left[1+\lambda^2 b e^{i\Theta}e^{i\gamma}\right],
\end{equation}
where 
\begin{equation}
b e^{i\Theta}\equiv\left(\frac{R_b}{1-\lambda^2}\right)\left[
\frac{\tilde A_{\rm P}^{u'}-\tilde A_{\rm P}^{t'}}{\tilde A_{\rm P}^{c'}-
\tilde A_{\rm P}^{t'}}\right].
\end{equation}
Consequently,  we obtain
\begin{equation}\label{xi-phiKS}
\xi_{\phi K_{\rm S}}^{(d)}=+e^{-i\phi_d}
\left[\frac{1+\lambda^2b e^{i\Theta}e^{-i\gamma}}{1+
\lambda^2b e^{i\Theta}e^{+i\gamma}}\right].
\end{equation}
The theoretical estimates of $b e^{i\Theta}$ 
suffer from large hadronic uncertainties. However, since this parameter enters 
(\ref{xi-phiKS}) in a doubly Cabibbo-suppressed way, we obtain the 
following expressions \cite{RF-EWP-rev}:
\begin{eqnarray}
{\cal A}_{\rm CP}^{\rm dir}(B_d\to \phi K_{\rm S})&=&0+
{\cal O}(\lambda^2)\label{BphiK-rel1}\\
{\cal A}_{\rm CP}^{\rm mix}(B_d\to \phi K_{\rm S})&=&-\sin\phi_d
+{\cal O}(\lambda^2),\label{BphiK-rel2}
\end{eqnarray}
where we made the plausible assumption that $b={\cal O}(1)$. On the other 
hand, the mixing-induced CP asymmetry of 
$B_d\to J/\psi K_{\rm S}$ measures also $-\sin\phi_d$, as we saw in
(\ref{Amix-BdpsiKS}). We arrive therefore at the following 
relation \cite{RF-EWP-rev,growo}:
\begin{equation}\label{Bd-phiKS-SM-rel}
-(\sin2\beta)_{\phi K_{\rm S}}\equiv
{\cal A}_{\rm CP}^{\rm mix}(B_d\to \phi K_{\rm S}) 
={\cal A}_{\rm CP}^{\rm mix}(B_d\to J/\psi K_{\rm S}) + 
{\cal O}(\lambda^2),
\end{equation}
which offers an interesting test of the SM. Since $B_d\to \phi K_{\rm S}$ is 
governed by penguin processes in the SM, this decay may well be affected by 
NP. In fact, if we assume that NP arises generically in the TeV regime, it can be 
shown through field-theoretical estimates that the NP contributions to  
$b\to s\bar s s$ transitions may well lead to sizeable violations of
(\ref{BphiK-rel1}) and (\ref{Bd-phiKS-SM-rel}) \cite{RF-Phys-Rep,FM-BphiK}. Moreover, 
this is also the case for several specific NP scenarios.

The experimental status can be summarized through the 
following averages \cite{HFAG}:
\begin{equation}
(\sin2\beta)_{\phi K_{\rm S}}=0.39\pm0.17, \quad
{\cal A}_{\rm CP}^{\rm dir}(B_d\to \phi K_{\rm S})=-0.01\pm0.12.
\end{equation}
During the recent years, the Belle results for $(\sin2\beta)_{\phi K_{\rm S}}$ 
\cite{Belle-BphiK} have 
moved quite a bit towards the SM reference value of $(\sin2\beta)_{\psi K_{\rm S}}$
given in (\ref{s2b-average}), and are now, within the errors, in agreement with 
the BaBar findings \cite{BaBar-BphiK}. Interestingly, 
the mixing-induced CP asymmetries of other $b\to s$ penguin modes show the 
same trend of having central values that are smaller than 0.681, as can be
seen in Fig.~\ref{fig:b-s} \cite{HFAG}. This
feature may in fact be due to the presence of NP contributions to the corresponding
decay amplitudes. However, the large uncertainties do not yet allow us to draw
definite conclusions.

\begin{figure}
   \centerline{
   \begin{tabular}{lc}
     {\small(a)} & \\
    &  \includegraphics[width=5.2truecm]{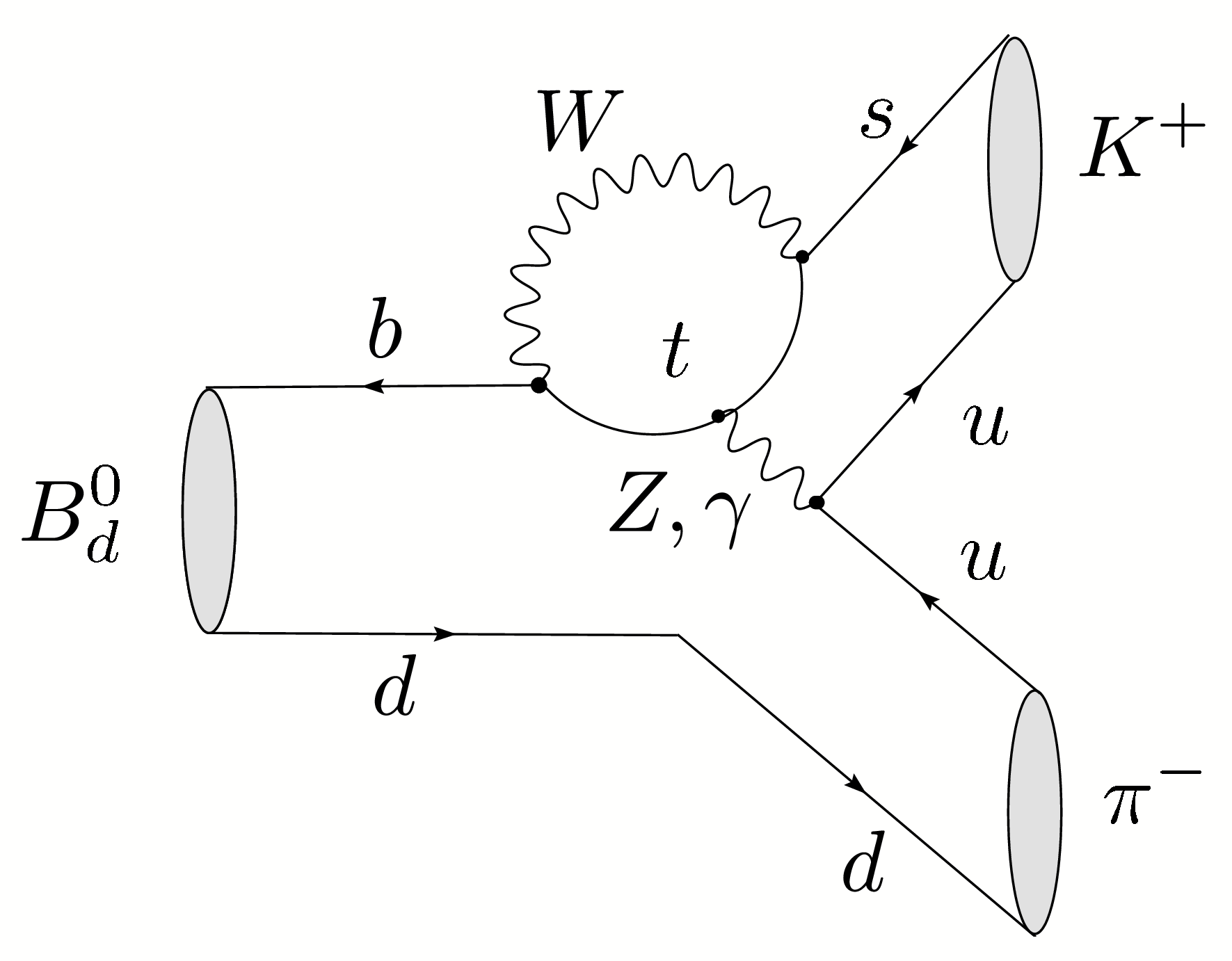}
     \hspace*{0.5truecm}
    \includegraphics[width=5.2truecm]{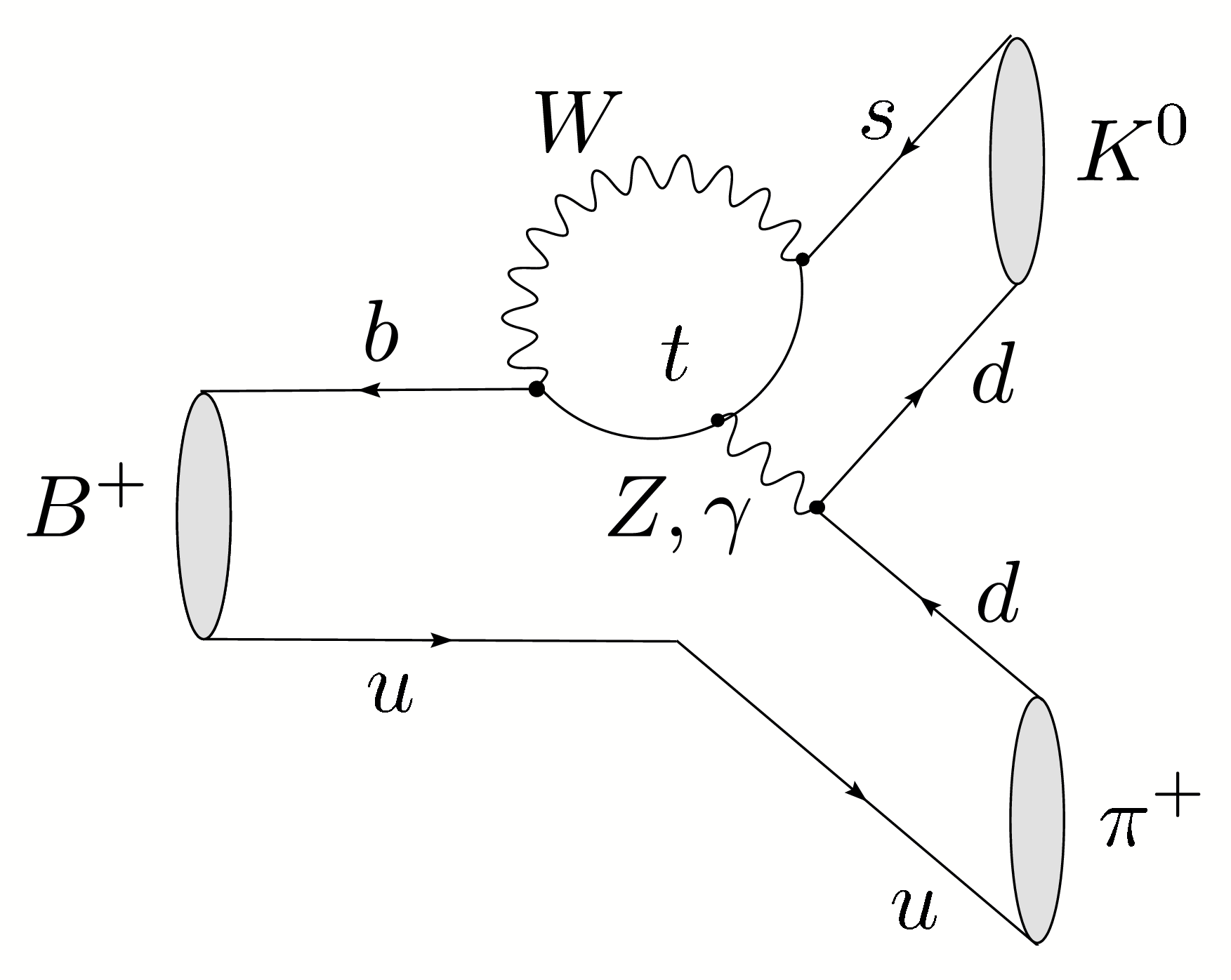} \\
     {\small(b)} & \\
    &     \includegraphics[width=5.2truecm]{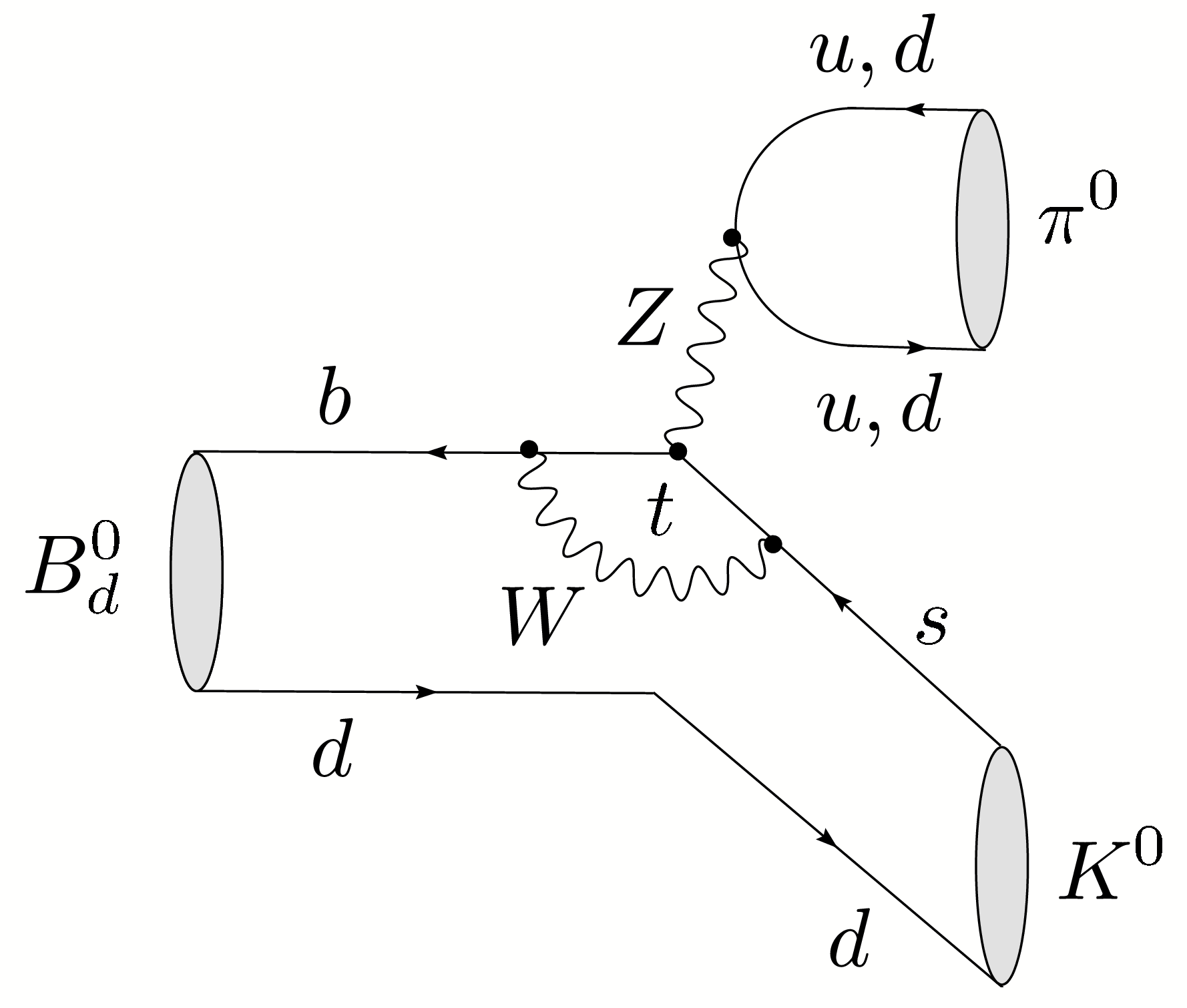} 
     \hspace*{0.5truecm}
    \includegraphics[width=5.2truecm]{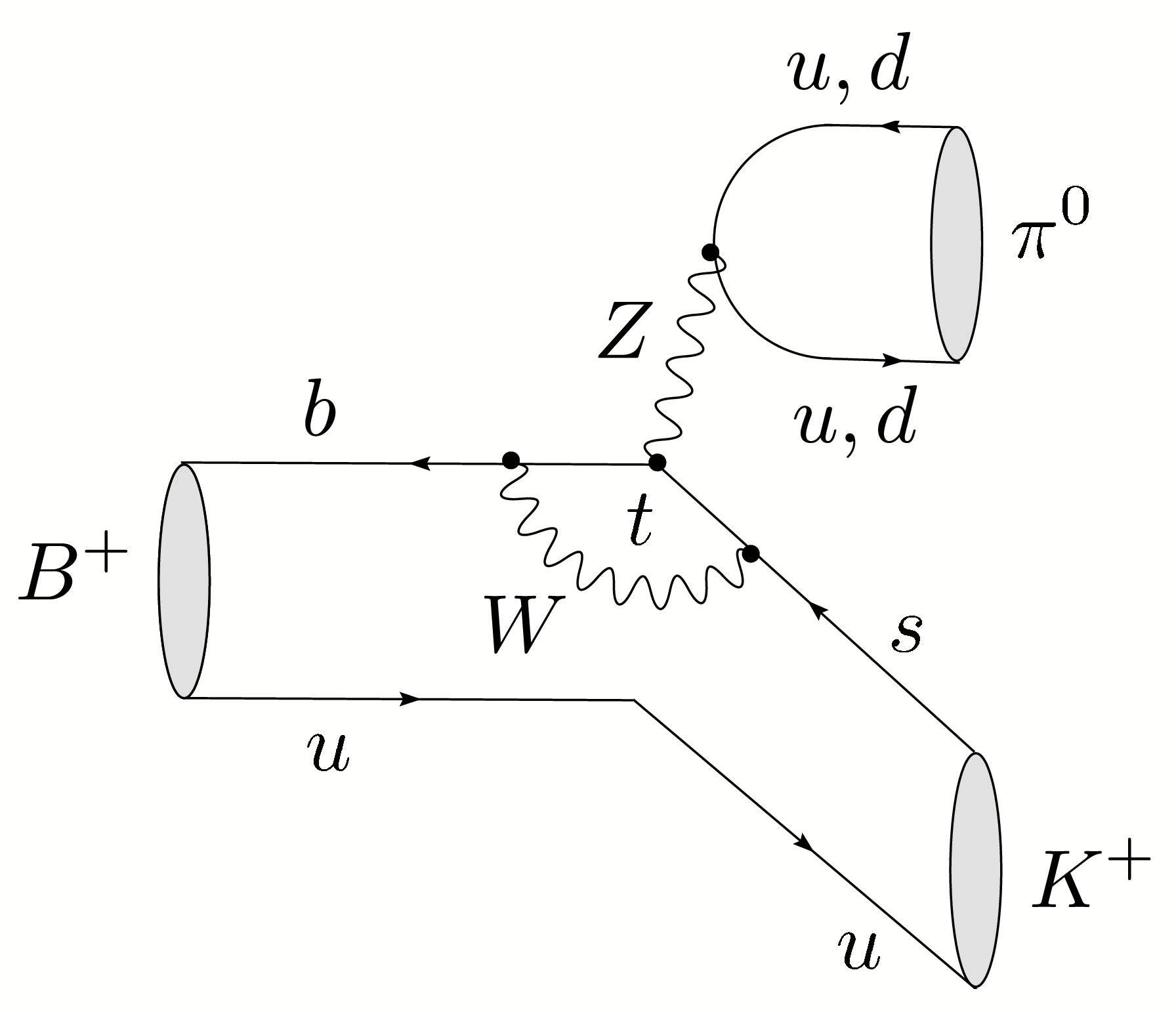} 
     \end{tabular}}
     \caption{Examples of the colour-suppressed (a) and colour-allowed (b) 
     EW penguin contributions to the $B\to\pi K$ system.}\label{fig:BpiK-EWP}
\end{figure}

\boldmath
\subsection{The $B\to \pi K$ Puzzle}\label{ssec:BpiK}
\unboldmath
Another hot topic is the exploration of $B\to\pi K$ decays, which are also
$b\to s$ transitions. Since tree amplitudes are suppressed by a CKM 
factor $\lambda^2 R_b\sim 0.02$ with respect to the penguin amplitudes,
these decays are actually dominated by QCD penguins. A classification of the
$B\to\pi K$ system is offered by their EW penguin contributions 
(see Fig.~\ref{fig:BpiK-EWP}):
\begin{itemize}
\item[(a)] In the $B^0_d\to\pi^-K^+$ and $B^+\to\pi^+K^0$ decays, EW penguins 
contribute in colour-suppressed form and are hence expected to play a minor r\^ole.
\item[(b)] In the $B^0_d\to\pi^0K^0$ and $B^+\to\pi^0K^+$ decays, EW penguins 
contribute in colour-allowed form and have therefore a significant impact on the decay 
amplitude, entering at the same order of magnitude as the tree contributions,
i.e.\ at the $20\%$ level. 
\end{itemize}
Interestingly, EW penguins offer an attractive avenue for NP to enter  
the $B\to\pi K$ system \cite{EWP-NP}, and the $B$-factory data for decays of 
class (b) raise indeed the possibility of having a  modified EW penguin sector through 
the impact of NP, which has received a lot of attention in the literature (see, e.g., 
Ref.~\cite{BpiK-papers}). 

Here we shall discuss key results of the recent analysis performed 
in Ref.~\cite{FRS-07}, following closely the strategy developed 
in Refs.~\cite{BFRS2,BFRS3,BFRS-5}. The starting point is given by $B\to\pi\pi$ modes.
Using the $SU(3)$ flavour symmetry of strong interactions and another plausible
dynamical assumption,\footnote{Consistency checks of these working assumptions
can be performed, which are all supported by the current data.} the data for the 
$B\to\pi\pi$ system can be converted into the hadronic parameters of the $B\to\pi K$ 
modes, thereby allowing us to calculate their observables in the SM. Moreover, also 
$\gamma$ can be extracted, with the result
\begin{equation}
\gamma=\left(70.0^{+3.8}_{-4.3}\right)^\circ,
\end{equation}
which is in agreement with the SM fits of the UT \cite{CKMfitter,UTfit}. 

\begin{figure}[t]
\begin{center}
\includegraphics[width=0.52\textwidth]{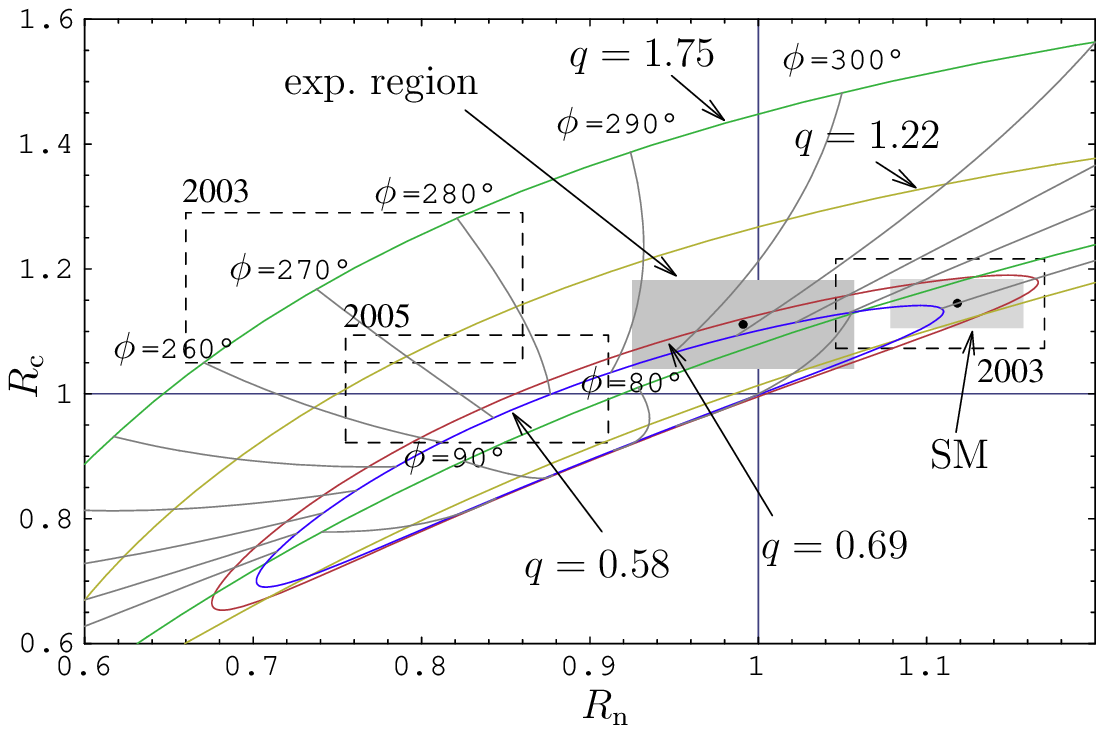}
\end{center}
\vspace*{-0.5truecm}
\caption{\label{fig:RnRc} The situation in the $R_{\rm n}$--$R_{\rm c}$ plane,
as discussed in the text.}
\end{figure}

As far as the $B\to\pi K$ observables with tiny EW penguin contributions are 
concerned, perfect agreement between the SM expectation and the experimental 
data is found. Concerning the $B\to\pi K$ observables receiving sizeable contributions 
from EW penguins, we distinguish between CP-conserving and CP-violating 
observables. In the former case, the key quantities are given by the following 
ratios of CP-averaged $B\to\pi K$ branching ratios \cite{BF98}:
\begin{eqnarray}
R_{\rm c}&\equiv&2\left[\frac{\mbox{BR}(B^+\to\pi^0K^+)+
\mbox{BR}(B^-\to\pi^0K^-)}{\mbox{BR}(B^+\to\pi^+ K^0)+
\mbox{BR}(B^-\to\pi^- \bar K^0)}\right]=1.11\pm0.07\\
R_{\rm n}&\equiv&\frac{1}{2}\left[
\frac{\mbox{BR}(B_d^0\to\pi^- K^+)+\mbox{BR}(\bar B_d^0\to\pi^+ 
K^-)}{\mbox{BR}(B_d^0\to\pi^0K^0)+\mbox{BR}(\bar B_d^0\to\pi^0\bar K^0)}
\right]=0.99\pm0.07,
\end{eqnarray}
where also the experimental averages are indicated \cite{HFAG}.
In these quantities, the EW penguin effects enter in colour-allowed form through 
the modes involving neutral pions, and are theoretically described by a parameter 
$q$, which measures the ``strength" of the EW penguin with respect to the tree 
contributions, and a CP-violating phase $\phi$. In the SM, the $SU(3)$ flavour 
symmetry allows a prediction of $q=0.60$ \cite{NR}, and $\phi$ 
{\it vanishes.} However, in the case of CP-violating NP effects in the EW penguin 
sector, $\phi$ would take a value different from zero. 
In Fig.~\ref{fig:RnRc}, we show the situation in the $R_{\rm n}$--$R_{\rm c}$ plane. 
Here the various contours correspond to different values of $q$, and the position on 
the contour is parametrized through the CP-violating phase $\phi$. We observe that 
the SM prediction (on the right-hand side) is very stable in time, having now significantly
reduced errors. On the other hand, the $B$-factory data have moved quite
a bit towards the SM, thereby reducing the ``$B\to\pi K$ puzzle" for the CP-averaged
branching ratios, which emerged already in 2000 \cite{BF00}. In comparison with the situation of the $B\to\pi K$ observables with tiny EW penguin 
contributions,  
the agreement between the new data for the $R_{\rm c,n}$ and their SM predictions 
is not as perfect. However, a case for a modified EW penguin sector cannot be made 
through the new measurements of these quantities. 

\begin{figure}
\vspace*{0.5truecm}
\begin{center}
\includegraphics[width=0.55\textwidth]{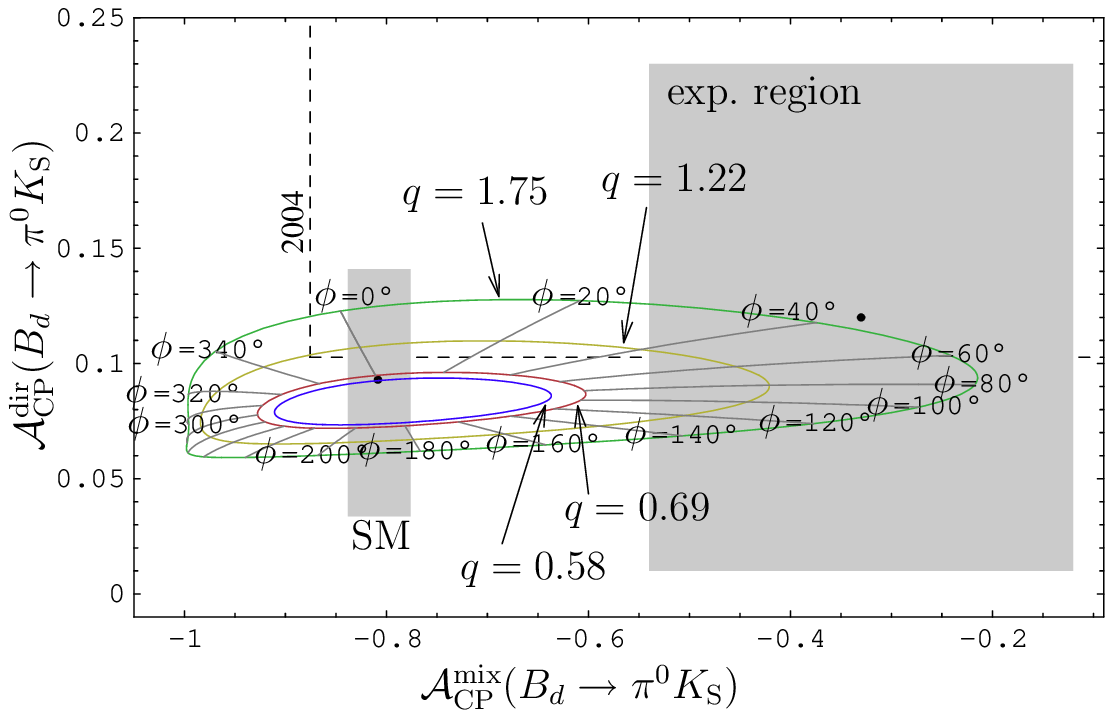}
\end{center}
\vspace*{-0.5truecm}
\caption{\label{fig:ACP}The situation in the 
${\cal A}_{\rm CP}^{\rm mix}(B_d\to\pi^0K_{\rm S})$--${\cal A}_{\rm CP}^{\rm 
dir}(B_d\to\pi^0K_{\rm S})$ plane, as discussed in the text.}
\end{figure}

Let us now have a closer look at the CP asymmetries of the 
$B^0_d\to\pi^0 K_{\rm S}$ and $B^\pm\to\pi^0K^\pm$ channels. 
As can be seen in Fig.~\ref{fig:ACP}, SM predictions for the CP-violating observables
of $B^0_d\to\pi^0K_{\rm S}$ are obtained that are much sharper than the current
$B$-factory data. In particular ${\cal A}_{\rm CP}^{\rm mix}(B_d\to\pi^0K_{\rm S})$
offers a very interesting quantity. We also see that the experimental central
values can be reached for large {\it positive} values of $\phi$. 
For the new input data, the non-vanishing difference
\begin{equation}\label{Delta-A}
\Delta A \equiv {\cal A}_{\rm CP}^{\rm dir}(B^\pm\to\pi^0K^\pm)-
{\cal A}_{\rm CP}^{\rm dir}(B_d\to\pi^\mp K^\pm)
\stackrel{\rm exp}{=}-0.140\pm0.030
\end{equation}
is likely to be generated through hadronic effects, i.e.\ not through the impact of 
physics beyond the SM. A similar conclusion was drawn in Ref.~\cite{GR-06}, where 
it was also noted that the measured  values of $R_{\rm c}$ and $R_{\rm n}$ are 
now in accordance with the SM.

Performing, finally, a simultaneous fit to $R_{\rm n}$, $R_{\rm c}$ and the 
CP-violating $B_d\to\pi^0K_{\rm S}$ asymmetries yields the following result:
\begin{equation}
q=1.7_{-1.3}^{+0.5},\quad \phi=+\left(73_{-18}^{+6}\right)^\circ.
\end{equation}
Interestingly, these parameters -- in particular the large {\it positive} phase -- 
would also allow us to accommodate the experimental values of
$(\sin2\beta)_{\phi K_{\rm S}}$ and the CP asymmetries of other 
$b\to s$ penguin modes with central values smaller than 
$(\sin2\beta)_{\psi K_{\rm S}}$. The large central value of $q$ would be excluded 
by constraints from rare decays in simple scenarios where NP enters only 
through $Z$ penguins, but could still be accommodated in other scenarios, 
e.g.\ in models with leptophobic $Z'$ bosons.

\subsection{Prospects for LHCb}\label{ssec:prosp}
Unfortunately, it is unlikely that the current $B$ factories will allow us to 
establish -- or rule out -- the tantalizing option of having NP in the $b\to s$ 
penguin processes. However, at LHCb, this exciting topic
can be explored with the help of the decay $B^0_s\to \phi\phi$ \cite{FG-1}. 
A handful of events have been observed in this mode a few 
years ago by the CDF collaboration at the Tevatron, corresponding to a 
branching ratio of  $(14^{+6}_{-5}\pm 6)\times 10^{-6}$~\cite{Acosta:2005eu}. 
A  proposal for studying time and angular dependence in this decay mode has been
made by the LHCb collaboration \cite{LHCb-Bsphiphi}. The proposal is based on 
an estimated sample of about 3100 events collected in one year of running. In order
to control hadronic uncertainties, the decay mode $B_s\to\phi\phi$ may be related
through the $SU(3)$ flavour symmetry to $B_s\to \phi \bar K^{*0}$ and plausible 
dynamical assumptions, which can be checked through experimental control 
channels \cite{FG-1}. The current $B$-factory data on the CP asymmetries
of the $b\to s$ penguin modes leave ample space for NP phenomena in
the $B^0_s\to \phi\phi$ decay to be discovered at LHCb. Let us next
have a closer look at other key targets of the physics programme of this experiment.

\begin{figure}[t]
 \centering
  \includegraphics[width=6.8truecm]{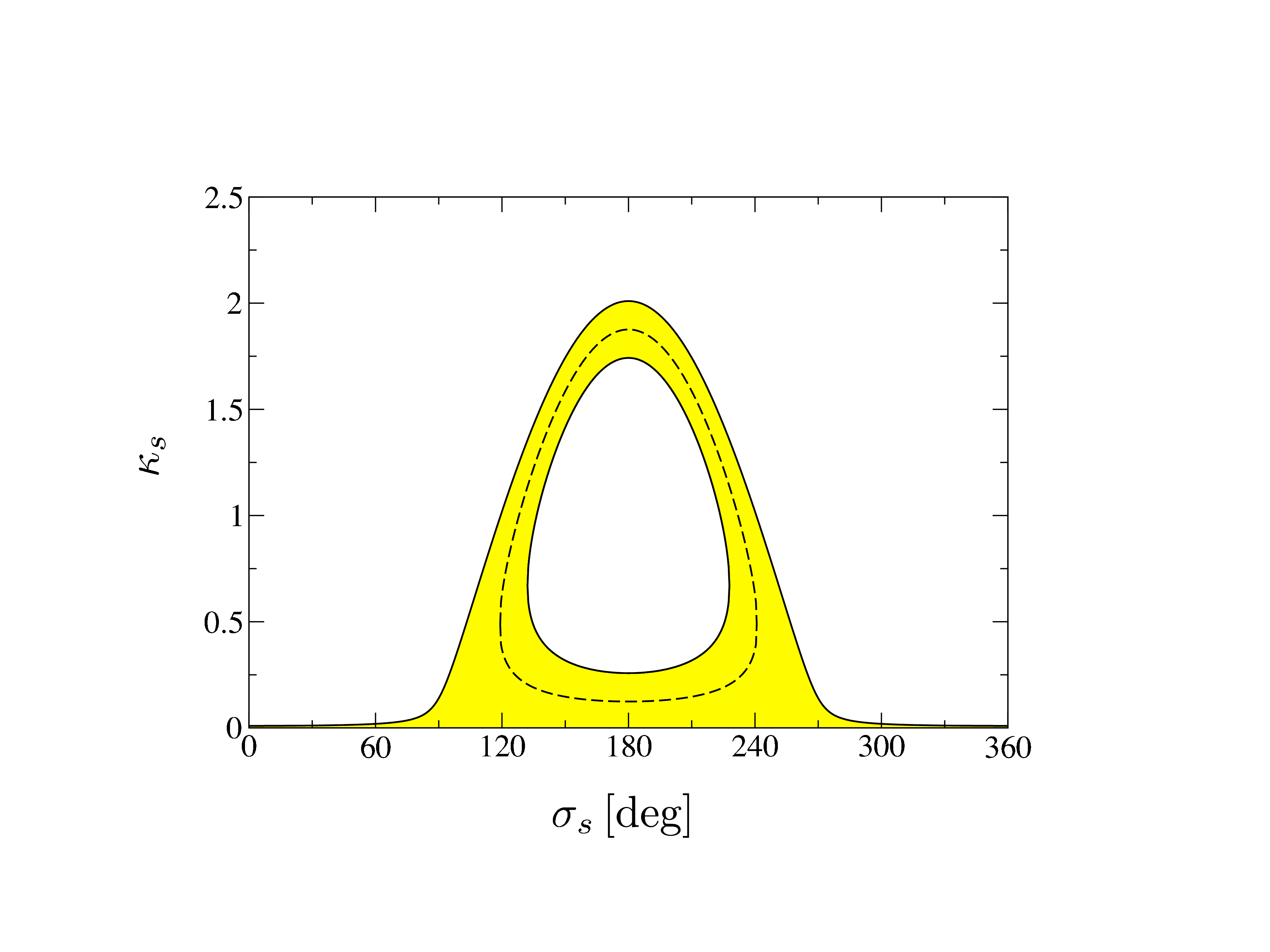} 
     \vspace*{-0.4truecm}
  \caption{The allowed region in the $\sigma_s$--$\kappa_s$ plane of
  NP parameters for $B^0_s$--$\bar B^0_s$ mixing.}\label{fig:MDs-NP}
\end{figure}

\section{HIGHLIGHTS OF \boldmath$B$\unboldmath-PHYSICS AT 
THE LHC}\label{sec:LHC}
\setcounter{equation}{0}
Since the current $e^+e^-$ $B$ factories run at the $\Upsilon(4S)$ resonance, 
which decays into $B_{u,d}$, but not into $B_s$ mesons, the $B_s$ system cannot 
be explored by the BaBar and Belle experiments.\footnote{The asymmetric $e^+e^-$ KEKB collider was recently also operated at the $\Upsilon(5S)$ resonance, 
allowing the Belle experiment to take first $B_s$ data \cite{Belle-U5S}.} 
However, plenty of $B_s$ mesons are produced at hadron colliders, i.e.\ at 
the Tevatron and soon at the LHC. The $B$-decay programme at the LHC
is characterized by its high statistics and the complementarity to the
studies at the  $e^+e^-$ $B$ factories; in particular, the physics potential
of the $B_s$-meson system, which offers various powerful strategies for 
the exploration of CP violation, can be fully exploited. 

\boldmath
\subsection{In Pursuit of New Physics with 
$\Delta M_s$}\label{ssec:Bs-prelim}
\unboldmath
As we discussed in Subsection~\ref{ssec:Mix-Par}, the mass difference 
$\Delta M_s$ of the $B_s$-meson system was recently measured at the 
Tevatron, with the results summarized in (\ref{MDs}). On the other hand, 
the HPQCD collaboration has reported the following lattice QCD prediction 
\cite{HPQCD-DMs}:
\begin{equation}\label{HPQCD-DMs}
\Delta M_s^{\rm SM}=20.3(3.0)(0.8)\,{\rm ps}^{-1}.
\end{equation}
In contrast to the case of $\Delta M_d$ discussed in Subsection~\ref{ssec:NP-mix}, 
the CKM factor entering this SM value does not require information on $\gamma$ 
and $|V_{ub}/V_{cb}|$, as 
\begin{equation}
|V_{ts}^*V_{tb}|=|V_{cb}|\left[1+{\cal O}(\lambda^2)\right],
\end{equation}
which is an important advantage. Using (\ref{rhoq-def}), we may convert the
experimental value of $\Delta M_s$ into the allowed region in the 
$\sigma_s$--$\kappa_s$ plane shown in Fig.~\ref{fig:MDs-NP} \cite{BF-DMs}. 
We see that  the measurement of $\Delta M_s$ leaves 
ample space for the NP parameters $\sigma_s$ and $\kappa_s$, which can
also be accommodated in specific scenarios (e.g.\ SUSY, extra $Z'$ and little 
Higgs models). It should be noted that the experimental errors are already 
significantly smaller than the theoretical lattice QCD uncertainties.
The experimental results on $\Delta M_s$ have immediately triggered 
a lot of theoretical activity (see, e.g., \cite{BF-DMs,DMs-papers,BBGT}). 

As in the case of the $B_d$-meson system, the allowed region in the 
$\sigma_s$--$\kappa_s$ plane will be dramatically
reduced as soon as measurements of CP violation in the $B_s$-meson 
system become available. The ``golden" channel in this respect is 
$B^0_s\to J/\psi \phi$, our next topic.

\boldmath
\subsection{The Decay $B^0_s\to J/\psi \phi$}\label{ssec:Bspsiphi}
\unboldmath
This mode is the counterpart of the $B^0_d\to J/\psi K_{\rm S}$ transition, where
we have just to replace the down quark by a strange quark. The structures of the
corresponding decay amplitudes are completely analogous to each other. However,
there is also an important difference with respect to $B^0_d\to J/\psi K_{\rm S}$,
since the final state of $B^0_s\to J/\psi \phi$ contains two vector mesons and is,
hence, an admixture of different CP eigenstates. Using the angular distribution of the 
$J/\psi [\to\ell^+\ell^-]\phi [\to\ K^+K^-]$ decay products, the CP eigenstates
can be disentangled \cite{DDLR} and the time-dependent decay rates calculated
\cite{DDF,DFN}. As in the case of $B^0_d\to J/\psi K_{\rm S}$, the
hadronic matrix elements cancel then in the mixing-induced observables. For the
practical implementation, a set of three linear polarization amplitudes is usually 
used: $A_0(t)$ and $A_\parallel(t)$ correspond to CP-even final-state configurations,
whereas $A_\perp(t)$ describes a CP-odd final-state configuration.

It is instructive to illustrate how this works by having a closer look at the
one-angle distribution, which takes the following form \cite{DDF,DFN}:
\begin{equation}
\frac{d\Gamma(B^0_s(t)\to J/\psi \phi)}{d\cos\Theta}\propto
\left(|A_0(t)|^2+|A_\parallel(t)|^2\right)
\frac{3}{8}\left(1+\cos^2\Theta\right)+|A_\perp(t)|^2\frac{3}{4}\sin^2\Theta.
\end{equation}
Here $\Theta$ is defined as the angle between the momentum of the $\ell^+$
and the normal to the decay plane of the $K^+K^-$ system in the $J/\psi$
rest frame. The time-dependent measurement of the angular dependence
allows us to extract the following observables:
\begin{equation}
P_+(t)\equiv |A_0(t)|^2+|A_\parallel(t)|^2, \quad
P_-(t)\equiv |A_\perp(t)|^2,
\end{equation}
where $P_+(t)$ and $P_-(t)$ refer to the CP-even and CP-odd final-state configurations,
respectively. If we consider the case of having an initially, i.e.\ at time $t=0$, present
$\bar B^0_s$ meson, the CP-conjugate quantities $\bar P_\pm(t)$ can be extracted
as well. Using an {\it untagged} data sample, the untagged rates
\begin{equation}
P_\pm(t)+\overline{P}_\pm(t)\propto
\left[(1\pm\cos\phi_s)e^{-\Gamma_{\rm L}t}+
(1\mp\cos\phi_s)e^{-\Gamma_{\rm H}t}\right]
\end{equation}
can be determined, while a {\it tagged} data sample allows us to measure
the CP-violating asymmetries
\begin{equation}
\frac{P_\pm(t)-\overline{P}_\pm(t)}{P_\pm(t)+\overline{P}_\pm(t)}=
\pm\left[\frac{2\,\sin(\Delta M_st)\sin\phi_s}{(1\pm\cos\phi_s)e^{+\Delta\Gamma_st/2}+
(1\mp\cos\phi_s)e^{-\Delta\Gamma_st/2}}\right].
\end{equation}
In the presence of CP-violating NP contributions to $B^0_s$--$\bar B^0_s$
mixing, we obtain
\begin{equation}
\phi_s=-2\lambda^2\eta+\phi_s^{\rm NP}\approx -2^\circ+
\phi_s^{\rm NP}\approx \phi_s^{\rm NP}.
\end{equation}
Consequently, NP of this kind would be indicated by the following features:
\begin{itemize}
\item The {\it untagged} observables depend on {\it two} exponentials;
\item {\it sizeable} values of the CP-violating asymmetries.
\end{itemize}

\begin{figure}[t] 
 \centering
\begin{tabular}{cc}
  \includegraphics[width=5.8truecm]{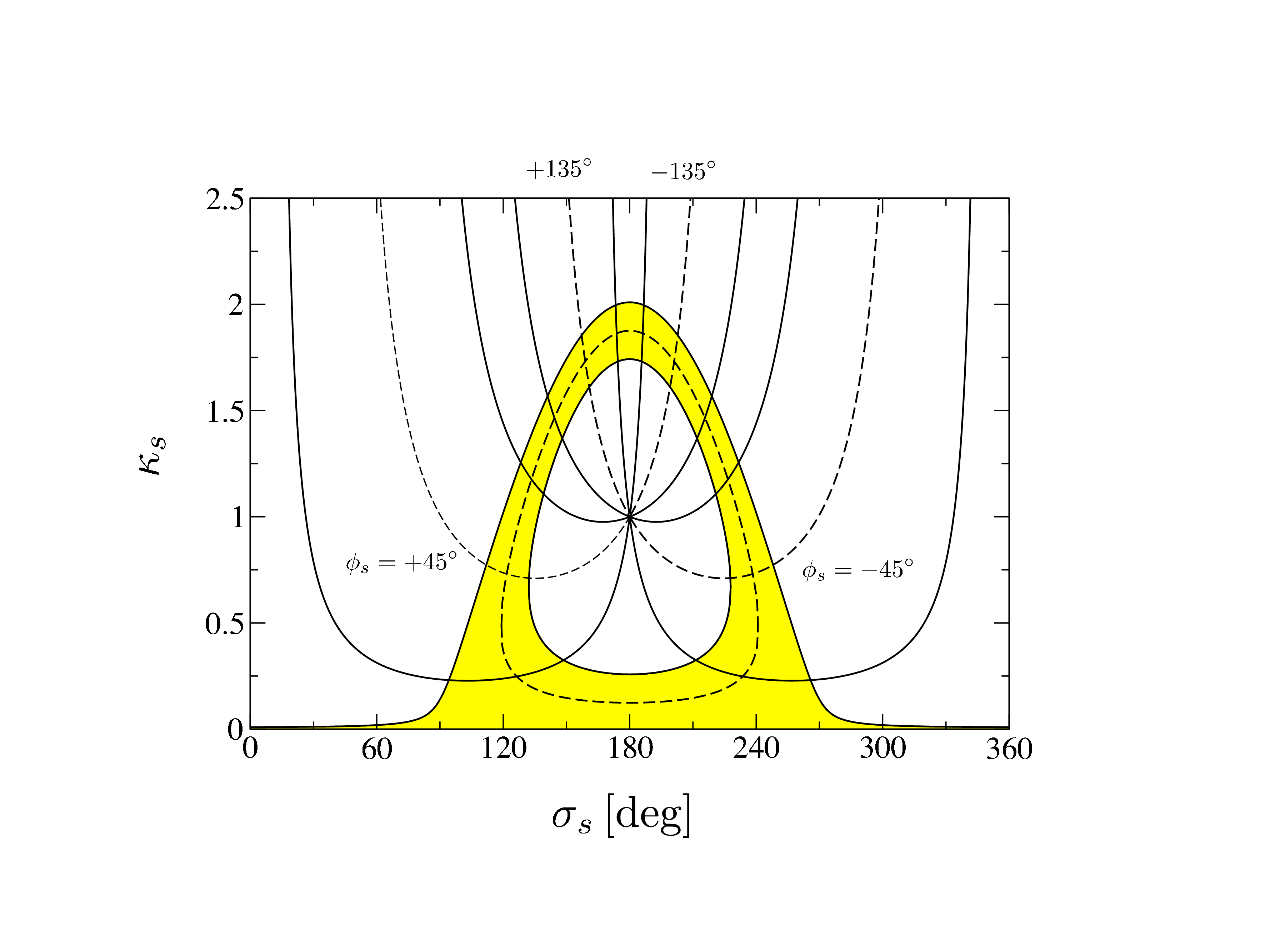} &
    \includegraphics[width=5.8truecm]{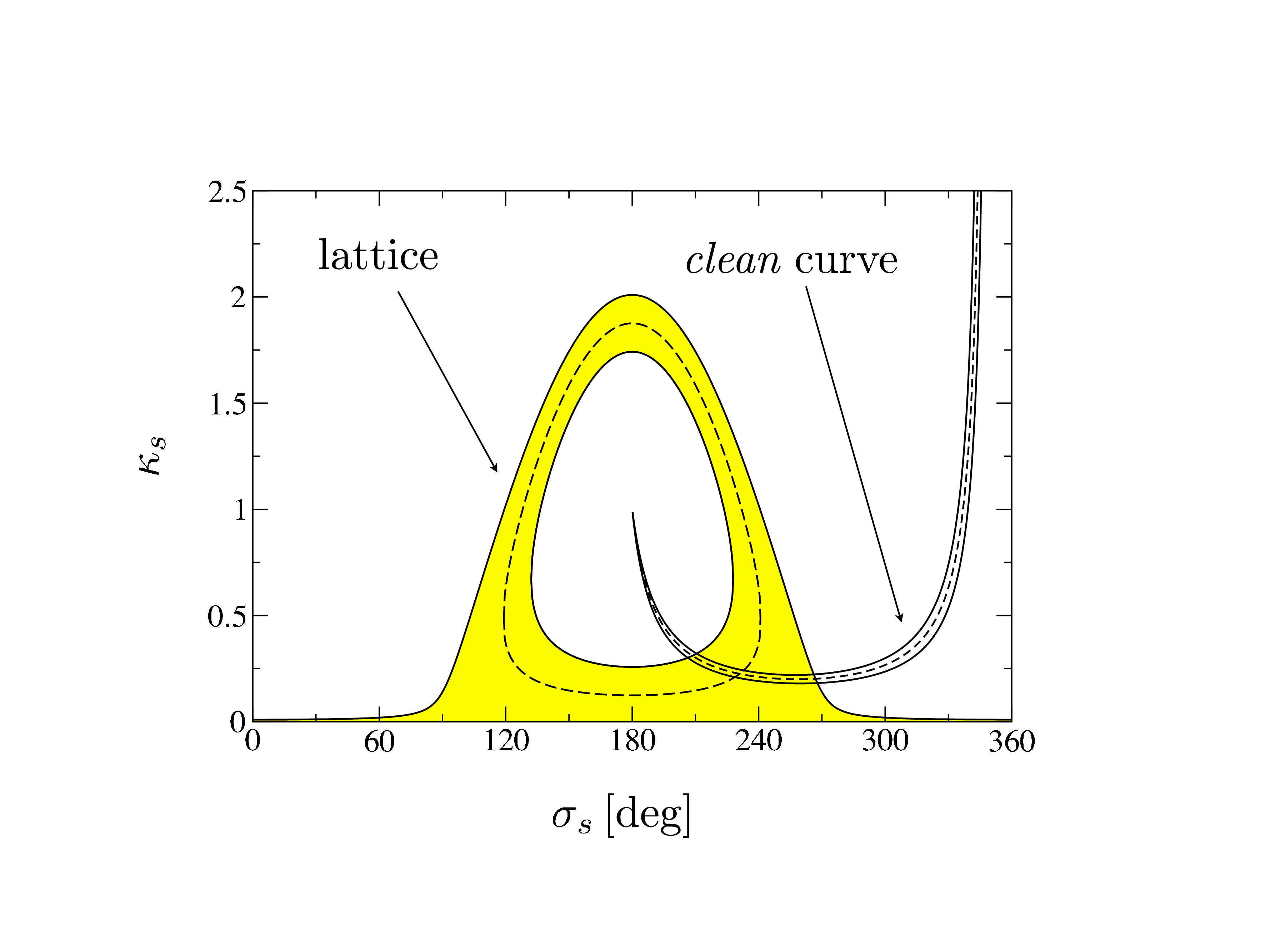} 
    \end{tabular}
    \vspace*{-0.4truecm}
   \caption[]{Impact of the measurement of CP violation in 
   $B^0_s\to J/\psi\phi$: current D0 data (left panel), and a NP scenario
   with $(\sin\phi_s)_{\rm exp}=-0.20\pm0.02$ (right panel).}\label{fig:sis-kas-CP}
\end{figure}

These general features hold also for the full three-angle distribution
\cite{DDF,DFN}: it is much more involved than the one-angle case, but
provides also additional information through interference terms of the
form 
\begin{equation}
\mbox{Re}\{A_0^\ast(t)A_\parallel(t)\}, \quad
\mbox{Im}\{A_f^\ast(t)A_\perp(t)\} \, (f\in\{0,\parallel\}).
\end{equation}
From an experimental point of view, there is no experimental draw-back with
respect to the one-angle case. Following these lines, $\Delta\Gamma_s$ 
(see (\ref{DG-det})) and $\phi_s$ can be extracted.  
Recently, the D0 collaboration has reported first results for the measurement 
of $\phi_s$ through the untagged, time-dependent three-angle 
$B^0_s\to J/\psi\phi$ distribution
\cite{D0-phis}:
\begin{equation}
\phi_s=-0.79\pm0.56\,\mbox{(stat.)} ^{+0.14}_{-0.01}\,\mbox{(syst.)}
=-(45\pm32^{+1}_{-8})^\circ,
\end{equation}
which is complemented by three additional mirror solutions. This phase is therefore not 
yet stringently constrained. Using (\ref{phiq-NP}), we then obtain the curves 
in the $\sigma_s$--$\kappa_s$ plane shown in the left panel of
Fig.~\ref{fig:sis-kas-CP}. Very recently, CDF reported first
bounds on $\phi_s$ from flavour-tagged $B^0_s\to J/\psi\phi$
decays \cite{cdf-tagged}.

Fortunately, $\phi_s$ will be very accessible at LHCb, where already the initial
$0.5\,\mbox{fb}^{-1}$ of data will give an uncertainty of
$\sigma(\phi_s)=0.046=2.6^\circ$ by the end of 2009, which will
be significantly improved further thanks to the $2\,\mbox{fb}^{-1}$ that should
be available by the end of 2010 \cite{nakada}. At some point, also in view of LHCb upgrade plans \cite{LHCb-up}, we have to include hadronic penguin uncert
ainties. 
This can be done with the help of the $B^0_d\to J/\psi \rho^0$ decay \cite{RF-ang}. In
order to illustrate the impact of the measurement of CP violation in 
$B^0_s\to J/\psi\phi$, we show in the right panel of Fig.~\ref{fig:sis-kas-CP}
the case corresponding to $(\sin\phi_s)_{\rm exp}=-0.20\pm0.02$. Such a 
measurement would give a NP signal at the $10\,\sigma$ level, which would
immediately rule out MFV models, and demonstrates 
the power of the $B_s$ system to search for NP \cite{BF-DMs}.  It should be
emphasized that the contour following from the measurement of
$\phi_s$ would be essentially clean, in contrast to the shaded region representing 
the constraint from the measured value of $\Delta M_s$, which suffers from 
lattice QCD uncertainties.

\subsection{Further Benchmark Decays for LHCb}
This experiment has a very rich physics programme. Besides many other 
interesting aspects, there are two major lines of research:
\begin{enumerate}
\item \underline{Precision measurements of $\gamma$:}\\
On the one hand, there are strategies using  tree decays: 
$B^0_s\to D_s^\mp K^\pm$ [$\sigma_\gamma\sim5^\circ$],
$B^0_d\to D^0K^{*}$ [$\sigma_\gamma\sim8^\circ$],
$B^\pm\to D^0K^\pm$ [$\sigma_\gamma\sim5^\circ$],
where we have also indicated the expected sensitivities for $10\,\mbox{fb}^{-1}$;
by 2013, a LHCb tree deterimation of $\gamma$ with 
$\sigma_\gamma=2^\circ\sim3^\circ$
should be available \cite{nakada}. This very impressive situation should be
compared with the current $B$-factory data, yielding the results summarized
in (\ref{gam-DK}). These extractions are essentially unaffected by NP effects.
On the other hand, $\gamma$ can also be determined through $B$ decays
with penguin contributions: $B^0_s\to K^+K^-$ and $B^0_d\to \pi^+\pi^-$
[$\sigma_\gamma\sim5^\circ$], $B^0_s\to D_s^+D_s^-$ and $B^0_d\to D_d^+D_d^-$.
The key question is whether discrepancies will arise in these determinations. 

\item \underline{Analyses of rare decays, which are absent at the SM tree level:}\\ 
prominent examples are $B^0_{s,d}\to\mu^+\mu^-$,
$B^0_d\to K^{*0}\mu^+\mu^-$ and $B^0_s\to \phi \mu^+\mu^-$. In order to 
complement the studies of CP violation in $b\to s$ penguin modes at the
$B$ factories, $B^0_s\to\phi\phi$ is a very interesting mode for LHCb, as
we noted in Subsection~\ref{ssec:prosp}.
\end{enumerate}
Let us next have a closer look at some of these decays.

\begin{figure}[t]
\centerline{
 \includegraphics[width=5.2truecm]{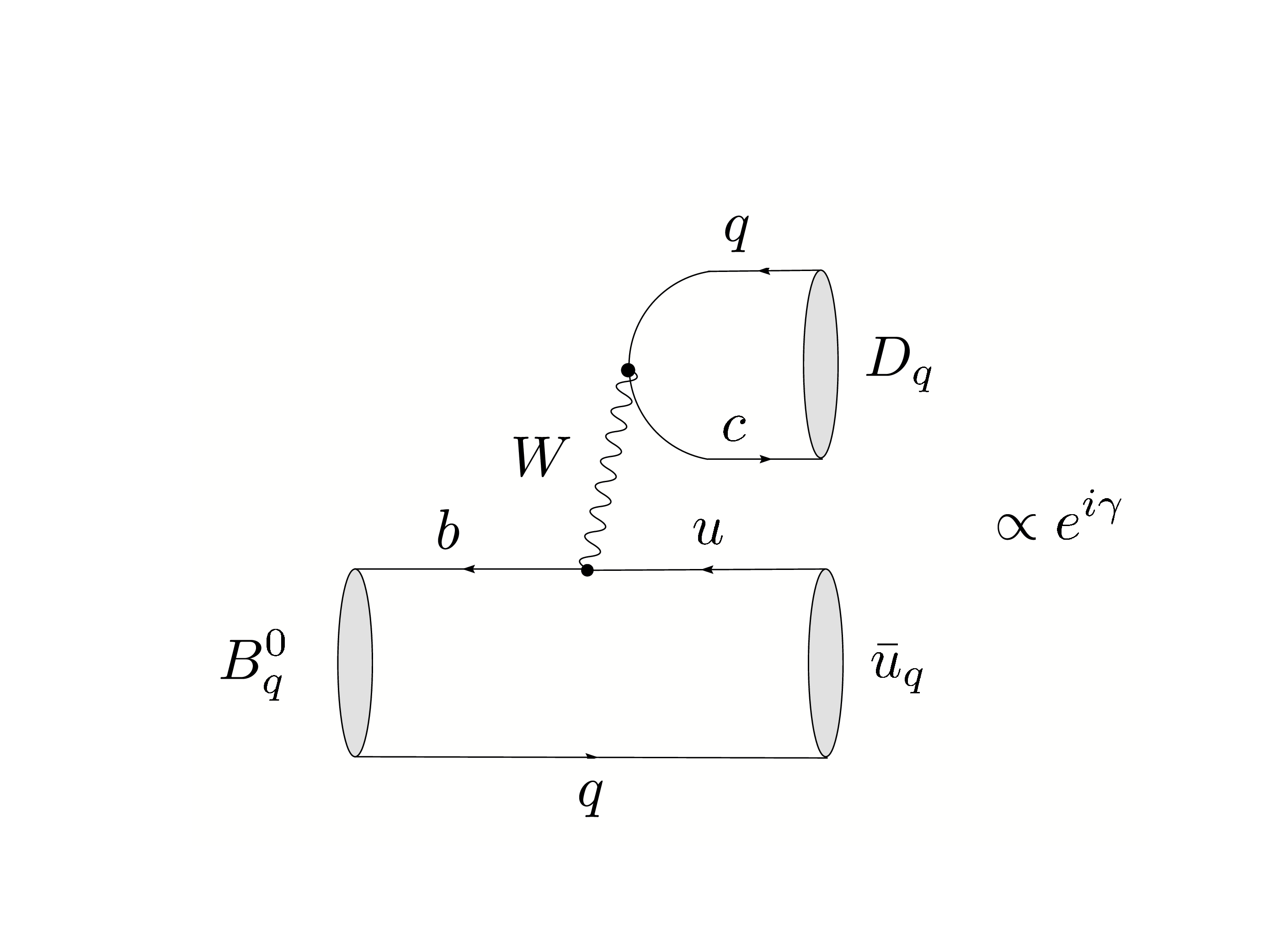}
 \hspace*{0.5truecm}
 \includegraphics[width=5.2truecm]{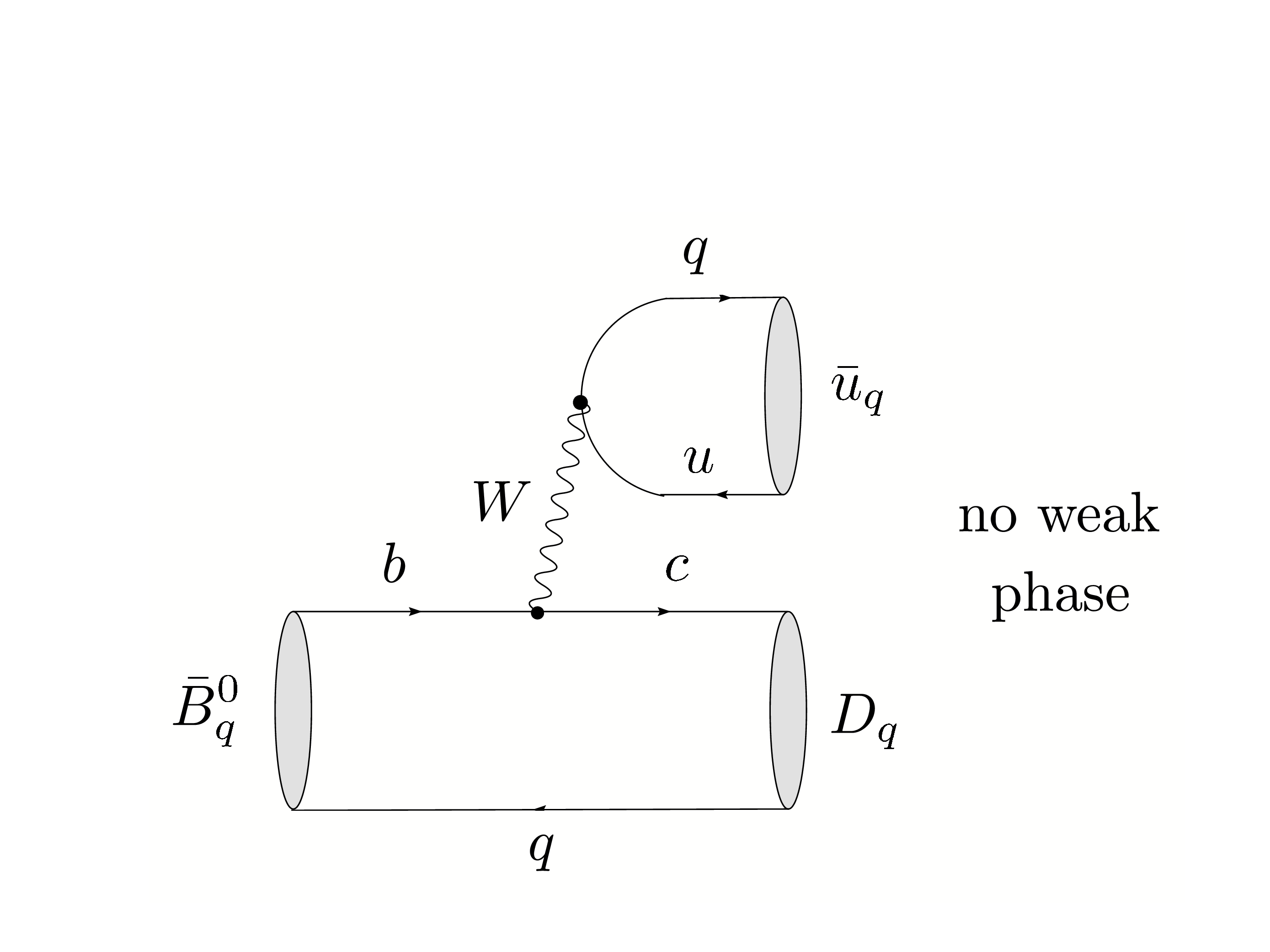}  
 }
 \vspace*{-0.3truecm}
\caption{Feynman diagrams contributing to $B^0_q\to D_q\bar u_q$
and $\bar B^0_q\to D_q \bar u_q$ 
decays.}\label{fig:BqDquq}
\end{figure}

\begin{figure}
\centerline{
 \includegraphics[width=3.0truecm]{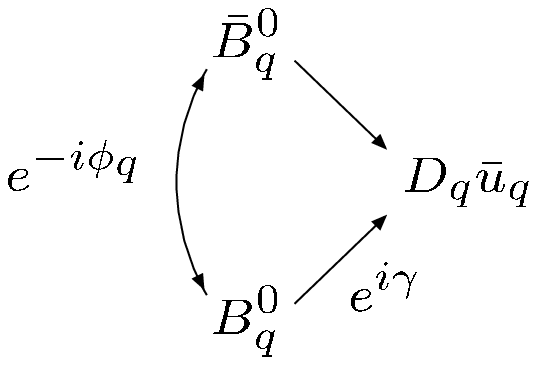} }
 \vspace*{-0.3truecm}
\caption{Interference effects between $B^0_q\to D_q\bar u_q$
and $\bar B^0_q\to D_q\bar u_q$ 
decays.}\label{fig:BqDquq-int}
\end{figure}

\subsubsection{$B_s\to D_s^\pm K^\mp$ and 
$B_d\to D^\pm \pi^\mp$}\label{ssec:BsDsK}
The decays $B_s\to D_s^\pm K^\mp$ \cite{BsDsK} and $B_d\to D^\pm \pi^\mp$
\cite{BdDpi} can be 
treated on the same theoretical basis, and provide new strategies to determine 
$\gamma$ \cite{RF-gam-ca}. Following this paper, we write these modes, which 
are pure ``tree" decays according to the classification of 
Subsection~\ref{sec:class}, generically as $B_q\to D_q \bar u_q$. 
As can be seen from the Feynman diagrams in Fig.~\ref{fig:BqDquq}, their
characteristic feature is that both a $B^0_q$ and a $\bar B^0_q$ meson may decay 
into the same final state $D_q \bar u_q$. Consequently,  as illustrated in 
Fig.~\ref{fig:BqDquq-int}, interference effects between $B^0_q$--$\bar B^0_q$ 
mixing and decay processes arise, which allow us to probe the weak phase 
$\phi_q+\gamma$ through measurements of the corresponding time-dependent
decay rates. 
 
 In the case of $q=s$, i.e.\ $D_s\in\{D_s^+, D_s^{\ast+}, ...\}$ and 
$u_s\in\{K^+, K^{\ast+}, ...\}$, these interference effects are governed 
by a hadronic parameter $X_s e^{i\delta_s}\propto R_b\approx0.4$, where
$R_b\propto |V_{ub}/V_{cb}|$ is the usual UT side, and hence are large. 
On the other hand, for $q=d$, i.e.\ $D_d\in\{D^+, D^{\ast+}, ...\}$ 
and $u_d\in\{\pi^+, \rho^+, ...\}$, the interference effects are described 
by $X_d e^{i\delta_d}\propto -\lambda^2R_b\approx-0.02$, and hence are tiny. 

Measuring the $\cos(\Delta M_qt)$ and $\sin(\Delta M_qt)$ terms of the
time-dependent $B_q\to D_q \bar u_q$ rates, a theoretically clean 
determination of $\phi_q+\gamma$ is possible \cite{BsDsK,BdDpi}. 
Since the $\phi_q$ can be determined separately, as we saw above, $\gamma$ 
can be extracted. However, in the practical implementation, there are problems:
we encounter an eightfold discrete ambiguity for $\phi_q+\gamma$, which
is very disturbing for the search of NP, and in the $q=d$ case, an additional input 
is required to extract $X_d$ since
${\cal O}(X_d^2)$ interference effects would otherwise have to be resolved,
which is impossible. Performing a combined analysis of the $B^0_s\to D_s^{+}K^-$
and $B^0_d\to D^+\pi^-$ decays, these problems can be solved \cite{RF-gam-ca}.
This strategy exploits the fact that these transitions are related to each other
through the $U$-spin symmetry of strong interactions,\footnote{The $U$-spin 
symmerty is an $SU(2)$ subgroup of the $SU(3)_{\rm F}$ flavour-symmetry 
group of QCD, connecting
$d$ and $s$ quarks in analogy to the isospin symmetry, which relates $d$ and
$u$ quarks to each other.} allowing us to simplify the hadronic sector. Following
these lines, an unambiguous value of $\gamma$ can be extracted from the
observables. To this end, $X_d$ has actually not to be fixed, and $X_s$ may only enter
through a $1+X_s^2$ correction, which is determined through untagged $B_s$
rates. The first studies for LHCb are very promising, and are 
currently further refined \cite{BDpi-Uspin}. 

\subsubsection{The $B_s\to K^+K^-$, $B_d\to \pi^+\pi^-$ System}
\begin{figure} 
\centerline{
 \includegraphics[width=4.6truecm]{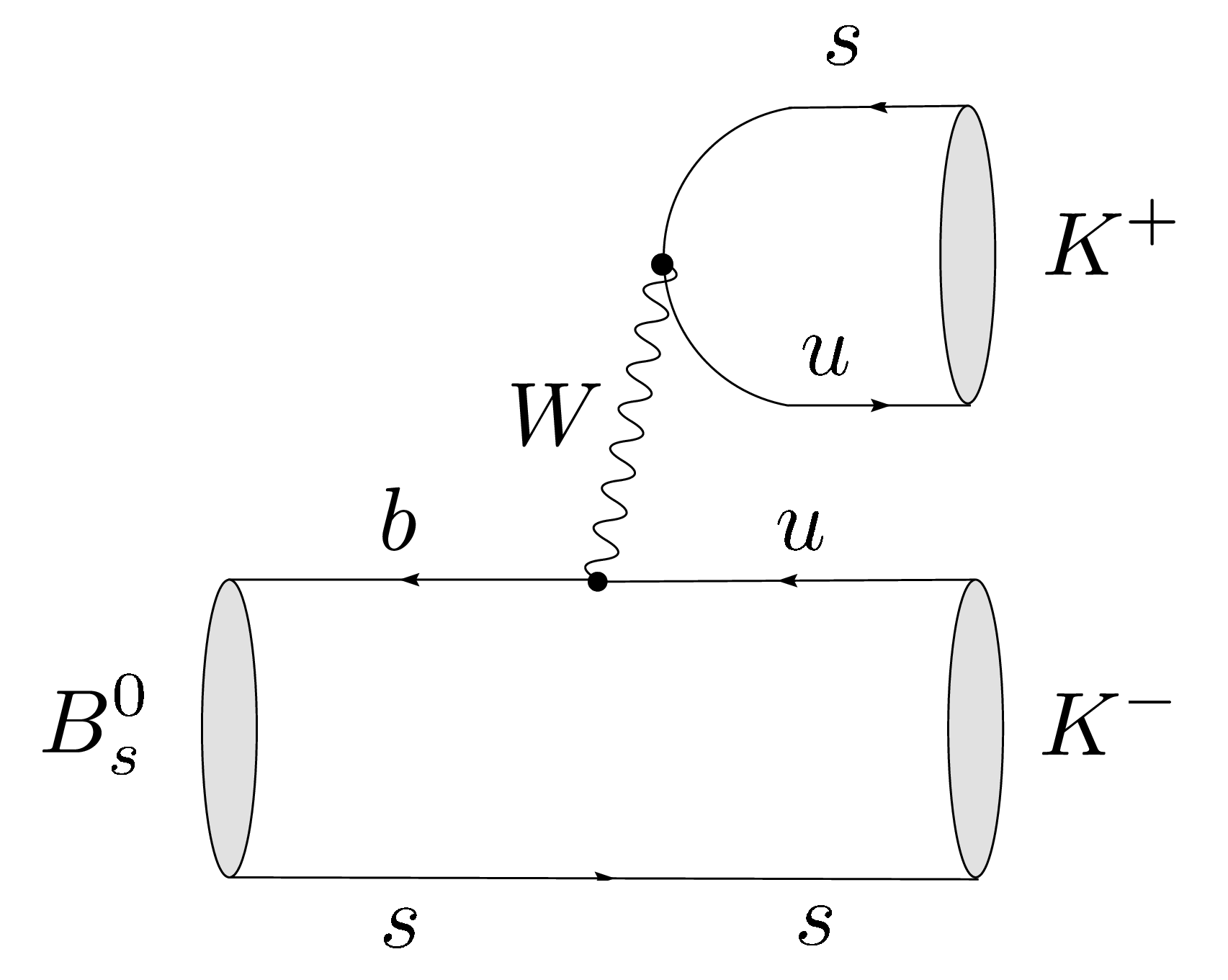}
 \hspace*{0.5truecm}
 \includegraphics[width=5.0truecm]{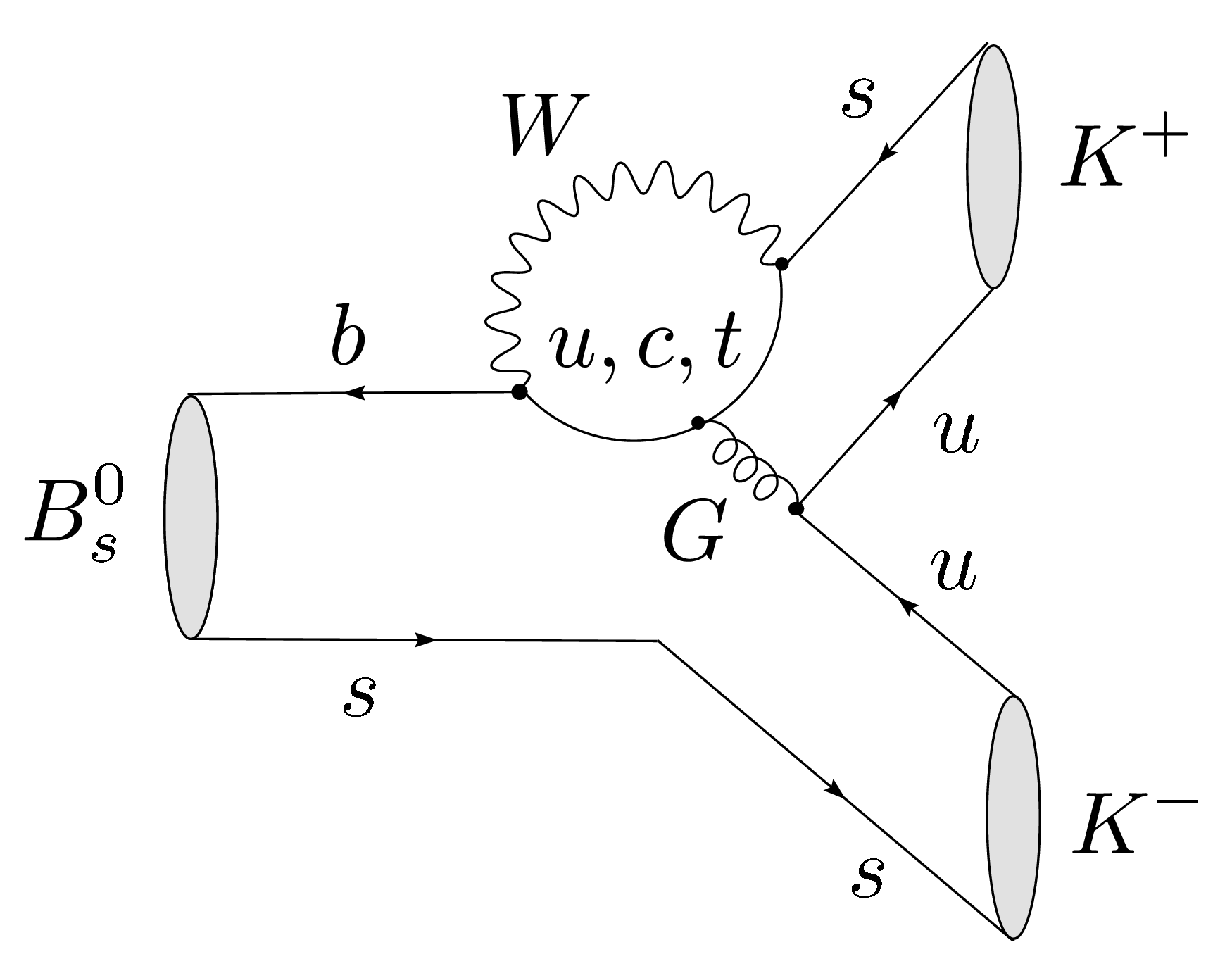}  
 }
 \vspace*{-0.5truecm}
\caption{Feynman diagrams contributing to the $B^0_s\to K^+K^-$
decay.}\label{fig:BsKK-diag}
\end{figure}

\begin{figure}
\centerline{
 \includegraphics[width=4.6truecm]{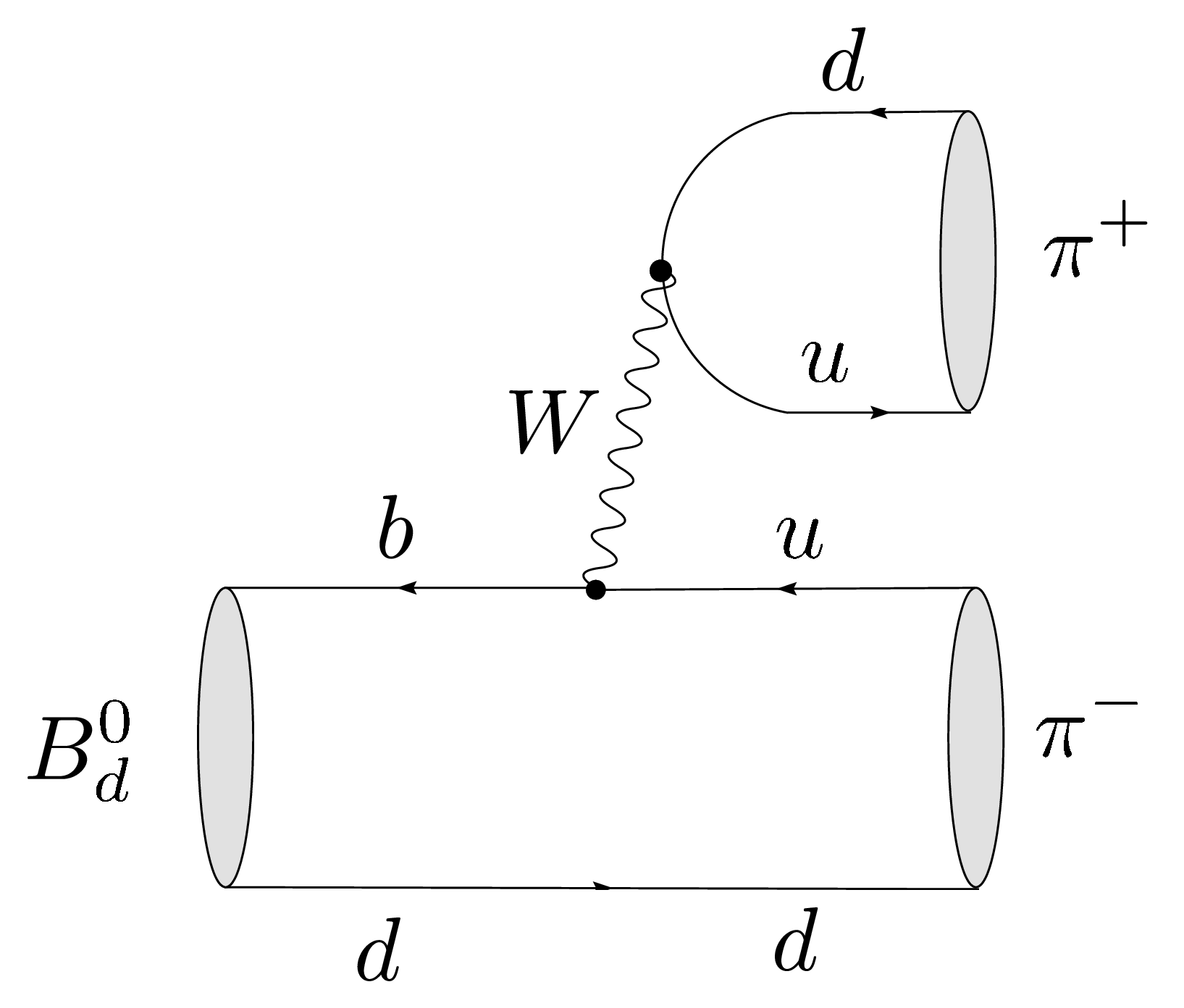}
 \hspace*{0.5truecm}
 \includegraphics[width=5.0truecm]{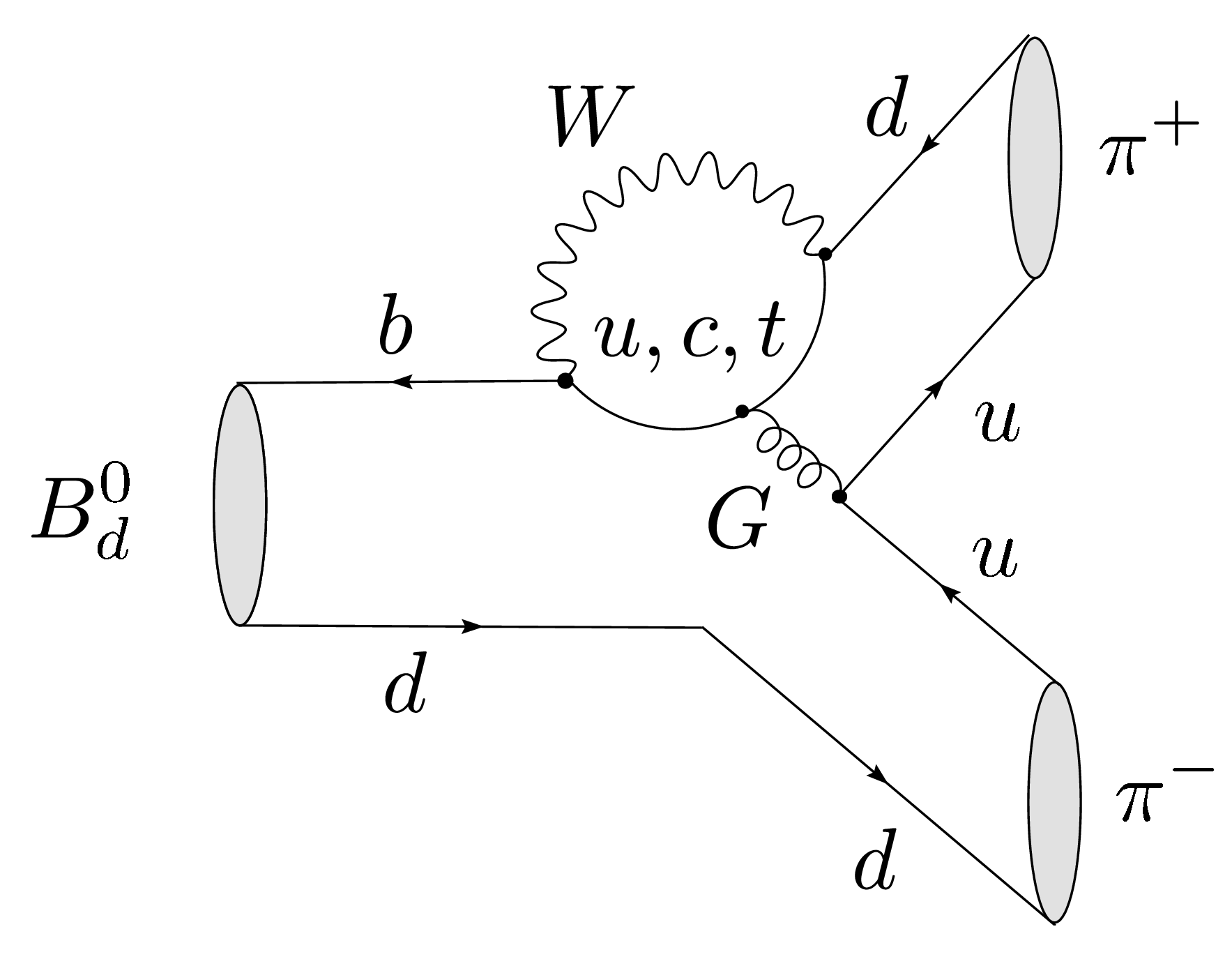}  
 }
 \vspace*{-0.5truecm}
\caption{Feynman diagrams contributing to the $B^0_d\to \pi^+\pi^-$
decay.}\label{fig:Bpipi-diag}
\end{figure}

\begin{figure}
   \centering
  \includegraphics[width=7.0truecm]{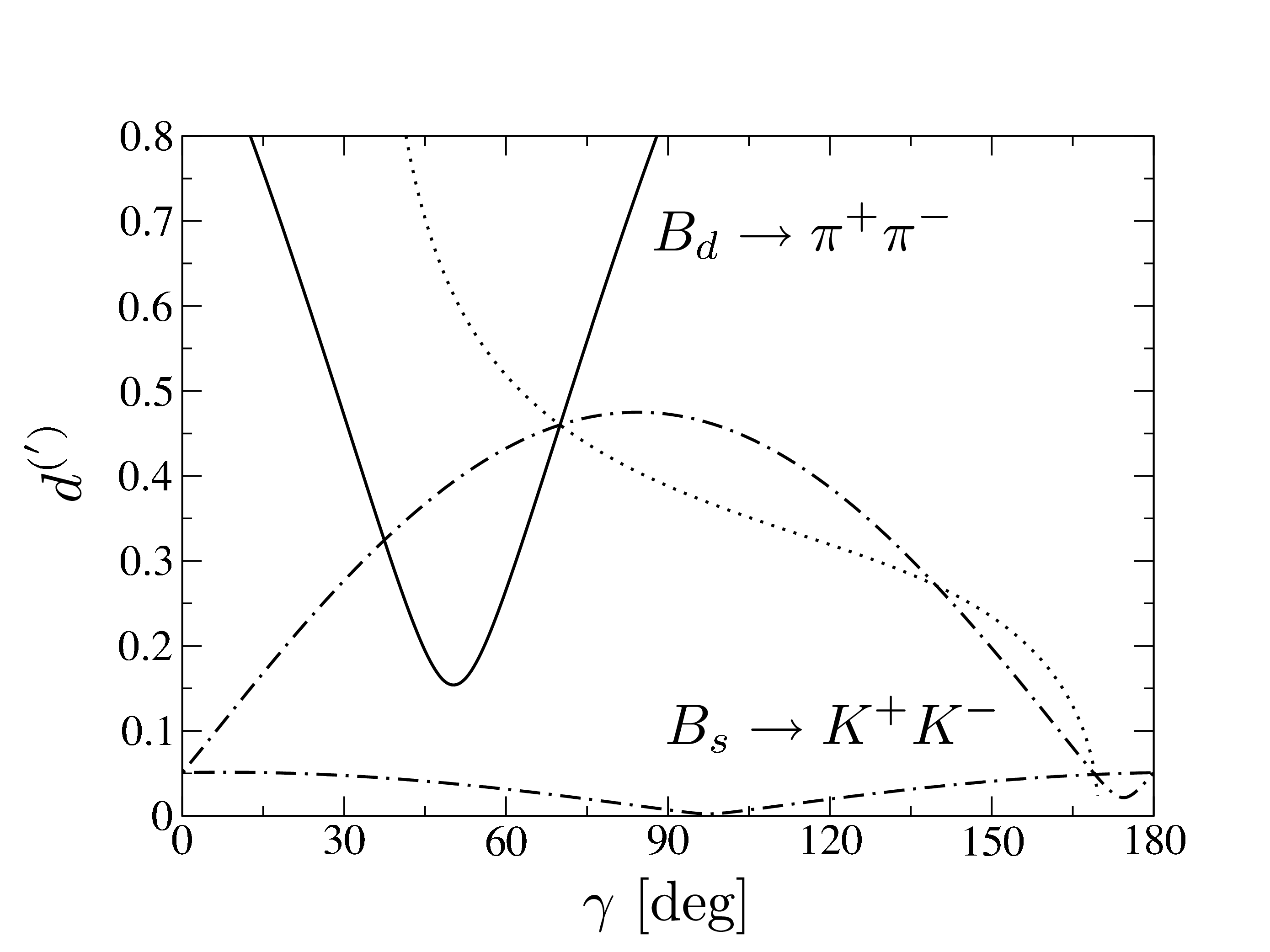} 
  \vspace*{-0.3truecm}
     \caption{The contours in the $\gamma$--$d^{(')}$ plane for an example 
     corresponding to the CP asymmetries
   ${\cal A}_{\rm CP}^{\rm dir}(B_d\to\pi^+\pi^-)=-0.24$ and 
   ${\cal A}_{\rm CP}^{\rm mix}(B_d\to\pi^+\pi^-)=+0.59$, as well as
   ${\cal A}_{\rm CP}^{\rm dir}(B_s\to K^+K^-)=+0.09$ and
   ${\cal A}_{\rm CP}^{\rm mix}(B_s\to K^+K^-)=-0.23$.}\label{fig:Bs-Bd-contours}
\end{figure}

As can be seen in Fig.~\ref{fig:BsKK-diag}, the decay $B^0_s\to K^+K^-$ is a 
$\bar b \to \bar s$ transition that involves tree and penguin contributions.
In analogy to the $B\to\pi K$ case discussed in Subsection~\ref{ssec:BpiK},  
the latter topologies play actually the dominant r\^ole in $B^0_s\to K^+K^-$.
If we replace the strange quarks in Fig.~\ref{fig:BsKK-diag} through down
quarks, we obtain the decay 
topologies for the $B^0_d\to\pi^+\pi^-$ channel shown in Fig.~\ref{fig:Bpipi-diag}.
However, because of the different CKM structure, the tree topologies play 
the dominant r\^ole in $B^0_d\to \pi^+\pi^-$, although the QCD penguins have
an important impact as well. Following the discussion of Subsections~\ref{ssec:BpsiK}
and \ref{ssec:BphiK}, we may write the corresponding decay amplitudes in
the SM as follows  \cite{RF-BsKK}:
\begin{eqnarray}
A(B^0_d\to\pi^+\pi^-)& \propto & \left[e^{i\gamma}-de^{i\theta}\right]\\
A(B_s^0\to K^+K^-)& \propto &
\left[e^{i\gamma}+\left(\frac{1-\lambda^2}{\lambda^2}\right)d'e^{i\theta'}\right],
\end{eqnarray}
where the CP-conserving hadronic parameters $de^{i\theta}$ and
$d'e^{i\theta'}$ descripe -- sloppily speaking -- the ratios of penguin to tree
contributions. The direct and mixing-induced CP asymmetries take then the 
following general form:
\begin{equation}
{\cal A}_{\rm CP}^{\rm dir}(B_d\to \pi^+\pi^-)=
G_1(d,\theta;\gamma), \quad
{\cal A}_{\rm CP}^{\rm mix}(B_d\to \pi^+\pi^-)=
G_2(d,\theta;\gamma,\phi_d)
\end{equation}
\begin{equation}
{\cal A}_{\rm CP}^{\rm dir}(B_s\to K^+K^-)=
G_1'(d',\theta';\gamma), \quad
{\cal A}_{\rm CP}^{\rm mix}(B_s\to K^+K^-)=
G_2'(d',\theta';\gamma,\phi_s).
\end{equation}
Since $\phi_d$ is already known  and  $\phi_s$ is negligibly small
in the SM -- or can be determined with the help of $B^0_s\to J/\psi \phi$ 
should CP-violating NP contributions to $B^0_s$--$\bar B^0_s$ mixing make 
it sizeable -- we may convert the measured values of 
${\cal A}_{\rm CP}^{\rm dir}(B_d\to \pi^+\pi^-)$, 
${\cal A}_{\rm CP}^{\rm mix}(B_d\to \pi^+\pi^-)$ and
${\cal A}_{\rm CP}^{\rm dir}(B_s\to K^+K^-)$, 
${\cal A}_{\rm CP}^{\rm mix}(B_s\to K^+K^-)$ into {\it theoretically clean}
contours in the $\gamma$--$d$ and $\gamma$--$d'$ planes, respectively.
In Fig.~\ref{fig:Bs-Bd-contours}, we show these contours (solid and dot-dashed) 
for an example, which is inspired by the current $B$-factory data.

Looking at the Feynman diagrams shown in Figs.~\ref{fig:BsKK-diag} and 
\ref{fig:Bpipi-diag}, we see that $B^0_d\to\pi^+\pi^-$ is actually related to 
$B^0_s\to K^+K^-$ through the interchange 
of all down and strange quarks. Consequently, each decay topology contributing
to $B^0_d\to\pi^+\pi^-$ has a counterpart in $B^0_s\to K^+K^-$ and vice versa, 
and the corresponding hadronic parameters can be related to each other
with the help of the $U$-spin flavour symmetry of strong interactions,
implying the following relations \cite{RF-BsKK}:
\begin{equation}\label{U-spin-rel}
d'=d, \quad \theta'=\theta.
\end{equation}
Applying the former, we may extract $\gamma$ and $d$ through the 
intersections of the theoretically clean $\gamma$--$d$ and $\gamma$--$d'$ 
contours. In the example of Fig.~\ref{fig:Bs-Bd-contours}, a twofold
ambiguity arises from the solid and dot-dashed curves. However, as 
discussed in Ref.~\cite{RF-BsKK}, it can be resolved with the help of the dotted 
contour, thereby leaving us with the ``true" solution of $\gamma=70^\circ$ in this 
case. Moreover, we may determine $\theta$ and $\theta'$, which allow an interesting 
internal consistency check of the second $U$-spin relation in (\ref{U-spin-rel}).

This strategy is very promising from an experimental point of view for LHCb, 
where an accuracy for $\gamma$ of a few degrees can be achieved 
\cite{LHCb-analyses}. As far as possible $U$-spin-breaking 
corrections to $d'=d$ are concerned, they enter the determination of $\gamma$ 
through a relative shift of the $\gamma$--$d$ and $\gamma$--$d'$ contours; 
their impact on the extracted value of $\gamma$ therefore depends on the form 
of these curves, which is fixed through the measured observables. In the examples discussed in Ref.~\cite{RF-BsKK} and Fig.~\ref{fig:Bs-Bd-contours}, the 
extracted value of $\gamma$ would be very stable with respect to such effects. 
It should also be noted
that the $U$-spin relations in (\ref{U-spin-rel}) are particularly robust since they 
involve only ratios of hadronic amplitudes, where all $SU(3)$-breaking decay constants
and form factors cancel in factorization and also chirally enhanced terms
would not lead to  $U$-spin-breaking corrections \cite{RF-BsKK}.

\begin{figure}
\centerline{
 \includegraphics[width=7.0truecm]{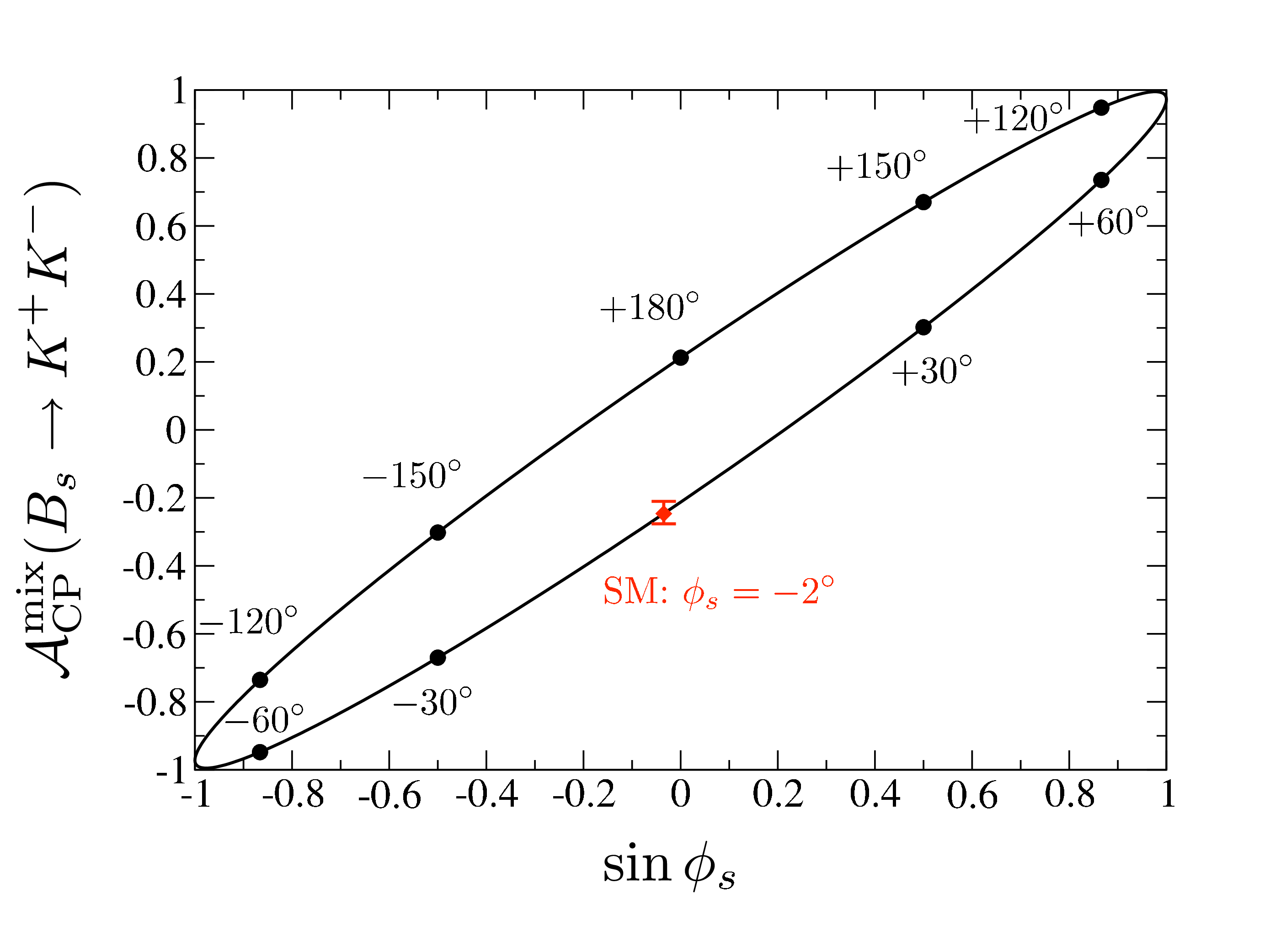}}
 \vspace*{-0.3truecm}
\caption{The correlation between $\sin\phi_s$, which can be determined 
through mixing-induced CP violation in $B^0_s\to J/\psi \phi$, and 
${\cal A}_{\rm CP}^{\rm mix}(B_s\to K^+K^-)$. Each point on the curve 
corresponds to a given value of $\phi_s$, as indicated by the numerical 
values \cite{RF-BsKK-07}.}\label{fig:Bs-NP-2}
\end{figure}

A detailed analysis of the status and prospects of the $B_{s,d}\to \pi\pi,\pi K, KK$ 
system in view of the first results on the $B_s$ modes from the Tevatron 
\cite{CDF-BsK+K-} was recently performed in Ref.~\cite{RF-BsKK-07}. Interestingly, 
the data for the  $B_d\to\pi^+\pi^-$, $B_s\to K^+K^-$ system favour the 
BaBar measurement of the direct CP violation in $B_d\to\pi^+\pi^-$, which 
results in a fortunate situation
for the extraction of $\gamma$, yielding $\gamma=(66.6^{+4.3+4.0}_{-5.0-3.0})^\circ$, 
where the latter errors correspond to a an estimate of $U$-spin-breaking effects. An important further step will be the measurement of the mixing-induced CP 
violation in $B_s\to K^+K^-$, which is predicted in the SM as 
${\cal A}_{\rm CP}^{\rm mix}(B_s\to K^+K^-)=-0.246^{+0.036+0.052}_{-0.030-0.024}$, 
where the second errors illustrate the impact of large non-factorizable 
$U$-spin-breaking corrections. In the case of CP-violating NP contributions to 
$B^0_s$--$\bar B^0_s$ mixing also this observable
would be sensitively affected, as can be seen in Fig.~\ref{fig:Bs-NP-2}, and
allows an unambiguous determination of  the $B^0_s$--$\bar B^0_s$ mixing 
phase with the help of $B_s\to J/\psi \phi$ at LHCb. Finally, the measurement 
of the direct CP violation in the $B_s\to K^+K^-$ channel will make the full exploitation 
of the physics potential of the $B_{s,d}\to \pi\pi, \pi K, KK$ modes possible.

\subsubsection{The Rare Decays $B^0_s\to\mu^+\mu^-$ and 
$B^0_d\to\mu^+\mu^-$}\label{ssec:Bmumu}
\begin{figure}[t]
\centerline{
 \includegraphics[width=4.6truecm]{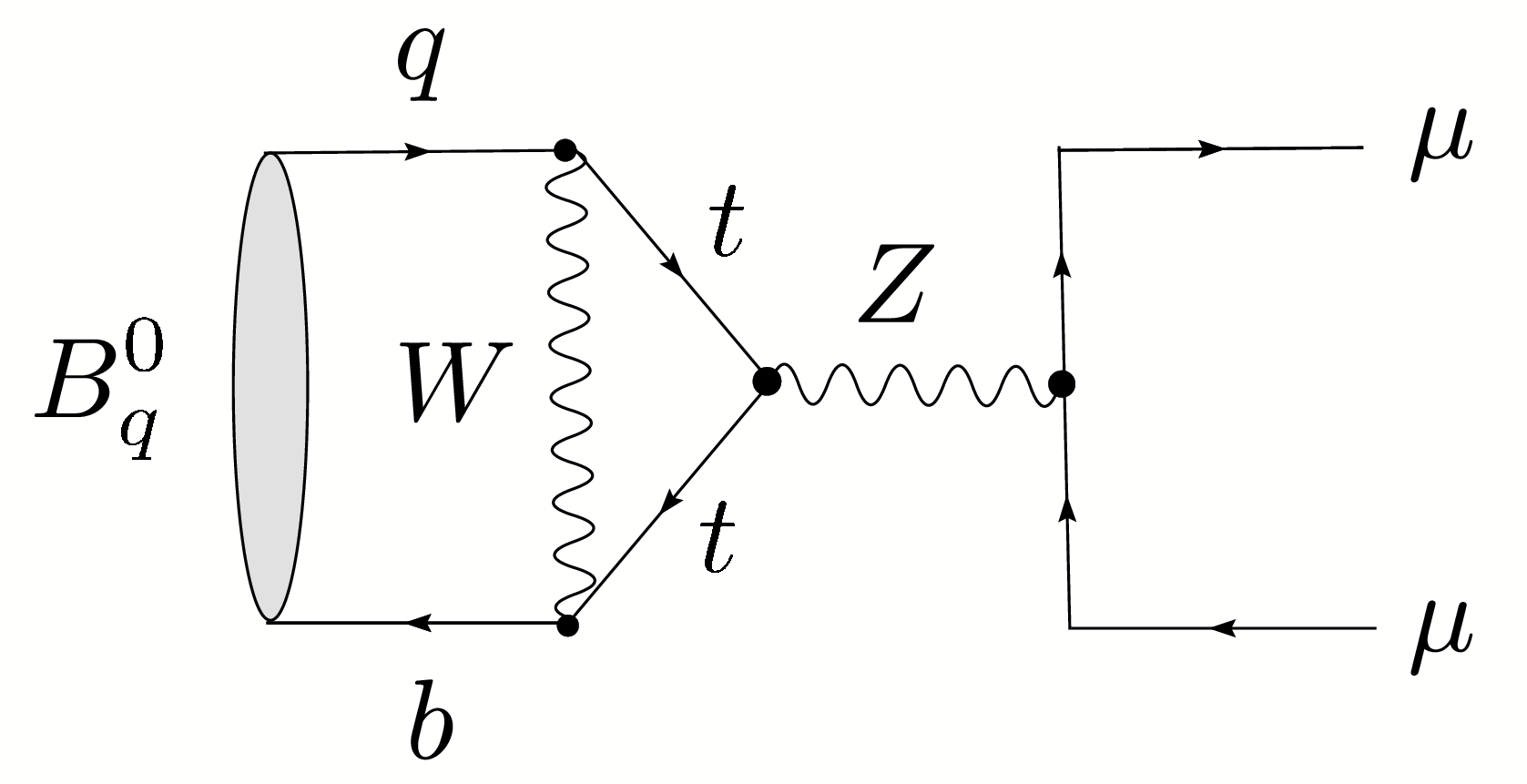}
 \hspace*{0.5truecm}
 \includegraphics[width=4.3truecm]{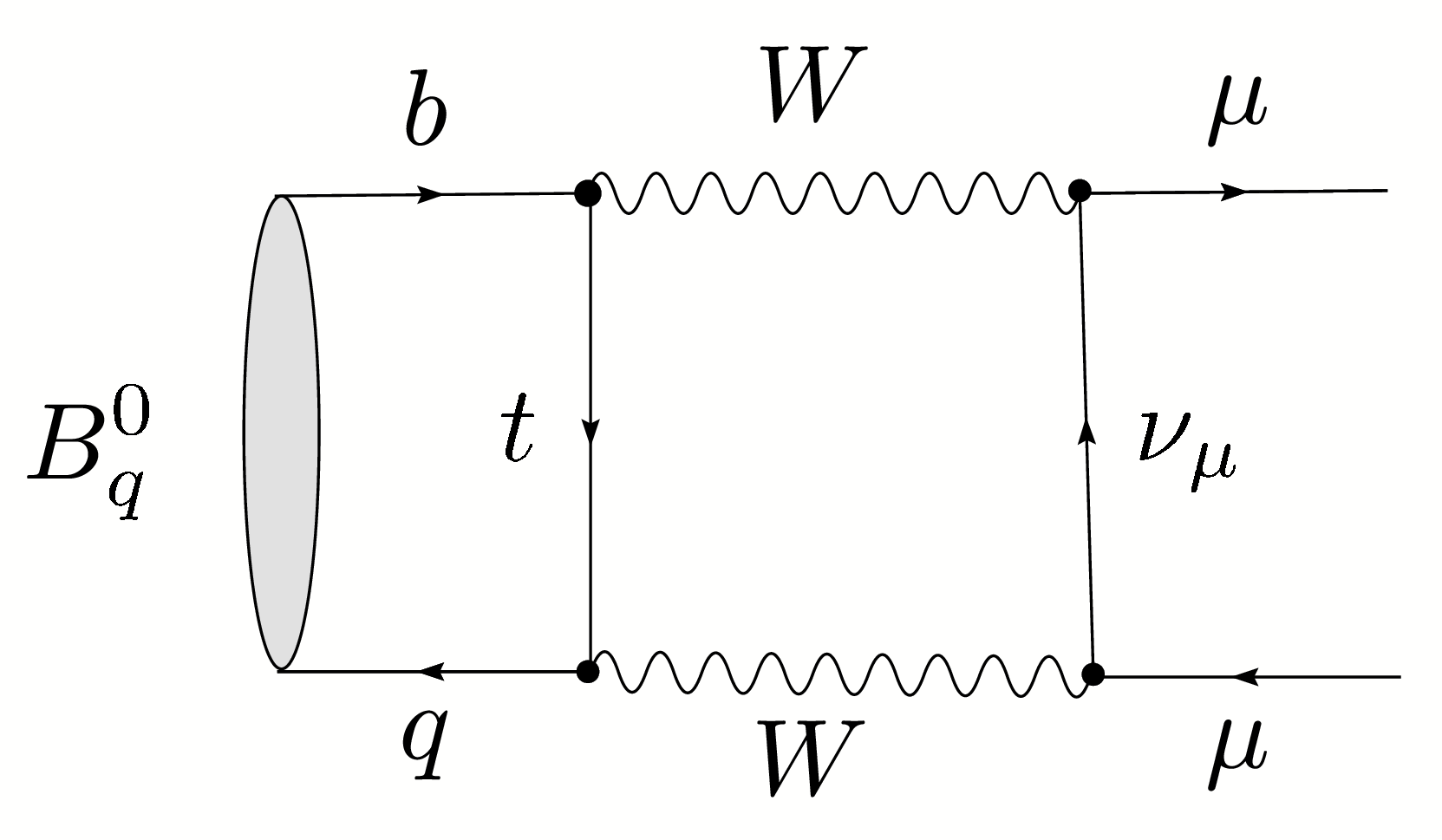}  
 }
\caption{Feynman diagrams contributing to 
$B^0_q\to \mu^+\mu^-$ ($q\in\{s,d\}$) decays.}\label{fig:Bqmumu}
\end{figure}

As can be seen in Fig.~\ref{fig:Bqmumu}, these decays originate from 
$Z^0$-penguin and box diagrams in the SM. The corresponding low-energy 
effective Hamiltonian is given as follows \cite{BBL-rev}:
\begin{equation}\label{Heff-Bmumu}
{\cal H}_{\rm eff}=-\frac{G_{\rm F}}{\sqrt{2}}\left[
\frac{\alpha}{2\pi\sin^2\Theta_{\rm W}}\right]
V_{tb}^\ast V_{tq} \eta_Y Y_0(x_t)(\bar b q)_{\rm V-A}(\bar\mu\mu)_{\rm V-A} 
\,+\, {\rm h.c.},
\end{equation}
where $\alpha$ denotes the QED coupling and $\Theta_{\rm W}$ is the
Weinberg angle. The short-distance physics is described by 
$Y(x_t)\equiv\eta_Y Y_0(x_t)$, where $\eta_Y=1.012$ is a perturbative 
QCD correction, and the Inami--Lim function
$Y_0(x_t)$ describes the top-quark mass dependence. We observe that
only the matrix element $\langle 0| (\bar b q)_{\rm V-A}|B^0_q\rangle$ 
is required. Since here the vector-current piece vanishes, as
the $B^0_q$ is a pseudoscalar meson, this matrix element is simply
given by the decay constant $f_{B_q}$. 
Consequently, we arrive at a very favourable 
situation with respect to the hadronic matrix elements. Since, moreover, 
NLO QCD corrections were calculated, and long-distance contributions are 
expected to play a negligible r\^ole \cite{Bmumu}, the $B^0_q\to\mu^+\mu^-$ 
modes belong to the cleanest rare $B$ decays.

Using also the data for the mass differences $\Delta M_q$ to reduce the hadronic 
uncertainties,\footnote{This input allows us to replace the decay constants $f_{B_q}$
through the bag parameters $\hat B_{B_q}$.} the following SM predictions were 
obtained in \cite{BBGT}:
\begin{eqnarray}
\mbox{BR}(B_s\to\mu^+\mu^-) &=& (3.35\pm0.32)\times 10^{-9}\\
\mbox{BR}(B_d\to\mu^+\mu^-) &=& (1.03\pm0.09)\times 10^{-10}.
\end{eqnarray}
Consequently, these branching ratios are extremely tiny. But things could actually
be much more exciting, as NP effects may significantly enhance 
$\mbox{BR}(B_s\to\mu^+\mu^-)$ \cite{WG2-rep}. The current upper 
bounds (95\% C.L.) from the CDF collaboration read as follows
\cite{CDF-Bmumu}:
\begin{equation}\label{Bmumu-exp-CDF}
\mbox{BR}(B_s\to\mu^+\mu^-)<5.8\times10^{-8}, \quad
\mbox{BR}(B_d\to\mu^+\mu^-)<1.8 \times10^{-8},
\end{equation}
while the D0 collaboration finds the following 90\% C.L. (95\% C.L.) upper limit 
\cite{D0-Bmumu}:
\begin{equation}\label{Bmumu-exp-D0}
\mbox{BR}(B_s\to\mu^+\mu^-)<7.5~(9.3) \times 10^{-8}.
\end{equation}
Consequently, there is still a long way to go within the SM. However, by the end
of 2009, with $0.5\,\mbox{fb}^{-1}$ data, LHCb can exclude or discover a 
NP contribution to $B_s\to\mu^+\mu^-$  at the level of the SM \cite{nakada}. 
This decay is also very interesting for ATLAS and CMS, where detailed signal
and background studies are currently in progress \cite{smsp}.

\subsubsection{The Rare Decay $B^0_d\to K^{*0}\mu^+\mu^-$}
The key observable for NP searches through this channel is the following
forward--backward asymmetry:
\begin{equation}
A_{\rm FB}(\hat s)=\frac{1}{{\rm d}\Gamma/{\rm d}\hat s}
\left[\int_0^{+1} {\rm d}(\cos\theta)\frac{{\rm d}^2\Gamma}{{\rm d} \hat s \, 
{\rm d}(\cos\theta)} - \int_{-1}^0{\rm d}(\cos\theta)
\frac{{\rm d}^2\Gamma}{{\rm d} \hat s \, {\rm d}(\cos\theta)}\right].
\end{equation}
Here $\theta$ is the angle between the $B^0_d$ momentum and that of the 
$\mu^+$ in the dilepton centre-of-mass system, and $\hat s \equiv s/M_B^2$ with 
$s=(p_{\mu^+}+p_{\mu^-})^2$. A particularly interesting kinematical point is
characterized by  
\begin{equation}
A_{\rm FB}(\hat s_0)|_{\rm SM}=0,
\end{equation}
as $\hat s_0$ is quite robust with respect to hadronic uncertainties 
(see, e.g., \cite{BKastll}). In SUSY extensions of the SM, $A_{\rm FB}(\hat s)$ 
could take opposite sign or take a dependence on $\hat s$ without a zero point 
\cite{ABHH}. The current $B$-factory data for the inclusive $b\to s\ell^+\ell^-$ 
branching ratios and the integrated forward--backward asymmetries are in 
accordance with the SM, but suffer still from large uncertainties. This situation will 
improve dramatically at the LHC. Here LHCb will measure the zero crossing
point by $\sim2013$ with $10\,\mbox{fb}^{-1}$ with 
$\sigma(s_0)=0.27 (\mbox{GeV}/c^2)^2$,
corresponding to 19k events \cite{nakada}. For other interesting observables 
provided by  $B^0_d\to K^{*0}\mu^+\mu^-$, see Ref.~\cite{matias-rare}. Also 
alternative $b\to s\mu^+\mu^-$ modes are currently under study, such as 
$B^0_s\to\phi\mu^+\mu^-$ and $\Lambda_b\to \Lambda\mu^+\mu^-$
\cite{nakada,smsp}.

\section{CONCLUSIONS AND OUTLOOK}\label{sec:concl}
\setcounter{equation}{0}
In this decade, we have seen tremendous progress in the exploration
of the phenomenology of flavour physics and CP violation that was made 
possible through a fruitful interplay between theory and experiment. Altogether, 
the $e^+e^-$ $B$ factories have already 
produced ${\cal O}(10^9)$ $B\bar B$ pairs, and the Tevatron has recently 
succeeded in observing $B^0_s$--$\bar B^0_s$ mixing. Thanks to 
these efforts, CP violation is now well established in the $B$-meson system, 
thereby complementing the neutral $K$-meson system, where this unexpected
effect was discovered almost 45 years ago. The $B$-factory data agree globally 
with the KM mechanism of CP violation in an impressive way, but we have 
also hints for possible discrepancies, which could be first footprints of NP in
the quark-flavour sector. Unfortunately, definite conclusions cannot yet be drawn 
as the uncertainties are still too large. 

Exciting new perspectives for $B$ physics and the exploration of CP violation 
will arise in the summer of 2008 through the start of the LHC, with its
dedicated $B$-decay experiment LHCb. Thanks to the large statistics that
can be collected there and the full exploitation of the physics potential of the
$B_s$-meson system, we will be able to enter a new territory in the investigation
of CP violation. The golden channel for the search of CP-violating NP 
contributions to $B^0_s$--$\bar B^0_s$ mixing is $B^0_s\to J/\psi \phi$, where the
recent measurement of $\Delta M_s$ still leaves ample space for such effects both in 
terms of the general NP parameters and in specific extensions of the SM. In contrast
to the theoretical interpretation of $\Delta M_s$, the corresponding CP asymmetries
have not to rely on non-perturbative lattice QCD calculations. Moreover, it will
be very interesting to search for CP-violating NP effects in $b\to s$ penguin
processes through the $B^0_s\to \phi\phi$ channel. These measurements will
be complemented by other key ingredients for the search of NP: precision 
measurements of the UT angle $\gamma$ by means of various processes with 
a very different dynamics, and powerful analyses of rare $B$ decays. 

In addition to $B$ physics, which was the focus of this lecture, there are 
other important flavour probes. An outstanding example is charm physics, 
where evidence for $D^0$--$\bar D^0$ mixing was found at the $B$ factories 
in the spring of 2007 \cite{D-mix}, and very recently also at CDF 
\cite{CDF-Dmix}. The mixing parameters are measured in the ball park of the 
SM predictions, which are unfortunately affected
by large long-distance effects. A striking NP signal would be given by CP-violating
effects (for recent theoretical analyses, see, e.g.\ Ref.\ \cite{D-th}). There is also a
powerful charm-physics programme at LHCb. As far as kaon physics is concerned,
the future lies on the extremely rare decays $K^+\to\pi^+\nu\bar\nu$ and 
$K_{\rm L}\to \pi^0\nu\bar\nu$: these are very clean from the theoretical point 
of view, but unfortunately hard to measure. Nevertheless, there is a proposal to 
take this challenge and to measure the former channel at the CERN SPS, and 
efforts to explore the latter -- even more difficult decay -- at J-PARC in Japan. 
Moreover, interesting flavour probes are offered by top physics and the flavour violation 
in the neutrino and charged lepton sectors; regarding the latter (for a recent overview,
see Ref.~\cite{WG3-rep}), an experimental investigation of the lepton flavour 
violating decay $\mu  \rightarrow e \gamma$  is the target of the MEG 
experiment at PSI, and studies of  $\mu \rightarrow e$ conversion are proposed at 
FNAL and J-PARC. Further studies in this direction using  $\tau$ decays at the 
LHC  and at a possible future super-$B$ factory will be important. Finally,  
continued searches of  electric dipole moments and measurements of the 
anomalous magnetic moment of the muon are  essential parts of the future 
experimental programme, providing also a strong interplay with theory. 

In view of the quickly approaching start of the LHC, there is a burning question: 
what is the synergy between the plenty of information following from analyses 
of the flavour sector with the high-$Q^2$ programme of ATLAS and CMS? 
The main goal of these experiments is to explore electroweak 
symmetry breaking, in particular the question of whether this is actually caused by 
the Higgs mechanism, to produce and observe new particles, which could themselves 
be the mediators of new flavour- and CP-violating interactions, and then to go
back to the deep questions of particle physics, such as the origin of dark matter
and the baryon asymmetry of the Universe. It is obvious that there should be a 
very fruitful interplay between these ``direct" studies of NP and the ``indirect" 
information provided by flavour physics \cite{CERN-WS}. I have no doubts 
that the next years will be extremely exciting!

\vskip1cm
\noindent

\section*{ACKNOWLEDGEMENTS}

I would like to thank the students for their interest in my lectures,
the discussion leaders for their efforts to complement them in the
discussion sessions, and the local organizers -- in particular Claudio Dib -- 
for hosting this exciting school in Vi\~na del Mar.


\begin{thebibliography}{999}


\bibitem{CP-obs}J.H. Christenson {\it et al.}, 
{Phys.\ Rev.\ Lett.}~{\bf 13} (1964) 138.

\bibitem{eps-prime}V.~Fanti {\it et al.}\  [NA48 Collaboration],
{ Phys.\ Lett.}~{\bf B465} (1999) 335;\\
A.~Alavi-Harati {\it et al.}\  [KTeV Collaboration],
{ Phys.\ Rev.\ Lett.}~{\bf 83} (1999) 22.

\bibitem{NA48-final}J.R.~Batley {\it et al.}\  [NA48 Collaboration],
{ Phys.\ Lett.}~{\bf B544} (2002) 97.

\bibitem{KTeV-final}A.~Alavi-Harati {\it et al.}\  [KTeV Collaboration],
{ Phys.\ Rev.}~{\bf D67} (2003) 012005.

\bibitem{CP-B-obs}B. Aubert {\it et al.}\ [BaBar Collaboration],
{ Phys.\ Rev.\ Lett.}~{\bf 87} (2001) 091801;\\
K. Abe {\it et al.}\ [Belle Collaboration],
{ Phys.\ Rev.\ Lett.}~{\bf 87} (2001) 091802.

\bibitem{superB} M.~Bona {\it et al.},
  arXiv:0709.0451 [hep-ex];
 A.~G.~Akeroyd {\it et al.},
  arXiv:hep-ex/0406071.
  
\bibitem{Belle-U5S}K.~Abe {\it et al.}  [Belle Collaboration],
  hep-ex/0610003;\\
  J.~Wicht  {\it et al.}\ [Belle Collaboration],
  arXiv:0712.2659 [hep-ex].
  
\bibitem{nakada}T.~Nakada, talk at 2nd Workshop on Flavour Dynamics,
Albufeira, Portugal, 3--10 November 2007 
[http://www.theorie.physik.uni-muenchen.de/lsfritzsch/albufeira/].

\bibitem{WG2-rep}G.~Buchalla {\it et al.},
  arXiv:0801.1833 [hep-ph],
Report of Working Group 2 of the CERN Workshop ``Flavour in the era of the LHC'', Geneva, Switzerland, November 2005 -- March 2007.
  
\bibitem{wagner}C. Wagner,  lecture given at this school.

\bibitem{barenboim}G. Barenboim,  lecture given at this school.

\bibitem{ellis}J. Ellis,  lecture given at this school.

\bibitem{sach}A.D. Sakharov,
 { JETP Lett.}~{\bf 5} (1967) 24.
  
\bibitem{shapos}V.A. Rubakov, M.E. Shaposhnikov, 
{ Usp.\ Fiz.\ Nauk} {\bf 166} (1996) 493; 
{ Phys.\ Usp.}~{\bf 39} (1996) 461;\\
A. Riotto and M. Trodden, { Annu.\ Rev.\ Nucl.\ Part.\ 
Sci.}~{\bf 49} (1999) 35.

\bibitem{LG-rev}For a recent review, see W. Buchm\"uller, R.D. Peccei 
and T. Yanagida,
  Ann.\ Rev.\ Nucl.\ Part.\ Sci.\  {\bf 55} (2005) 311.

\bibitem{cab}N. Cabibbo,
  { Phys.\ Rev.\ Lett.}~{\bf 10} (1963) 531.

\bibitem{KM}M. Kobayashi and T. Maskawa,
{ Prog.\ Theor.\ Phys.}~{\bf 49} (1973) 652.

\bibitem{CKM-book}M. Battaglia {\it et al.}, CERN 2003-002-corr,
{\it The CKM matrix and the unitarity triangle}
(CERN, Geneva, 2003) [hep-ph/0304132].

\bibitem{epsp-rev}For a recent review, see A.J. Buras and M. Jamin,
  { JHEP} {\bf 0401} (2004) 048.

\bibitem{superweak}L. Wolfenstein,
  { Phys.\ Rev.\ Lett.}~{\bf 13} (1964) 562.
  
  \bibitem{BF-rev}A.J. Buras and R. Fleischer,
Adv.\ Ser.\ Direct.\ High Energy Phys.\  {\bf 15} (1998) 65.

\bibitem{BLS-textbook}G. Branco, L. Lavoura 
and J. Silva,  {\it CP Violation}, International Series of Monographs on 
Physics 103, Oxford Science Publications (Clarendon Press, Oxford, 1999).

\bibitem{BiSa-textbook}I.I. Bigi and A. I. Sanda, {\it CP Violation},
Cambridge Monographs on Particle Physics, Nuclear Physics and 
Cosmology (Cambridge University Press, Cambridge, 2000).

\bibitem{mannel-book}T.~Mannel,
  Springer Tracts Mod.\ Phys.\  {\bf 203} (2004) 1.

\bibitem{buras-spain}A.J. Buras,
  lectures given at 2004 European School of High-Energy Physics, 
  Sant Feliu de Guixols, Barcelona, Spain, 30 May -- 12 June 2004
  [hep-ph/0505175]. 

\bibitem{bigi-lecture}I.I.~Bigi,
  lectures given at 2007 European School of High-Energy Physics, Aronsborg, 
  Sweden, 18 June -- 1 July 2006. 
  [hep-ph/0701273].
  
\bibitem{nir-lecture}Y.~Nir,  arXiv:0708.1872 [hep-ph],
lectures given at 2nd Workshop on Monte Carlo Tools for Beyond the Standard Model Physics (MC4BSM), Princeton, New Jersy, 23--24 March 2007 and at the 2nd Joint Fermilab-CERN Hadron Collider Physics Summer School, CERN, Geneva, Switzerland, 6--15 June 2007. 
    
\bibitem{ali-rev}A.~Ali,
  arXiv:0712.1022 [hep-ph].
  
\bibitem{gronau-rev}M.~Gronau,
  Int.\ J.\ Mod.\ Phys.\  {\bf A22} (2007) 1953.

\bibitem{HoLi-rev}A.~H\"ocker and Z.~Ligeti,
  Ann.\ Rev.\ Nucl.\ Part.\ Sci.\  {\bf 56} (2006) 501.
    
\bibitem{SM}S.L. Glashow, { Nucl.\ Phys.}~{\bf 22} (1961) 579;\\
S. Weinberg, { Phys.\ Rev.\ Lett.}~{\bf 19} (1967) 1264;\\ 
A. Salam, in {\it Elementary Particle Theory}, ed.\ N. Svartholm 
(Almqvist and Wiksell, Stockholm, 1968).

\bibitem{pich}A.~Pich,  lecture given at this school,
  arXiv:0705.4264 [hep-ph].
 
\bibitem{GIM}S.L. Glashow, J. Iliopoulos and L. Maiani,
{ Phys.\ Rev.}~{\bf D2} (1970) 1285.

\bibitem{PDG}S. Eidelman {\it et al.}\ [Particle Data Group],
{  Phys.\ Lett.}~{\bf B592}, 1 (2004).

\bibitem{jarlskog}C. Jarlskog, { Phys.\ Rev.\ Lett.}~{\bf 55}
(1985) 1039; { Z. Phys.}~{\bf C29} (1985) 491.

\bibitem{BBG}J. Bernabeu, G. Branco and M. Gronau, { Phys.\ 
Lett.}~{\bf B169} (1986) 243.

\bibitem{PDG-n} W.M.~Yao {\it et al.}  [Particle Data Group],
 { J.\ Phys.}~{\bf G33} (2006) 1.
  
\bibitem{wolf}L. Wolfenstein, { Phys.\ Rev.\ Lett.}~{\bf 51} (1983)
1945.

\bibitem{blo}A.J. Buras, M.E. Lautenbacher and G. Ostermaier, 
{ Phys.\ Rev.}~{\bf D50} (1994) 3433. 

\bibitem{Brev01}A.J. Buras, hep-ph/0101336,
lectures given at Erice International School 
of Subnuclear Physics: Theory and Experiment Heading for New Physics, 
Erice, Italy, 27 August -- 5 September 2000.

\bibitem{AKL}R. Aleksan, B. Kayser and D. London, { Phys.\ Rev.\ 
Lett.}~{\bf 73} (1994) 18.

\bibitem{JS}C. Jarlskog and R. Stora, { Phys.\ Lett.}~{\bf B208} (1988) 268.

\bibitem{ut}L.L. Chau and W.-Y. Keung, { Phys.\ Rev.\ 
Lett.}~{\bf 53} (1984) 1802.

\bibitem{CKMfitter}J.~Charles {\it et al.}~[CKMfitter Group], 
{Eur.\ Phys.\ J.}~C {\bf 41} (2005) 1; for the most recent updates, see
http://ckmfitter.in2p3.fr/.

\bibitem{UTfit}M.~Bona {\it et al.}~[UTfit Collaboration],
  {JHEP} {\bf 0507} (2005) 028; for the most recent updates, see
http://utfit.roma1.infn.it/.
    
\bibitem{kop}B. Kopeliovich, lectures given at this school.

\bibitem{khod}A.~Khodjamirian,
  lectures given at the 2003 European School on High-Energy Physics, Tsakhkadzor, Armenia, 24 August -- 6 September 2003 [hep-ph/0403145].

\bibitem{luscher}M. L\"uscher,
Annales Henri Poincare {\bf 4} (2003) S197
[hep-ph/0211220];
  PoS {\bf LAT2005} (2006) 002.
  
\bibitem{sach-lat}C.T.~Sachrajda,
  { AIP Conf.\ Proc.}~{\bf 842} (2006) 198.
  
\bibitem{delmo} M.~Della Morte,
  PoS {\bf LAT2007}, 008 (2007).

\bibitem{fulvia}F. De Fazio,
hep-ph/0010007.

\bibitem{Belle-leptonic} K.~Ikado {\it et al.},
  { Phys.\ Rev.\ Lett.}~{\bf 97} (2006) 251802.
  
  \bibitem{Babar-leptonic}B.~Aubert {\it et al.}~[BaBar Collaboration],
  hep-ex/0608019.
  
\bibitem{hou}W.S.~Hou,
  { Phys.\ Rev.}~{\bf D48} (1993) 2342.
  
\bibitem{browder}T.E.~Browder,
 {  Nucl.\ Phys.\ Proc.\ Suppl.}~{\bf 163} (2007) 117.
  
\bibitem{CLEO-c}D.G. Cassel,
eConf {\bf C0304052} (2003) WG501 [hep-ex/0307038].

\bibitem{davies-eps}C. Davies, plenary talk at HEP2005 Europhysics Conference, 
Lisbon, Portugal, 21--27 July 2005, http://www.lip.pt/events/2005/hep2005/.
  
\bibitem{fD-lat} C.~Aubin {\it et al.},
  Phys.\ Rev.\ Lett.\  {\bf 95} (2005) 122002.

\bibitem{fD-cleoc}M.~Artuso {\it et al.}\  [CLEO Collaboration],
  Phys.\ Rev.\ Lett.\  {\bf 95} (2005) 251801.

\bibitem{ro-st} J.L.~Rosner and S.~Stone,
  arXiv:0802.1043 [hep-ex].

\bibitem{IW}N. Isgur and M.B. Wise,
{ Phys.\ Lett.}~{\bf B232} (1989) 113 and
{\bf B237} (1990) 527.

\bibitem{neubert-rev}M. Neubert, { Phys.\ Rep.}~{\bf 245} (1994) 259.

\bibitem{BaBar-book}{\it The BaBar Physics Book}, eds.\ P. Harrison and
H.R. Quinn, SLAC-R-504 (1998).

\bibitem{neu-BDast}M.~Neubert,
{ Phys.\ Lett.}~{\bf B264} (1991) 455.

\bibitem{luke}M.E. Luke,
{ Phys.\ Lett.}~{\bf B252} (1990) 447.

\bibitem{Gambino}
  P.~Gambino and N.~Uraltsev,
  { Eur.\ Phys.\ J.} {\bf C34} (2004) 181.

\bibitem{OBuchmuller}
  O.~Buchm\"uller and H.~Fl\"acher,
 Phys.\ Rev.\ {\bf D73} (2006) 073008.
  
\bibitem{BF-DMs}P.~Ball and R.~Fleischer,
  { Eur.\ Phys.\ J.}~{\bf C48} (2006) 413.
 
\bibitem{HFAG}Heavy Flavour Averaging Group [E. Barberio {\it et al.}], 
  arXiv:0704.3575 [hep-ex];
for online updates, see  http://www.slac.stanford.edu/xorg/hfag. 
  
\bibitem{Vublatt}
M. Okamoto {\it et al.},
 {Nucl.\ Phys.\ Proc.\ Suppl.} {\bf 140} (2005) 461;\\
E.~Gulez {\it et al.},
{ Phys.\ Rev.} {\bf D73} (2006) 074502.

\bibitem{LCSR}A.~Khodjamirian {\it et al.},
{  Phys.\ Rev.} {\bf D62} (2000) 114002;\\
  P.~Ball and R.~Zwicky,
{  JHEP} {\bf 0110} (2001) 019;
{  Phys.\ Rev.} {\bf D71} (2005) 014015;
{ Phys.\ Rev.} {\bf D71} (2005) 014029;
{  Phys.\ Lett.} {\bf B625} (2005) 225.

\bibitem{blucher}E.~Blucher {\it et al.},
  hep-ph/0512039.

\bibitem{HEFF-TREE}F.J. Gilman and M.B. Wise,
{ Phys.\ Rev.}~{\bf D20} (1979) 2392;\\
G. Altarelli, G. Curci, G. Martinelli and S. Petrarca,
{ Phys.\ Lett.}~{\bf B99} (1981) 141;\\
A.J. Buras and P.H. Weisz,
{ Nucl.\ Phys.}~{\bf B333} (1990) 66.

\bibitem{BBL-rev}G. Buchalla, A.J. Buras and M.E. Lautenbacher,
{ Rev.\ Mod.\ Phys.}~{\bf 68} (1996) 1125.

\bibitem{BSS}M. Bander, D. Silverman and A. Soni,
{ Phys.\ Rev.\ Lett.}~{\bf 43} (1979) 242.

\bibitem{RF-DIPL}R. Fleischer,
{ Z.\ Phys.}~{\bf C58} (1993) 483.

\bibitem{BF-PEN}A.J. Buras and R. Fleischer,
{ Phys.\ Lett.}~{\bf B341} (1995) 379.

\bibitem{CHARM-PEN}M. Ciuchini, E. Franco, G. Martinelli, M. Pierini 
and L. Silvestrini,
{ Phys.\ Lett.}~{\bf B515} (2001) 33;\\
C. Isola, M. Ladisa, G. Nardulli, T.N. Pham and P. Santorelli,
{ Phys.\ Rev.}~{\bf D65} (2002) 094005;\\
C.W. Bauer, D. Pirjol, I.Z. Rothstein and I.W. Stewart,
Phys.\ Rev.\  {\bf D70} (2004) 054015.

\bibitem{RF-EWP}R. Fleischer,
{ Z.\ Phys.}~{\bf C62} (1994) 81;
{ Phys.\ Lett.}~{\bf B321} (1994) 259 and
{\bf B332} (1994) 419.

\bibitem{RF-rev}R. Fleischer,
{ Int.\ J.\ Mod.\ Phys.}~{\bf A12} (1997) 2459.

\bibitem{EWP-BpiK}N.G. Deshpande and X.-G. He,
{ Phys.\ Rev.\ Lett.}~{\bf 74} (1995) 26 [E: ibid., p.\ 4099];\\
M. Gronau, O.F. Hernandez, D. London and J.L. Rosner,
{ Phys.\ Rev.}~{\bf D52} (1995) 6374.

\bibitem{PAPIII}R. Fleischer,
 { Phys.\ Lett.}~{\bf B365} (1996) 399.

\bibitem{Neu-Ste}M. Neubert, B. Stech,
{ Adv.\ Ser.\ Direct.\ High Energy Phys.}~{\bf 15} (1998) 294,
and references therein.

\bibitem{largeN}A.J. Buras and J.-M. G\'erard,
{ Nucl.\ Phys.}~{\bf B264} (1986) 371;\\
A.J. Buras, J.-M. G\'erard and R. R\"uckl,
{ Nucl.\ Phys.}~{\bf B268} (1986) 16.

\bibitem{BBNS}M. Beneke, G. Buchalla, M. Neubert and
C. Sachrajda,
{ Phys.\ Rev.\ Lett.}~{\bf 83} (1999) 1914;
{ Nucl.\ Phys.}~{\bf B591} (2000) 313;
{Nucl.\ Phys.}~{\bf B606} (2001) 245.

\bibitem{QCDF-old}J.D. Bjorken, { Nucl.\ Phys.\ (Proc.\ Suppl.)}
{\bf B11} (1989) 325;\\
M. Dugan and B. Grinstein, { Phys.\ Lett.}~{\bf B255} (1991) 583;\\
H.D. Politzer and M.B. Wise, { Phys.\ Lett.}~{\bf B257} (1991) 399.

\bibitem{PQCD}H.-n.~Li and H.L.~Yu,
{Phys.\ Rev.}~{\bf D53} (1996) 2480;\\
Y.Y.~Keum, H.-n.~Li and A.I.~Sanda,
{Phys.\ Lett.}\ {\bf B504} (2001) 6;\\
Y.Y.~Keum and H.-n.~Li,
{Phys.\ Rev.}~{\bf D63} (2001) 074006;\\
A.~Ali {\it et al.}, 
  Phys.\ Rev.~{\bf D76} (2007) 074018.

\bibitem{SCET}C.W.~Bauer, D.~Pirjol and I.W.~Stewart,
{Phys.\ Rev.\ Lett.}~{\bf 87} (2001) 201806;\\
C.W.~Bauer, B.~Grinstein, D.~Pirjol and I.W.~Stewart,
{Phys.\ Rev.}~{\bf D67} (2003) 014010.

\bibitem{sum-rules}A. Khodjamirian,
{ Nucl.\ Phys.}~{\bf B605} (2001) 558;\\
A. Khodjamirian, T. Mannel and B. Melic,
{ Phys.\ Lett.}~{\bf B571} (2003) 75.

\bibitem{BFRS2}A.J. Buras, R. Fleischer, S. Recksiegel and F. Schwab,
{ Phys.\ Rev.\ Lett.}~{\bf 92} (2004) 101804.

\bibitem{BFRS3}A.J. Buras, R. Fleischer, S. Recksiegel and F. Schwab,
  { Nucl.\ Phys.}~{\bf B697} (2004) 133.

\bibitem{ALP-Bpipi}A. Ali, E. Lunghi and A.Y. Parkhomenko,
{ Eur.\ Phys.\ J.}~{\bf C36} (2004) 183.

\bibitem{CGRS}C.W. Chiang, M. Gronau, J.L. Rosner and D.A. Suprun,
  { Phys.\ Rev.}~{\bf D70} (2004) 034020.

\bibitem{CP-B-dir}B. Aubert {\it et al.}\ [BaBar Collaboration],
  { Phys.\ Rev.\ Lett.}~{\bf 93} (2004) 131801;\\
  Y. Chao {\it et al.}\  [Belle Collaboration],
  { Phys.\ Rev.\ Lett.}~{\bf 93} (2004) 191802.
  
\bibitem{gw}M. Gronau and D. Wyler,
{ Phys.\ Lett.}~{\bf B265} (1991) 172.

\bibitem{ADS}D. Atwood, I. Dunietz, A. Soni,
{ Phys.\ Rev.\ Lett.}~{\bf 78} (1997) 3257;
{ Phys.\ Rev.}~{\bf D63} (2001) 036005.

\bibitem{fw}R. Fleischer and D. Wyler,
{ Phys.\ Rev.}~{\bf D62} (2000) 057503.

\bibitem{GHLR}M. Gronau, J.L. Rosner and D. London,
  { Phys.\ Rev.\ Lett.}~{\bf 73} (1994) 21;\\
 M. Gronau, O.F. Hernandez, D. London and J.L. Rosner,
  { Phys.\ Rev.}~{\bf D50} (1994) 4529.
  
\bibitem{RF-BsKK}R. Fleischer,
  { Phys.\ Lett.}~{\bf B459} (1999) 306.

\bibitem{bisa}A.B. Carter and A.I. Sanda,
{ Phys.\ Rev.\ Lett.}~{\bf 45} (1980) 952;
{ Phys.\ Rev.}~{\bf D23} (1981) 1567;\\
I.I. Bigi and A.I. Sanda,
{ Nucl.\ Phys.}~{\bf B193} (1981) 85.

\bibitem{WG-sum}G.~Cavoto, R.~Fleischer, K.~Trabelsi and J.~Zupan,
  arXiv:0706.4227 [hep-ph].
  
\bibitem{RF-gam-ca}R. Fleischer,
 {  Nucl.\ Phys.}~{\bf B671} (2003) 459.

\bibitem{BFRS-5}A.J. Buras, R. Fleischer, S. Recksiegel and F. Schwab,
  Eur.\ Phys.\ J.\  {\bf C45} (2006) 701.
  
\bibitem{CDF-Bc}F. Abe {\it et al.}  [CDF Collaboration],
{ Phys.\ Rev.\ Lett.}~{\bf 81} (1998) 2432.

\bibitem{D0-Bc}D0 Collaboration, D0 Note 4539-CONF (August 2004).

\bibitem{CDF-Bc-nl}D.~Acosta {\it et al.}\  [CDF Collaboration],
  Phys.\ Rev.\ Lett.\  {\bf 96} (2006) 082002.

\bibitem{masetti}M. Masetti, { Phys.\ Lett.}~{\bf B286} (1992) 160.

\bibitem{IKP}M.A. Ivanov, J.G. K\"orner and O.N. Pakhomova,
{ Phys.\ Lett.}~ {\bf B555} (2003) 189.

\bibitem{IL}T. Inami and C.S. Lim,
{ Prog.\ Theor.\ Phys.}~{\bf 65} (1981) 297
[E: ibid., p.\ 1772].

\bibitem{LLNP}S. Laplace, Z. Ligeti, Y. Nir and G. Perez,
{Phys.\ Rev.}~{\bf D65} (2002) 094040.

\bibitem{BBLN-CFLMT}M. Beneke, G. Buchalla, A. Lenz and U. Nierste,
{ Phys.\ Lett.}~{\bf B576} (2003) 173;\\
M.~Ciuchini, E.~Franco, V.~Lubicz, F.~Mescia and C.~Tarantino,
{ JHEP} {\bf 0308} (2003) 031.

\bibitem{LEPBOSC}$B$ Oscillations Working Group:
{\tt http://lepbosc.web.cern.ch/LEPBOSC/}.

\bibitem{DMs-obs}V.M.~Abazov {\it et al.}\  [D0 Collaboration],
   Phys.\ Rev.\ Lett.\  {\bf 97} (2006) 021802;\\
   A.~Abulencia {\it et al.}  [CDF Collaboration],
  Phys.\ Rev.\ Lett.\  {\bf 97} (2006) 062003. 

\bibitem{D0-DMs}The D0 Collaboration, D0note 5474-conf (2007)
[http://www-d0.fnal.gov].
   
\bibitem{CDF-DMs}A.~Abulencia {\it et al.}  [CDF Collaboration],
  Phys.\ Rev.\ Lett.\  {\bf 97} (2006) 242003. 

\bibitem{lenz}A.~Lenz,
  arXiv:0710.0940 [hep-ph].
  
\bibitem{DDF}A.S. Dighe, I. Dunietz and R. Fleischer,
{ Eur.\ Phys.\ J.}~{\bf C6} (1999) 647.

\bibitem{D0-DG}V.M.~Abazov {\it et al.}  [D0 Collaboration],
  Phys.\ Rev.\ Lett.\  {\bf 98} (2007) 121801.

\bibitem{CDF-DG}T.~Aaltonen {\it et al.}\ [CDF Collaboration],
  arXiv:0712.2348 [hep-ex].

\bibitem{smsp}A.~Krasznahorkay,
  ``Outlook for B physics at the LHC in ATLAS and CMS,'' in the proceedings of
  DIS 2007, Munich, Germany, 16--20 April 2007.

\bibitem{dun}I. Dunietz,
{ Phys.\ Rev.}~{\bf D52} (1995) 3048.

\bibitem{FD-CP}R. Fleischer and I. Dunietz,
{ Phys.\ Rev.}~{\bf D55} (1997) 259.

\bibitem{FD-NCP}R. Fleischer and I. Dunietz,
{ Phys.\ Lett.}~{\bf B387} (1996) 361.

\bibitem{DFN}I. Dunietz, R. Fleischer and U. Nierste,
  { Phys.\ Rev.}~{\bf D63} (2001) 114015.

\bibitem{RF-BdsPsiK}R. Fleischer,
  { Eur.\ Phys.\ J.}~{\bf C10} (1999) 299.

\bibitem{FM-BpsiK}R. Fleischer and T. Mannel,
  { Phys.\ Lett.}~{\bf B506} (2001) 311.

\bibitem{BMR}H. Boos, T. Mannel and J. Reuter,
  { Phys.\ Rev.}~{\bf D70} (2004) 036006.
  
\bibitem{CPS}M. Ciuchini, M. Pierini and L. Silvestrini,
  Phys.\ Rev.\ Lett.\  {\bf 95} (2005) 221804.
  
\bibitem{LHC-Book}P. Ball {\it et al.}, hep-ph/0003238,
in CERN Report on {\it Standard Model physics (and more) at
the LHC} (CERN, Geneva, 2000), p.\ 305.

\bibitem{FlMa}R. Fleischer and J. Matias,
  { Phys.\ Rev.}~{\bf D66} (2002) 054009.
  
\bibitem{FIM}R. Fleischer, G. Isidori and J. Matias,
  { JHEP} {\bf 0305} (2003) 053.

\bibitem{buras-NP}A.J.~Buras,
  Springer Proc.\ Phys.\  {\bf 98} (2005) 315.
  
\bibitem{MFV-1}A.J. Buras, P. Gambino, M. Gorbahn, S. J\"ager and L. Silvestrini,
  { Phys.\ Lett.}~{\bf B500} (2001) 161.

\bibitem{MFV-2}G. D'Ambrosio, G.F. Giudice, G. Isidori and A. Strumia,
  { Nucl.\ Phys.}~{\bf B645} (2002) 155.


\bibitem{RF-Phys-Rep}R. Fleischer,
  { Phys.\ Rep.}~{\bf 370} (2002) 537.

\bibitem{rare}For reviews, see, for instance, 
A. Ali,
  hep-ph/0412128;\\
  G. Isidori,
  { AIP Conf.\ Proc.}~{\bf 722} (2004) 181;\\
  M. Misiak,
  { Acta Phys.\ Polon.}~{\bf B34} (2003) 4397;\\
  T.~Hurth,
  Rev.\ Mod.\ Phys.\  {\bf 75} (2003) 1159.
  
\bibitem{FM-BphiK}R. Fleischer and T. Mannel,
 { Phys.\ Lett.}~{\bf B511} (2001) 240.
 
 \bibitem{RF-JPHYSG} R.~Fleischer,
  {J.\ Phys.}~{\bf G32} (2006) R71.

\bibitem{JLQCD}S.~Aoki {\it et al.}\  [JLQCD coll.],
{ Phys.\ Rev.\ Lett.} {\bf 91} (2003) 212001.

\bibitem{HPQCD}A.~Gray {\it et al.}\  [HPQCD coll.],
{  Phys.\ Rev.\ Lett.} {\bf 95} (2005) 212001.

\bibitem{Okamoto}M.~Okamoto,
 { PoS {\bf LAT2005}} (2005) 013.

\bibitem{BphiK-old}D. London and R.D. Peccei,
{ Phys.\ Lett.}~{\bf B223} (1989) 257;\\
N.G. Deshpande and J. Trampetic,  
{ Phys.\ Rev.}~{\bf D41} (1990) 895 and 2926;\\
J.-M. G\'erard and W.-S. Hou, { Phys.\ Rev.}~{\bf D43} (1991) 2909; 
{ Phys.\ Lett.}~{\bf B253} (1991) 478.


\bibitem{DH-PhiK}N.G. Deshpande and X.-G. He, { Phys.\ Lett.}~{\bf B336} (1994)
471.

\bibitem{RF-EWP-rev}R. Fleischer,
  { Int.\ J.\ Mod.\ Phys.}~{\bf A12} (1997) 2459.

\bibitem{growo}Y. Grossman and M.P. Worah,
  { Phys.\ Lett.}~{\bf B395} (1997) 241.

\bibitem{Belle-BphiK}K.F.~Chen {\it et al.}  [Belle Collaboration],
  { Phys.\ Rev.\ Lett.}~{\bf 98} (2007) 031802.

\bibitem{BaBar-BphiK}B.~Aubert {\it et al.}  [BaBar Collaboration],
  { Phys.\ Rev.\ Lett.}~{\bf 99} (2007) 161802.

  \bibitem{EWP-NP}R.~Fleischer and T.~Mannel,
  hep-ph/9706261;\\
Y.~Grossman, M.~Neubert and A.L.~Kagan,
  JHEP {\bf 9910} (1999) 029.

\bibitem{BpiK-papers}T.~Yoshikawa,
  { Phys.\ Rev.}~{\bf D68} (2003) 054023;\\
  M.~Gronau and J.L.~Rosner,
  { Phys.\ Lett.}~{\bf B572} (2003) 43;\\
  M.~Beneke and M.~Neubert,
  { Nucl.\ Phys.}~{\bf B675} (2003) 333;\\
  V.~Barger, C.W.~Chiang, P.~Langacker and H.S.~Lee,
  Phys.\ Lett.\  {\bf B598} (2004) 218;\\
   Y.L.~Wu and Y.F.~Zhou,
  Phys.\ Rev.\ {\bf D72} (2005) 034037.

\bibitem{FRS-07}R.~Fleischer, S.~Recksiegel and F.~Schwab,
  { Eur.\ Phys.\ J.}~{\bf C51} (2007) 55.

\bibitem{BF98}A.J.~Buras and R.~Fleischer,
  Eur.\ Phys.\ J.\   {\bf C11} (1999) 93.
  
\bibitem{NR}M.~Neubert and J.L.~Rosner,
  Phys.\ Rev.\ Lett.\  {\bf 81} (1998) 5076.
  
\bibitem{BF00}A.J.~Buras and R.~Fleischer,
  {\it Eur.\ Phys.\ J.}~{\bf C16} (2000) 97.

\bibitem{GR-06}M.~Gronau and J.L.~Rosner,
  Phys.\ Rev.\   {\bf D74} (2006) 057503; 
  Phys.\ Lett.\  {\bf B644} (2007) 237.

\bibitem{FG-1}R.~Fleischer and M.~Gronau,
 { Phys.\ Lett.}~{\bf B660} (2008) 212.
  
\bibitem{Acosta:2005eu}D.~Acosta {\it et al.}  [CDF Collaboration],
Phys.\ Rev.\ Lett.\  {\bf 95} (2005) 031801.

\bibitem{LHCb-Bsphiphi}S. Amato {\it et al.}, CERN-LHCb-2007-047. 

\bibitem{HPQCD-DMs}E.~Dalgic {\it et al.} [HPQCD Collaboration],
  Phys.\ Rev.\   {\bf D76} (2007) 011501.

\bibitem{DMs-papers}M.~Carena {\it et al.}, 
   Phys.\ Rev.\   {\bf D74} (2006) 015009;\\
  M.~Ciuchini and L.~Silvestrini,
  Phys.\ Rev.\ Lett.\  {\bf 97} (2006) 021803;\\
 M.~Endo and S.~Mishima,
  Phys.\ Lett.\   {\bf B640} (2006) 205;\\
  Z.~Ligeti, M.~Papucci and G.~Perez,
  Phys.\ Rev.\ Lett.\  {\bf 97} (2006) 101801;\\
  J.~Foster, K.I.~Okumura and L.~Roszkowski,
   Phys.\ Lett.\   {\bf B641} (2006) 452;\\
 Y.~Grossman, Y.~Nir and G.~Raz,
  Phys.\ Rev.\ Lett.\  {\bf 97} (2006) 151801;\\
S.~Baek, J.H.~Jeon and C.S.~Kim,
  Phys.\ Lett.\  {\bf B641} (2006) 183;\\
  M.~Blanke and A.J.~Buras,
  JHEP {\bf 0705} (2007) 061.
  
\bibitem{BBGT}M.~Blanke, A.J.~Buras, D.~Guadagnoli and C.~Tarantino,
  JHEP {\bf 0610} (2006)  003.


\bibitem{DDLR}A.S. Dighe, I. Dunietz, H. Lipkin and J.L. Rosner,
  { Phys.\ Lett.}~{\bf B369} (1996) 144.


\bibitem{D0-phis}V.~M.~Abazov {\it et al.}  [D0 Collaboration],
  Phys.\ Rev.\ Lett.\  {\bf 98} (2007) 121801.

\bibitem{cdf-tagged}T.~Aaltonen {\it et al.}  [CDF Collaboration],
  arXiv:0712.2397 [hep-ex].

\bibitem{LHCb-up}F.~Muheim,
  Nucl.\ Phys.\ Proc.\ Suppl.\  {\bf 170} (2007) 317.

\bibitem{RF-ang}R.~Fleischer,
  Phys.\ Rev.\   {\bf D60} (1999) 073008.

\bibitem{BsDsK}R. Aleksan, I. Dunietz and B. Kayser,
{ Z.\ Phys.}\  {\bf C54} (1992) 653.

\bibitem{BdDpi}I.~Dunietz and R.G.~Sachs,
{ Phys.\ Rev.}\  {\bf D37} (1988) 3186  [E: {\bf D39} (1989) 3515];\\
I.~Dunietz,
{ Phys.\ Lett.}\  {\bf B427} (1998) 179;\\
D.A.~Suprun, C.W.~Chiang and J.L.~Rosner,
{ Phys.\ Rev.}\  {\bf D65} (2002) 054025.

\bibitem{BDpi-Uspin}G. Wilkinson, in G.~Cavoto {\it et al.},
  Proceedings of CKM 2005 (WG5), San Diego, California, 15--18
   March 2005 [hep-ph/0603019];
V. Gligorov and G. Wilkinson, work in progress.
  
\bibitem{LHCb-analyses}G. Balbi {\it et al.}, CERN-LHCb/2003-123 and
124; R. Antunes Nobrega {\it et al.}\ [LHCb Collaboration], {\it Reoptimized
LHCb Detector, Design and Performance}, Technical Design Report 9, 
CERN/LHCC 2003-030; A. Carbone, J. Nardulli, S. Pennazzi, A.~Sarti and
V. Vagnoni, CERN-LHCb-2007-059.

\bibitem{CDF-BsK+K-}A.~Abulencia {\it et al.}  [CDF Collaboration],
  { Phys.\ Rev.\ Lett.}~{\bf 97} (2006) 211802;
  CDF Note 8579 (2006).

\bibitem{RF-BsKK-07}R.~Fleischer,
  Eur.\ Phys.\ J.\   {\bf C52} (2007)  267.

\bibitem{Bmumu}G.~Buchalla and A.J.~Buras,
{ Nucl.\ Phys.}\ {\bf B400} (1993) 225 
and  {\bf B548} (1999) 309;\\
M.~Misiak and J.~Urban, 
{ Phys.\ Lett.}\  {\bf B451} (1999) 161.

\bibitem{CDF-Bmumu}T. Aaltonen {\it et al.}\  [CDF Collaboration],
  arXiv:0712.1708 [hep-ex].

\bibitem{D0-Bmumu}D0 Collaboration, D0 Note 5344-CONF (2007) 
[http://www-d0.fnal.gov].

\bibitem{BKastll}G.~Burdman,
  Phys.\ Rev.\   {\bf D57} (1998) 4254;\\
  M.~Beneke, T.~Feldmann and D.~Seidel,
  Eur.\ Phys.\ J.\   {\bf C41} (2005) 173;\\
A.~Ali, G.~Kramer and G.~h.~Zhu,
  Eur.\ Phys.\ J.\  C {\bf 47} (2006) 625.
  
\bibitem{ABHH}A.~Ali, P.~Ball, L.T.~Handoko and G.~Hiller,
  Phys.\ Rev.\   {\bf D61} (2000) 074024.
  
\bibitem{matias-rare}F.~Kr\"uger and J.~Matias,
  Phys.\ Rev.\  {\bf D71} (2005) 094009;\\
    E.~Lunghi and J.~Matias,
  JHEP {\bf 0704} (2007) 058.
  
\bibitem{D-mix}B.~Aubert {\it et al.}  [BABAR Collaboration],
  Phys.\ Rev.\ Lett.\  {\bf 98} (2007) 211802;\\
M.~Staric {\it et al.}  [Belle Collaboration],
  Phys.\ Rev.\ Lett.\  {\bf 98} (2007) 211803.
  
\bibitem{CDF-Dmix}T.~Aaltonen {\it et al.}  [CDF Collaboration],
        arXiv:0712.1567 [hep-ex].

\bibitem{D-th}Y.~Nir,
  JHEP {\bf 0705} (2007) 102 (2007);\\
  E.~Golowich, J.~Hewett, S.~Pakvasa and A.A.~Petrov,
  Phys.\ Rev.\  {\bf D76} (2007) 095009.
  
  \bibitem{WG3-rep}M.~Raidal {\it et al.},
  arXiv:0801.1826 [hep-ph],
Report of Working Group 3 of the CERN Workshop ``Flavour in the era of the LHC'', Geneva, Switzerland, November 2005 -- March 2007.

\bibitem{CERN-WS}These topics were addressed at the CERN Workshop 
``Flavour in the era of the LHC", November 2005 -- March 2007 
[http://flavlhc.web.cern.ch/flavlhc/].

\end{thebibliography}
\end{document}